\pdfoutput=1
\documentclass[cernpreprint,UKenglish,texlive=2011,txfonts]{latex/atlasdoc}
\usepackage[siunitx=false,biblatex=false]{latex/atlaspackage}

\usepackage[process]{latex/atlasphysics}
\usepackage{units}
\usepackage{bm}

\graphicspath{{logos/}{figures/}}

\usepackage{paper-defs}

\AtlasTitle{Performance of $b$-Jet Identification in the ATLAS Experiment}
\author{The ATLAS Collaboration}
\AtlasRefCode{PERF-2012-04}
\PreprintIdNumber{CERN-PH-EP-2015-216}
\AtlasJournal{JINST}
\AtlasAbstract{%
The identification of jets containing $b$ hadrons is important for the 
  physics programme of the ATLAS experiment at the Large Hadron Collider.
  Several algorithms to identify jets containing $b$ hadrons are described, 
  ranging from those based on the reconstruction of an inclusive secondary vertex
  or the presence of tracks with large impact parameters to combined tagging algorithms
  making use of multi-variate discriminants.
  An independent $b$-tagging algorithm based on the reconstruction of muons inside jets 
  as well as the $b$-tagging algorithm used in the online trigger are also presented.

  The $b$-jet tagging efficiency, the $c$-jet tagging efficiency and the mistag rate 
  for light flavour jets in data
  have been measured with a number of complementary methods.
  The calibration results are presented as scale factors defined as the ratio 
  of the  efficiency (or mistag rate) in data to that in simulation.
  In the case of $b$ jets, where more than one calibration method exists, the results from the various 
  analyses have been combined taking into account the statistical correlation as
  well as the correlation of the sources of systematic uncertainty.
}

\hypersetup{pdftitle={Performance of b-Jet Identification in the ATLAS Experiment},pdfauthor={The ATLAS Collaboration}}

\begin{document}

\maketitle
\tableofcontents

\newpage

\section{Introduction}
\label{sec:intro}

The identification of jets containing $b$ hadrons 
is an important tool used in a spectrum of measurements comprising the 
Large Hadron Collider (LHC) physics programme.
In precision measurements in the top quark sector as well as in the 
search for the Higgs boson and new phenomena,
the suppression of background processes that contain predominantly 
light-flavour jets using $b$-tagging is of great use. It may also 
become critical to achieve an understanding of the flavour structure of any new physics 
(e.g. supersymmetry) revealed at the LHC.

Several algorithms to identify jets containing $b$ hadrons have been developed, 
exploiting the long lifetime, high mass and decay multiplicity of $b$ hadrons and the 
hard $b$-quark fragmentation function.
They range from an algorithm that uses the signed
significance of the decay length with respect to the proton-proton collision
location, in the following referred to as the primary vertex,
of an inclusively reconstructed secondary vertex to more refined algorithms
using both secondary vertex properties and the significance of the transverse and 
longitudinal impact parameters of the charged particle tracks. The most
discriminating observables resulting from these algorithms are combined in
artificial neural networks.
An independent $b$-tagging algorithm based on reconstructed muons inside jets, exploiting the 
relatively large fraction of $b$-hadron decays with muons in the final state, about 20\%,
and the $b$-tagging algorithm used for the online trigger selection have also been developed.

The performance of the tagging algorithms has been characterised in simulated events,
including the dependence on additional proton-proton
interactions in the same bunch crossing, referred to as pile-up in the following.
A first comparison between data and simulation focuses on the basic ingredients for $b$-tagging,
namely the track properties, including the impact parameter distributions.
A second comparison focuses more specifically on tracks in $b$ jets, and is made
possible by fully reconstructing the $b$-hadron decay $B^{\pm} \rightarrow \Jpsi K^{\pm}$.

To use $b$-tagging in physics analyses, the efficiency $\epsilon_b$ with
which a jet containing a $b$ hadron is tagged by a $b$-tagging
algorithm needs to be measured. Other necessary pieces of information are 
the probability of mistakenly tagging a jet containing a $c$ hadron (but not a $b$ hadron) 
or a light-flavour parton ($u$-, $d$-, $s$-quark or gluon $g$) jet as a $b$ jet.
In the following, these are referred to as the $c$-jet tagging efficiency and mistag
rate, respectively. 

Several methods have been developed to measure the $b$-jet tagging efficiency, the $c$-jet tagging efficiency
and the mistag rate in data. The $b$-jet tagging efficiency has been measured in an inclusive 
sample of jets with muons inside and in samples of $t\bar t$ events with one or two 
leptons in the final state. The $c$-jet tagging efficiency has been measured in an inclusive 
sample of jets associated to $D^{\star +}$ mesons as well as in a sample of $W+c$ events. 
The mistag rate has been measured in an inclusive jet sample.
The calibration results are presented as data-to-simulation scale factors, derived from the ratio of the efficiency or mistag rate measured in data to that obtained in simulated events.
Where more than one calibration method exists the results from the various analyses
have been combined taking into account the statistical and systematic correlation.

This paper is intended to provide a complete description of almost all the $b$-tagging developments in ATLAS during Run 1 
of the LHC in the years 2010 -- 2012. The results are illustrated with data taken in the year 2011 at a centre-of-mass energy of 7~\TeV{}.
As these developments extended over a period of years, there is some variation
between the simulated samples and systematic uncertainties used for the data
efficiency measurements depending on the chronology. Also, several of the methods
developed to measure the tagging efficiency of $b$ jets on the small samples
available at the start of Run 1 have meanwhile been abandoned in favour of more
precise calibration methods developed later; this is reflected in the choice of
results used in the combination of $b$-jet efficiency measurements made to
achieve the ultimate precision.
In those methods used previously, quoted values and uncertainties for
parameters entering the analysis do reflect the best knowledge at the time. They
have not been updated since to benefit from the improved present knowledge
on some of the analysis ingredients.
Section~\ref{sec:samples} starts with a discussion of the data and simulated
samples used throughout this paper, along with a description of the corrections
applied to the simulated samples to reproduce the experimental conditions
present in the data.
The various $b$-tagging algorithms are described in Sections~\ref{sec:lifetime_tagging}, \ref{sec:smt_tagger}, and~\ref{sec:trigger}.
Section~\ref{sec:pu} discusses the effects of pile-up, while
Section~\ref{sec:btag_inputs} provides a comparison between data and simulated
samples of distributions of selected quantities important for $b$-tagging.
Calibrations of the $b$-jet tagging efficiency and their combination are discussed in Sections~\ref{sec:beff_mubased}, \ref{sec:beff_ttbarbased}, \ref{sec:combination}, and~\ref{tandp}.
Calibrations of the $c$-jet tagging efficiency are covered in Sections~\ref{sec:ceff_Wc} and~\ref{sec:ceff_dstarbased}, while the mistag rate calibration is discussed in Sections~\ref{sec:mistag} and~\ref{smt:mtag}.

\section{Data and simulation samples, object selection}
\label{sec:samples}

The studies presented in this paper are generally based on a data sample
corresponding to approximately 4.7~fb$^{-1}$ of 7~\TeV{} proton-proton collision data,
after requiring the data to be of good quality; slight differences exist due to
variations in data quality requirements.
The data have been collected in 2011 using the ATLAS experiment. The ATLAS
detector is a large, general-purpose collider detector and is described in detail
elsewhere~\cite{PERF-2007-01}. Its most prominent features, as relevant to $b$-jet
identification and its performance estimation, are:
\begin{itemize}
\item An Inner Detector (ID)~\cite{IDET-2013-01}, providing tracking and vertexing capabilities for
  $|\eta| < 2.5$.\footnote{
    ATLAS uses a right-handed coordinate system with its origin at the nominal
    interaction point in the centre of the detector, and the $z$ axis along the beam line.
    The $x$ axis points to the centre of the LHC ring, and the $y$ axis points upwards.
    Cylindrical coordinates $(r,\phi)$ are used in the transverse plane, with $\phi$
    being the azimuthal angle around the beam line. The pseudorapidity $\eta$ is defined
    in terms of the polar angle $\theta$ as $\eta=-\ln\tan(\theta/2)$.}
  It is immersed in an axial 2~T magnetic field and features three subdetectors
  employing different techniques.
  A pixel detector consisting of three layers of silicon pixel sensors is
  located closest to the beam line. It is followed by a silicon microstrip
  detector (SCT), consisting of eight (eighteen) layers of silicon microstrip
  sensors arranged in cylinders (disks) in its barrel (endcap) region, and by a straw tube tracker
  providing of order 36 measurements for track reconstruction as well as causing
  high-energy electrons to generate transition radiation.
  Especially the pixel and microstrip layers are essential for the purpose of a
  precise reconstruction of tracks and of displaced vertices.
\item A fine-grained lead and liquid argon sampling calorimeter, providing
  electromagnetic calorimetry up to $|\eta| < 3.2$.
\item A hermetic hadronic calorimeter covering the range $|\eta| < 4.9$. Its
  central part is a steel and scintillating tile sampling calorimeter; its
  forward parts are again sampling calorimeters, using a liquid argon detection
  medium and copper and tungsten absorbers.
\item A large air-core Muon System (MS), providing stand-alone precision muon
  momentum reconstruction in the range $|\eta| < 2.7$ using a combination of
  drift tube and resistive plate chamber technologies, and equipped with
  dedicated detectors for triggering and precise timing. A system of one barrel
  and two endcap magnet toroids provides a bending power ranging between 1 Tm
  and 7.5 Tm, lowest in the transition region between the toroids.
\end{itemize}
A three-level trigger system was used to reduce the event rate from the 20 MHz
bunch crossing rate to $\sim$~200 Hz.
The trigger selections used in the different studies are described 
in the corresponding sections.

The key objects for $b$-tagging are the calorimeter jets, the tracks
reconstructed in the Inner Detector 
and the signal primary vertex of the hard-scattering collision of interest which is selected from the set of all 
reconstructed primary vertices. 
Each vertex is required to have two or more tracks.
Tracks are reconstructed from clusters of signals in the silicon pixel and
microstrip sensors, and drift circles in the straw tube tracker (collectively
referred to as ``hits'' in the following).
They are associated with the calorimeter jets based on their angular separation
$\Delta R(\mbox{track},\mbox{jet})  \equiv \sqrt{(\Delta\eta)^2+(\Delta\phi)^2}$. 
The association $\Delta R$ cut varies as a function of the jet $\pt$, resulting
in a narrower cone for jets at high $\pt$\ which are more collimated.
At 20 \GeV, it is 0.45 while for more energetic jets with a $\pt$ of 150~\GeV{} the 
$\Delta R$ cut is 0.26.
Any given track is associated with at most one jet; if it satisfies the association
criterion with respect to more than one jet, the jet with the smallest $\Delta
R$ is chosen.
The track selection criteria depend on the $b$-tagging algorithm, and
are detailed in Section~\ref{sec:lifetime_tagging}. 

Jets used in this paper are reconstructed from topological clusters~\cite{PERF-2007-01} formed from energy deposits in the 
calorimeters using the anti-$k_t$ algorithm with a radius parameter of $0.4$~\cite{antikt, antikt2, antikt3}.
The jet reconstruction is done at the electromagnetic scale and then a scale factor is 
applied in order to obtain the jet energy at the hadronic scale.
In the studies based on jets with associated muons, the jet energy is further corrected for the 
energy of the muon and the average energy of  the corresponding neutrino in
simulated events, to arrive at the jet energy scale of an inclusive $b$-jet sample.
The measurement of the jet energy and the specific cuts used to reject jets of bad quality are described in Ref.~\cite{PERF-2012-01}. 
The jets are generally required to have $|\eta|<2.5$ and transverse momentum $\pt > 20\GeV$.
Furthermore, the jet vertex fraction (JVF) is computed as the summed transverse momentum of the tracks associated 
with a jet consistent with originating from the selected primary vertex
(defined as having a longitudinal impact parameter with respect to it less than 1~mm)
divided by the summed transverse momentum of all tracks associated with a jet,
where only tracks with transverse impact parameters less than 1.5~mm are considered;
it is required to be larger than 0.75. The selection of the primary vertex is described in Section~\ref{sec:lifetime_ingredients}.
Some measurements of the $b$-jet tagging efficiency make use of soft muons ($\pt\ > 4 \gev$)
associated with jets, using a spatial matching of $\Delta R{\rm (jet, \mu)} < 0.4$.

Multiple Monte Carlo (MC) simulated samples are used throughout this paper.
The properties and performance of the tagging algorithms are mostly studied using simulated samples of
\ttbar{} events, which unless otherwise stated are generated with
MC@NLO v3.41~\cite{bib:mcatnlo} interfaced to HERWIG v6.520~\cite{Corcella:2000bw};
for several studies and performance measurements, multijet samples generated
using \Pythia{} v6.423~\cite{pythia2} are used.
To reproduce the pile-up conditions in the data, extra collisions have been superimposed on the simulated events.
To simulate the detector response,
the generated events are processed through a GEANT4~\cite{geant} simulation of the ATLAS detector, 
and then reconstructed and analysed in the same way as the data.
The simulated detector geometry corresponds to a perfectly aligned Inner Detector and the majority of the disabled 
silicon detector (pixel and strip)
modules and front-end chips present in data are masked in the simulation.
The ATLAS simulation infrastructure is described in more detail in Ref.~\cite{SOFT-2010-01}.

To bring the simulation into agreement with data for distributions where discrepancies are known to 
be present, corrections have been applied to some of the simulated samples.
The average number of interactions per bunch crossing, denoted
$\langle \mu \rangle$, ranged between~4 and~20~\cite{DAPR-2011-01}.
Its distribution in simulated events has been reweighted 
to ensure a good agreement in the distribution of the number of 
reconstructed primary vertices between data and simulation.
The fraction of pile-up interactions leading to visible signatures
(\emph{reconstructible interactions}) in the region 2.09 < |\eta| < 3.84 is
computed from Refs.~\cite{STDM-2010-11,Antchev:2011vs}, and is used
to scale the $\langle \mu \rangle$ values prior to the reweighting described
above, to bring the numbers of reconstructible interactions in agreement between
data and simulated events.
Applying this scaling has been verified to lead to a good agreement between data
and simulated events also in the average number of reconstructed primary
vertices as a function of $\langle \mu \rangle$.
When appropriate, the \pT{} spectrum of the simulated jets has also been reweighted to match the
spectrum in data, to account e.g. for the fact that the prescale factors of low threshold jet triggers
present in data are not activated in the simulation.

The labelling of the flavour of a jet in simulation is done by spatially matching the 
jet with generator level partons~\cite{CSCbook}: if a $b$ quark with a
transverse momentum of more than 5~\GeV{} is found within $\Delta R(b,\mbox{jet}) < 0.3$ of the jet direction,
the jet is labelled as a $b$ jet.
If no $b$ quark is found the procedure is repeated for $c$ quarks and $\tau$ leptons. 
A jet for which no such association can be made is labelled as a light-flavour jet.

\section{Lifetime-based tagging algorithms}
\label{sec:lifetime_tagging}

The lifetime-based tagging algorithms take advantage of the relatively long lifetime of hadrons
containing a $b$ quark, of the order of 1.5 ps ($c\tau\approx 450\ \mu$m).
A $b$ hadron with $\pt=50 \GeV$ will have a significant mean flight path length
$\langle l \rangle = \beta\gamma c\tau$, travelling on average about 3 mm in the transverse direction
before decaying and therefore leading to topologies with at least one vertex displaced from the 
point where the hard-scatter collision occurred. 
Two classes of algorithms aim at identifying such topologies.
An inclusive approach consists of using the impact parameters of the charged-particle tracks from the $b$-hadron decay products.
The transverse impact parameter, $d_0$, is the distance of closest approach of the track to the primary vertex point, in the $r$--$\phi$ projection.
The longitudinal impact parameter, $z_0$, is the difference between the $z$
coordinates of the primary vertex position and of the track at this point of closest approach in $r$--$\phi$.
The tracks from $b$-hadron decay products tend to have large impact parameters which can be
distinguished from tracks stemming from the primary vertex.
Two tagging algorithms exploiting these properties
are discussed in this article: JetProb, used mostly for early data, and IP3D for high-performance tagging.
The second approach is to reconstruct explicitly the displaced vertices. 
Two algorithms make use of this technique: the SV algorithm 
attempts to reconstruct an inclusive secondary vertex; while the JetFitter algorithm aims at 
reconstructing the complete $b$-hadron decay chain.
Finally, the results of several of these algorithms are combined in the MV1
tagger to improve the light-flavour-jet rejection and to increase the range 
of $b$-jet tagging efficiency for which the algorithms can be applied.
These algorithms are discussed in detail in Sections~\ref{sec:ip_algo}--\ref{sec:comb-vertex-algos}.

\subsection{Key ingredients}
\label{sec:lifetime_ingredients}

The determination on an event-by-event basis of the primary vertex~\cite{ATLAS-CONF-2010-069} is particularly important for $b$-tagging, since it
defines the reference point with respect to which impact parameters and vertex displacements are expressed.
The precision of the reconstructed vertex positions improves with increasing associated track multiplicity.
For example, in minimum bias events it improves from approximately 300~$\mu$m (600~$\mu$m) in the $x$ and $y$ ($z$) directions for two-track vertices to 20~$\mu$m (35~$\mu$m) for vertices with 70 associated tracks.
The vertex resolution depends strongly on the event topology, and significantly better resolutions can be achieved in events with high-\pt{} jets or leptons.
The number of reconstructed primary vertices is substantially larger than one in the presence of pile-up interactions:
during the highest instantaneous luminosity of the 2011 data taking period, six primary vertex candidates were reconstructed on average.
The adopted strategy is to choose the primary vertex candidate that maximises the sum of the associated tracks' $\pt^{2}$.
The performance of this algorithm depends on the final state and on the pile-up conditions (as will be discussed further in Section~\ref{sec:pu});
simulation studies indicate that the probability to choose the correct primary vertex in $\ttbar$ events is higher than 98\%,
while in lower-multiplicity final states it can be considerably lower.

The actual tagging is performed on the sub-set of tracks in the event that are associated with the jet.
Once associated with a jet, tracks are subject to specific requirements designed to select well-measured tracks and to reject 
so-called fake tracks (in which not all hits used for the track reconstruction originate
from a single charged particle) and tracks from long-lived particles ($K_s$, $\Lambda$ and other hyperon decays) 
or material interactions (photon conversions or hadronic interactions).
The $b$-tagging baseline quality level requires at least seven precision hits (pixel or micro-strip hits) on the track, and
at least two of these in the pixel detector, one of which must be in the innermost pixel layer.
Only tracks with $\pt>1\GeV$ are considered.
The transverse and longitudinal impact parameters defined with respect to the primary vertex must fulfil 
$|d_0|<1$ mm and $|z_0|\sin\theta<1.5$~mm, where $\theta$ is the track polar angle
(the factor $\sin\theta$ serves to make the efficiency for tracks to pass these selection
criteria less dependent on their polar angles).
This selection is used by all the tagging algorithms relying on the impact parameters of tracks.
The average number of $b$-tagging quality tracks associated to a jet with $\pt=50\GeV$ 
(200~\GeV) is 3.5 (7).
In typical $t\bar{t}$ events, the average number of selected tracks per light-flavour ($b$ quark) jet is 3.7 (5.5) and their
average $\pt$ is $6.6\GeV$ (6.3~\GeV), respectively.
The SV and JetFitter algorithms use looser track selection criteria, in particular to maximise the efficiency to 
identify tracks originating from material interactions or decays of long-lived
particles; these tracks are subsequently removed 
for $b$-tagging purposes. The main differences in the selection cuts for the SV algorithm are:
$\pt>400\MeV$, $|d_0|<3.5$ mm (no cut on $z_0$).
The corresponding cuts used by the JetFitter algorithm are:
$\pt>500\MeV$, $|d_0|<7$ mm, $|z_0|\sin\theta<10$~mm.
Both algorithms make a requirement of 
at least one hit in the pixel detector (with no requirement on the innermost pixel layer).

\subsection{Impact parameter-based algorithms}
\label{sec:ip_algo}

For the tagging itself, the impact parameters of tracks are computed with
respect to the selected primary vertex.
Given that the decay point of the $b$ hadron must lie along its flight path,
the transverse impact parameter is signed to further discriminate the tracks from $b$-hadron decay from
tracks originating from the primary vertex.
The sign is defined as positive if the track intersects the jet axis 
 in front of the primary vertex, and as negative if the intersection lies behind the primary vertex.
The jet axis is defined by the calorimeter-based jet direction. 
However if an inclusive secondary vertex is found in the jet (cf. Section~\ref{sec:vertex-algos}), the jet direction is replaced by the direction 
of the line joining the primary and the secondary vertices.
The experimental resolution generates a random sign for the tracks originating from the
primary vertex, while tracks from the $b$-/$c$-hadron decay normally have a positive sign.
Decays of e.g. $K_s^0$ and $\Lambda^0$ as well as interactions in the detector material also
produce tracks with positively signed impact parameters, enhancing the probability to identify
light flavour jets as $b$-quark jets.

JetProb~\cite{jetprob_aleph} 
is an implementation of a simple algorithm extensively used at LEP and later at the Tevatron.
It uses the track impact parameter significance $S_{d_0}\equiv d_0/\sigma_{d_0}$,
where $\sigma_{d_0}$ is the uncertainty on the reconstructed $d_{0}$.
The $S_{d_0}$ value 
of each selected track in a jet, $i$,
is compared to a pre-determined resolution function ${\cal R}(S_{d_{0}})$ for
prompt tracks, in order to measure the probability that the track originates
from the primary vertex, ${\cal P}_{\text{trk},i}$, as

\begin{equation}
  {\cal P}_{\text{trk},i} = \int_{-\infty}^{-|S_{d_0}^{i}|} {\cal R}(x) dx.
\end{equation}
The resolution function is determined from experimental data using the negative side of the signed impact 
parameter distribution, assuming that the contribution from heavy-flavour particles is negligible.
The individual track probabilities ${\cal P}_{\text{trk},i}$ for the $N$ tracks with positive $d_0$ are then combined 
as follows:

\begin{equation} {\cal P}_{\text{jet}} = {\cal P}_0 \sum_{j=0}^{N-1}\frac{ (-\ln{\cal P}_0)^j } {j!}, 
\;\;\mbox{where}\;\;
 {\cal P}_0 = \prod_{i=1}^{N} {\cal P}_{\text{trk},i} .
\end{equation}
For light-flavour jets and a perfect suppression of tracks resulting from decays
of long-lived hadrons or from material interactions, the distribution of
${\cal P}_{\text{jet}}$ should be uniform, while it should peak around zero for $b$ jets.
This robust algorithm with no dependence on simulation was mostly used for data taken before 2011, 
and is still used for online $b$-tagging (this is discussed in Section~\ref{sec:trigger}).

IP3D is a more powerful algorithm relying on both the transverse and longitudinal impact parameters, as well as their 
correlations. 
It is based on a log-likelihood ratio (LLR) method in which for each track the measurement 
$S \equiv (d_0/\sigma_{d_0}, z_0/\sigma_{z_0})$ is compared 
to pre-determined two-dimensional probability density functions (PDFs) obtained from simulation for both the $b$- and light-flavour-jet hypotheses.
The ratio of probabilities defines the track weight.
The jet weight is the sum of the logarithms of the individual track weights.
The LLR formalism allows track categories to be used by defining different dedicated PDFs for each of them.
Currently two exclusive categories are used:
the tracks that share a hit in the pixel detector or more than one hit in the
silicon strip detector with another track, and those that do not.

\subsection{Vertex-based algorithms}
\label{sec:vertex-algos}

To further increase the discrimination between $b$ jets and light-flavour jets, an inclusive three dimensional vertex
formed by the decay products of the $b$ hadron, including the products of
the possible subsequent charm hadron decay, can be sought.
The algorithm starts from all tracks that are significantly displaced from the primary vertex\footnote{$d_{3D}/\sigma_{d_{3D}}>2$,
where $d_{3D}$ is the three dimensional distance between the primary vertex and the point of closest approach of the track to this vertex,
and $\sigma_{d_{3D}}$ its uncertainty.} and associated with the jet, and forms vertex candidates for
track pairs with vertex fit $\chi^{2}<4.5$. 
Vertices compatible with long-lived particles or material interactions are rejected:
the invariant mass of the charged-particle track four-momenta is used to reject vertices
that are likely to originate from $K_s$, $\Lambda$ decays and photon conversions,
while the position of the vertex in the $r$--$\phi$ projection is compared to a simplified description of
the innermost pixel layers to reject secondary interactions in the detector material.
All tracks from the remaining two-track vertices are combined into a single
inclusive vertex, using an iterative procedure to remove the track yielding the
largest contribution to the $\chi^2$ of the vertex fit until this contribution passes a predefined threshold.

A simple discriminant between $b$ jets and light-flavour jets is the flight length 
significance $L_{3D}/\sigma_{L_{3D}}$, i.e., the distance between the primary vertex 
and the inclusive secondary vertex divided by the measurement uncertainty.
The significance is 
signed with respect to the jet direction, in the same way as the transverse impact parameter of 
tracks is.
The flight length significance is the discriminating observable on which the SV0 tagging algorithm relies.
As is typical for secondary vertex tagging algorithms, the mistag rate is much smaller than for 
impact parameter-based algorithms, 
but the limited secondary vertex finding efficiency, of approximately 70\%, can be a drawback.

\begin{figure}
  \subfloat[]{\label{fig:perf:sv_mass}\includegraphics[width=0.32\textwidth]{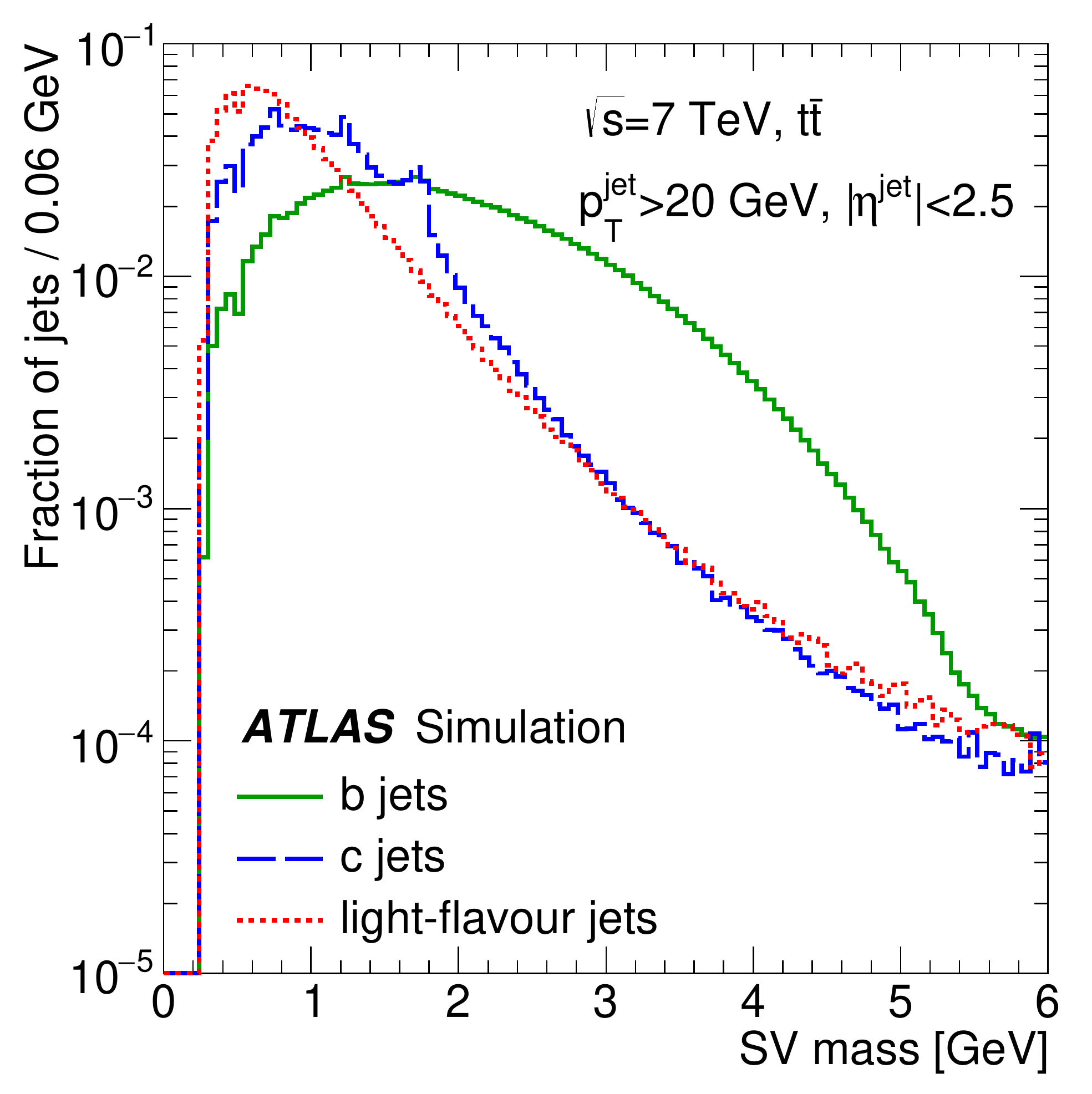}}
  \subfloat[]{\label{fig:perf:sv_efrc}\includegraphics[width=0.32\textwidth]{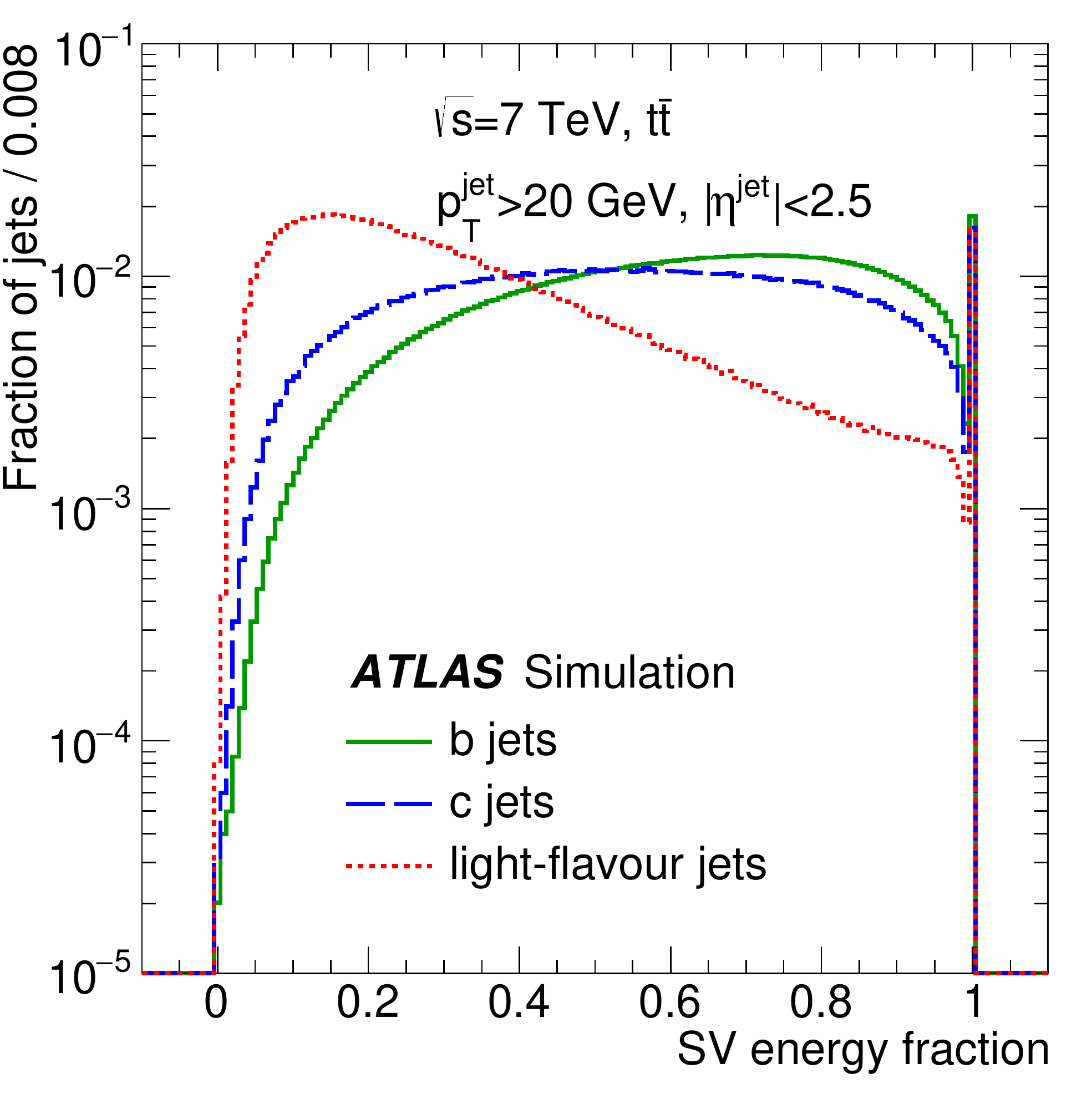}}
  \subfloat[]{\label{fig:perf:sv_effi}\includegraphics[width=0.32\textwidth]{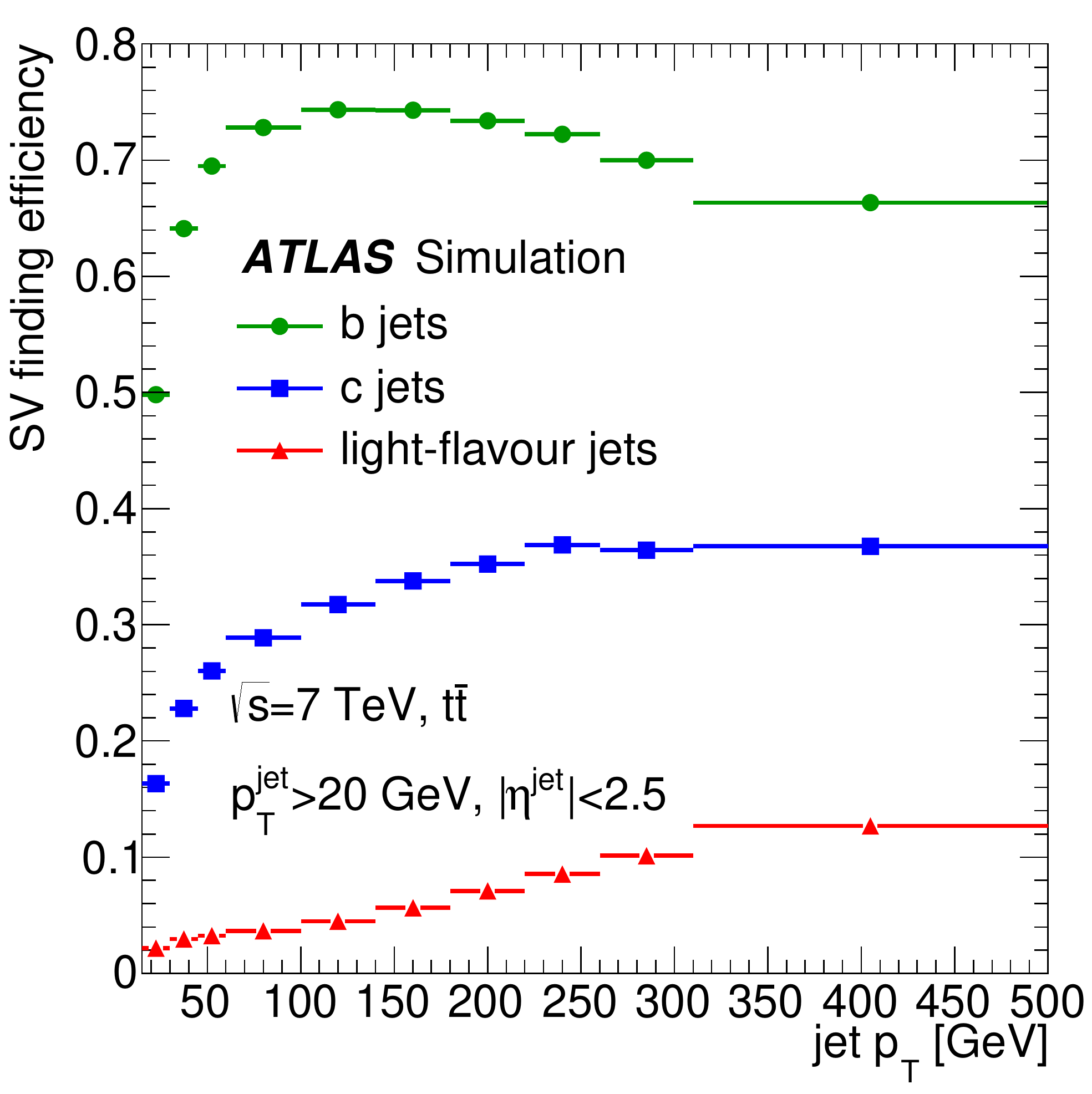}}
  \caption{The vertex mass (a), energy fraction (b) and vertex finding efficiency
    (c) of the inclusive secondary vertices found by the SV1 algorithm,
    for three different flavours of jets.}
  \label{fig:perf:sv_perf}
\end{figure}

SV1 is another tagging algorithm based on the same secondary vertex finding infrastructure, 
but it provides a better performance 
as it is based on a likelihood ratio formalism, like the one explained previously for the IP3D algorithm.
Three of the vertex properties are exploited:
the \emph{vertex mass} (i.e., the invariant mass of all charged-particle tracks used to
reconstruct the vertex, assuming that all tracks are pions), 
the ratio of the sum of the energies of these tracks 
to the sum of the energies of all tracks in the jet, and
the number of two-track vertices.
In addition, the $\Delta R$ between the jet direction and the direction of the line joining the 
primary vertex and the secondary vertex is used in the LLR.
Some of these properties are illustrated in Fig.~\ref{fig:perf:sv_perf} for $b$ jets, $c$ jets and light-flavour jets in simulated $\ttbar$ events.
SV1 relies on a two-dimensional distribution of the first two
variables and on two one-dimensional distributions of the latter variables.
The secondary vertex finding efficiency depends in particular on the event topology.
SV1 requires an a priori knowledge of $\epsilon_b^{SV}$ and the corresponding
efficiency for light-flavour jets, $\epsilon_l^{SV}$, obtained from simulation.
This efficiency is shown as a function of the jet $\pt$ in Fig.~\ref{fig:perf:sv_effi}.

\begin{figure}
  \subfloat[]{\label{fig:perf:jf_mass_b}
    \includegraphics[width=0.32\textwidth]{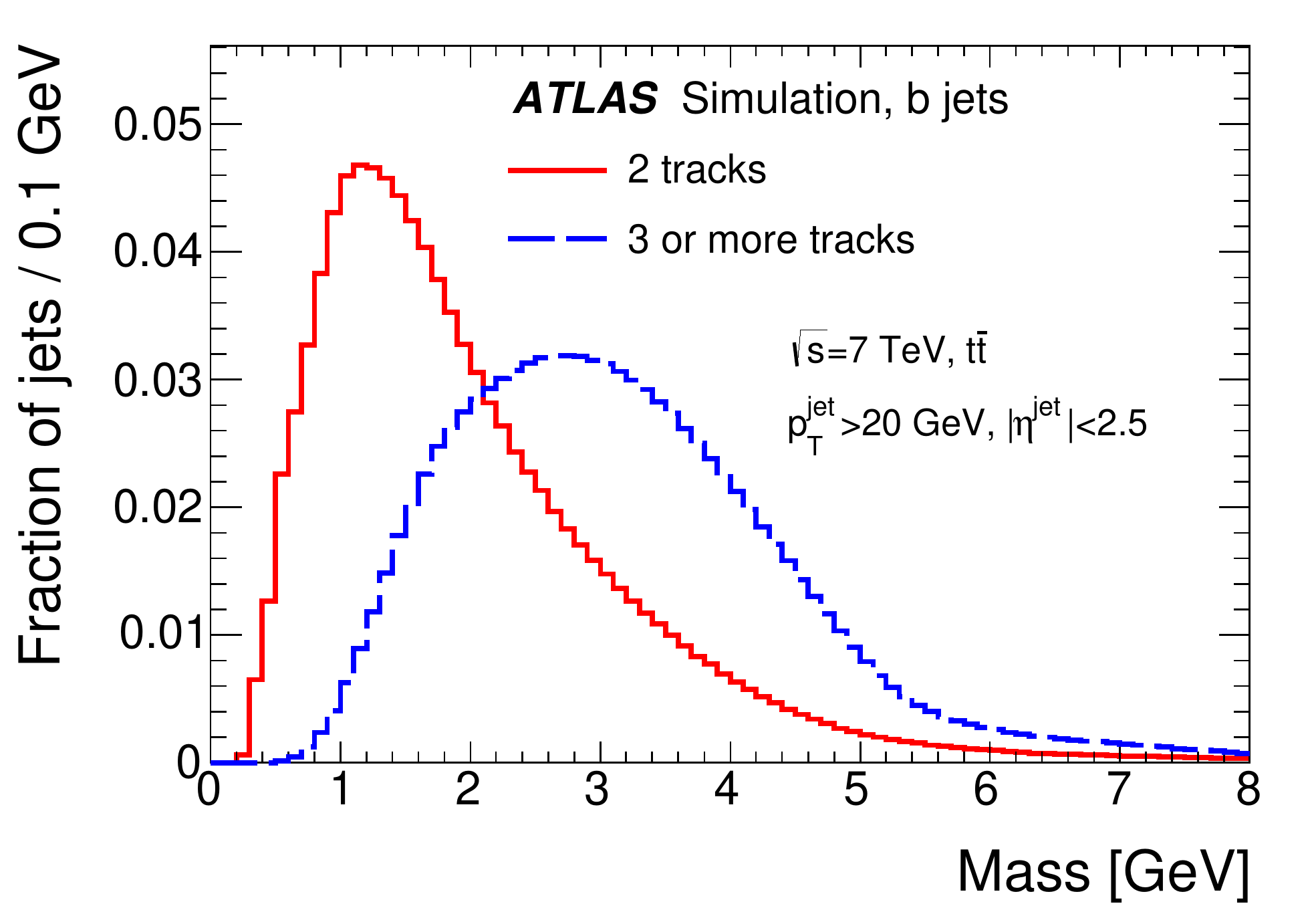}}
  \subfloat[]{\label{fig:perf:jf_mass_c}
    \includegraphics[width=0.32\textwidth]{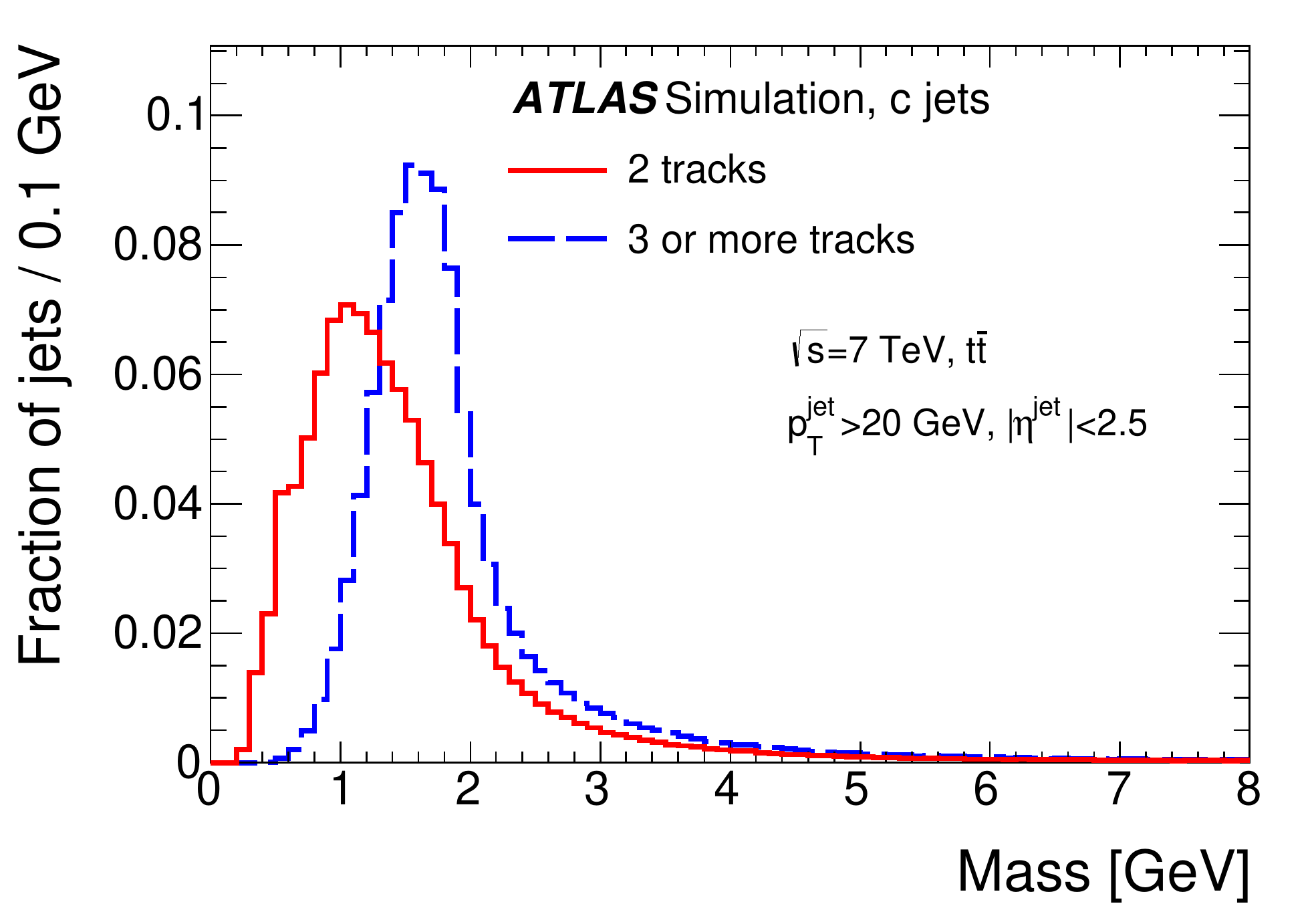}}
  \subfloat[]{\label{fig:perf:jf_mass_l}
    \includegraphics[width=0.32\textwidth]{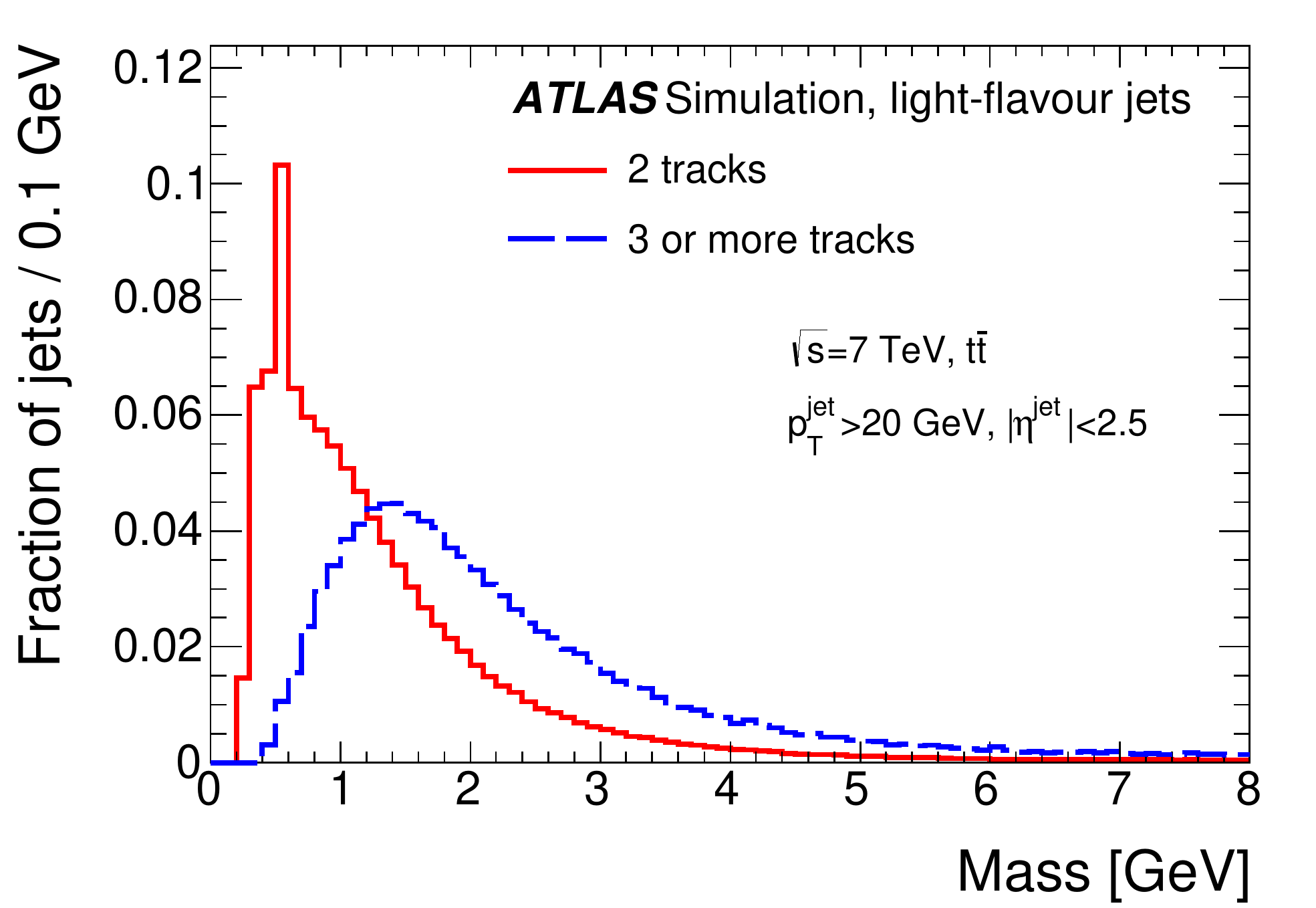}}
  \\
  \subfloat[]{\label{fig:perf:jf_efrc_b}
    \includegraphics[width=0.32\textwidth]{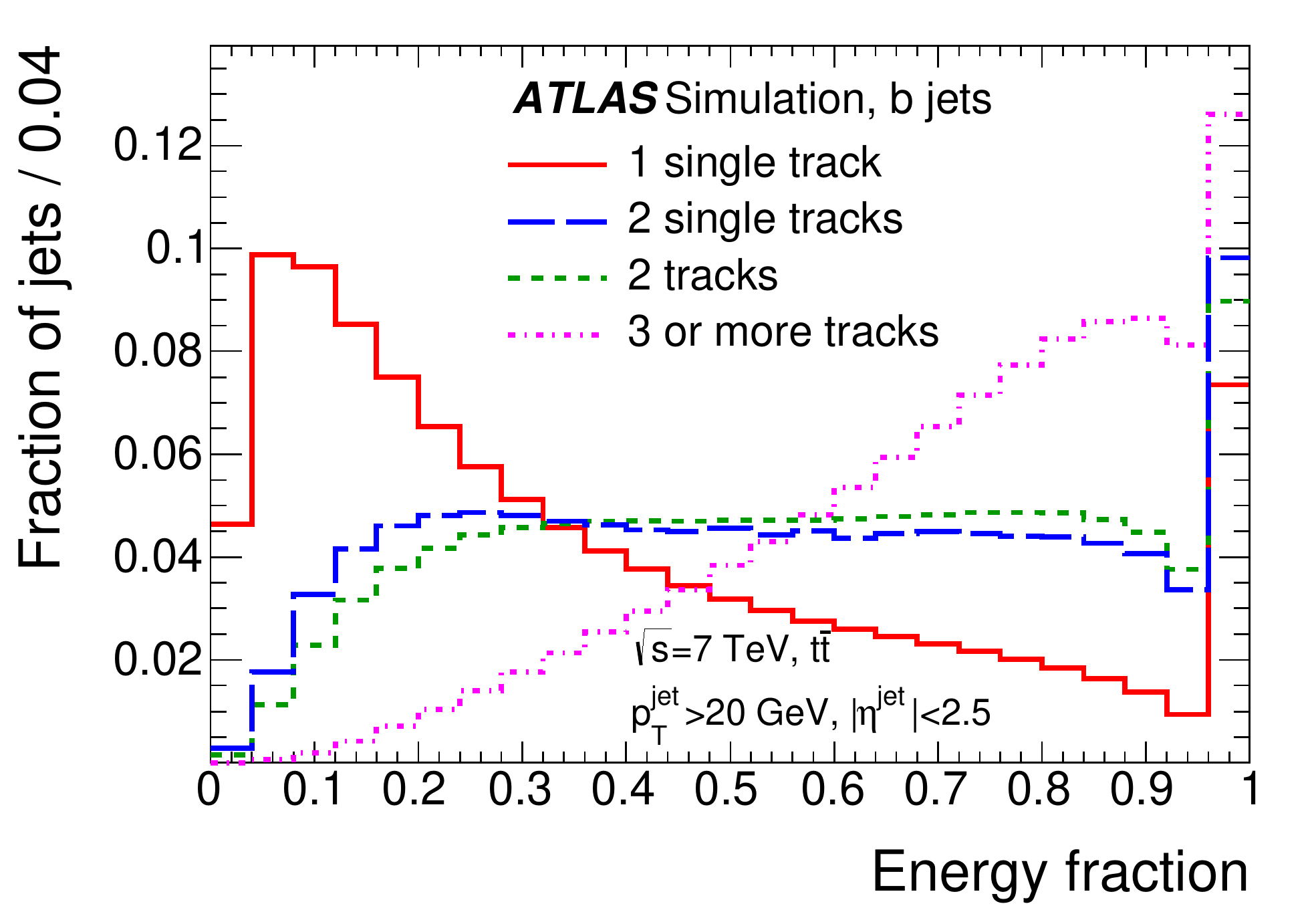}}
  \subfloat[]{\label{fig:perf:jf_efrc_c}
    \includegraphics[width=0.32\textwidth]{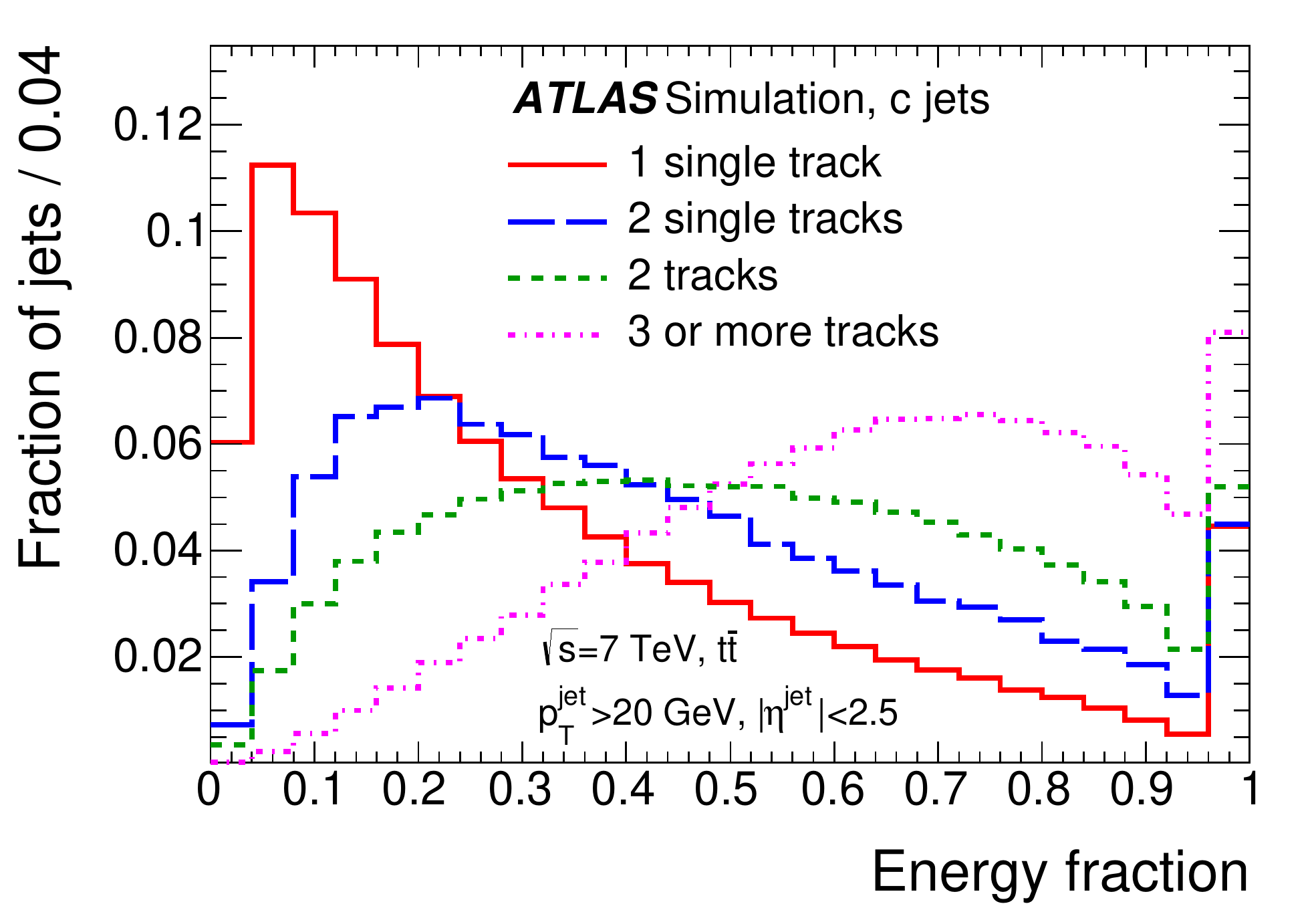}}
  \subfloat[]{\label{fig:perf:jf_efrc_l}
    \includegraphics[width=0.32\textwidth]{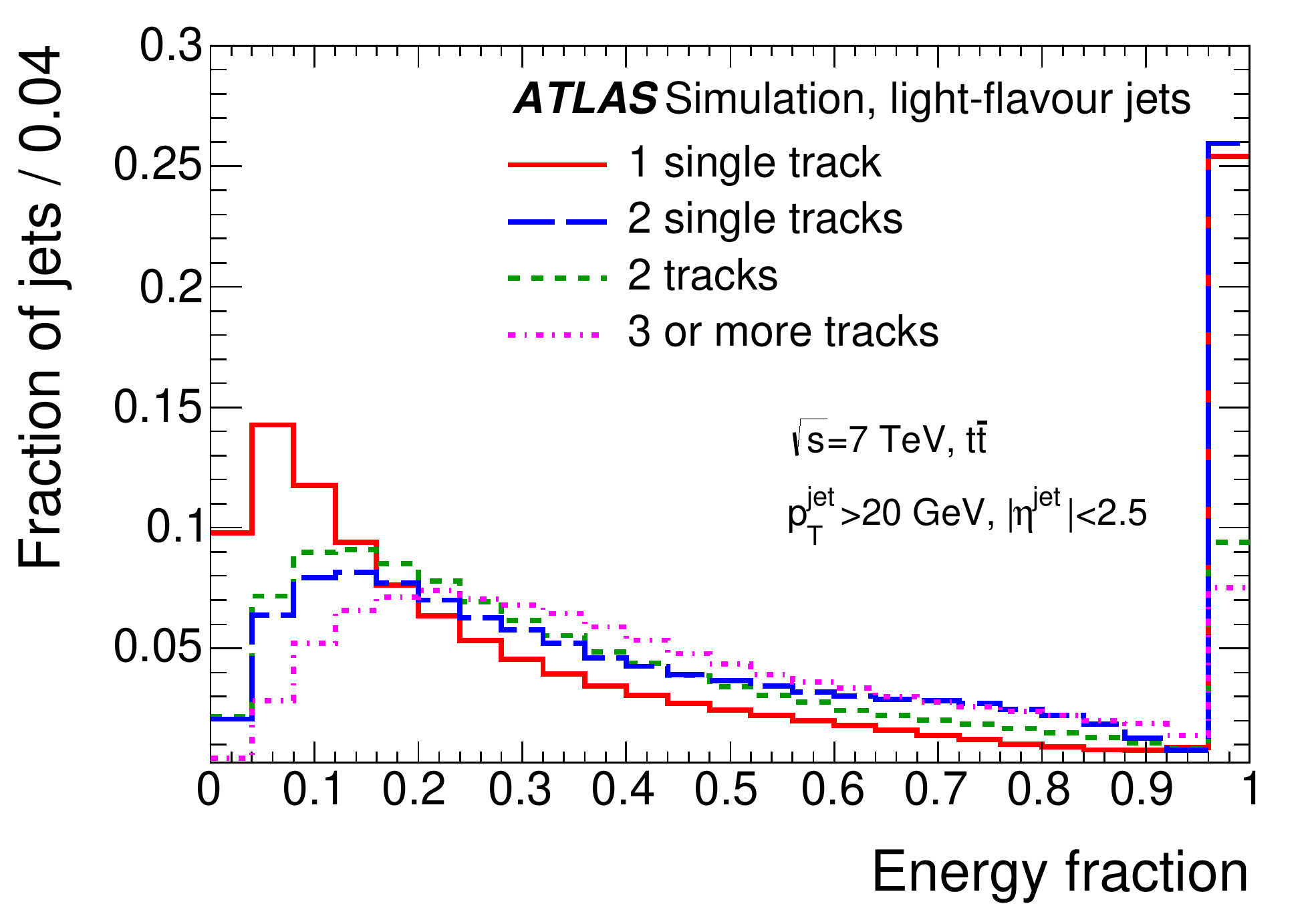}}
  \\
  \subfloat[]{\label{fig:perf:jf_sig3d_b}
    \includegraphics[width=0.32\textwidth]{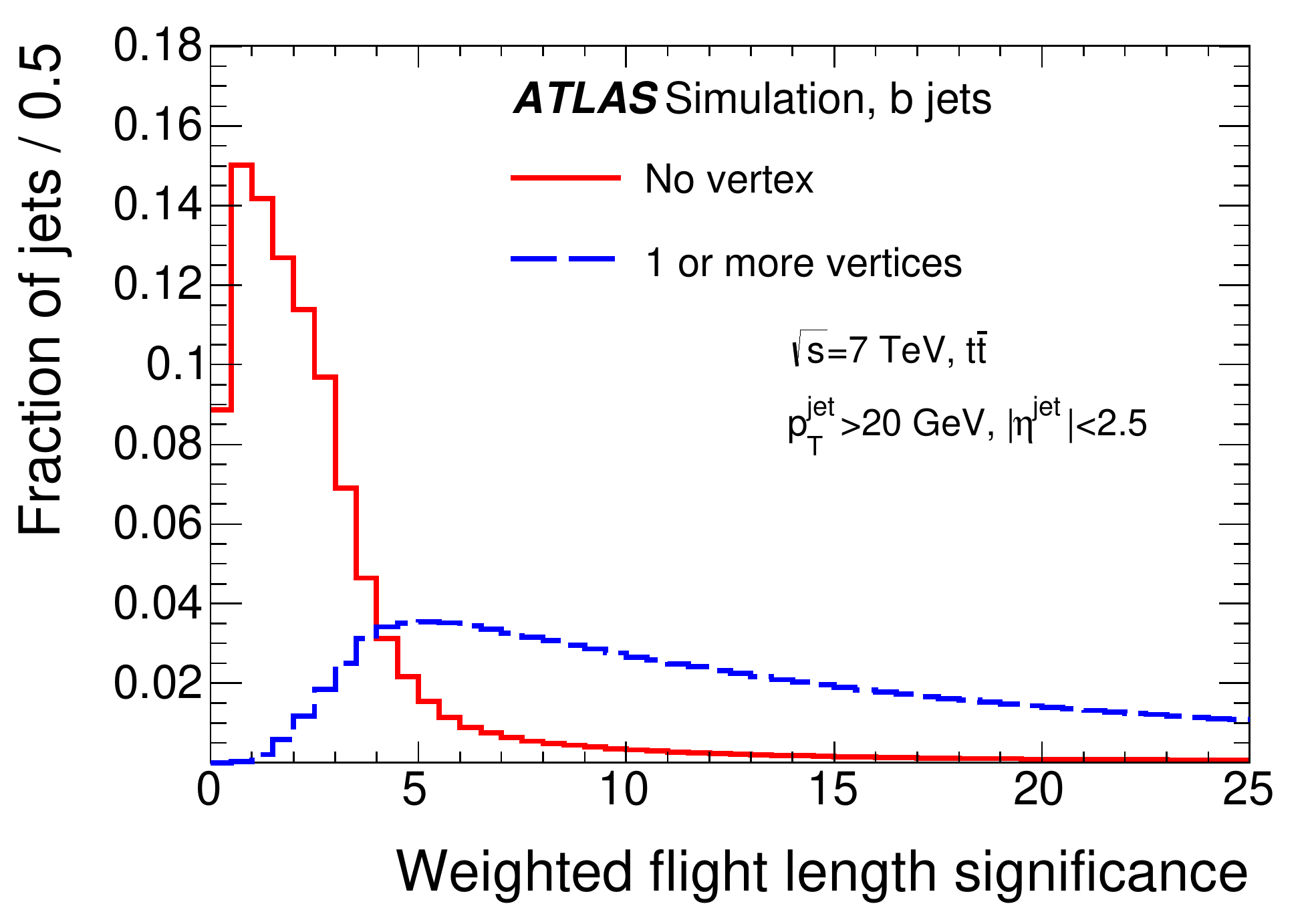}}
  \subfloat[]{\label{fig:perf:jf_sig3d_c}
    \includegraphics[width=0.32\textwidth]{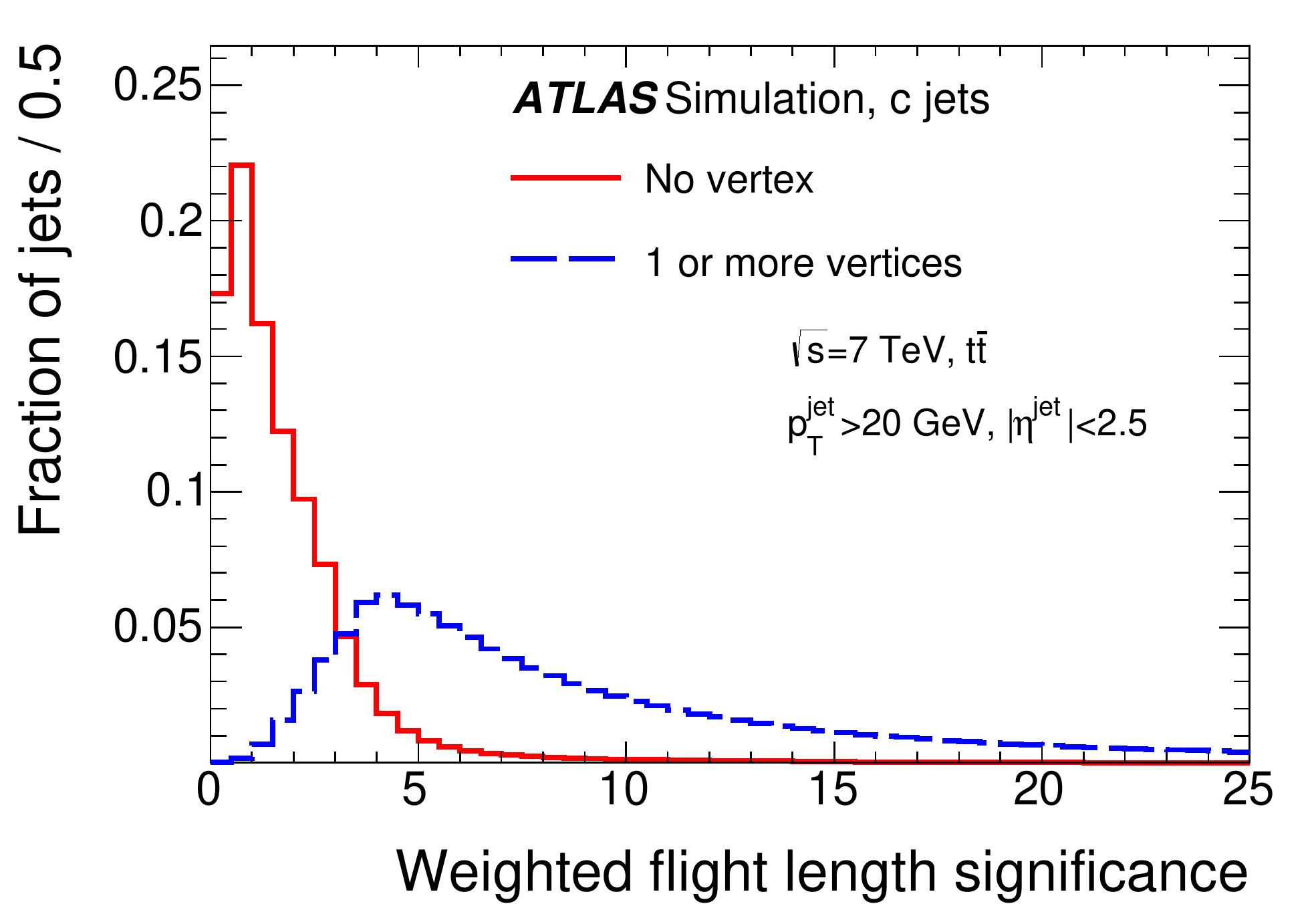}}
  \subfloat[]{\label{fig:perf:jf_sig3d_l}
    \includegraphics[width=0.32\textwidth]{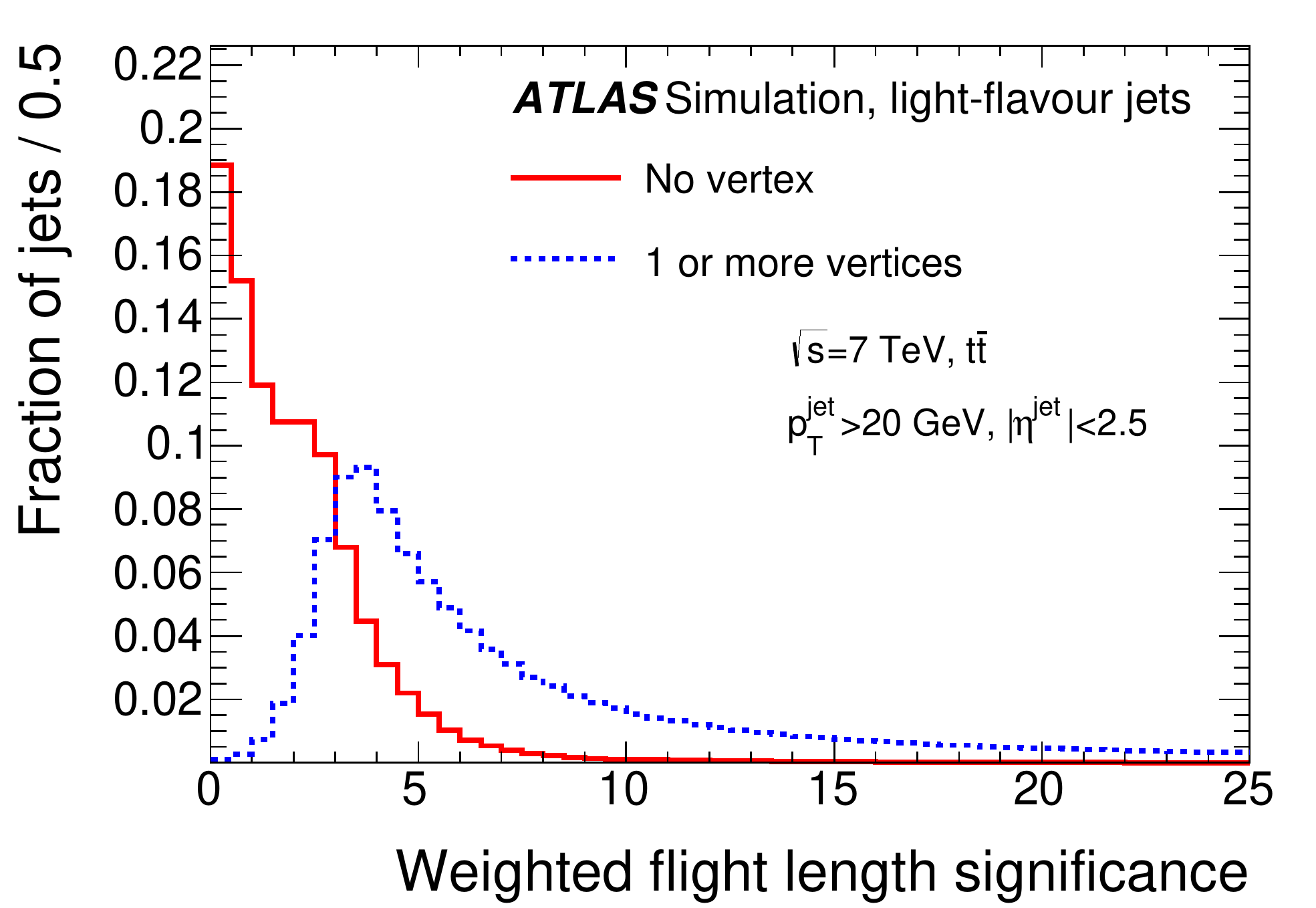}}
  \caption{The vertex mass (top), energy fraction (middle) and flight length significance (bottom) for $b$ jets (left), 
    $c$ jets (middle) and light-flavour jets (right), split according to the decay chain topology found by JetFitter. 
    In the case that no vertex with at least two outgoing tracks has been reconstructed, these quantities
    are computed from reconstructed single track vertices as explained in the text.
    The distributions are obtained from a simulated sample of \ttbar\ events
    generated with \Powheg~\cite{bib:powheg,bib:powheg2} interfaced to \Pythia.
    \label{fig:perf:jetfitter}}
\end{figure}

A very different algorithm, JetFitter~\cite{CSCbook}, exploits the topological structure
of weak $b$- and $c$-hadron decays inside the jet.
A Kalman filter is used to find a common line in three dimensions on which the primary vertex and the bottom and charm vertices lie,
as well as their positions on this line approximating the $b$-hadron flight path.
With this approach, the $b$- and $c$-hadron vertices are not merged,
even when only a single track is attached to each of them.
In the JetFitter algorithm, the decay topology is described by the following discrete variables:
the number of vertices with at least two tracks, the total number of tracks at
these vertices, and the number of additional single track vertices on the $b$-hadron flight axis.
The vertex information is condensed in the following observables, shown in Fig.~\ref{fig:perf:jetfitter}:
the vertex mass (the invariant mass of all charged particle tracks attached to the decay chain),
the energy fraction (the energy of these charged particles divided by the sum
of the energies of all charged particles associated to the jet), and the
flight length significance $L/\sigma_{L}$ (the average displaced vertex
decay length divided by its uncertainty; the individual reconstructed vertices
contribute to the average decay length weighted by the inverse square of their
decay length uncertainties).
The six JetFitter variables defined above are used as input nodes in an artificial neural network.
As the input variable distributions depend on the \pt\ and $|\eta|$ of the jets,
these kinematic variables are included as two additional input nodes.
To ensure that the jet \pt{} and $|\eta|$ spectra of the $b$, $c$ and light-flavour
jets in the training sample are not used by the neural network to separate the
different jet flavours, a two-dimensional reweighting yielding flat kinematic
distributions for all three jet flavours is performed prior to the neural network
training. A coarse two-dimensional binning with seven bins in \pt{} and three bins 
in $|\eta|$ is used for the reweighting.
The JetFitter neural network has three output nodes, corresponding to the 
$b$-, $c$- and light-flavour-jet hypotheses, referred to as $P_b$, $P_c$ and $P_l$.
The network topology includes two hidden layers, with 12 and 7 nodes, respectively.
A discriminating variable to select $b$ jets and reject light-flavour jets is then
defined from the values of the corresponding output nodes: $w_{\rm JetFitter} = \ln(P_b/P_l)$.

\subsection{Combined tagging algorithms}
\label{sec:comb-vertex-algos}

The vertex-based algorithms exhibit much lower mistag rates than the impact parameter-based ones, but their efficiency 
for actual $b$ jets is limited by the secondary vertex finding efficiency.
Both approaches are therefore combined to define versatile and powerful tagging algorithms.
The LLR-based IP3D and SV1 algorithms are combined in a straightforward manner by summing their respective weights: 
this is the so-called IP3D+SV1 algorithm.
Another combination technique is the use of an artificial neural network, which can take advantage of complex correlations between 
the input values.
Two tagging algorithms are defined in this way,
IP3D+JetFitter and MV1.

The IP3D+JetFitter algorithm is defined in the same way as the JetFitter algorithm itself, with the only
difference being that the output weight of the IP3D algorithm is used as an additional input node, and 
that the number of nodes in the two intermediate hidden layers is increased to 9 and 14, respectively.
The discriminating variable to select $b$ jets and reject light-flavour jets is
defined as $w_{\rm IP3D+JetFitter} = \ln(P_b/P_l)$.
A specific tuning of the IP3D+JetFitter algorithm to provide a better discrimination 
between $b$ and $c$ jets uses 
$w_{\rm IP3D+JetFitter(c)} = \ln(P_b/P_c)$ as a discriminant.

\begin{figure}
  \subfloat[]{\includegraphics[width=0.32\textwidth]{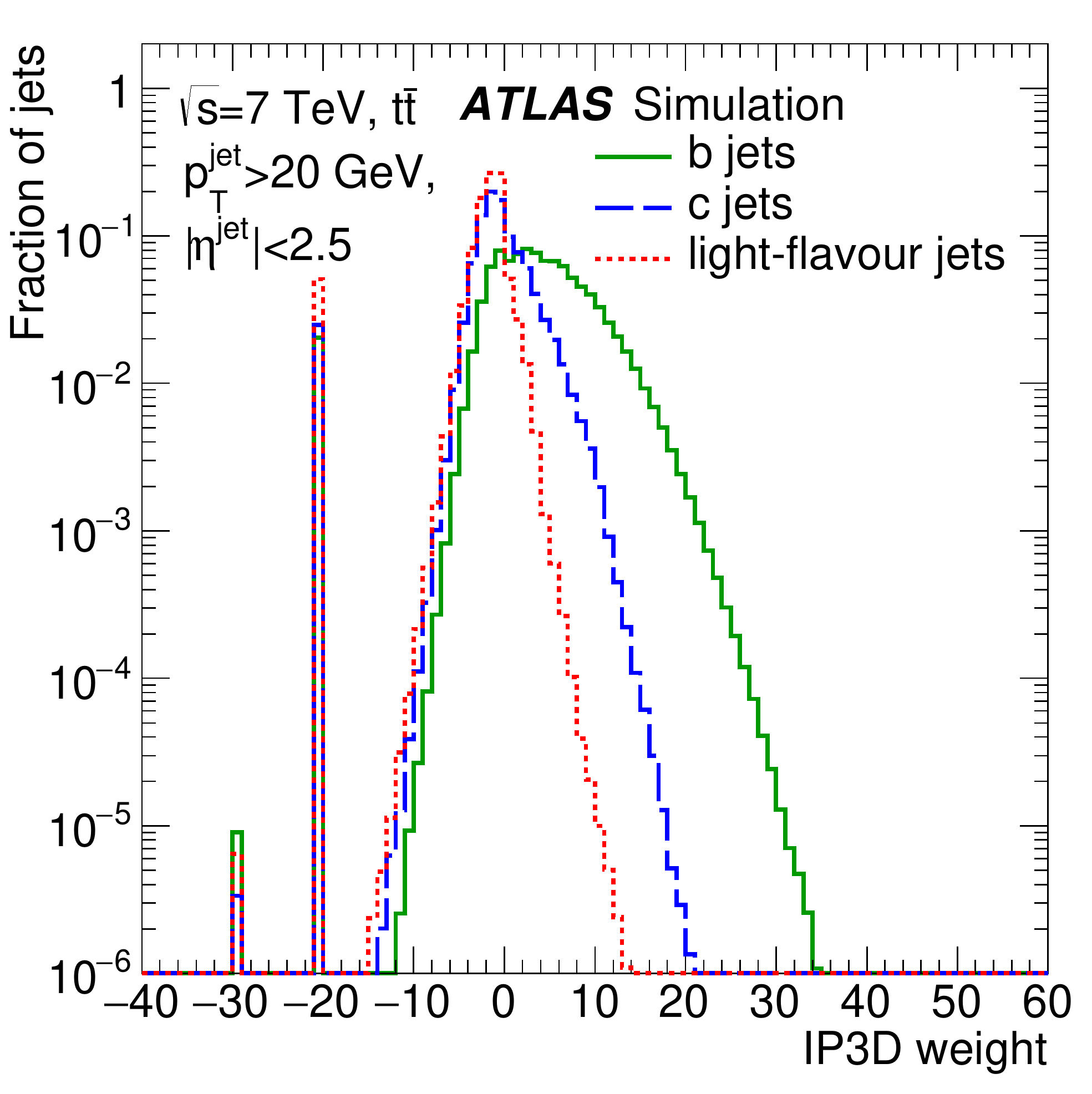}}
  \subfloat[]{\includegraphics[width=0.32\textwidth]{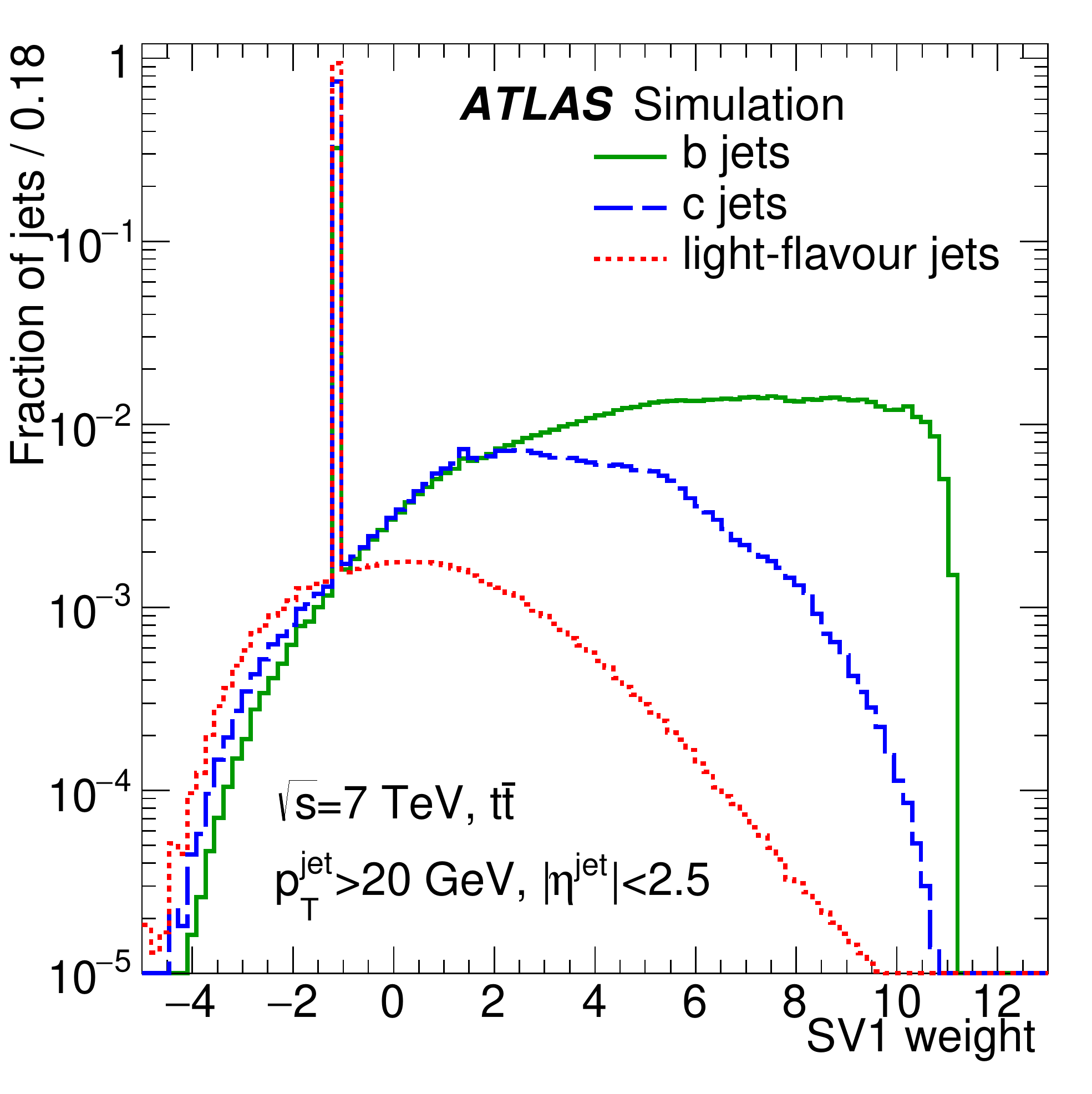}}
  \subfloat[]{\includegraphics[width=0.32\textwidth]{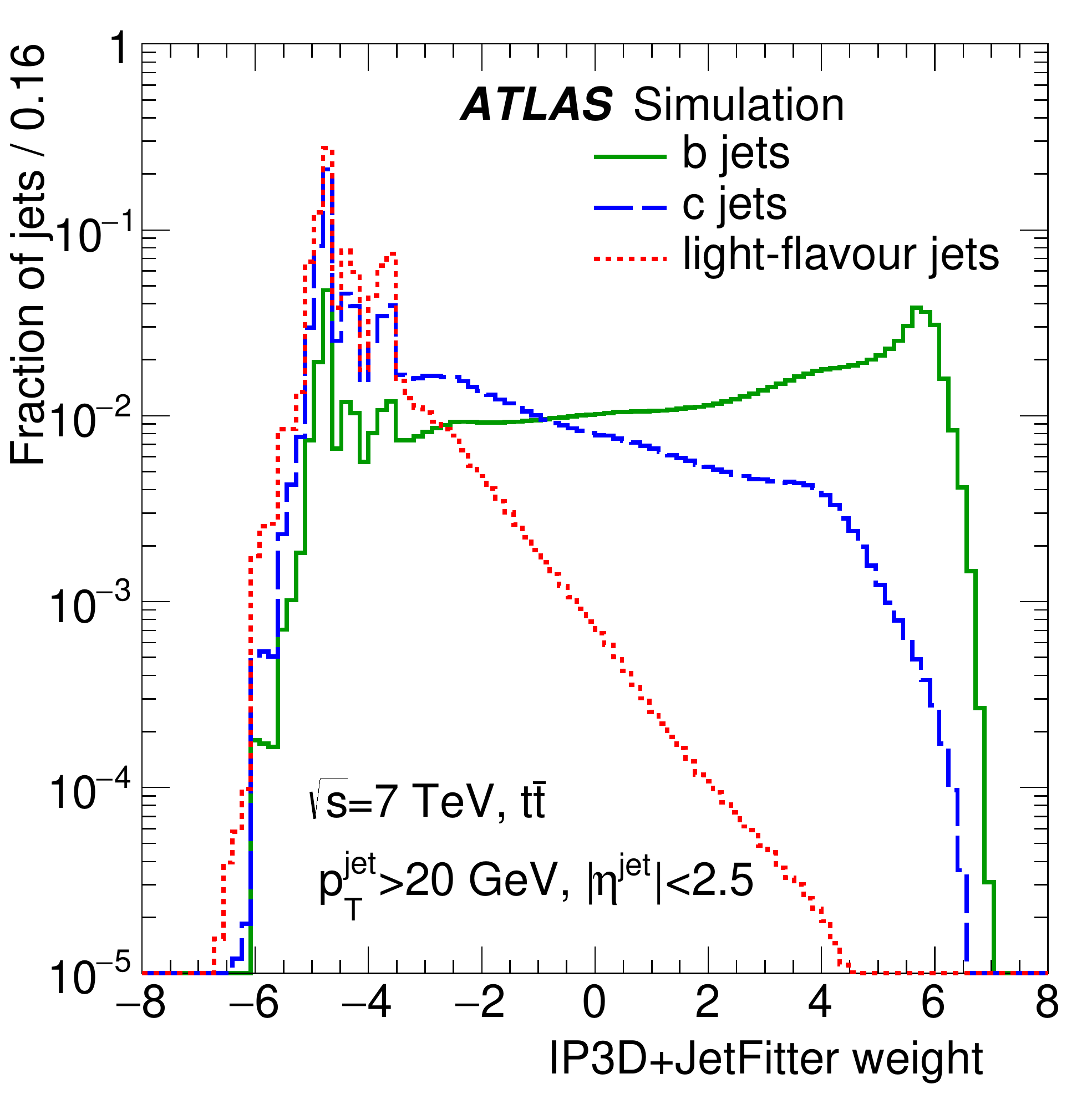}}
  \caption{Distribution of the IP3D (a), SV1 (b) and IP3D+JetFitter (c) weights, 
    for $b$, $c$ and light-flavour jets. 
    These three weights are used as inputs for the MV1 algorithm.
    The spikes at $w_{\rm IP3D}\approx -20$ and $\approx -30$ correspond to pathological cases where the IP3D weight 
    could not be computed, due to the absence of good-quality tracks.
    The spike at $w_{\rm SV1}\approx -1$ corresponds to jets in which no secondary vertex could be 
    reconstructed by the SV1 algorithm, and where discrete probabilities for a $b$
    and light-flavour jet not to have a vertex are assigned.
    The irregular behaviour in $w_{\rm IP3D+JetFitter}$ arises because both the
    $w_{\rm IP3D}$ and the $w_{\rm JetFitter}$ distribution (not shown) exhibit several spikes.
    \label{fig:perf:mv1inputs}}
\end{figure}

\begin{figure}[htbp]
  \subfloat[]{\includegraphics[width=0.32\textwidth]{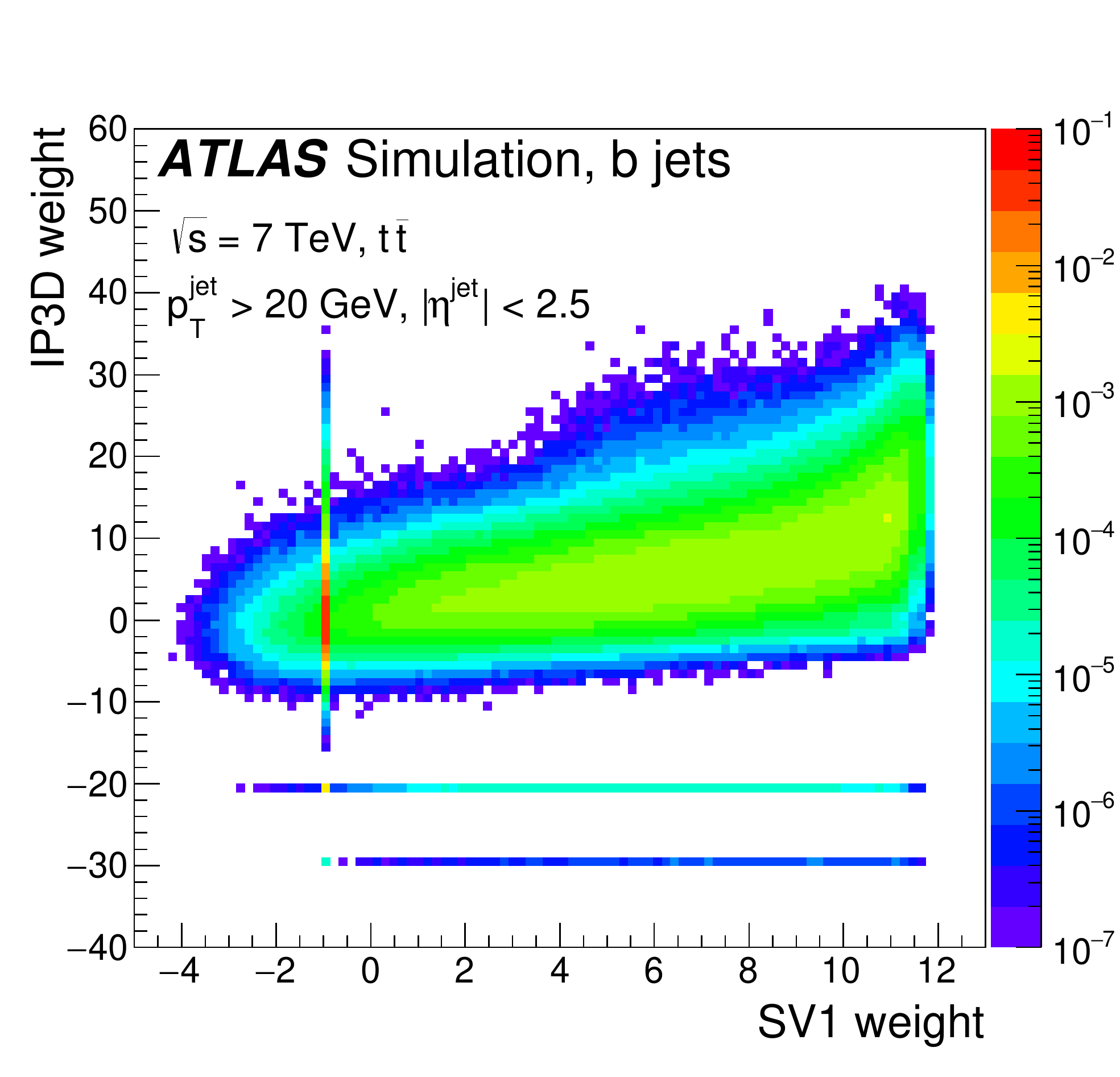}}
  \subfloat[]{\includegraphics[width=0.32\textwidth]{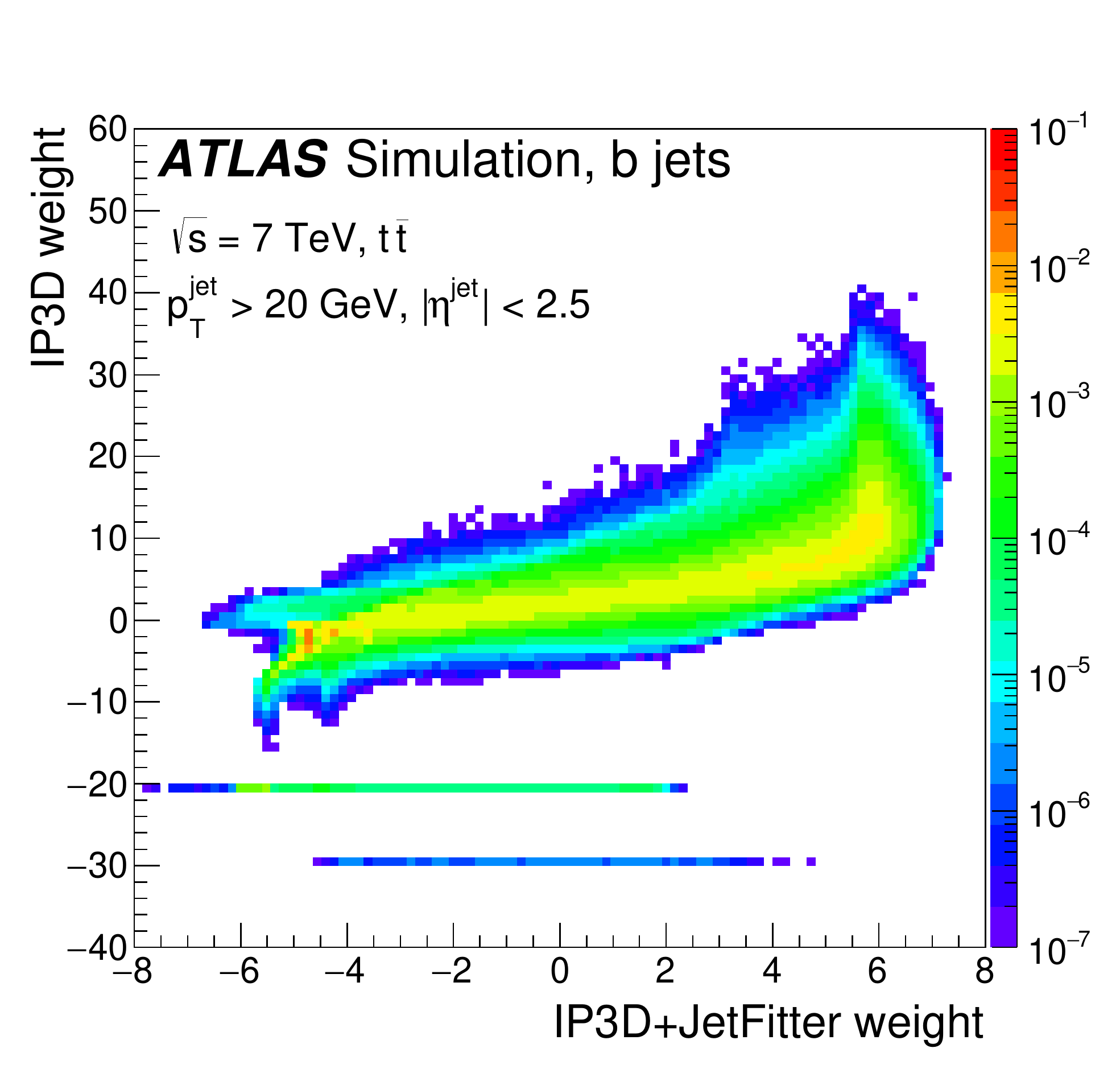}}
  \subfloat[]{\includegraphics[width=0.32\textwidth]{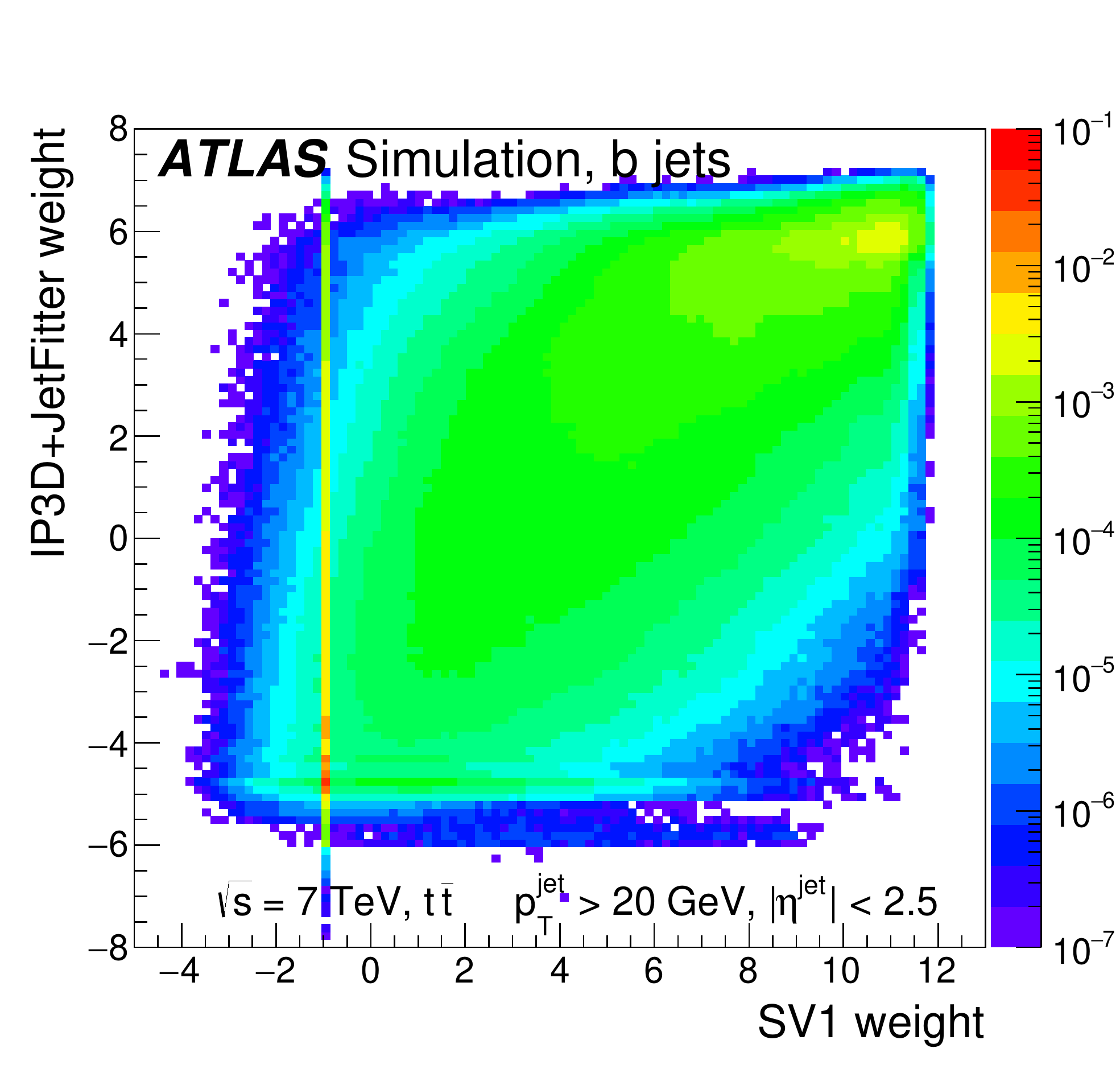}}
  \hfill
  \subfloat[]{\includegraphics[width=0.32\textwidth]{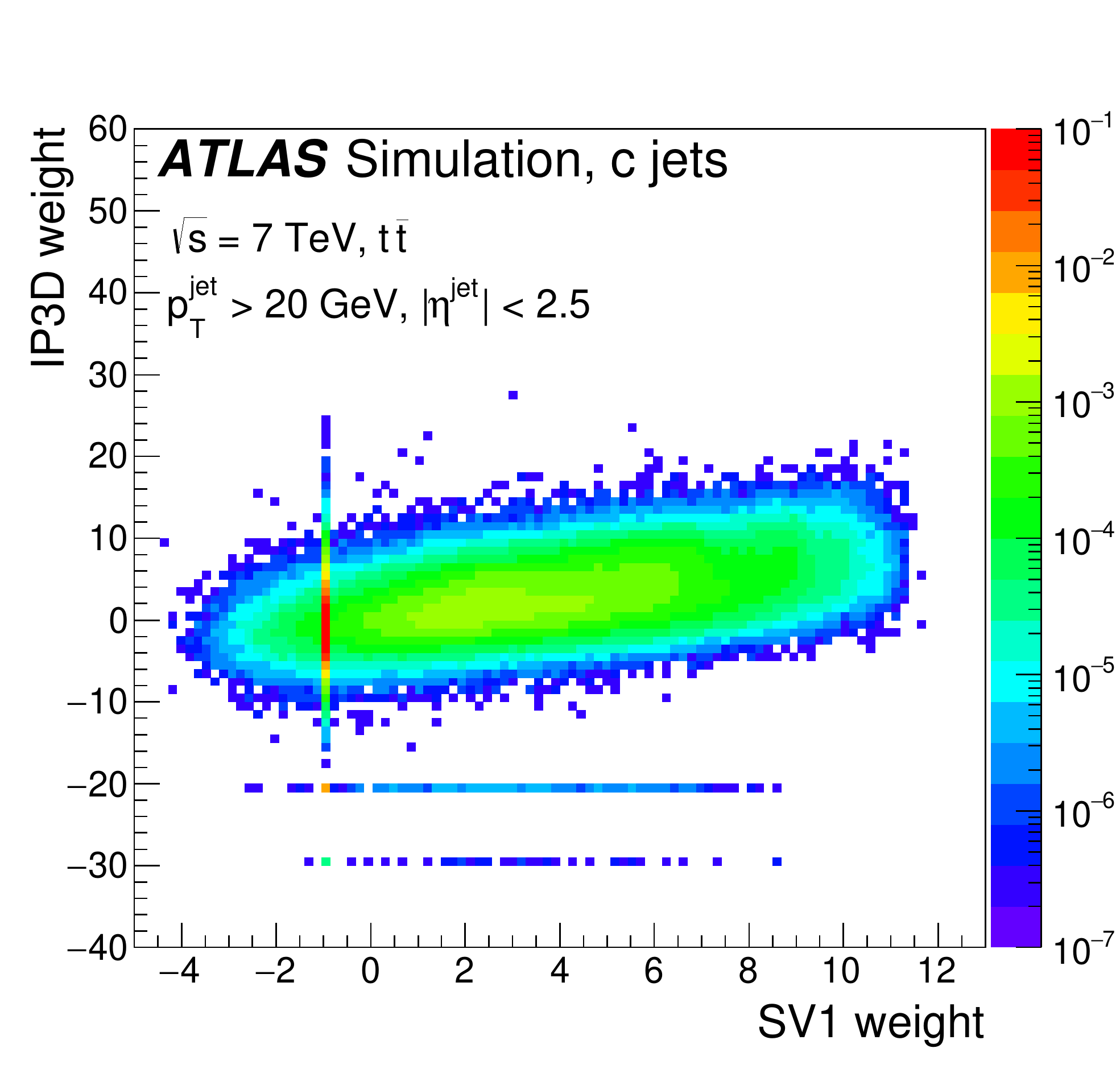}}
  \subfloat[]{\includegraphics[width=0.32\textwidth]{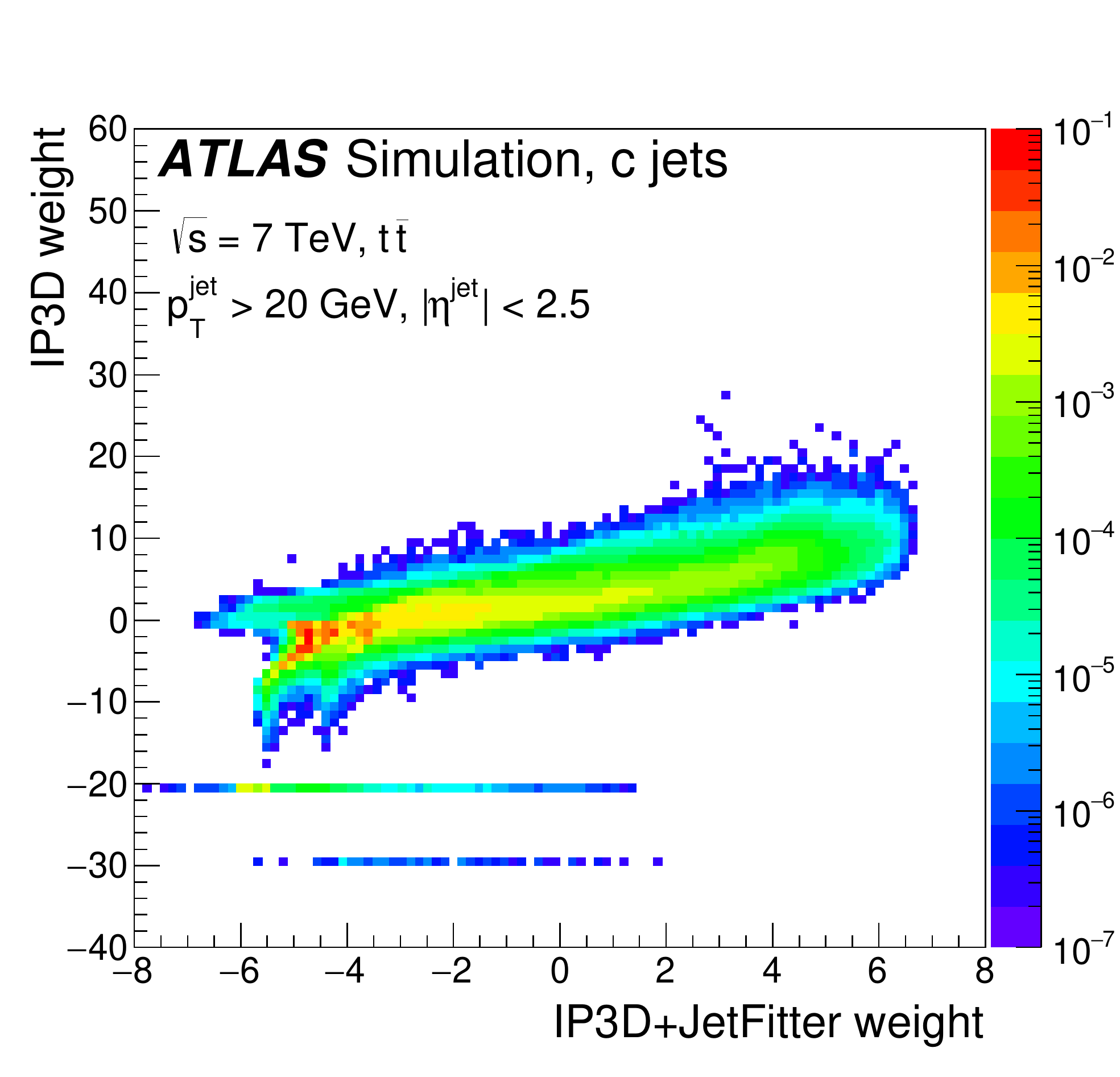}}
  \subfloat[]{\includegraphics[width=0.32\textwidth]{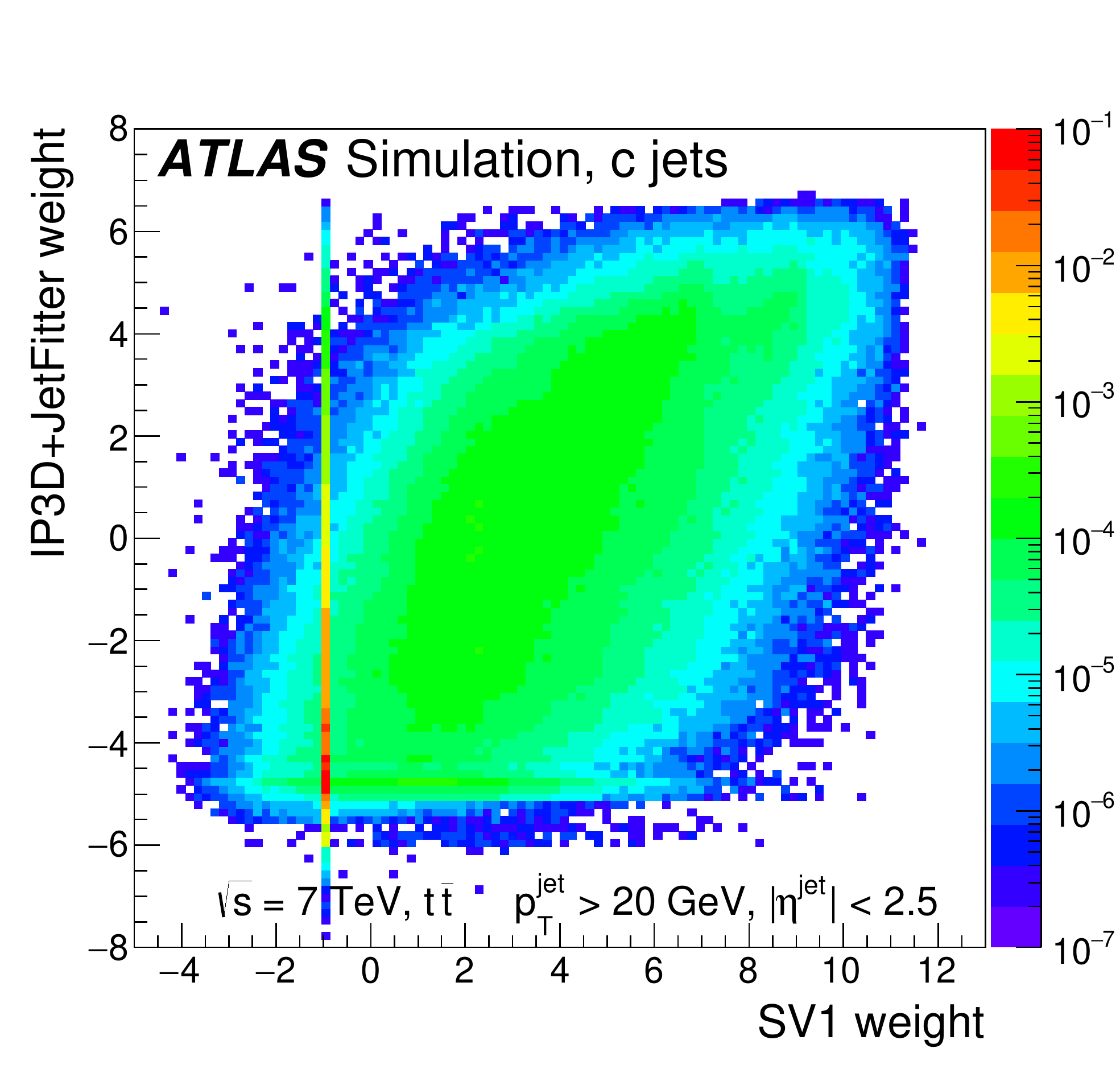}}
  \hfill
  \subfloat[]{\includegraphics[width=0.32\textwidth]{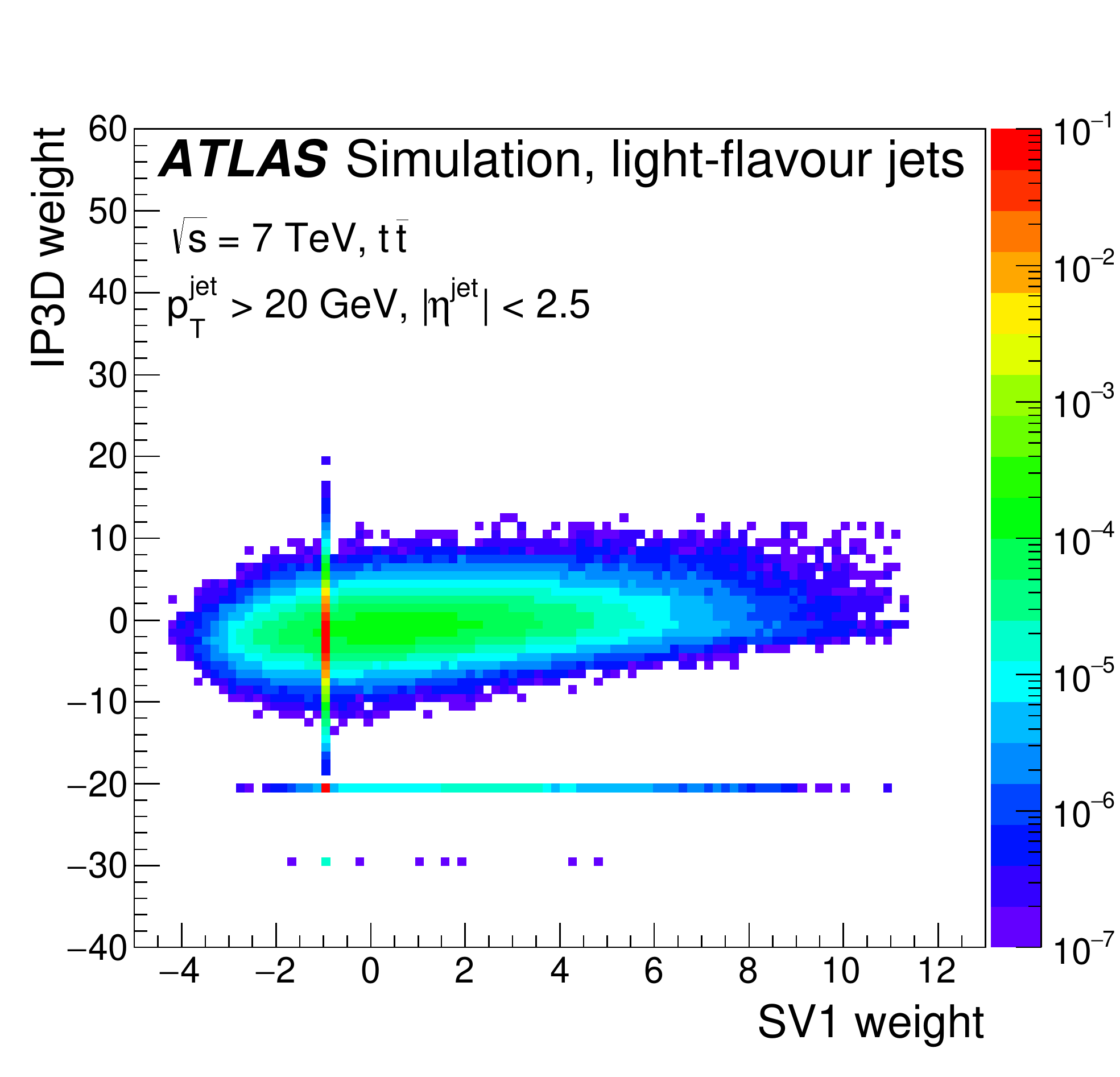}}
  \subfloat[]{\includegraphics[width=0.32\textwidth]{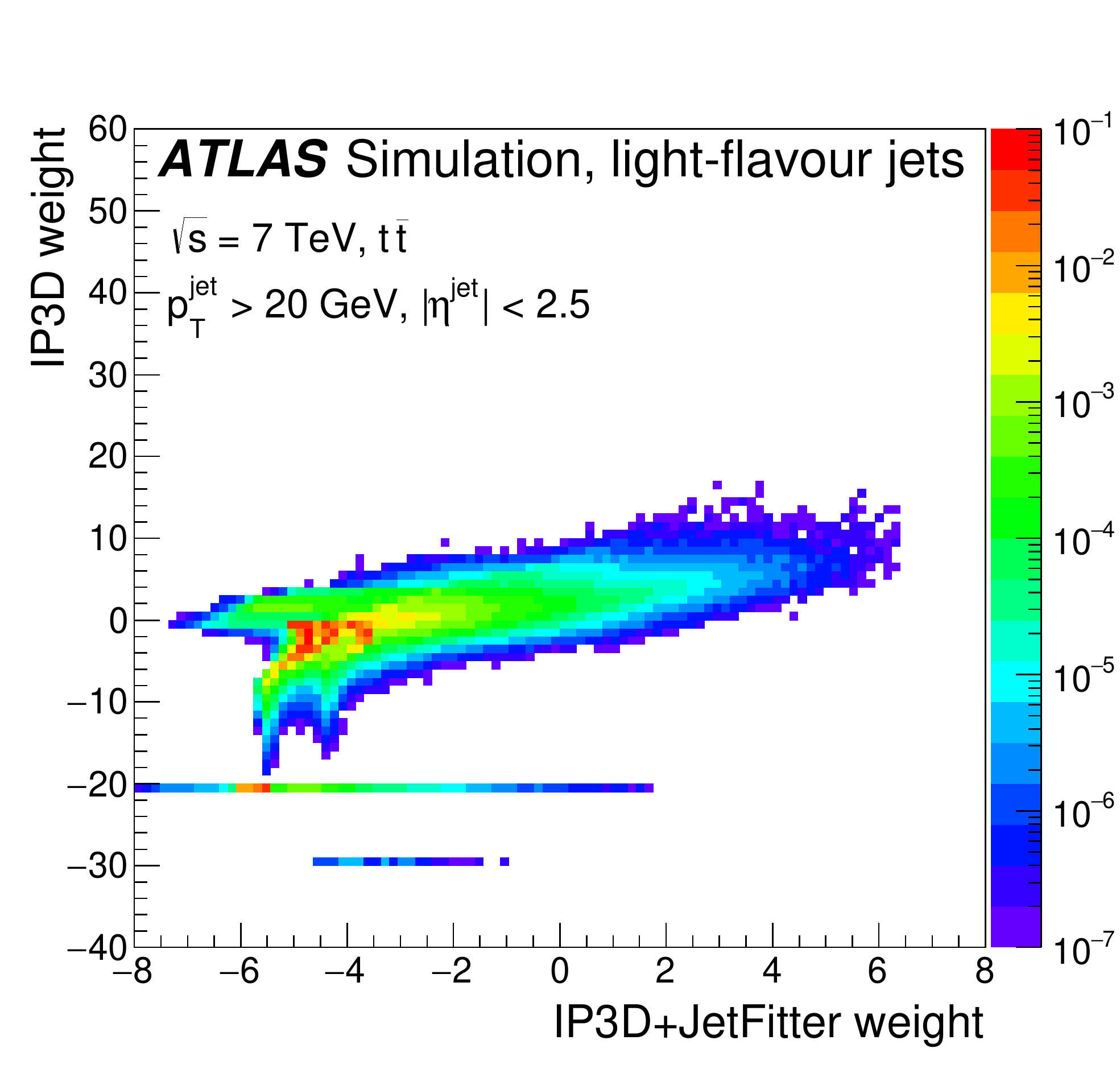}}
  \subfloat[]{\includegraphics[width=0.32\textwidth]{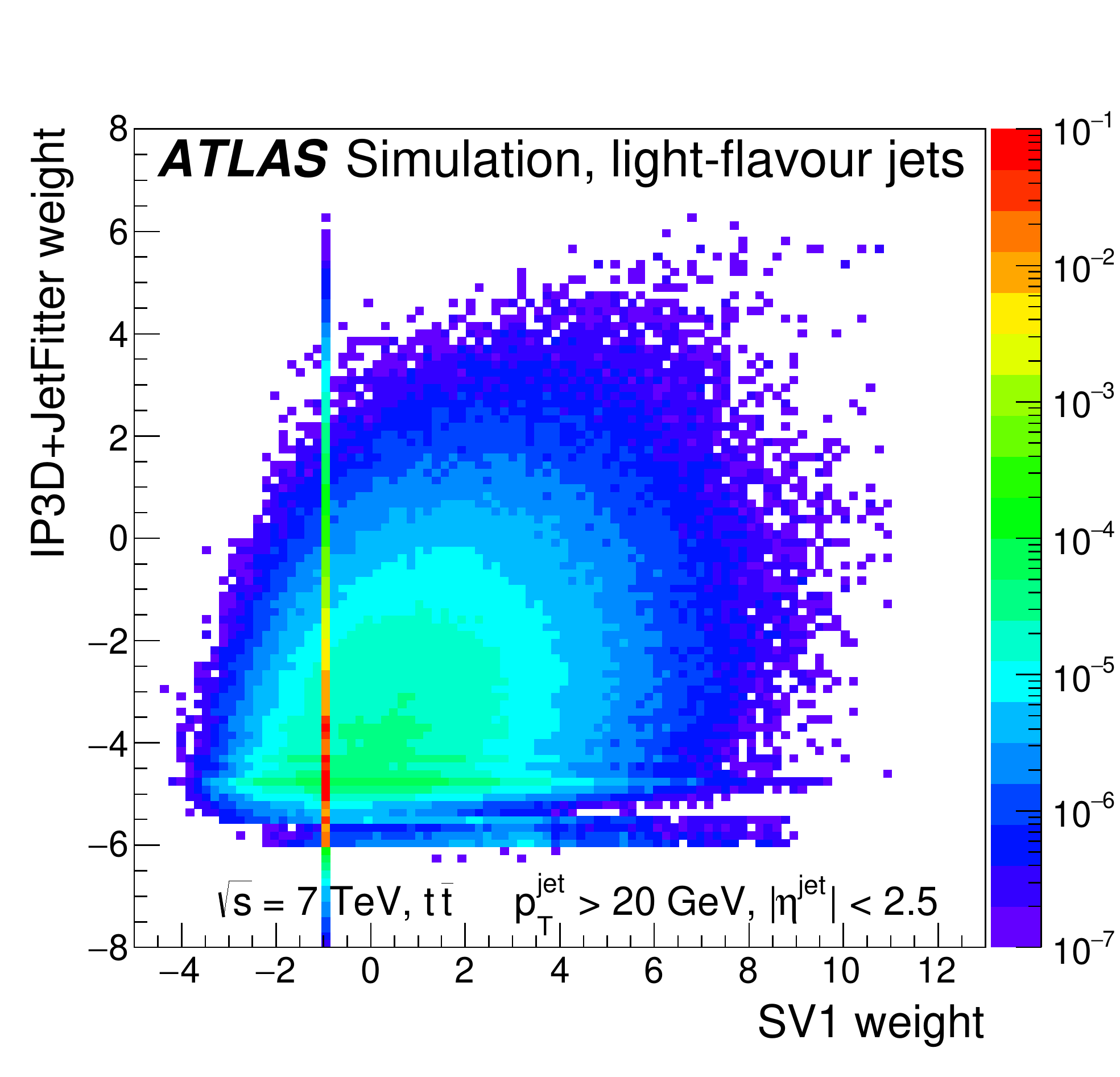}}
  \caption{Distributions of the correlations between the IP3D, SV1 and IP3D+JetFitter weights, 
    for $b$ jets (top), $c$ jets (middle) and light-flavour jets (bottom). 
    The spikes at $w_{\rm IP3D}\approx -20$ and $\approx -30$ correspond to pathological cases where the IP3D weight 
    could not be computed, due to the absence of good-quality tracks.
    The spike at $w_{\rm SV1}\approx -1$ corresponds to jets in which no secondary vertex could be 
    reconstructed by the SV1 algorithm, and where discrete probabilities for a $b$
    and light-flavour jet not to have a vertex are assigned.
    \label{fig:perf:mv1corr}}
\end{figure}

MV1 is an algorithm used widely in ATLAS physics analyses.
Distributions of the three MV1 input variables (the IP3D and SV1 discriminants as well as the sum of the IP3D and JetFitter discriminants)
are shown in Fig.~\ref{fig:perf:mv1inputs}, for $b$ jets, $c$ jets, and light-flavour jets in simulated $t\bar{t}$ events.
The distributions of the correlations between the three input weights are also shown 
in Fig.~\ref{fig:perf:mv1corr}, for $b$ jets, $c$ jets and light-flavour jets. 
These distributions illustrate the potential gain in combining the three weights: 
it can be seen that the IP3D weight has only limited correlations with the secondary vertex-based weights,  while naturally 
SV1 and IP3D+JetFitter weights are more correlated but the correlation is different in the $b$-jet, $c$-jet and light-flavour-jet samples.
The MV1 neural network is a perceptron with two hidden layers consisting of three and two nodes, respectively,
and an output layer with a single node which holds the final discriminant variable.
The implementation used is the MLP code from the TMVA package~\cite{tmva}.
The training relies on a back-propagation algorithm and is based on two simulated samples of $b$ jets (signal hypothesis) 
and light-flavour jets (background hypothesis).
Most of the jets are obtained from simulated $t\bar{t}$ events and their average transverse momentum is around 60 \gev. 
To provide jets with higher \pt{} for the training, 
simulated dijet events with jets in the $200\GeV<\pt<500\GeV$ range are also included.
As in the case of the JetFitter neural network, since the tagging performance depends strongly on the $\pt$ and, to a lesser extent, on the $\eta$ of the jet, 
biases may arise from the 
different kinematic spectra of the two training samples (of light-flavour and $b$ jets).
To reduce this effect, weighted training events are used.
Each jet is assigned to a category defined by a coarse two-dimensional grid in $(\pt,\eta)$ with four bins in $\eta$ and ten bins in \pt. 
Jets in the same category are given the same weight,
 defined according to the overall 
fraction of all jets in this category, and the jet category 
is fed to the network as an additional input variable.
The MV1 output weight distribution is shown in Fig.~\ref{fig:perf:mv1Weight} for $b$, $c$, and light-flavour jets in
simulated $t\bar{t}$ events. 
The spike around 0.15 corresponds mostly to jets for which no secondary vertex could be found.

\subsection{Performance in simulation}

\begin{figure}
  \begin{minipage}{0.49\textwidth}
    \includegraphics[width=\textwidth]{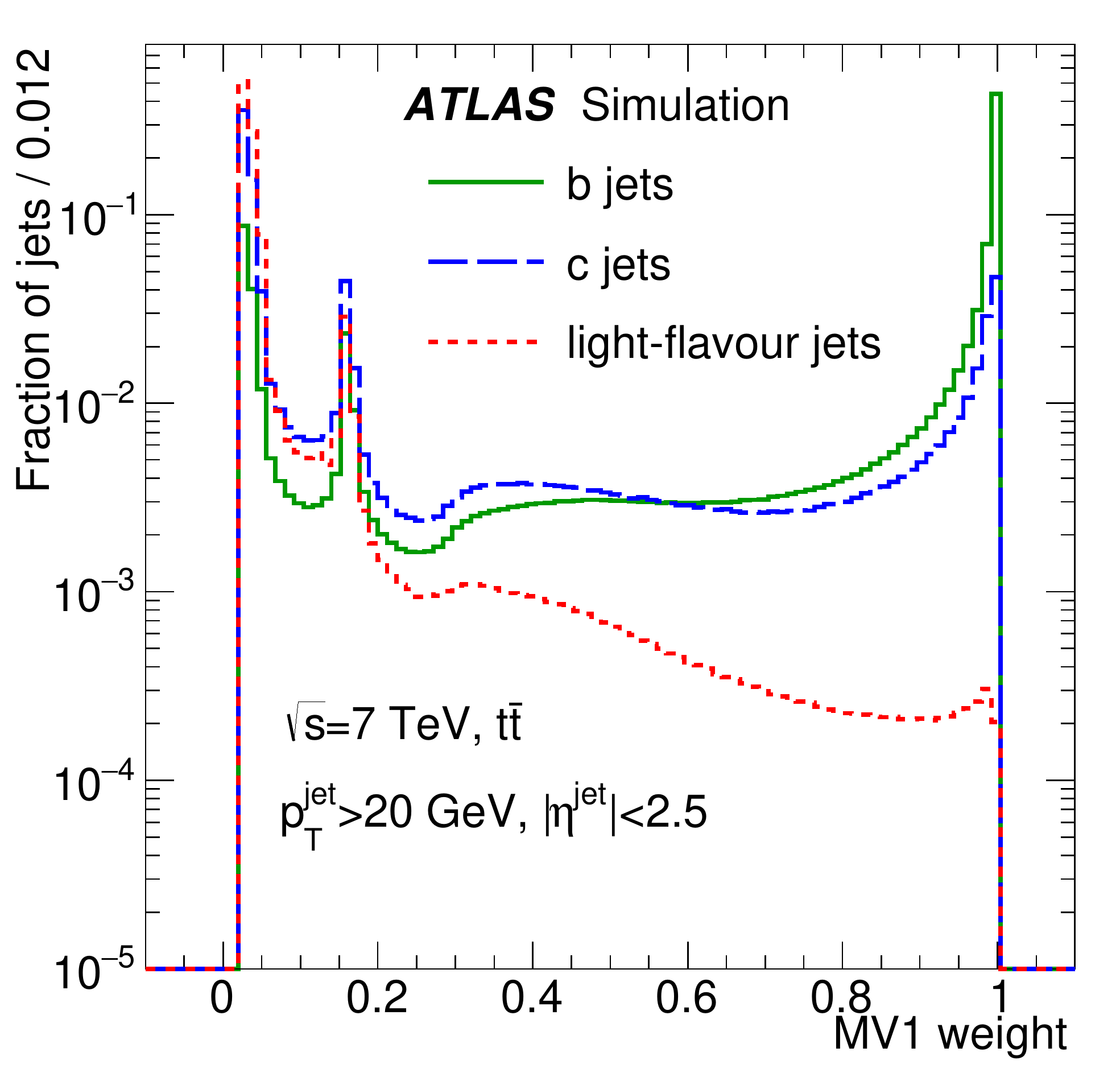}
    \caption{Distribution of the tagging weight obtained with the MV1 algorithm, for three different 
      flavours of jets.}
    \label{fig:perf:mv1Weight}
  \end{minipage}\hfill
  \begin{minipage}{0.49\textwidth}
    \vspace*{-3.5mm}
    \includegraphics[width=\textwidth]{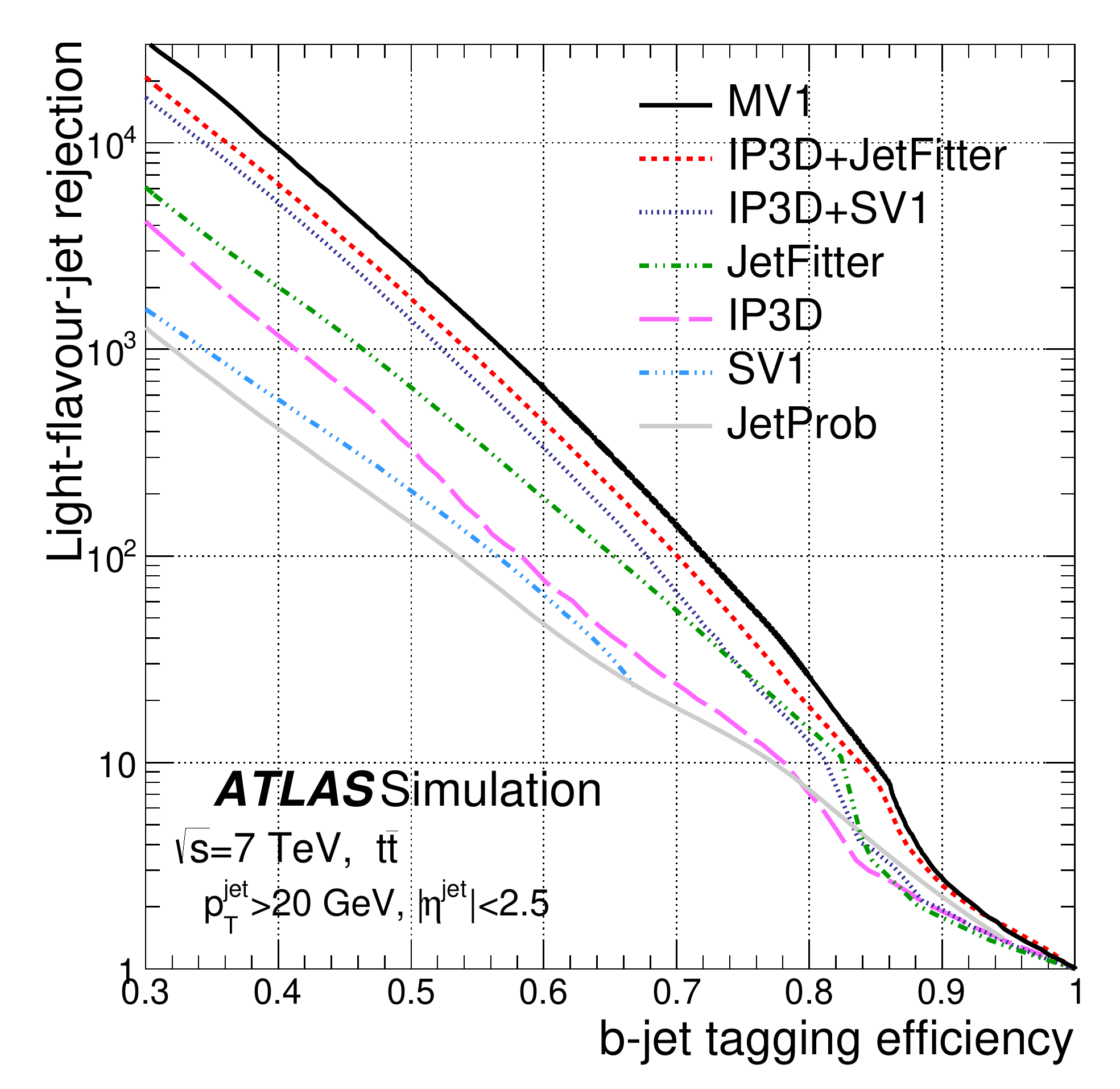}
    \caption{Light-flavour-jet rejection versus $b$-jet tagging efficiency, 
      for various tagging algorithms.}
    \label{fig:perf:lrejVSeff}
  \end{minipage}
\end{figure}

\begin{figure}
  \begin{minipage}{0.49\textwidth}
    \includegraphics[width=\textwidth]{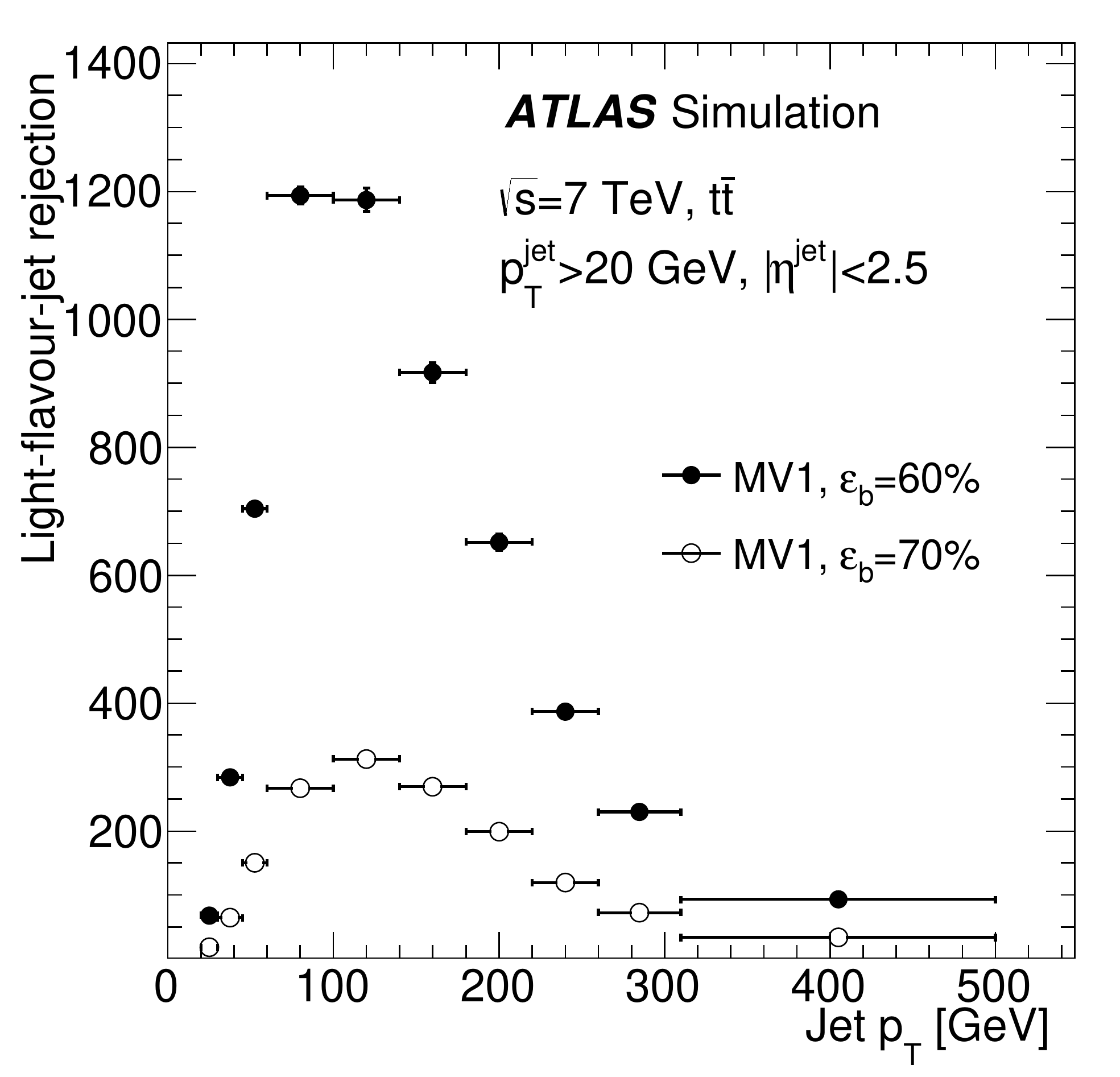}
    \caption{Light-flavour-jet rejection versus jet $\pt$, for the MV1 algorithm.
      In each bin the cut on the $b$-tagging weight is adjusted to maintain an average 60\% (70\%) $b$-jet tagging efficiency.}
    \label{fig:perf:lrejVSpt}
  \end{minipage}\hfill
  \begin{minipage}{0.49\textwidth}
    \includegraphics[width=\textwidth]{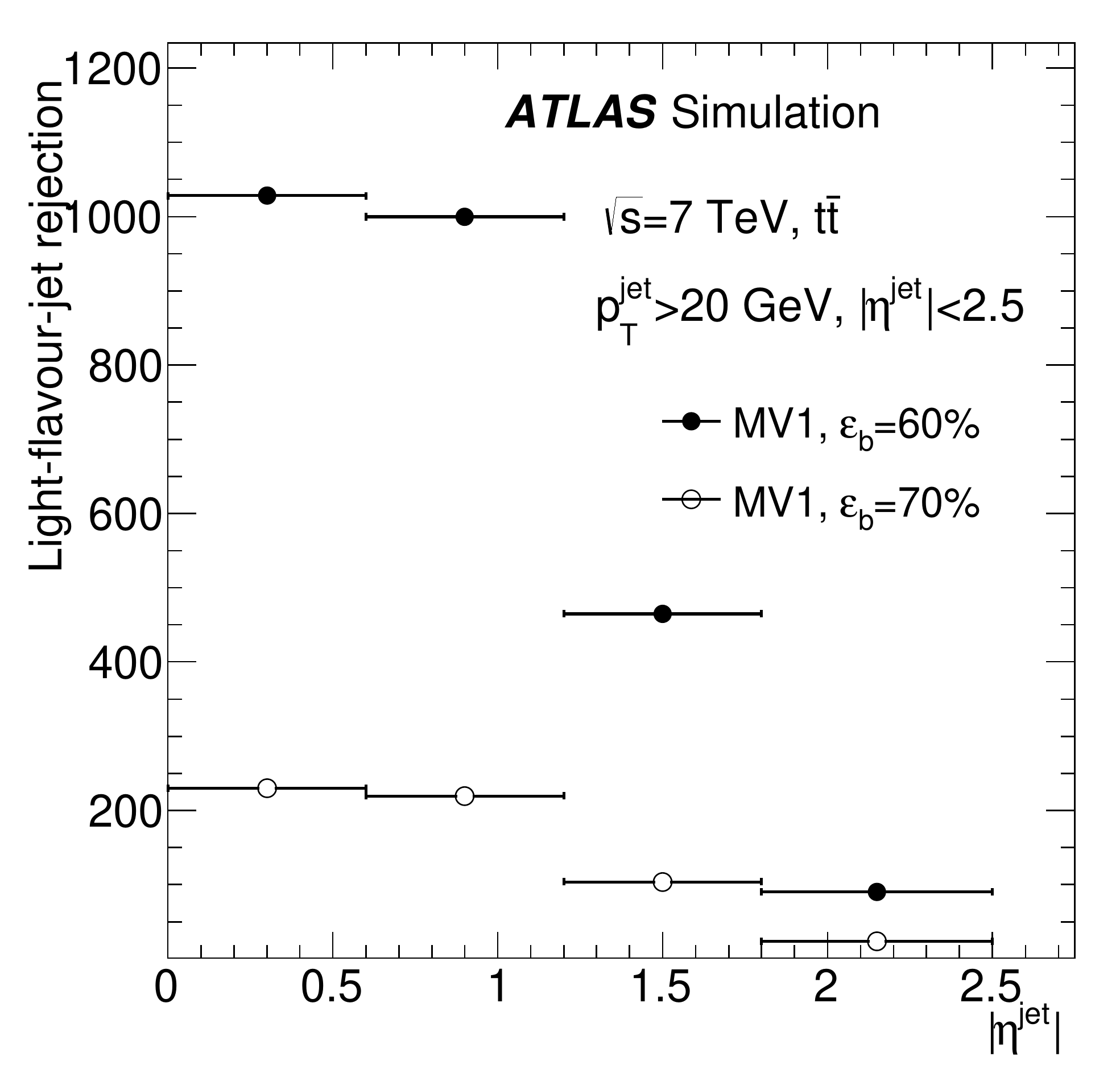}
    \caption{Light-flavour-jet rejection versus jet $|\eta|$, for the MV1 algorithm.
      In each bin the cut on the $b$-tagging weight is adjusted to maintain an average 60\% (70\%) $b$-jet tagging efficiency.}
    \label{fig:perf:lrejVSeta}
  \end{minipage}
\end{figure}

\begin{figure}
  \begin{minipage}{0.49\textwidth}
    \includegraphics[width=\textwidth]{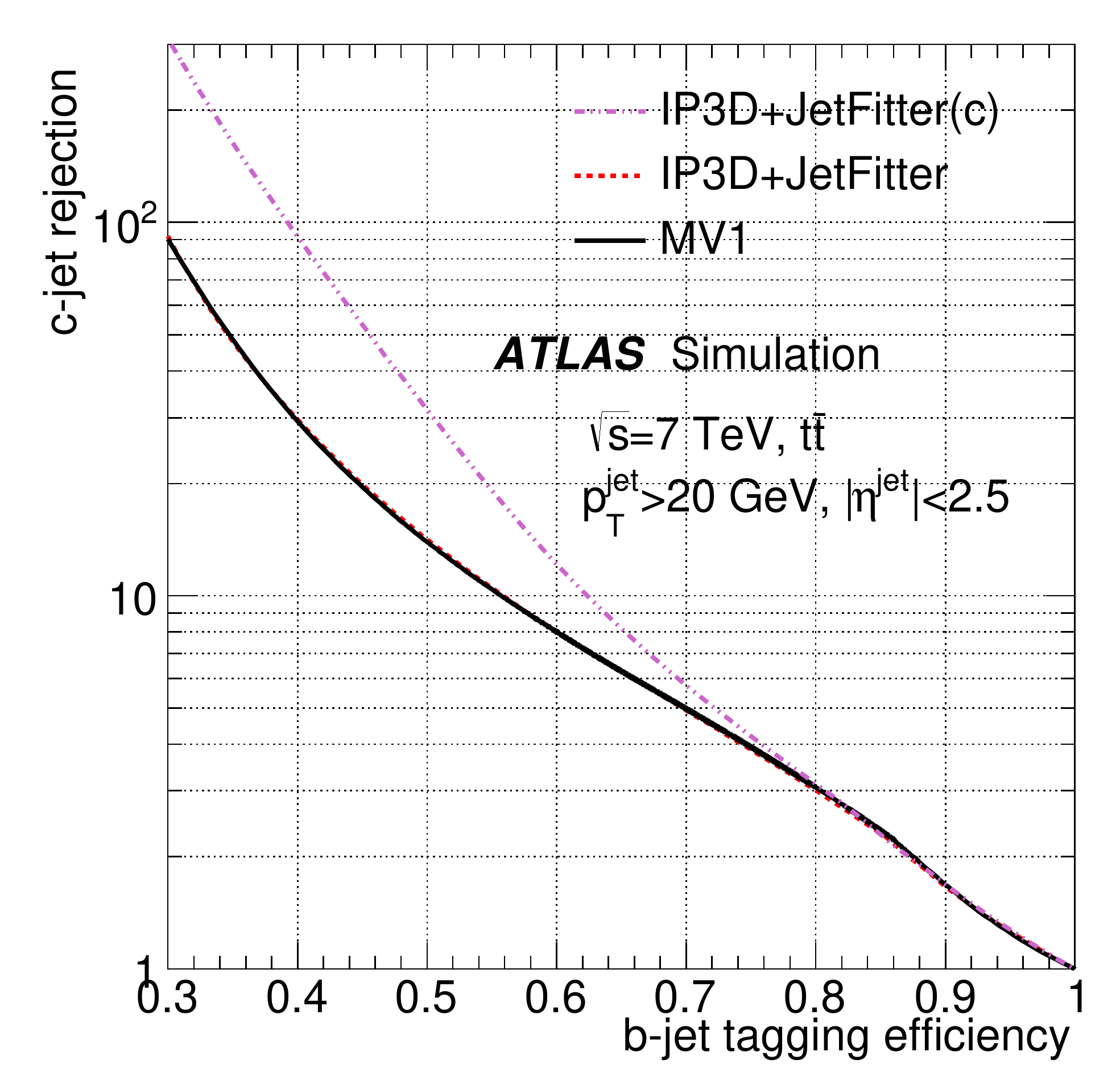}
    \caption{$c$-jet rejection versus $b$-jet tagging efficiency, for three tagging algorithms.}
    \label{fig:perf:crejVSeff}
  \end{minipage}\hfill
  \begin{minipage}{0.49\textwidth}
    \includegraphics[width=\textwidth]{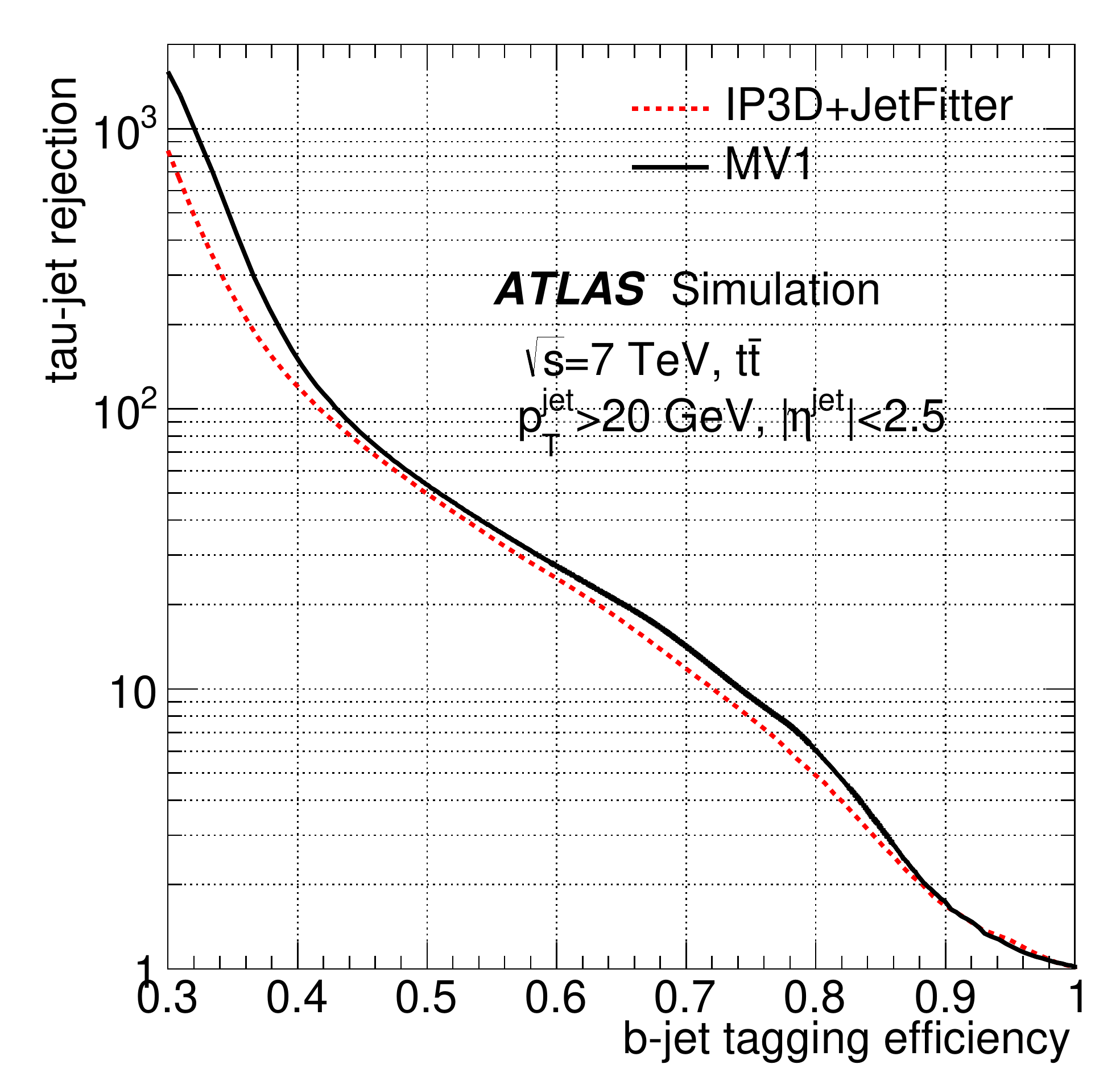}
    \caption{$\tau$-jet rejection versus $b$-jet tagging efficiency, for two tagging algorithms.}
    \label{fig:perf:trejVSeff}
  \end{minipage}
\end{figure}

The performance of the tagging algorithms is estimated in large samples of simulated $t\bar{t}$ events.
Figure~\ref{fig:perf:lrejVSeff} shows the light-flavour-jet rejection as a function of $b$-jet tagging efficiency.
As expected, a clear hierarchy between the standalone and combined algorithms is observed. 
In particular, the use of a combined tagging algorithm can improve the rejection by a factor 4 to 10 compared to JetProb in the 60--80\% efficiency range.

For physics analyses it is important to understand the light-flavour-jet rejection as a function of kinematic variables.
Figures~\ref{fig:perf:lrejVSpt} and~\ref{fig:perf:lrejVSeta} show the dependence on jet \pt{} and $\eta$, respectively.
The rejection is best at intermediate $\pt^{\text{jet}}$ values and in the central region.
At low~\pT\ and/or high~\abseta, the performance is degraded mostly
because of the increase of multiple scattering and secondary
interactions. For \pT{} greater than about $200\GeV$, some dilution arises because the fraction of
fragmentation tracks increases, and more $b$ hadrons fly beyond the
first pixel layer. In addition, a further performance degradation results
from pattern recognition issues in the core of very dense jets.

As mentioned in the previous section, algorithms such as IP3D+JetFitter can be tuned to achieve a better charm rejection.
For high-performance $b$-tagging algorithms, the ability to reject $c$ jets also becomes important.
Charm hadrons have sufficiently long lifetimes to also lead to reconstructible secondary vertices.
Since JetFitter relies not only on the long lifetimes of $b$ and $c$ hadrons but also on the full decay topology, 
it can help to discriminate $b$ jets and $c$ jets, 
for instance by separating $b$ jets with cascade charm decays ({\em i.e.} at least 2 vertices) from single-vertex $c$ jets.
The neural network used for the IP3D+JetFitter combination has three output neurons: one for each of the light-quark, $b$ and $c$ hypotheses. The usual IP3D+JetFitter algorithm is built using the LLR of the light-flavour-jet and $b$-jet outputs.
Figure~\ref{fig:perf:crejVSeff} shows the $c$-jet rejection versus the $b$-jet tagging efficiency.
On the other hand, the figure also shows that merely adding the SV1 and IP3D discriminants does not help to improve the performance with respect to IP3D+JetFitter.

Since hadronic decays of $\tau$ leptons can be reconstructed as jets which can mimic $b$ jets, it is useful to know the discrimination power between $\tau$ jets and $b$ jets. This is shown in Fig.~\ref{fig:perf:trejVSeff} for two tagging algorithms.

\section{Muon-based tagging algorithm}
\label{sec:smt_tagger}

Decays of $b$ hadrons to muons, either direct, $b \rightarrow \mu^-$, or through the 
cascade, $b \rightarrow c \rightarrow \mu^+$ 
(or, with significantly smaller rate, $b \rightarrow \bar{c} \rightarrow \mu^-$), 
can be exploited to identify $b$ jets.\footnote{Charge-conjugate decay modes are
  implied throughout this paper.}
The intrinsic efficiency of muon-based tagging algorithms is typically lower
than that of lifetime-based tagging algorithms due to the limited 
branching fraction of $b$ hadrons to muons ($\approx 20\%$, including both direct and cascade decays). 
The {\it Soft Muon Tagger} (SMT), which is described in this section, is a muon-based tagging algorithm 
that does not use any lifetime information. This makes it complementary to the lifetime-based techniques
and subject to significantly different sources of systematic uncertainties.

\subsection{Muon selection} 
\label{sec:smt_selection}

The muons considered for tagging in the SMT algorithm are required to be reconstructed both in the ID and the MS, so-called combined muons~\cite{PERF-2007-01}. Such muons must satisfy track quality requirements on the number of hits in the different ID sub-detectors, aimed at reducing the number of light-flavour hadron decays in flight. 
Candidate muons also have to be loosely compatible with the reconstructed primary vertex, in order to reject charged particles 
from additional proton collisions, especially at high LHC instantaneous luminosities, or
from nuclear interactions of the hard collision products with the detector material.
A candidate muon is associated with a jet if $\Delta R(\textrm{jet},\mu) < 0.5$.
If more than one jet fulfils this requirement, the muon is associated with the nearest jet only. 
The candidate muon must further fulfil a set of selection criteria, referred to as SMT selection criteria
in the following: $|d_{0}|<$ 3 mm, $|z_{0}\cdot \sin\theta|<$ 3 mm and $\pt >4\GeV$.

Light charged mesons ($\pi^{\pm}$, $K^{\pm}$) decay predominantly into muons
and thus contribute significantly 
to a sample of jets with associated muons.
Given the long lifetimes of light charged mesons, 
a small fraction of those mesons decay between the end of the ID volume and the
entrance of the muon system.
While in those cases the ID measures the track parameters for the meson, 
the MS is sensitive to the track of the muon produced in the decay, 
giving rise to an enlarged $\chi^2$ for the combination of both measurements.
In order to discriminate between $b$ and light-flavour jets the SMT therefore uses the $\chi^2$ of the statistical combination of the track parameters of muons reconstructed in the ID and MS, $\chi^2_{\rm match}$, normalised to the number of degrees of freedom. 
The momentum imbalance and kink from the decay between the light charged meson and daughter muon will result in $\chi^2_{\rm match}$ values larger on average than for decays of heavy-flavoured hadrons.
The $\chi^2_{\rm match}$ is defined as

\begin{equation}
\chi^2_{\rm match} = \frac{1}{5}(\vec{P}_{\rm ID}-\vec{P}_{\rm MS})^{\mathrm{T}} (W_{\rm ID}+W_{\rm MS})^{-1} (\vec{P}_{\rm ID}-\vec{P}_{\rm MS}),
\label{eqn:chi2m}
\end{equation}
where $\vec{P}_{\rm ID}$ and $\vec{P}_{\rm MS}$ are the 5-dimensional vectors of the trajectory helix parameters measured in the ID and MS, respectively, 
and $W_{\rm ID}$ and $W_{\rm MS}$ are their associated $5\times5$ covariance matrices.
  
\begin{figure}[!htbp]
  \centering \includegraphics[width=0.6\textwidth]{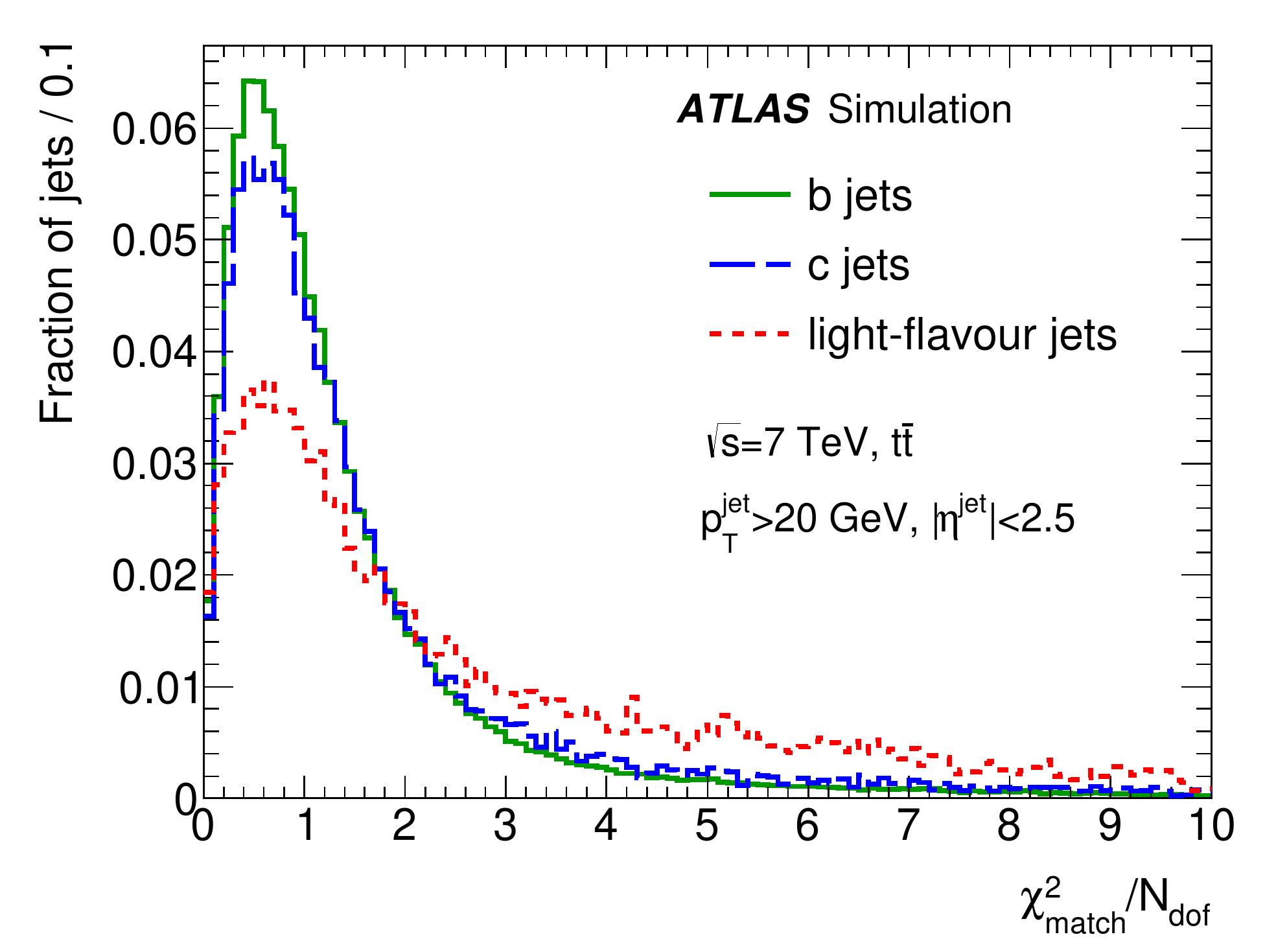}
  \caption{Distribution of the $\chi^2_{\rm match}$ variable for $b$ (solid green),
    $c$ (long-dashed blue) and light-flavour (dashed red) jets.}
  \label{fig:smt_chi2}
\end{figure}

The $\chi^2_{\rm match}$ distribution for the different flavour sources in simulated \ttbar{} events
is shown in Fig.~\ref{fig:smt_chi2}.
Compared to $b$ or $c$ jets, light-flavour jets indeed show a significantly broader $\chi^2_{\rm match}$ distribution.
A jet is considered tagged by the SMT if it has an associated candidate muon passing the
SMT selection criteria, which also include the requirement $\chi^2_{\rm match}<$ 3.2.

\subsection{Performance in simulation} 
\label{sec:smt_mc}

Various aspects of the performance of the SMT algorithm have been studied in simulated events of different physics processes.

An inclusive sample of di-muon events from $J/\psi$ meson and $Z$ boson
decays has been used to provide a clean source of genuine muons spanning
a wide transverse momentum range. This allows studies of the efficiency
of the SMT selection criteria for isolated muons, including the $\chi^2_{\text{match}}$ cut.
This efficiency, which is found to be on average around 95\%, has been studied as a function 
of the muon transverse momentum and pseudorapidity. It is found not to depend significantly on the
transverse momentum, and exhibit only a mild dependence on the pseudorapidity.

\begin{figure}[hbtp]
  \begin{minipage}{0.49\textwidth}
    \includegraphics[width=\textwidth]{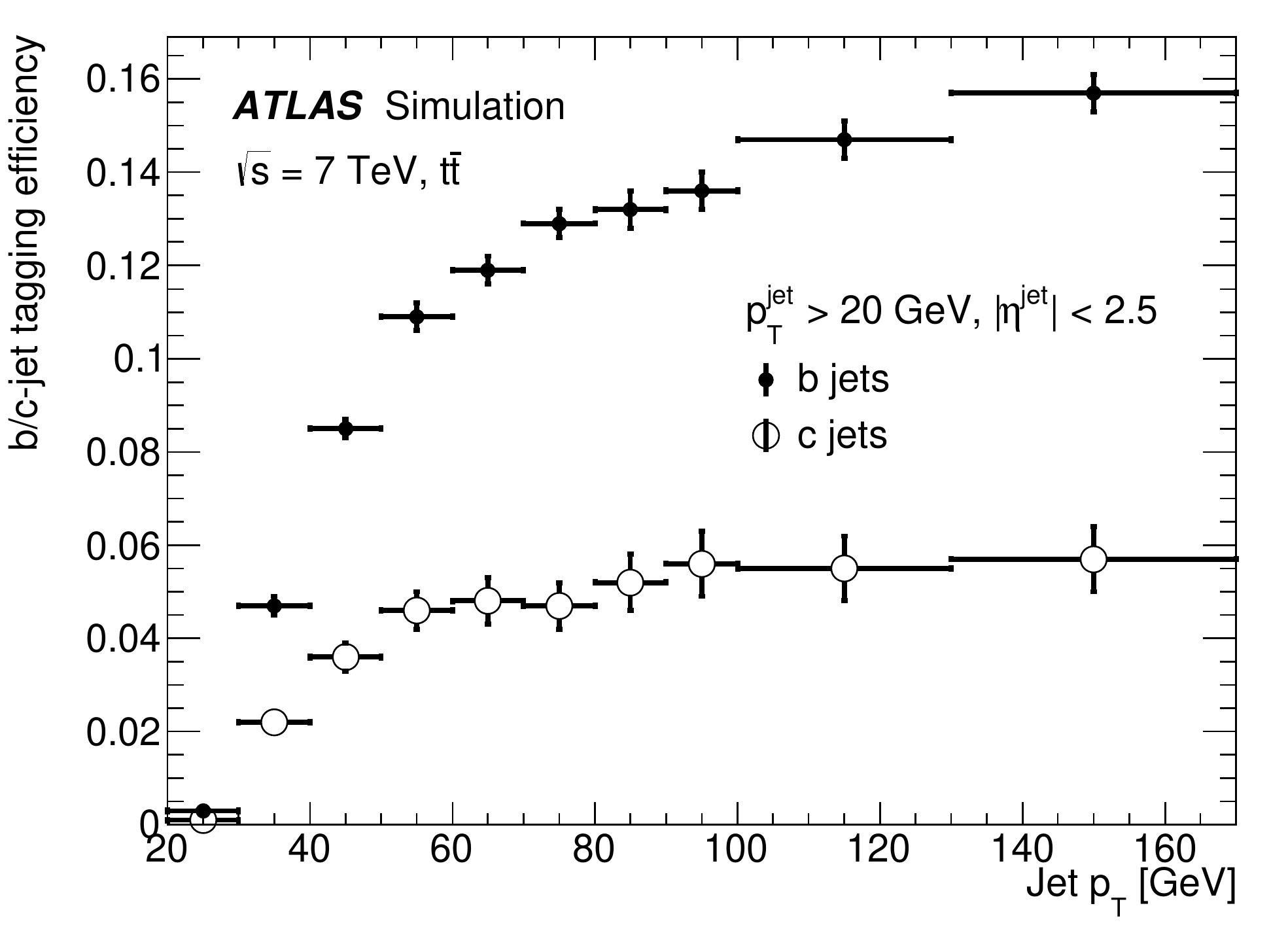}
    \caption{$b$- and $c$-jet tagging efficiencies of the SMT algorithm, and associated statistical uncertainties.}
    \label{fig:smt_mc_beff}
  \end{minipage}\hfill
  \begin{minipage}{0.49\textwidth}
    \includegraphics[width=\textwidth]{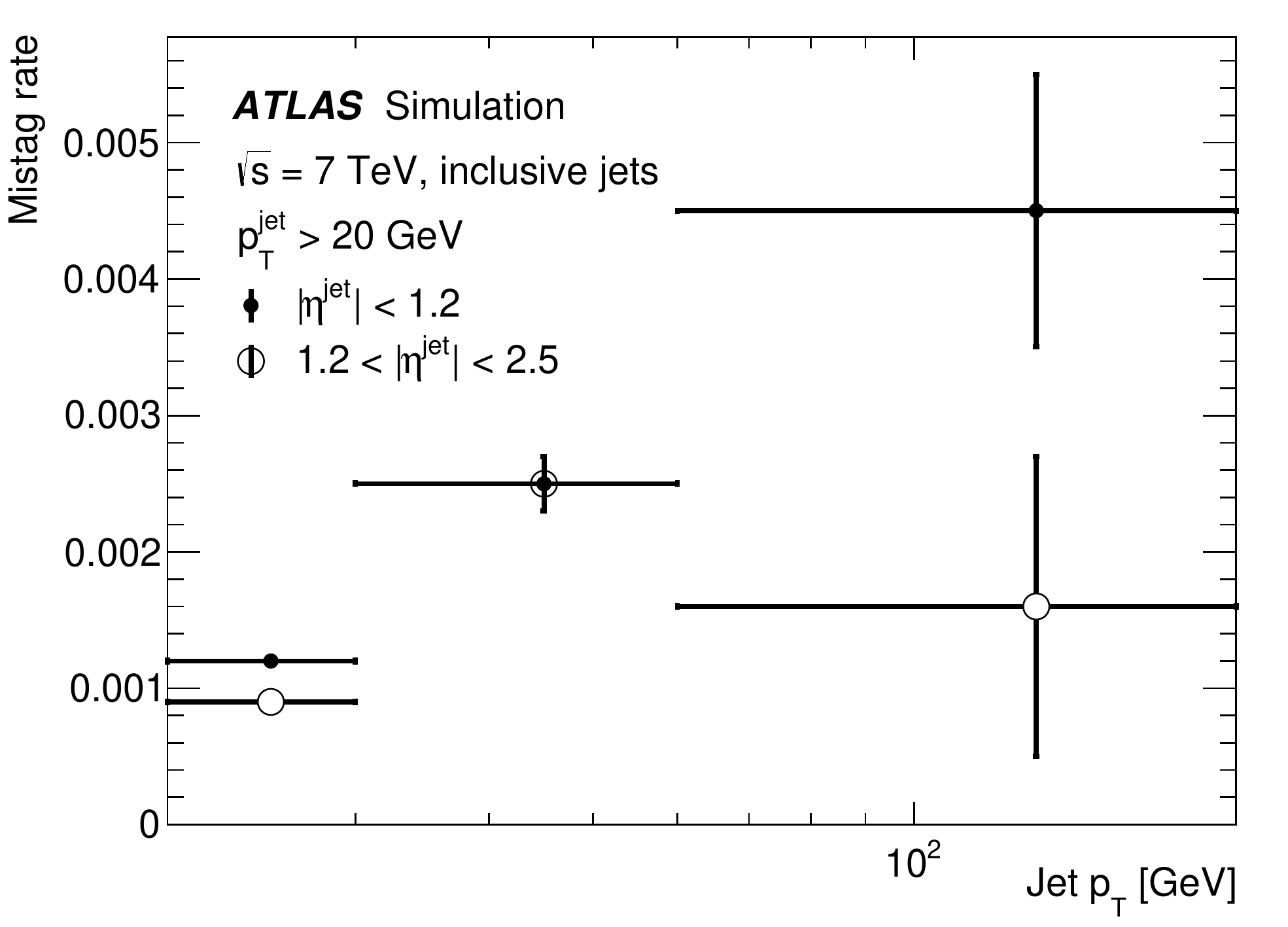}
    \caption{Mistag rate of the SMT algorithm, and associated statistical uncertainties, for light-flavour jets.}
    \label{fig:smt_mc_mistag}
  \end{minipage}
\end{figure}

The efficiency of the SMT algorithm to identify $b$ and $c$ jets has been evaluated using a sample 
of simulated $t\bar{t}$ events. The average $b$- and $c$-jet tagging efficiencies in this sample are found to be
11.1\% and 4.4\%, respectively. The efficiencies as a function of jet \pt{} are given in Fig.~\ref{fig:smt_mc_beff}.
As expected, the tagging efficiencies are significantly lower than what is typically found
for lifetime-based tagging algorithms, due to the limited branching ratio of
muonic $b$- and $c$-hadron decays. A dependence on the jet \pt{} is observed,
whereby a lower efficiency is found for lower \pt{}: softer jets originate from
decays of $b$ hadrons with lower transverse momentum, which in turn produce less
energetic tagging muons.
The latter are more likely to fail the SMT pre-selection requirement on the muon \pt{} (\pt{}$>4$ GeV). 
The efficiency becomes almost flat
when jets attain a \pt{} range where they produce high transverse momentum muons.

The mistag rate, i.e. the efficiency to falsely identify a light-flavour jet as a $b$ jet, has been estimated
using a sample of simulated inclusive jet events, generated with {\sc PYTHIA}.
As mentioned before, mistagging of light-flavour jets as $b$ jets is mainly caused by decays in flight of charged pions and kaons,
$\pi^+,K^+\to\mu^+\nu_{\mu}$.
Another source is instrumental effects like punch-through of hadrons through the 
calorimeters and nuclear interactions of particles within a jet with the 
material in the calorimeters, mimicking muons in the MS. 
The values of the mistag rate, determined as a function of jet \pt{} and $|\eta|$, are summarised in Fig.~\ref{fig:smt_mc_mistag}. 
They are found to be very low, demonstrating the power of the SMT tagging algorithm.

\section{$b$-jet trigger algorithm}
\label{sec:trigger}

The possibility to identify $b$ jets at trigger level is crucial for physics processes with purely hadronic final states containing $b$ jets
because the absence of leptons and the huge inclusive jet background make other trigger selections very challenging.

\subsection{Trigger selection}
\label{sec:trigger-selection}

The $b$-jet trigger selection  starts from the calorimetric jet candidates, reconstructed by
the hardware-based first level trigger (LVL1); the corresponding charged-particle tracks, reconstructed 
by the two subsequent software-based trigger levels, the second level trigger (LVL2) and the Event Filter (EF), are then analysed with 
lifetime-based algorithms.
For a detailed description of the ATLAS trigger scheme, including the detailed descriptions of tracking,
vertexing and beamspot determination in the trigger, see Ref.~\cite{PERF-2011-02}.

During the 2011 data taking, the $b$-tagging trigger selection was based on the impact parameter significance of the reconstructed tracks.
The tagging algorithm adopted for the primary physics trigger was an online version of the JetProb algorithm described in Section~\ref{sec:ip_algo},
applied to jet candidates identified by the LVL1 trigger.
To maximise the acceptance for different physics channels, various $b$-jet trigger selections were deployed during 2011 data taking,
differing in LVL1 jet requirements as well as in $b$-tagging requirements.
The trigger selections required either a single or multiple $b$-tagged jets, and the $b$ jets were selected at three working points.
These working points, referred to as \emph{tight}, \emph{medium} and \emph{loose}, correspond respectively to
approximately 40\%, 55\% and 70\% identification efficiency for selecting jets corresponding to true offline $b$ jets,
measured on a $t\bar{t}$ simulated sample. 
The $b$-tagging triggers also exploited a refined jet reconstruction at LVL2 and EF,
which offers a better correlation between online and offline jet \pt, to
reduce further the rate without compromising the jet trigger efficiency plateau of the LVL1 selection. 
The rate reduction provided solely by the request of one {\emph{tight}} (two {\emph{medium}})
$b$-tagged jet(s) is a factor of 6 (13) at LVL2 and  2 (4) at EF.

\begin{figure}[h!]
  \subfloat[]{\label{fig:BjetTriggerBasicPlots_a}\includegraphics[width=0.32\textwidth]{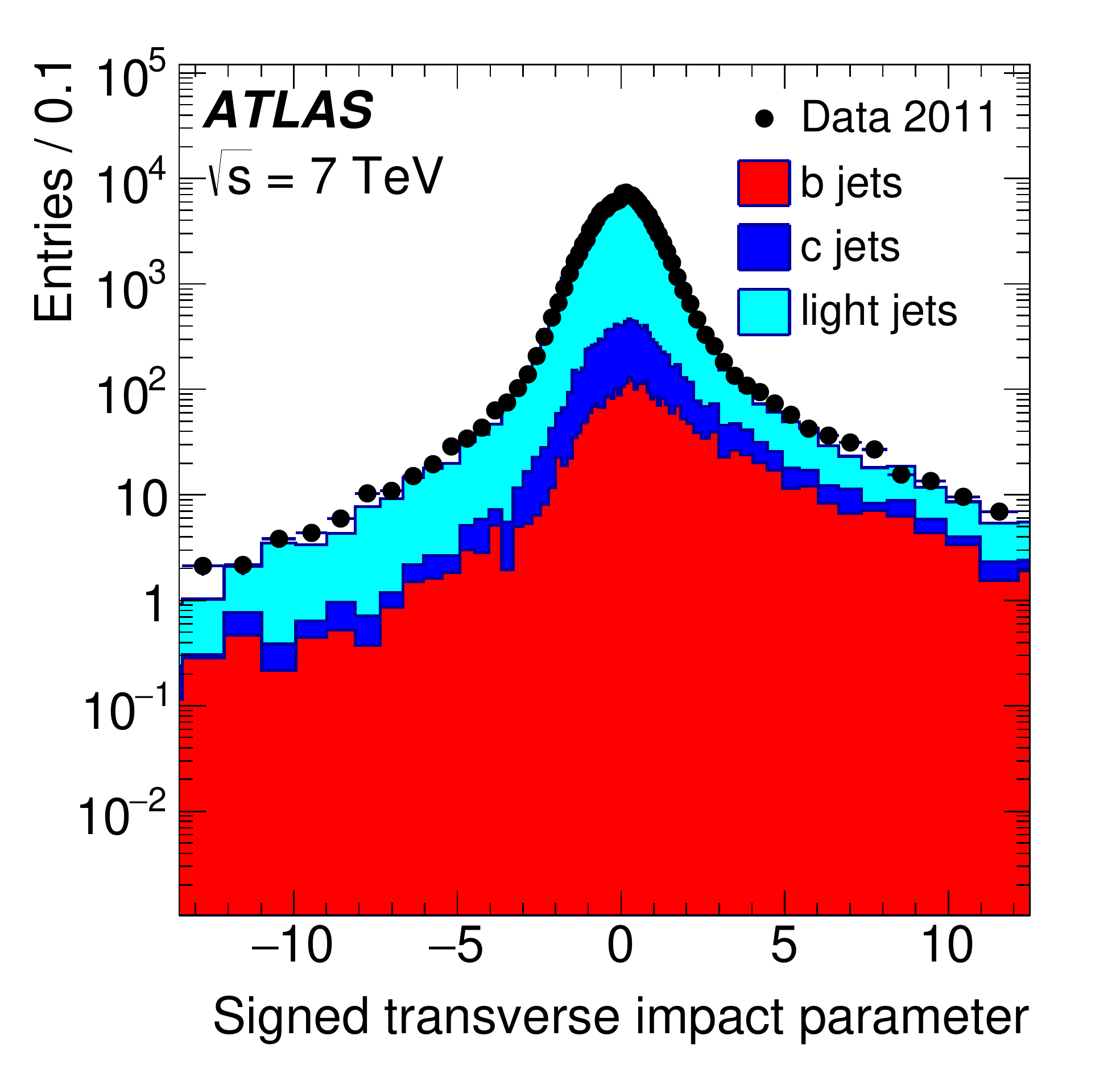}}
  \subfloat[]{\label{fig:BjetTriggerBasicPlots_b}\includegraphics[width=0.32\textwidth]{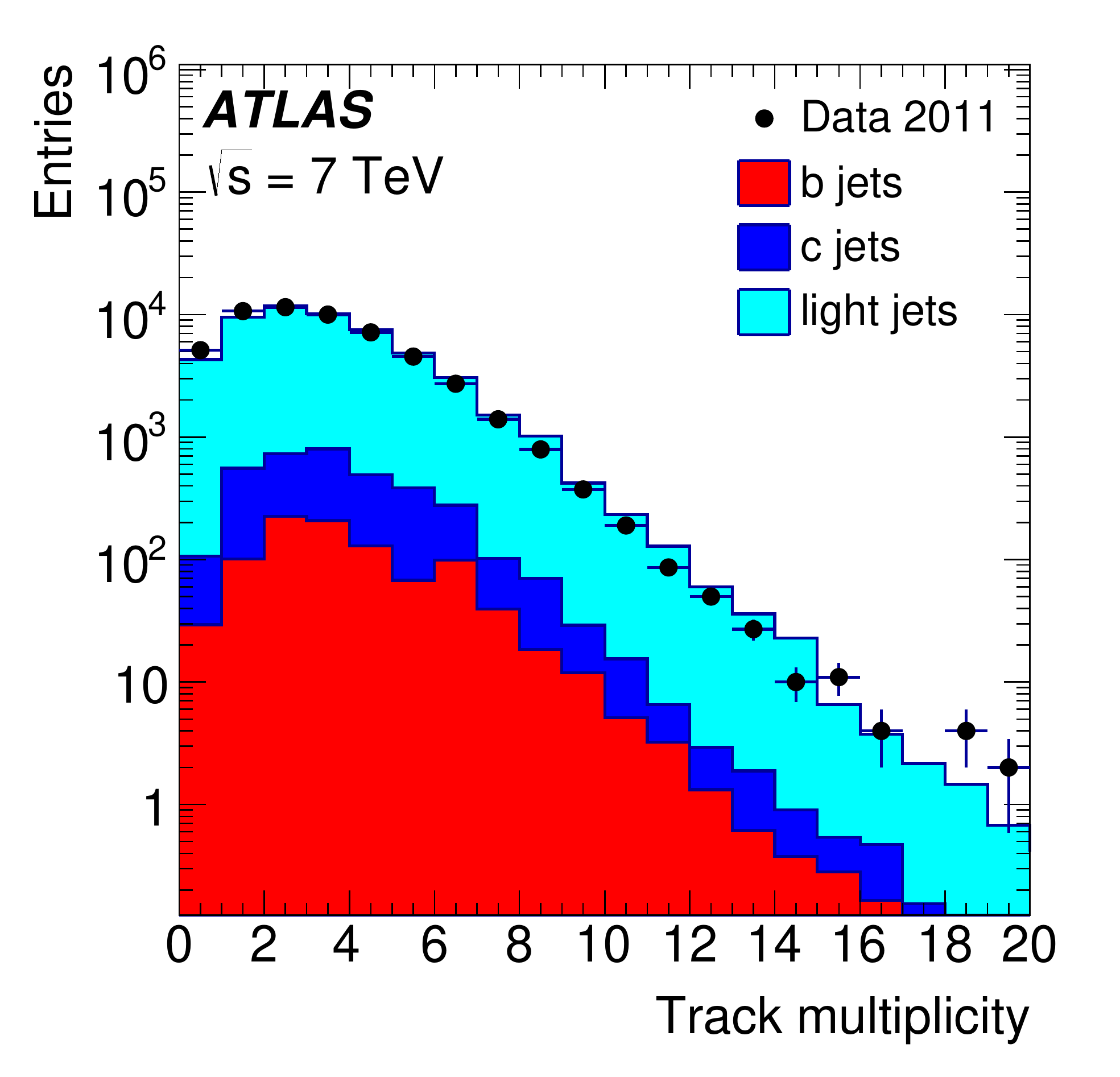}}
  \subfloat[]{\label{fig:BjetTriggerBasicPlots_c}\includegraphics[width=0.32\textwidth]{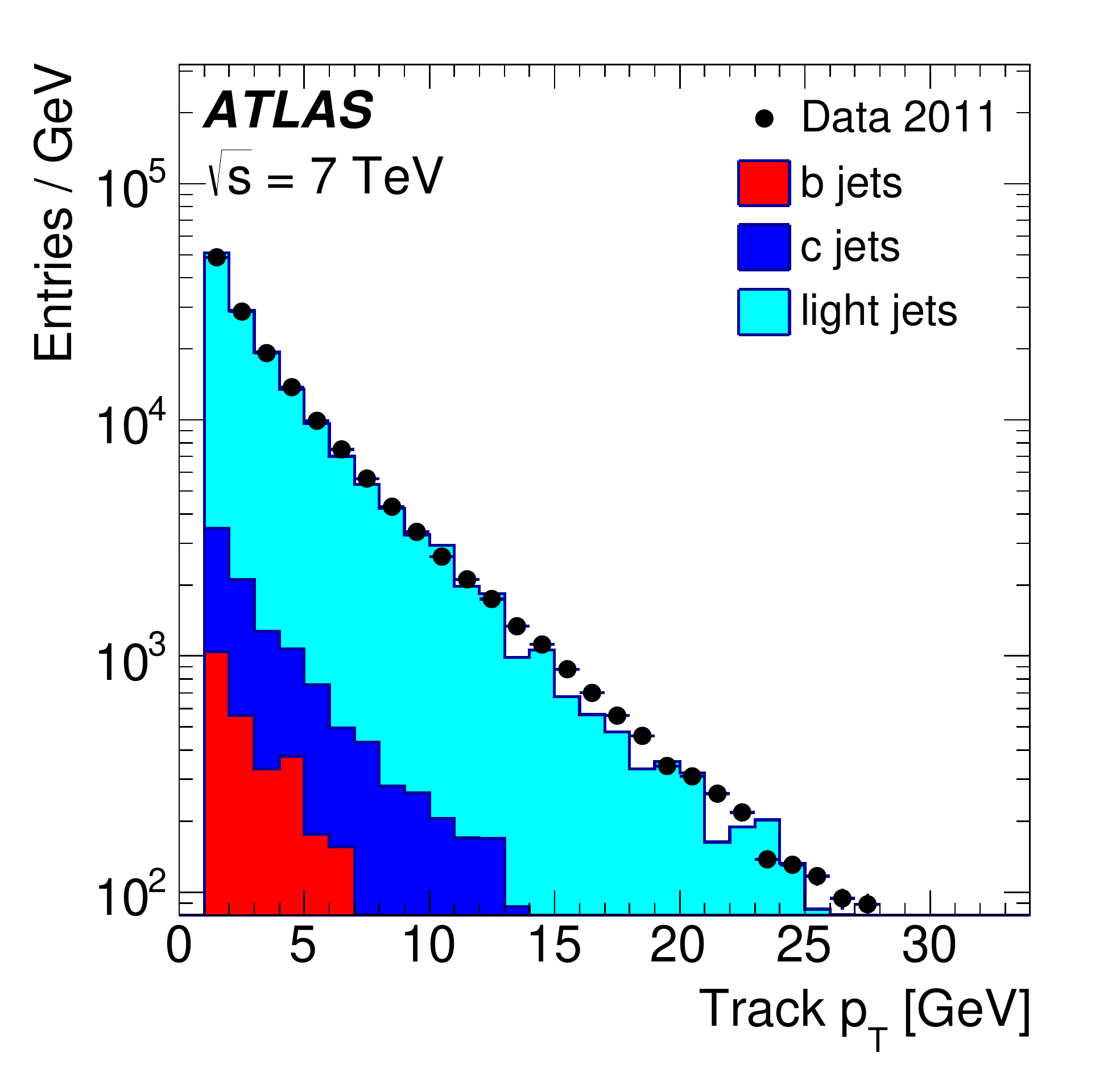}}
  \caption{Signed transverse impact parameter significance (a), multiplicity (b) and transverse momentum (c) of EF tracks that are
    reconstructed starting from a low-\pt{} jet identified by the LVL1 and are requested to satisfy online $b$-tagging criteria.
    The $b$, $c$ and light-flavour distributions are derived from a simulated 
dijet sample.}
  \label{fig:BjetTriggerBasicPlots}
\end{figure}

The data collected in 2011 are compared to a PYTHIA generated dijet sample,
and distributions of basic ingredients for the
$b$-jet triggers are shown in Fig.~\ref{fig:BjetTriggerBasicPlots}.
The overall agreement is good but to take into account deviations in the simulation,
especially in the impact parameter tails, data-driven techniques will be employed to derive data-to-simulation scale factors,
as described in Sections~\ref{sec:beff_mubased}, \ref{sec:ceff_dstarbased} and~\ref{sec:mistag}.

\subsection{Performance in simulation}

The different tagging methods are characterised, at each trigger level,
as a curve showing the light-flavour-jet rejection ($R_l$) versus the efficiency to select $b$ jets ($\epsilon_b$). 
The characterisation of trigger selections also involves studying
the bias that each trigger level imposes on the next one and on the final recorded sample. 
In particular, for the $b$-jet triggers, this can be derived as an additional rejection versus
efficiency curve for offline tagging algorithms, measured on a sample selected by a single $b$-jet trigger.

\begin{figure}[h!]
  \subfloat[]{\label{fig:BjetTriggerOfflineCorrelation_a}\includegraphics[width=0.49\textwidth]{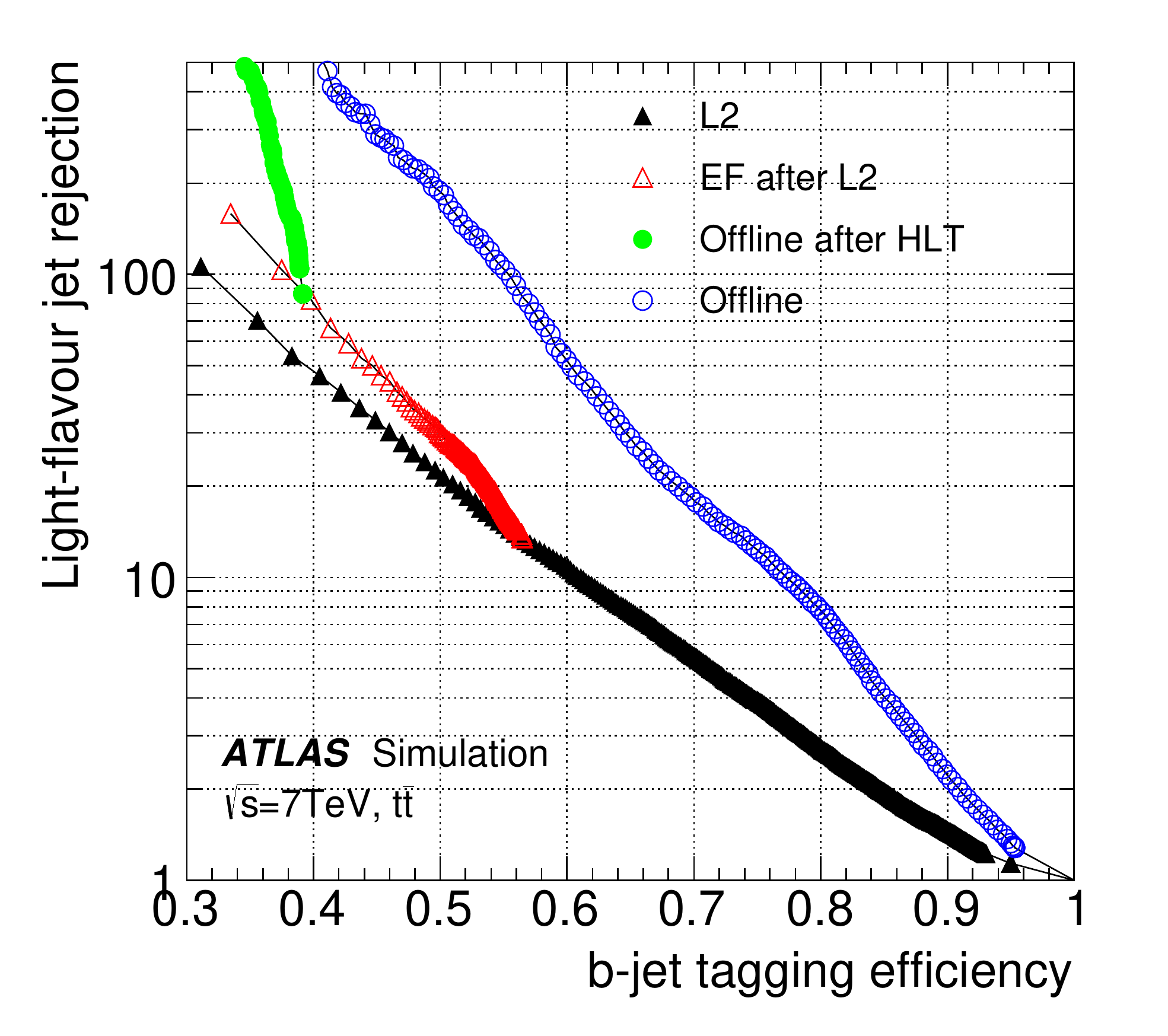}}
  \subfloat[]{\label{fig:BjetTriggerOfflineCorrelation_b}\includegraphics[width=0.49\textwidth]{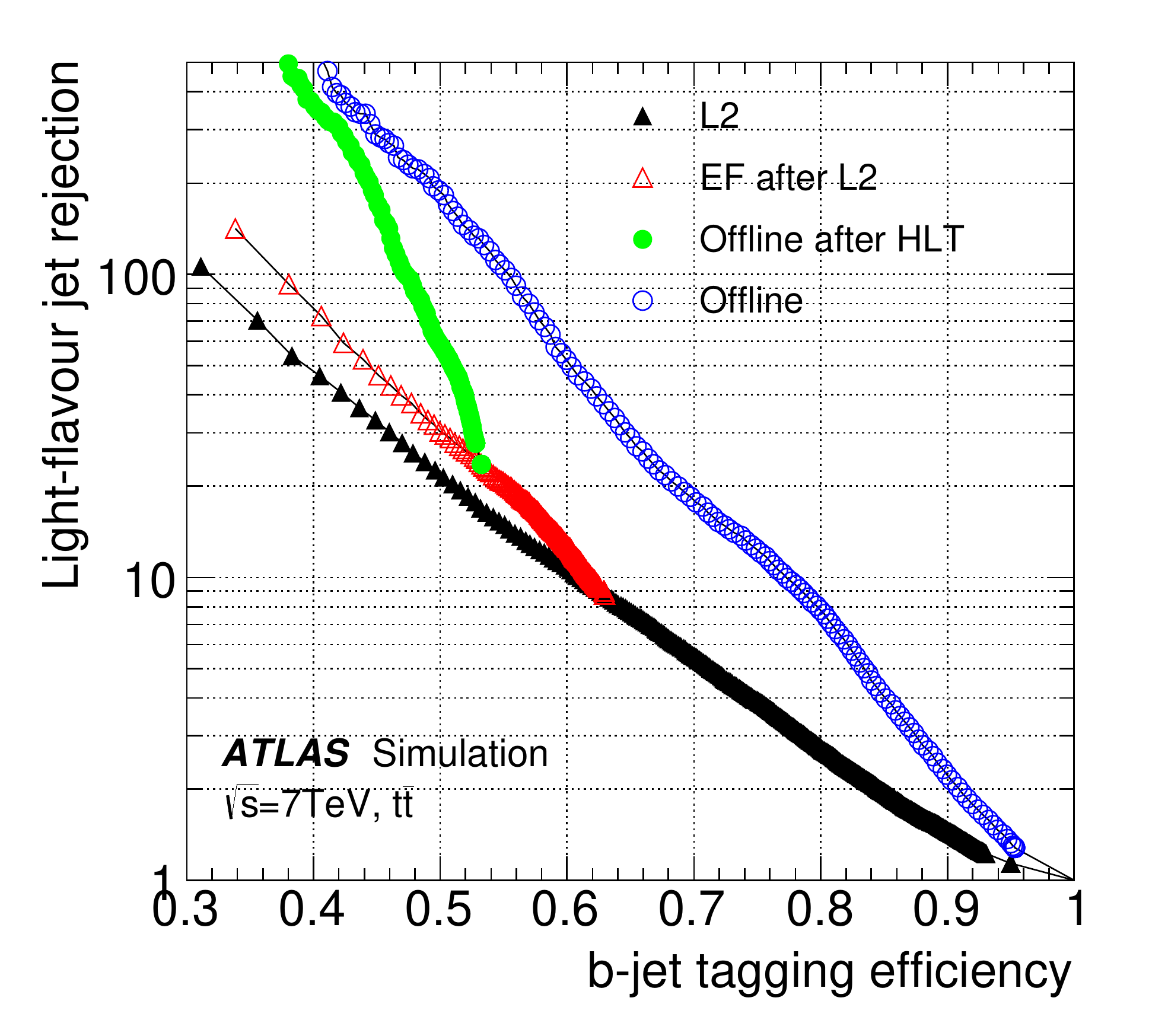}}
  \caption{The combined rejection versus efficiency curves for the LVL2, EF and offline JetProb tagging algorithms
    for the tight (a) and medium (b) trigger working points.
    The offline rejection versus efficiency curve, measured on an unbiased sample,
    is superimposed, providing an estimate of the correlation between online and offline selections.
    The offline jets are required to have $|\eta|<2.5$ and transverse momentum $\pt > 50\GeV$ (corresponding to
    the full acceptance of the jet trigger).}
  \label{fig:BjetTriggerOfflineCorrelation}
\end{figure}

The combined rejection versus efficiency curves for the LVL2, EF and offline selections based on JetProb and 
measured in a sample of {\sc HERWIG} generated $t\bar{t}$ events 
are shown in Fig.~\ref{fig:BjetTriggerOfflineCorrelation}; 
the EF (offline) performance is shown starting from the \emph{tight} and \emph{medium} L2 (EF)  working points.

When compared with the same curve measured on an unbiased sample, the curve
describing the offline rejection on jets selected by a single $b$-jet trigger also provides an estimate of
the correlation between the tagging algorithms used in the different selection stages.
In each plot an offline curve, which is obtained on an unbiased sample, is drawn to provide an estimate of this correlation.
For instance, Fig.~\ref{fig:BjetTriggerOfflineCorrelation_b} shows that a sample of jets selected by the offline $b$-tagging is not biased by the
$b$-jet trigger ``medium'' selection if the offline selection operates at an average efficiency of about 40\%.
However the use of the $b$-jet trigger is not limited to this unbiased offline sample since data-to-simulation
efficiency scale factors are derived for trigger selection and for combined trigger and offline selections.

\section{Dependence of the $b$-tagging performance on pile-up}
\label{sec:pu}

With the increasing instantaneous luminosity of the LHC during 2011 data taking,
the rate of pile-up interactions increased substantially with an average of 12
interactions per bunch crossing in the later data taking periods,
reaching maximum values of more than 20 interactions.
These additional interactions can potentially affect the $b$-tagging performance through several effects:
\begin{itemize}
\item The hard scatter primary vertex has to be identified among the reconstructed primary interaction vertices along
  the beam line (see Section \ref{sec:samples}).
  Identifying the wrong primary vertex as the signal vertex typically results 
  in rejecting tracks for the signal jets when applying the quality criteria for
  $b$-tagging tracks, consequently losing the power to tag these jets as $b$ jets. 
  This effect is less pronounced in final states containing jets and/or charged
  leptons with large transverse momenta, such as $t\bar{t}$ events.
  However, it can play an important role in topologies with lower transverse
  momenta of the final state objects or if some high transverse momentum objects
  are not reconstructed.
  Pile-up effects on vertex reconstruction can also lead to a worsening of
  the $z$-resolution of the primary vertex due to contamination from tracks 
  from nearby interactions. This will translate into a worsening of the
  longitudinal impact parameter resolution which constitutes an important input
  to $b$-tagging algorithms. Furthermore, the fraction of tracks in the tails of the
  longitudinal impact parameter distribution is increased, which also degrades the $b$-tagging performance.
  Studies in Ref.~\cite{IBLTDR} have shown that the fraction of $t\bar{t}$ events with a misidentified primary vertex
  is below 2\% for the number of additional interactions as present during data taking in 2011. The resolution of 
  the $z$ coordinate of the signal vertex degrades by about 10\% for an average of 12 additional interactions as 
  in the later data taking periods in 2011.
  As explained in Section~\ref{sec:samples}, a requirement on the jet vertex fraction has been applied to jets
  selecting only jets for the $b$-tagging analyses that are compatible with the selected primary event vertex.
  As a result, jets from the hard scatter interaction that are lost when the wrong primary vertex is
  selected as signal vertex do not enter into the determination of the performance of $b$-tagging algorithms.
  The consequences of this depend strongly on the specific analysis considered and are not discussed in
  detail in this paper. 
\item The increased density of charged particle tracks in the inner tracking detectors makes track reconstruction
  more challenging. 
  An increased rate of falsely associated hits or hits shared with other tracks, as well as an increased rate
  of fake tracks are the most important consequences. Furthermore, misassociated hits can lead to tails in the
  impact parameter resolutions for these tracks.
  These aspects have been studied in Refs.~\cite{IBLTDR,ATLAS-CONF-2012-042}. 
  It has been found that for the pile-up conditions in the 2011 data, 
  there is no significant degradation of the track reconstruction efficiency and the track impact parameter 
  resolution in the transverse plane. However, there is some increase of the rate of fake tracks and a slight
  worsening of the track impact parameter resolution along the $z$ direction.
\item Pile-up interactions can create additional jets reconstructed in the detector. If the
  corresponding interaction vertex is close to the primary vertex of the hard scatter process of interest, 
  charged particle tracks stemming from the pile-up interaction might be falsely associated to the hard
  scatter primary vertex and mimic lifetime signatures leading to an increased misidentification rate
  of non-$b$ jets.
  If the pile-up jet overlaps with a signal jet, tracks from the pile-up interaction might be misassociated
  with the signal jet, diluting the $b$-tagging performance.
  Studies in Ref.~\cite{IBLTDR} have shown that this is the main source of an increased multiplicity of
  tracks in signal jets in the presence of pile-up.   
  If the pile-up vertex is sufficiently displaced from the hard scatter vertex, the 
  corresponding tracks will be rejected by the selection criteria, typically not causing false identification
  of the pile-up jets.
\end{itemize}

\begin{figure}[!htb]
  \begin{center}
    \subfloat[]{\label{fig:MV1PileUp_a}\includegraphics[width=0.49\textwidth]{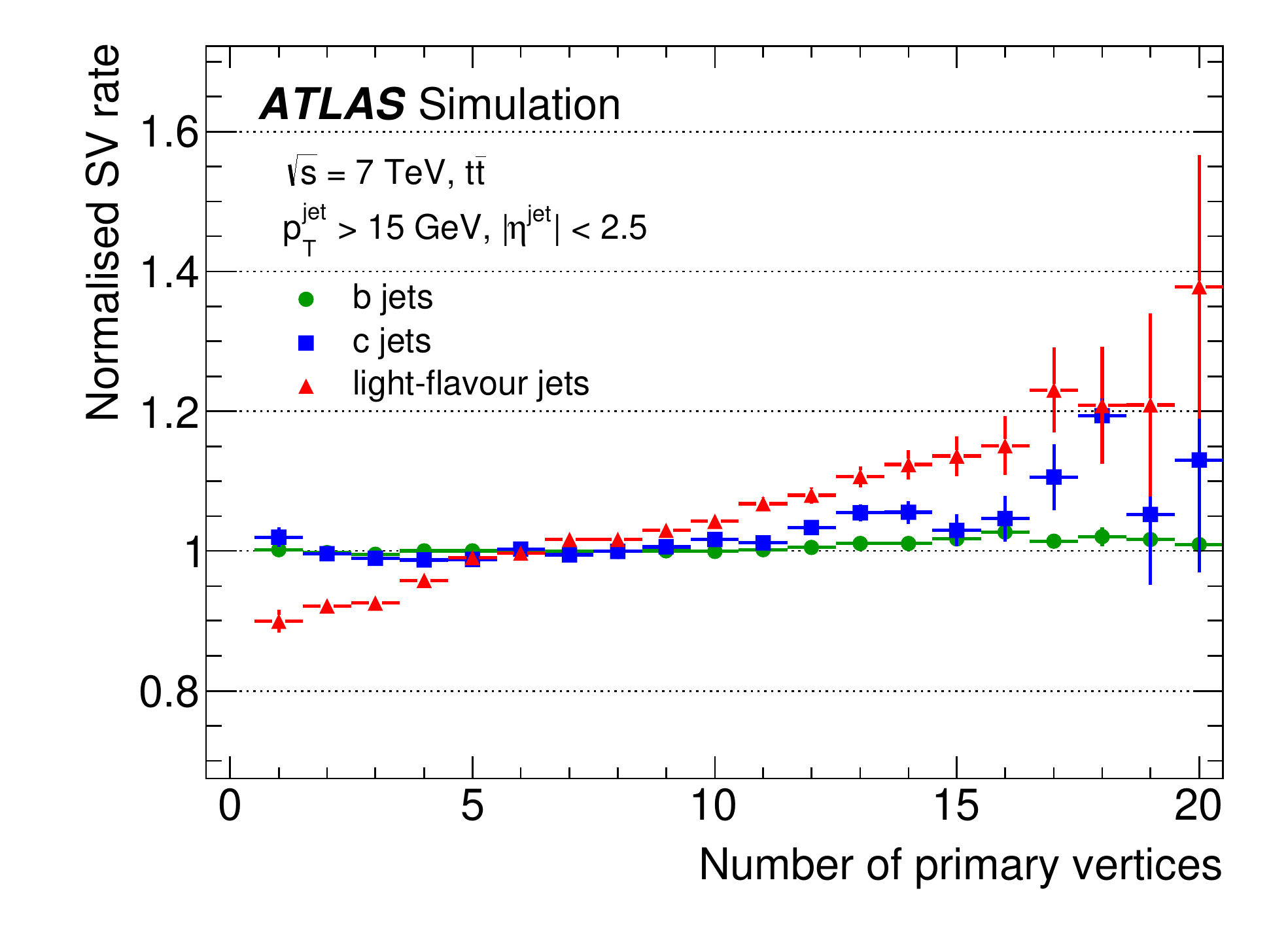}}
    \hfill
    \subfloat[]{\label{fig:MV1PileUp_b}\includegraphics[width=0.49\textwidth]{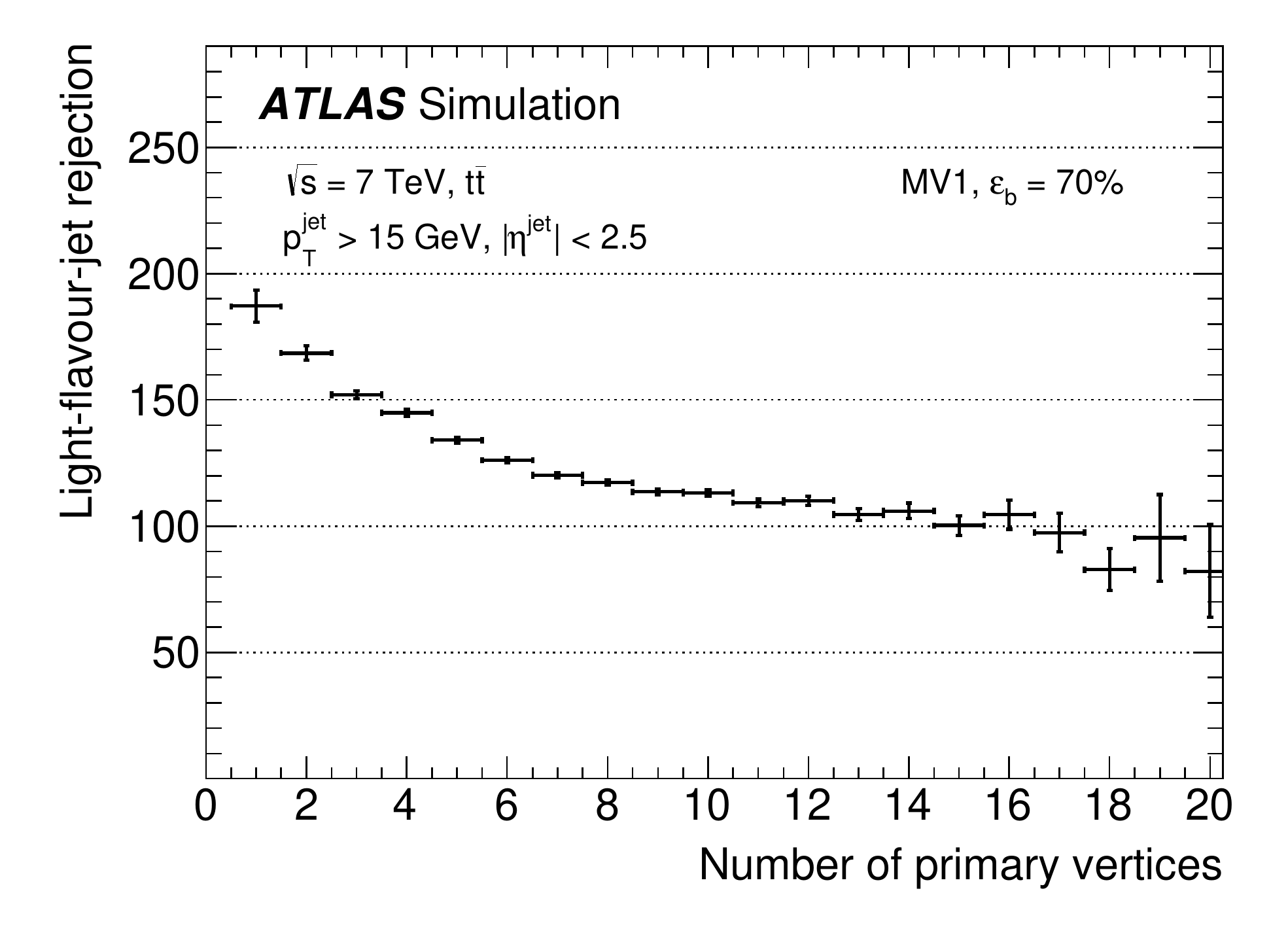}}
    \caption{The dependence of the secondary vertex reconstruction rate of the SV1 algorithm (a) and
      the light-flavour jet rejection of the MV1 algorithm (b) on the number of reconstructed primary vertices,
      estimated from simulated $t\overline{t}$ events.
      The secondary vertex reconstruction rate has been normalised to the rate in the inclusive sample.}
    \label{fig:MV1PileUp}
  \end{center}
\end{figure}

The dependence of the $b$-tagging performance on the number of reconstructed primary vertices has been studied
using simulated \ttbar{} events.
An important input to the $b$-tagging algorithms is the information from the reconstruction of 
inclusive secondary decay vertices in jets. The secondary vertex reconstruction can be affected by additional 
tracks from pile-up vertices.
Figure \ref{fig:MV1PileUp} shows the rate with which secondary vertices are reconstructed by the SV1 algorithm 
in jets of different flavour, normalised to the average secondary vertex reconstruction rate.
It can be seen that for $c$ and $b$ jets, where the reconstructed secondary vertices are mainly 
real vertices from decays of long-lived heavy hadrons, the secondary vertex rate is nearly independent
of the number of pile-up interactions. For light-flavour jets on the other hand an almost linear dependence can be observed,
leading to an increased misidentification rate of light-flavour jets.       
Figure \ref{fig:MV1PileUp} also shows the rejection of light-flavour jets for a $b$-jet tagging efficiency of 70\%
versus the number of reconstructed primary vertices for the MV1 algorithm for $t\overline{t}$ events. 
It can be seen that the light-flavour-jet rejection degrades with increasing
number of pile-up interactions, resulting in a light-flavour-jet rejection rate that is reduced by a factor 
of almost two for the highest level of pile-up as present in the year 2011.

\section{Simulation modelling of $b$-tagging input observables}
\label{sec:btag_inputs}

An acurate modelling of the $b$-tagging performance in the simulation is based on a correct description of the underlying 
quantities, such as the reconstruction efficiency and fake rate of tracks and vertices, and the properties of the reconstructed objects.
In this section, a comparison between data and simulation is presented for a number of $b$-tagging input observables.

\subsection{Measurement of the impact parameter resolution of charged particles}
\label{sec:ipres}

Two key ingredients for discriminating between tracks originating from displaced vertices and those originating from the primary vertex are:
the transverse impact parameter (IP) of a track, $d_{0}$, and $z_{0}\sin\theta$, the longitudinal impact parameter projected onto the direction perpendicular to the track.
Both of these quantities can be measured with respect to the primary vertex in an unbiased way ($d_{\rm PV}$ and $z_{\rm PV}\sin\theta$):
if the track under consideration was used for the primary vertex determination, it is first removed from the primary vertex which is subsequently refitted,
and the impact parameters are computed with respect to this new vertex.

Due to the fact that the primary interaction point has a spread itself (of approximately 25 $\mu$m in the $x$ and $y$ directions),
it is not possible to measure the impact parameter resolution IP$^{\rm track}$ directly.
Hence relating the impact parameter distributions to the purely track-based IP$^{\rm track}$ resolution is not straightforward
since it is convolved with the resolution on the primary vertex position:
$\sigma^{2}_{\rm IP} = \sigma^2_{\rm IP^{\rm track}}+\sigma^{2}_{\rm PV}$,
where $\sigma^{2}_{\rm PV}$ is the projection of the primary vertex uncertainty along the axis of closest approach
of the track to the primary vertex on the transverse or longitudinal plane.  

In this section, a measurement of the impact parameter resolution in data is presented.  
Since the measurement does not require a high-luminosity sample, and to limit pile-up effects, only the first runs of the collected data in 2011 are used.

The data were required to satisfy standard ID data quality requirements.
The simulation samples considered are PYTHIA generated dijet samples.
Events passing a logical OR of inclusive jet triggers, with at least 10 tracks used in the primary vertex reconstruction are retained for this study.  

Tracks fulfilling the following basic track quality selection are used:
\begin{itemize}
\item The track must be included in the primary vertex reconstruction.
\item $\pt>500\MeV$.
\item $|\eta|< 2.5$.
\item $\geqslant 2$ hits in the pixel detector.
\item $\geqslant 7$ hits in the combined pixel and SCT detectors.
\end{itemize}
In order to extract the correct impact parameter resolution from data it is important to understand how to subtract the contribution from the resolution on the position of the reconstructed primary vertex.
Since the primary vertex fit uses the beam spot constraint, the beam spot size is already included in the estimated uncertainty on the primary vertex position.
The tracks are divided into different categories of $\eta$, $\pt\sqrt{\sin\theta}$, and the number of innermost pixel layer hits to ensure an almost constant resolution within a single category.
Both the $d_{0}$ and $z_{0}\sin\theta$ resolution have been measured for each track category.
The pseudorapidity $\eta$ is chosen as it reflects the kinematics of the particle production mechanism while $\theta$ is more suitable for parametrising detector-related effects. Finally, $\pt\sqrt{\sin\theta}$ has been chosen instead of $\pt$ itself because it is directly linked to the multiple scattering contribution to the impact parameter resolution in the case that the material traversed by charged particles follows a cylindrical geometry.
The resolution is modelled as

\begin{equation}
  \sigma_{d_{0}}=\sqrt{a^{2}+\frac{b^{2}}{p^{2}\sin\theta^{3}}} .
\end{equation} 

The method used to subtract the primary vertex reconstruction contribution to the IP resolution, $\sigma_{\rm PV}$, relies on an iterative deconvolution procedure.
For each iteration it is possible to obtain the deconvolved distribution by multiplying the measured impact parameter of each track by a correction factor.
For example, for the transverse impact parameter with respect to the primary vertex:

\begin{equation}
  d_{\rm PV}\rightarrow d_{\rm PV}\sqrt{\frac{(K\sigma_{d_{0}})^{2}}{(K\sigma_{d_{0}})^{2}+\sigma^{2}_{\rm PV}}} ,
\end{equation}
where $K$ is a correction factor that depends on the iteration index. For the first iteration $K$ is equal to one.
For each iteration, $\sigma_{d_{\rm PV}}$ can be evaluated by fitting each $d_{\rm PV}$ distribution and for the $i$-th iteration it should be:

\begin{equation}
  (\sigma_{d_{\rm PV}})_{i}=K_{i}\sigma_{d_{0}}\sqrt{\frac{(K_{i+1}\sigma_{d_{0}})^{2}+\sigma^{2}_{\rm PV}}{(K_{i}\sigma_{d_{0}})^{2}+\sigma^{2}_{\rm PV}}} ,
\end{equation}
which can then be used to calculate $K_{i+1}$.
To evaluate the width of the core of the $d_{\rm PV}$ distributions, and hence estimate the impact parameter resolution,
a Gaussian fit is first applied to the whole distribution, and a temporary mean and width are obtained.
A new fit range, of width four times the temporary fit width, is then centred around the temporary mean;
finally the distribution is refitted within this new range.
The iterative procedure ends when the fitted $\sigma_{d_{\rm PV}}$ is stable within approximately $0.01\%$.
About five iterations are needed to make the $K$ factor converge to stable values that range between $~0.8$ and $~1.2$.
This iterative procedure was verified on Monte Carlo simulation;
the impact parameter resolutions derived from reconstructed tracks in simulated events converges well, especially at high $\pt$,
to the values derived from the tracks reconstructed directly from the simulated hits in the ID.

\begin{figure}[!htp]
  \begin{center}
    \subfloat[]{\label{fig:IP_a}\includegraphics[width=0.49\textwidth]{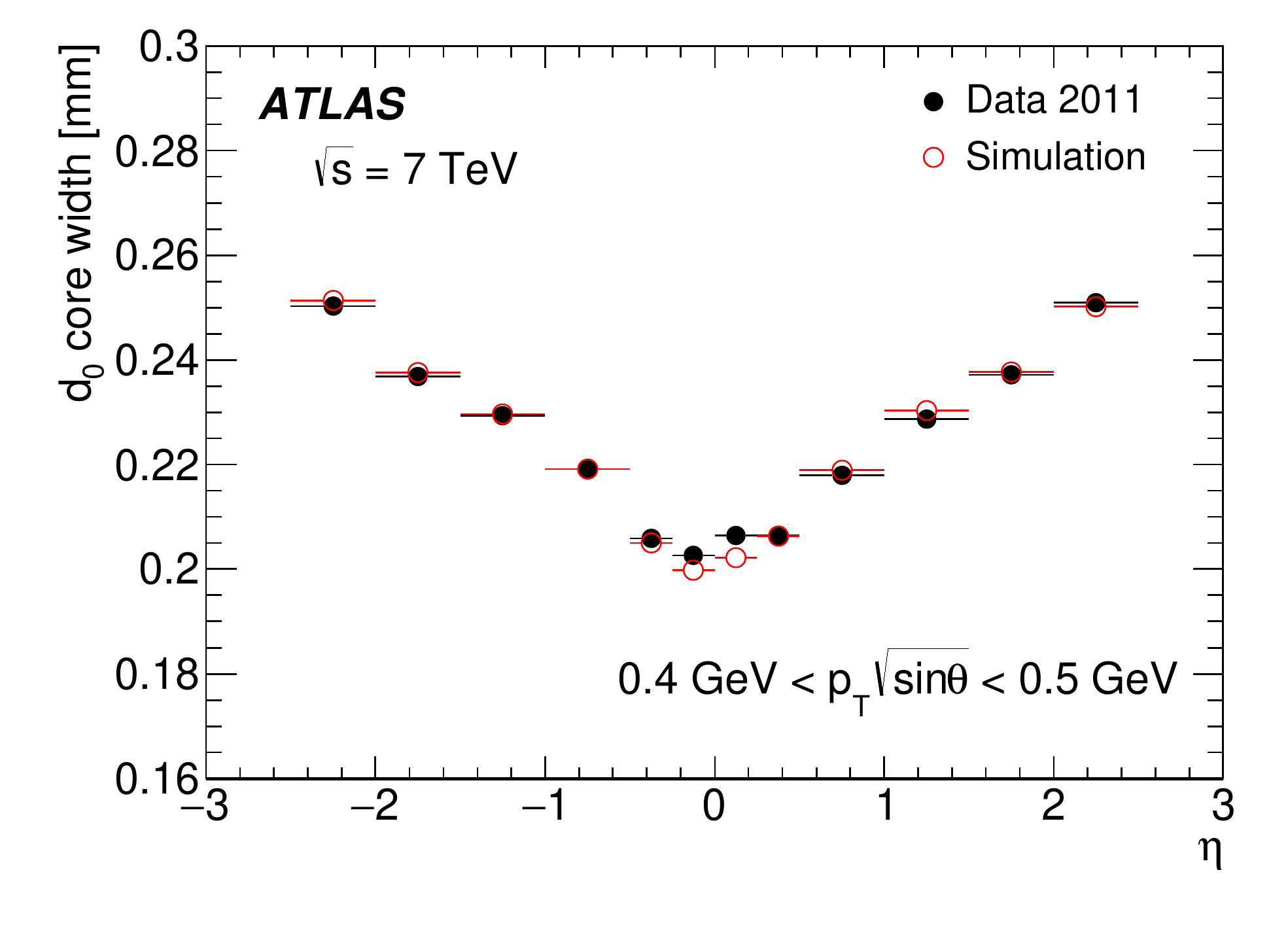}}
    \subfloat[]{\label{fig:IP_b}\includegraphics[width=0.49\textwidth]{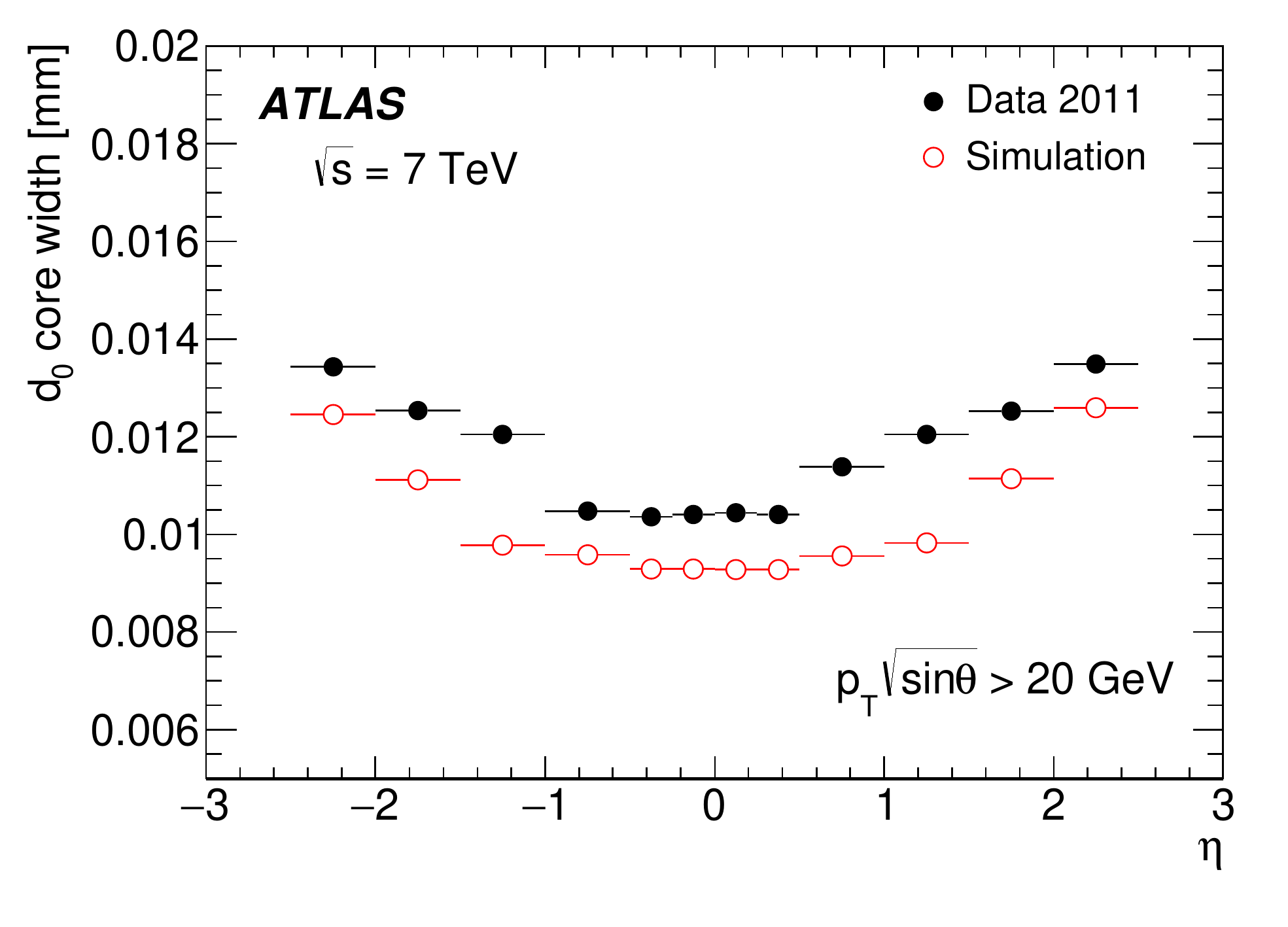}}
    \hfill 
    \subfloat[]{\label{fig:IP_c}\includegraphics[width=0.49\textwidth]{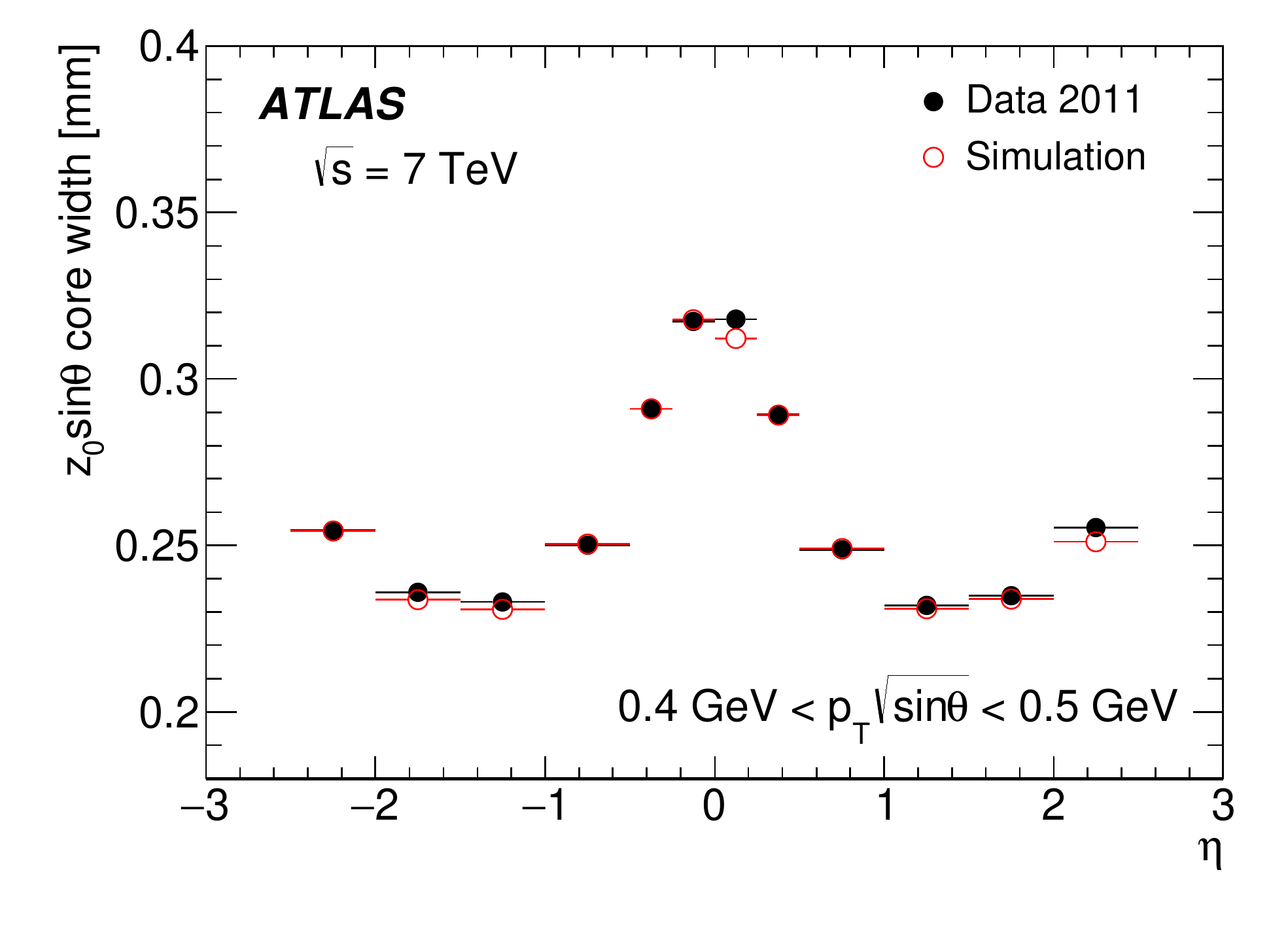}}
    \subfloat[]{\label{fig:IP_d}\includegraphics[width=0.49\textwidth]{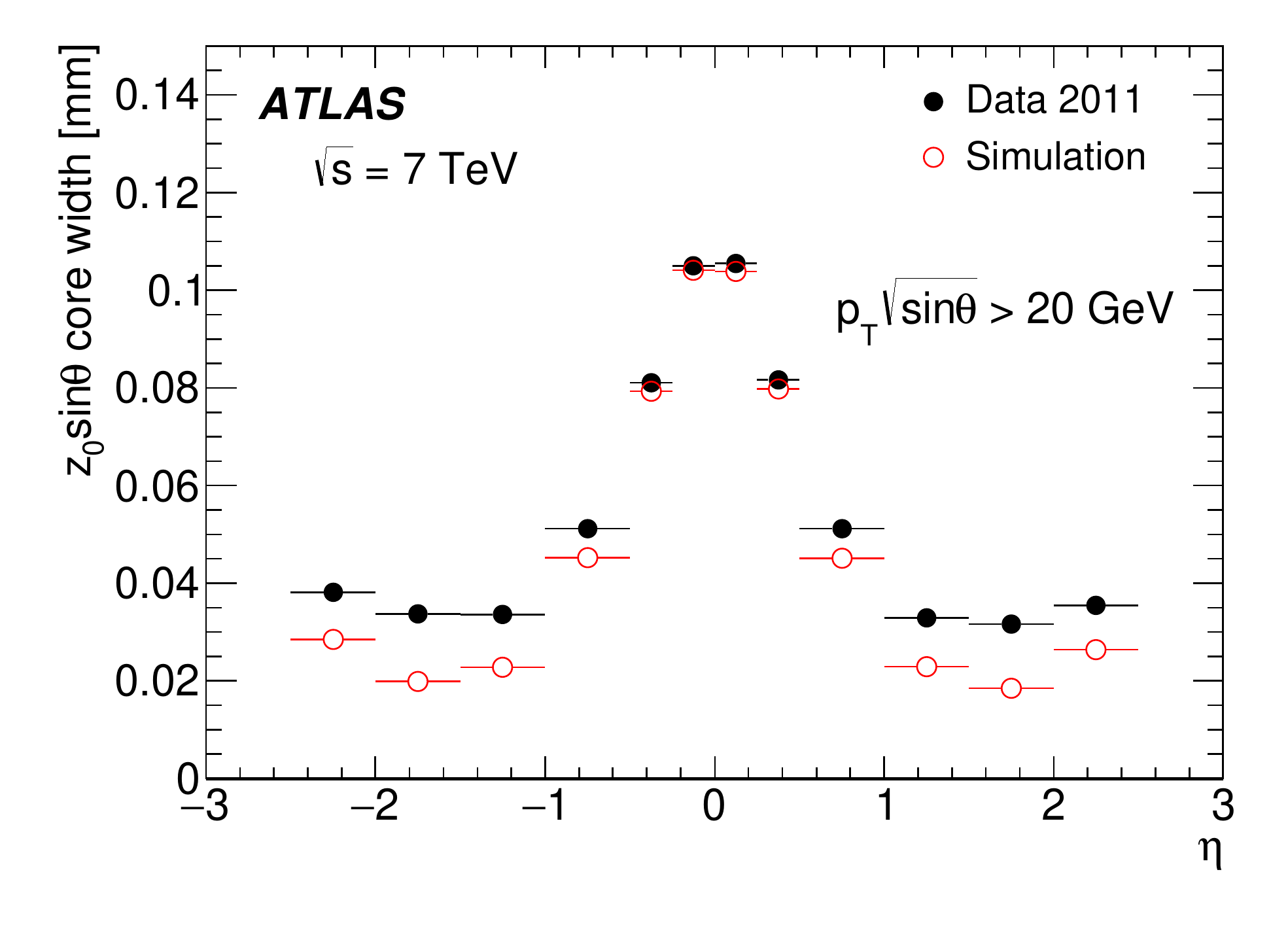}}
    \caption{Comparison between data and simulation of the transverse (top) and longitudinal (bottom) impact parameter resolutions measured with
      respect to the primary vertex as a function of $\eta$, for tracks with one hit in the innermost pixel detector layer, and for
      $0.4\,\GeV < \pt \sqrt{\sin{\theta}} < 0.5\,\GeV$ (left) and $\pt \sqrt{\sin{\theta}} > 20\,\GeV$ (right).}
    \label{fig:IP}
  \end{center}
\end{figure}

Figure \ref{fig:IP} shows the comparison between data and simulation for both the transverse and longitudinal impact parameter resolutions, measured with respect to the primary vertex as a function of $\eta$ for tracks with one hit in the innermost pixel detector layer, for two different $\pt\sqrt{\sin\theta}$ regions 
($0.4\GeV<\pt\sqrt{\sin\theta}<0.5\GeV$ and $\pt\sqrt{\sin\theta}>20\GeV$).
The $\eta$ dependence of the transverse impact parameter resolution is shown in the upper plots for low- and high-$\pt$ tracks. The low-$\pt$ tracks of the first 
region show a rise in resolution versus $|\eta|$ because of the increase in the multiple scattering contribution dominating the resolution in this momentum interval. At high $\pt$, the tracks of the second region, the hit resolution and potential residual misalignments of the silicon detectors are dominating, leading only to a moderate $\eta$ dependence in $d_0$. The lower plots show the resolution of the projected longitudinal impact parameter $z_0 \sin\theta$. Because of this projection and of the variation of the average pixel hit's cluster size with $\eta$, a strong dependence is seen both at low and high $\pt$. In both cases, $d_0$ and $z_0 \sin\theta$, the low-$\pt$ regime is well modelled in simulation thanks to the excellent description of the material in the beam pipe and the first layers of the pixel detector. The high-$\pt$ regime exhibits a significantly better resolution in simulation compared to data. These differences are attributed to residual alignment uncertainties in data not present in simulation, as well as to imperfections in the cluster modelling in the pixel sensors in simulation.

\subsection{Input variable comparisons using fully reconstructed $b$ hadrons}

In this section, a comparison between data and simulation is presented of observables entering $b$-tagging algorithms.
The main goal of this comparison is to validate the description in simulation of the $b$ jets and quantify possible differences.

A pure $b$-jet sample is obtained by exploiting an invariant mass based selection of fully reconstructed $b$ hadrons in an inclusive decay channel. 
It is possible to isolate a very pure $b$-jet sample to be used for the comparison
by matching those candidates to jets, albeit at the expense of the sensitivity to the modelling of heavy flavour lifetimes and decay processes.

\subsubsection{Comparison procedure and sample selection}
\label{subsec:procedure}

Although there are a reasonable number of decay channels of $b$ hadrons that
would be suitable for the selection of the $b$-jet-enriched sample, in practice only the decay mode 
$B^{\pm} \rightarrow \Jpsi(\mu^+\mu^-) K^{\pm} $ 
is chosen for this analysis.
This decay is characterised by both a clear signature and a high branching fraction
($\approx 10^{-3}$), compared to other decays involving a $\Jpsi$.

A logical OR of $\Jpsi$ triggers has been used for the event selection, applied
to the full 2011 dataset and to all simulated samples.
In the simulated signal sample, which is generated using {\sc PYTHIA},
true $B^\pm\rightarrow \Jpsi \; K^\pm$ decays are required, matched in $\Delta R$ to the reconstructed candidate.
The simulated jet kinematic (\pt, $\eta$) spectra are reweighted to match those in the data.

\begin{figure}
  \begin{minipage}{0.32\textwidth}
    \includegraphics[width=1.0\textwidth]{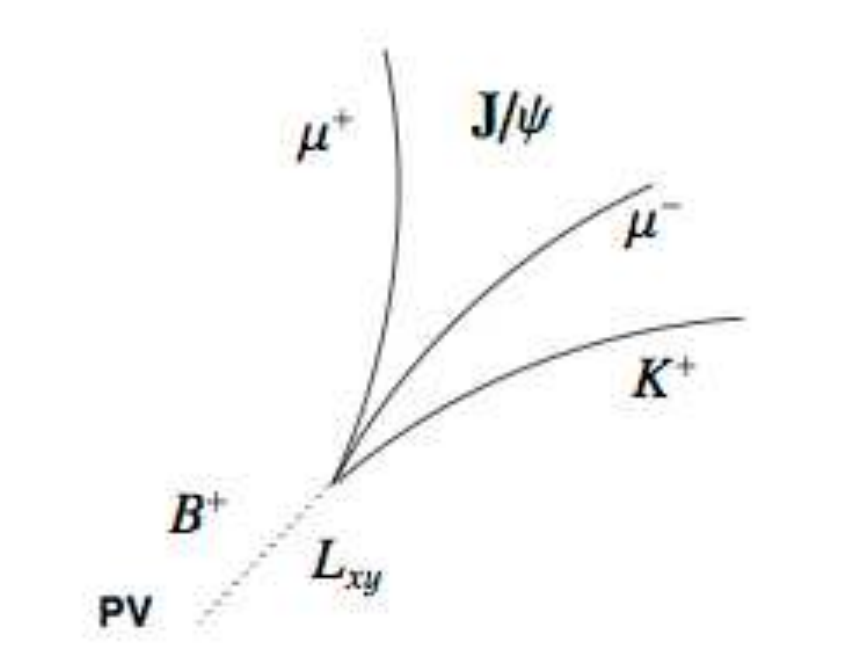}
    \caption{Schematic view of a $B^\pm\rightarrow \Jpsi \; K^\pm$ decay.}
    \label{fig:Bpm_graph}
  \end{minipage}\hfill
  \begin{minipage}{0.6\textwidth}
    \includegraphics[width=1.0\textwidth]{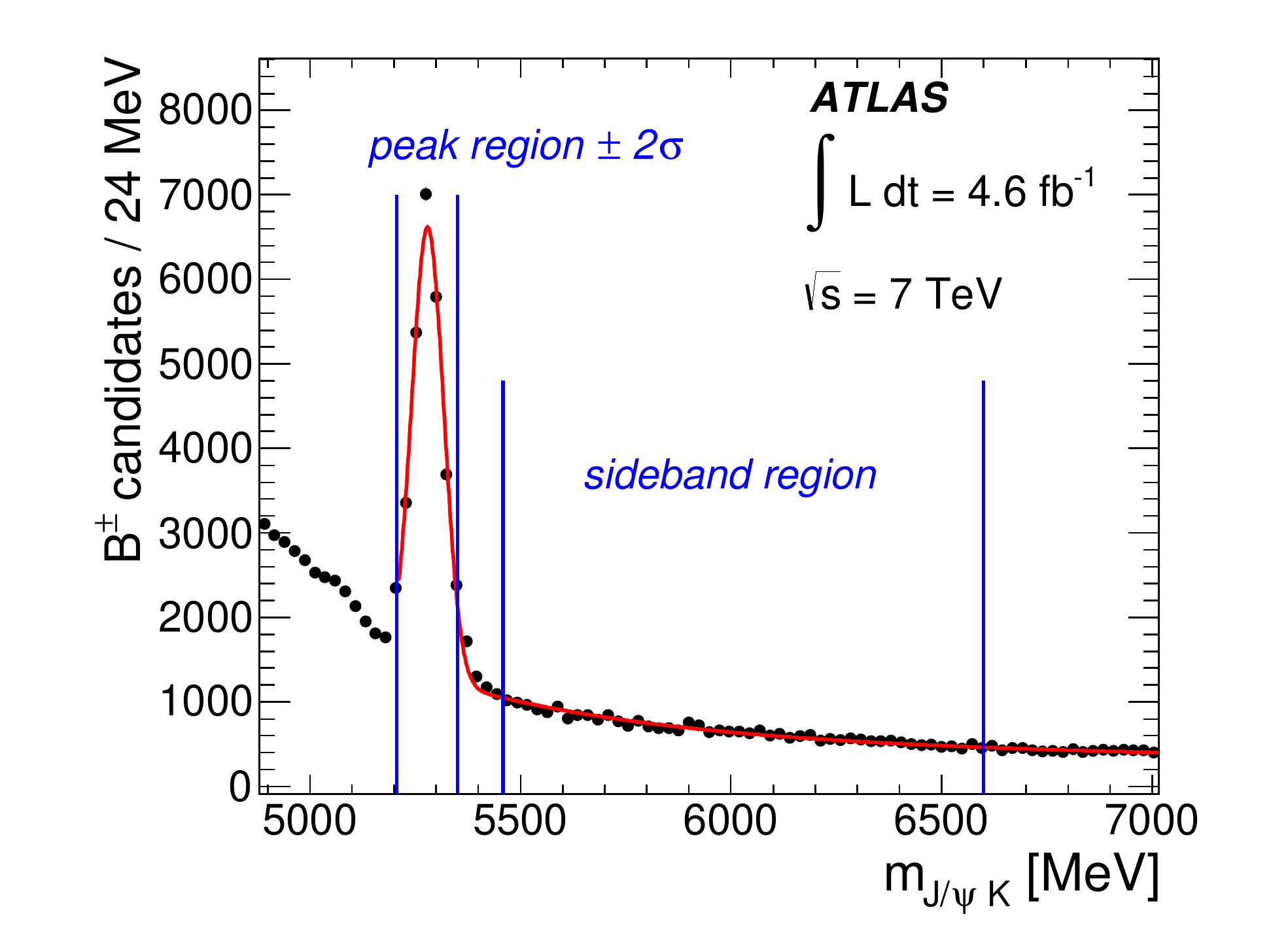}
    \caption{\textit \small Invariant mass distribution of $B^{\pm}$ candidates,
      in the $J/\psi K^{\pm}$ decay mode and on the full 2011 dataset.
      The red curve shows the result of the fit (Gaussian plus double exponential distributions).}
    \label{fig:Bpm_fit}
  \end{minipage}
\end{figure}

The $\Jpsi$ candidate is selected requiring two muons with $\pt > 4 \GeV$
and invariant mass within $200 \MeV$ from the $\Jpsi$ mass.
Secondly, a fully reconstructed $B^{\pm}$ candidate is selected following the scheme
shown in Fig.~\ref{fig:Bpm_graph}.
In the $B^{\pm}$ selection procedure all tracks that fulfil minimal quality requirements, and have a
transverse momentum greater than $2.5\GeV$, are refitted to a common vertex together with the selected muons. 
If more than one candidate is found in the event, the one with the lowest vertex fit $\chi^2$ is selected. 

Finally, the $B^{\pm}$ candidate is matched to a jet satisfying the selection criteria 
used in this paper ($\pt > 20 \GeV$, $|\eta|  < 2.5$) by means of an angular matching. 
No JVF requirement is imposed.
It should have $\Delta R(B,\rm jet) < 0.4$, where the candidate $B^{\pm}$
direction is estimated by summing the momenta of the muons and the third charged-particle track.
If more than one jet is found compatible with the $B^{\pm}$ candidate, only the jet having the
smallest $\Delta R$ is considered.

The obtained mass spectrum of all $B^{\pm}$ candidates is shown in Fig.~\ref{fig:Bpm_fit},
together with a fit to a Gaussian signal on top of a falling combinatorial background.
In order to separate the signal from the combinatorial background,
a sideband-based background subtraction procedure is adopted.
The sideband region is defined as the mass region between $3\sigma$ above the
signal peak and $6.6\GeV$, where $\sigma$ is the width of the Gaussian;
masses below the signal peak are not used since they
have a large contribution from partially reconstructed other $b$-hadron decays.
The main assumption is that the distributions in background events 
of variables under study are the same for the events in the sideband
region and under the resonance peak. This assumption has been tested
in simulated samples for each variable, discarding the ones showing correlation
with the invariant mass of the $B$ meson. 

In the following, the $B^{\pm}$ signal region is defined as the mass region
within two standard deviations from the signal peak.
For each variable, its distribution in the sideband region is subtracted from
that in the signal region, after proper normalisation.
This normalisation is evaluated from the fit to the invariant mass spectrum
using a double exponential function for the combinatorial background and
a Gaussian for the signal. Systematic effects on the background subtraction
have been estimated by replacing the double exponential with a single 
exponential. The statistical uncertainty on the estimated fraction
of combinatorial background is found to be negligible.

\subsubsection{Comparison of variables}

For the sake of clarity the investigated observables are divided into three
categories (although this procedure is not completely rigourous):
detector specific variables, variables sensitive to the hadronisation physics,
and tagging algorithm performance variables.

\begin{figure}
  \begin{center}
    \subfloat[]{\label{g:d0BL_a}\includegraphics[width=0.49\textwidth]{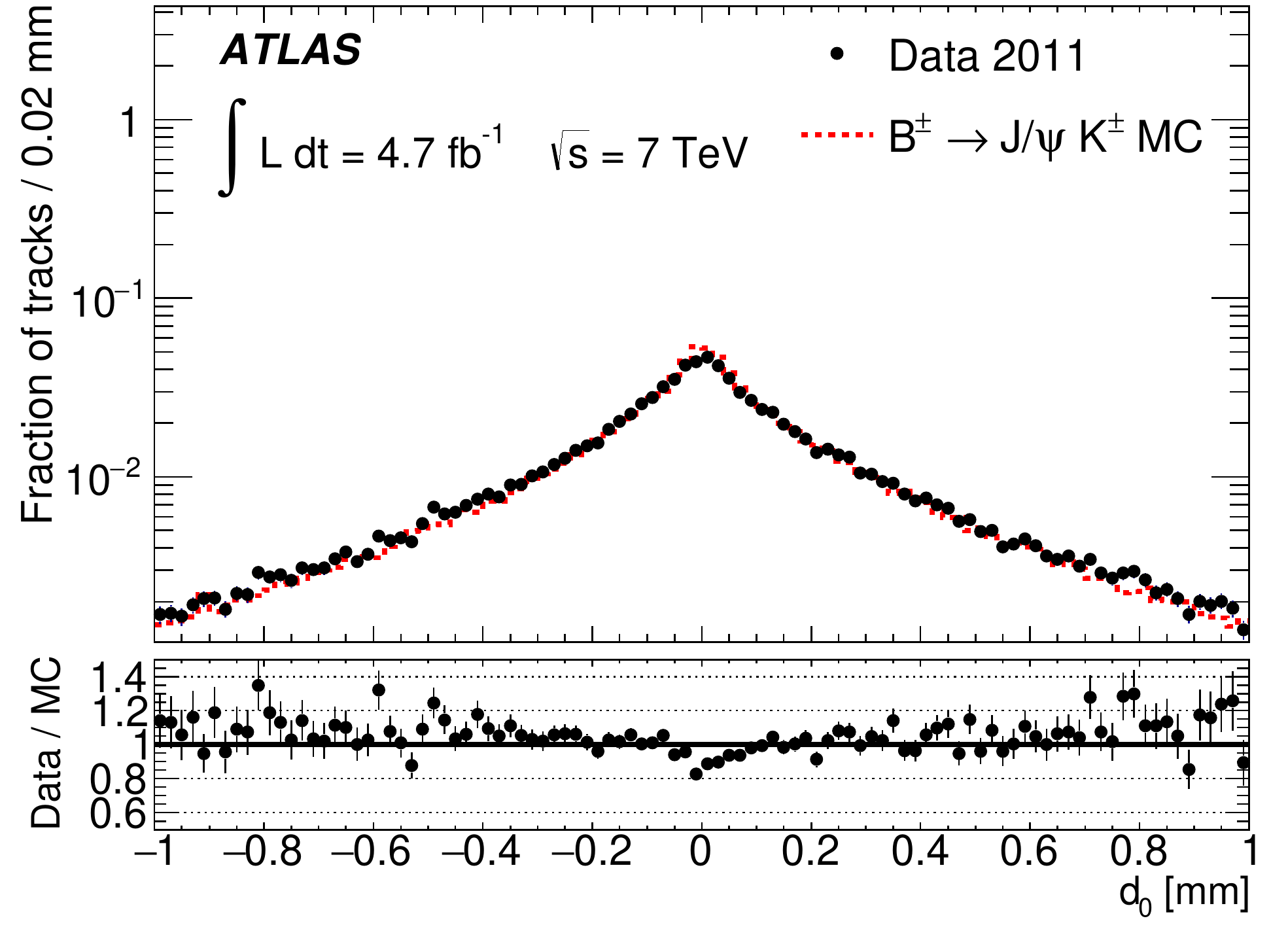}}
    \subfloat[]{\label{g:d0BL_b}\includegraphics[width=0.49\textwidth]{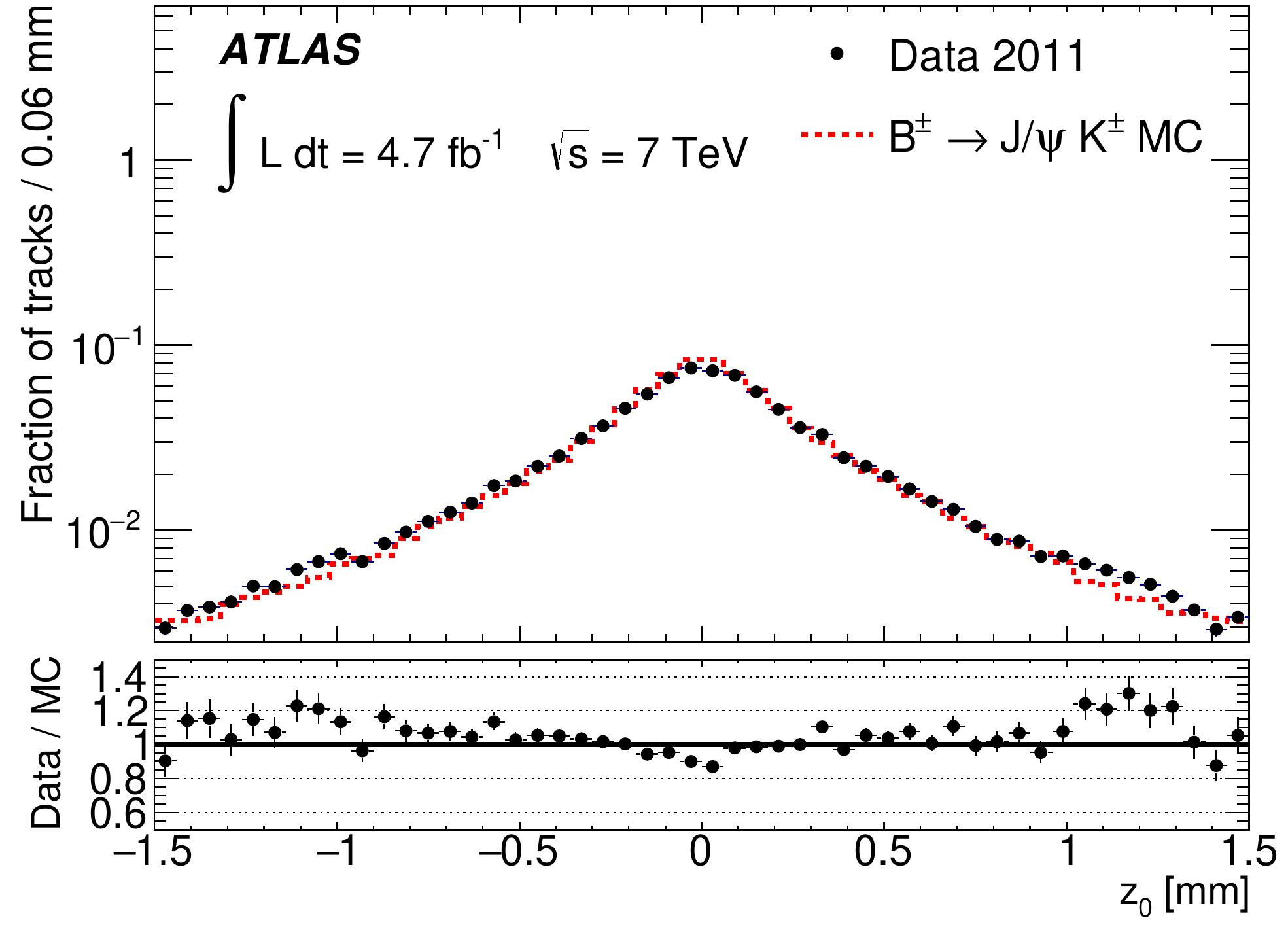}}
    \hfill
    \subfloat[]{\label{g:d0BL_c}\includegraphics[width=0.49\textwidth]{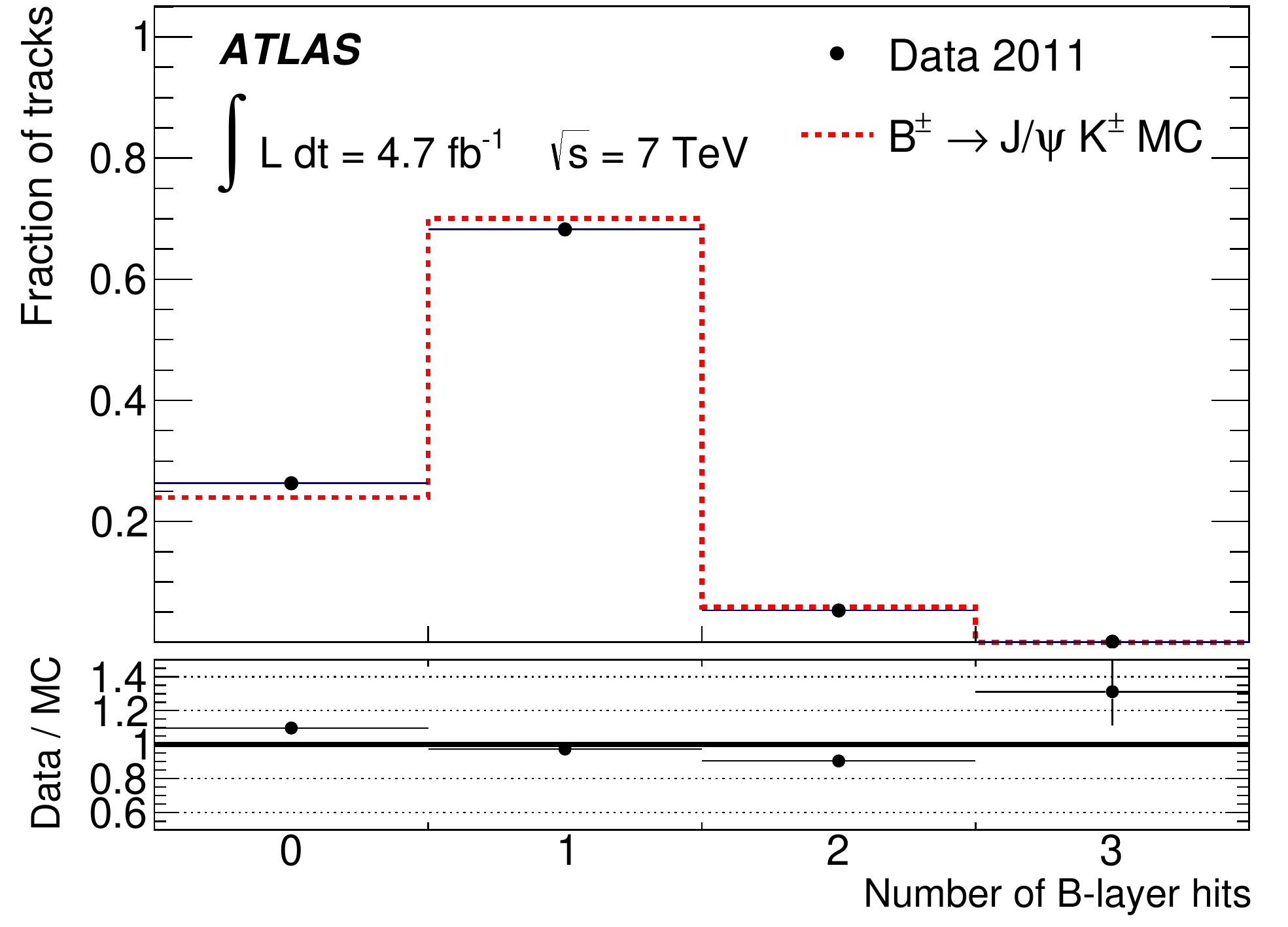}}
    \subfloat[]{\label{g:d0BL_d}\includegraphics[width=0.49\textwidth]{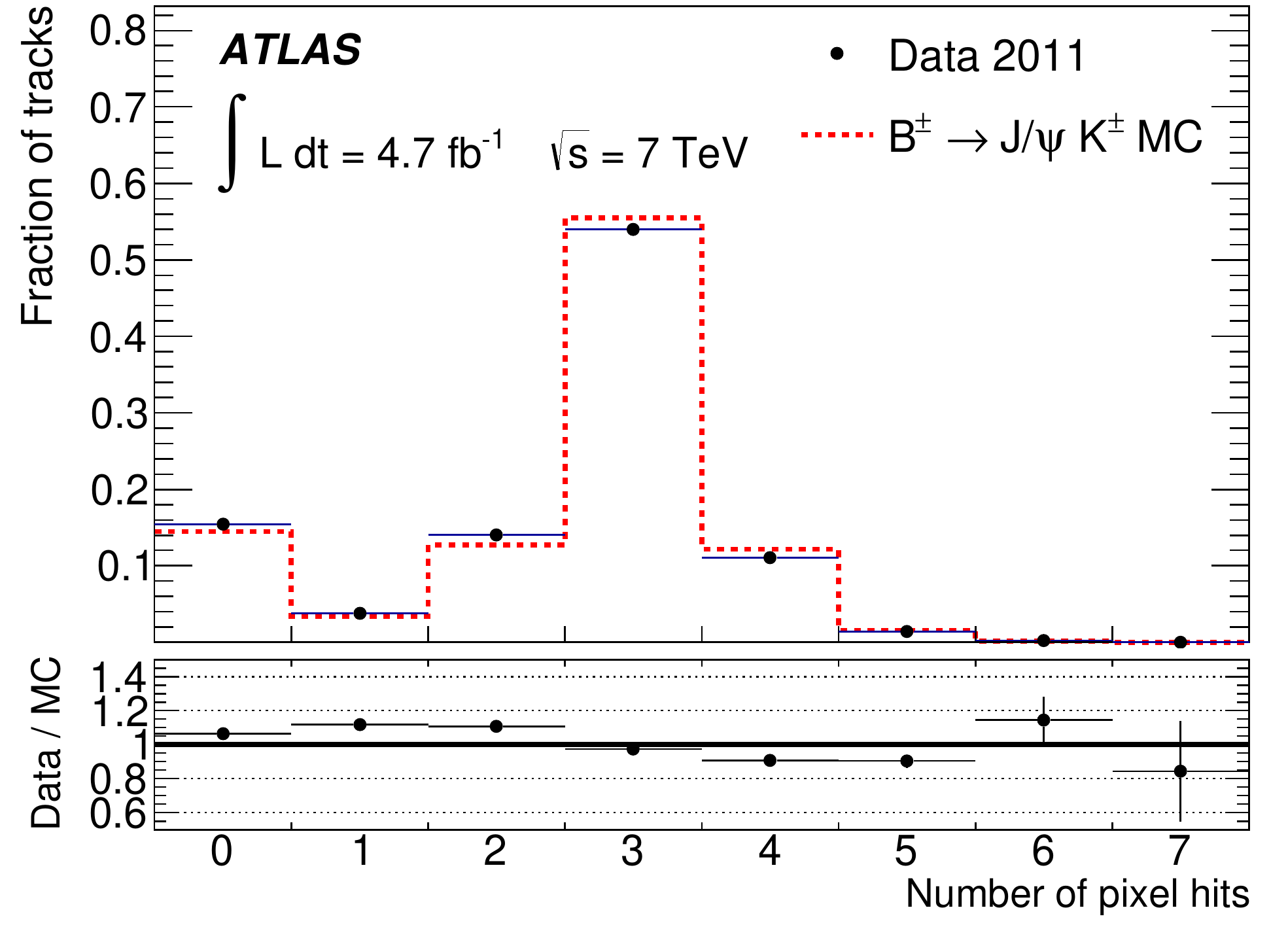}}
    \caption{Comparisons between data and simulation of the transverse (a) and longitudinal (b) impact parameters,
      the number of hits in the innermost pixel layer (c) and in all pixel detector layers (d) 
      for hadronisation tracks associated to the matched $b$ jet.}
    \label{g:d0BL}
  \end{center}
\end{figure}

The first group of variables that have been compared are those mostly related
to detector reconstruction effects.
The $B^{\pm}$ decay tracks as well as the hadronisation tracks, defined as those tracks that
have been associated to the matched jet but not identified as the $B^{\pm}$ 
decay products, have been used for this comparison.
Inner Detector hits and impact parameters 
in the $xy$ plane ($d_0$) and along the
beam axis ($z_0$) with respect to the primary vertex and their errors
for $B^{\pm}$ decay tracks have been studied.
Hadronisation tracks have been used to study
the associated number of innermost pixel detector layer and other pixel detector hits,
which are of utmost importance for tagging algorithm performances.
The distributions of these quantities in data and simulation are shown in Fig.~\ref{g:d0BL}.
The impact parameter distributions are slightly wider in data than in simulated events,
consistent qualitatively with the slightly worse impact parameter resolutions observed in Fig.~\ref{fig:IP},
and the numbers of innermost layer and total pixel hits associated to each
track are slightly lower in data than in simulation.

\begin{figure}
  \subfloat[]{\label{g:HaddRn_a}\includegraphics[width=0.49\textwidth]{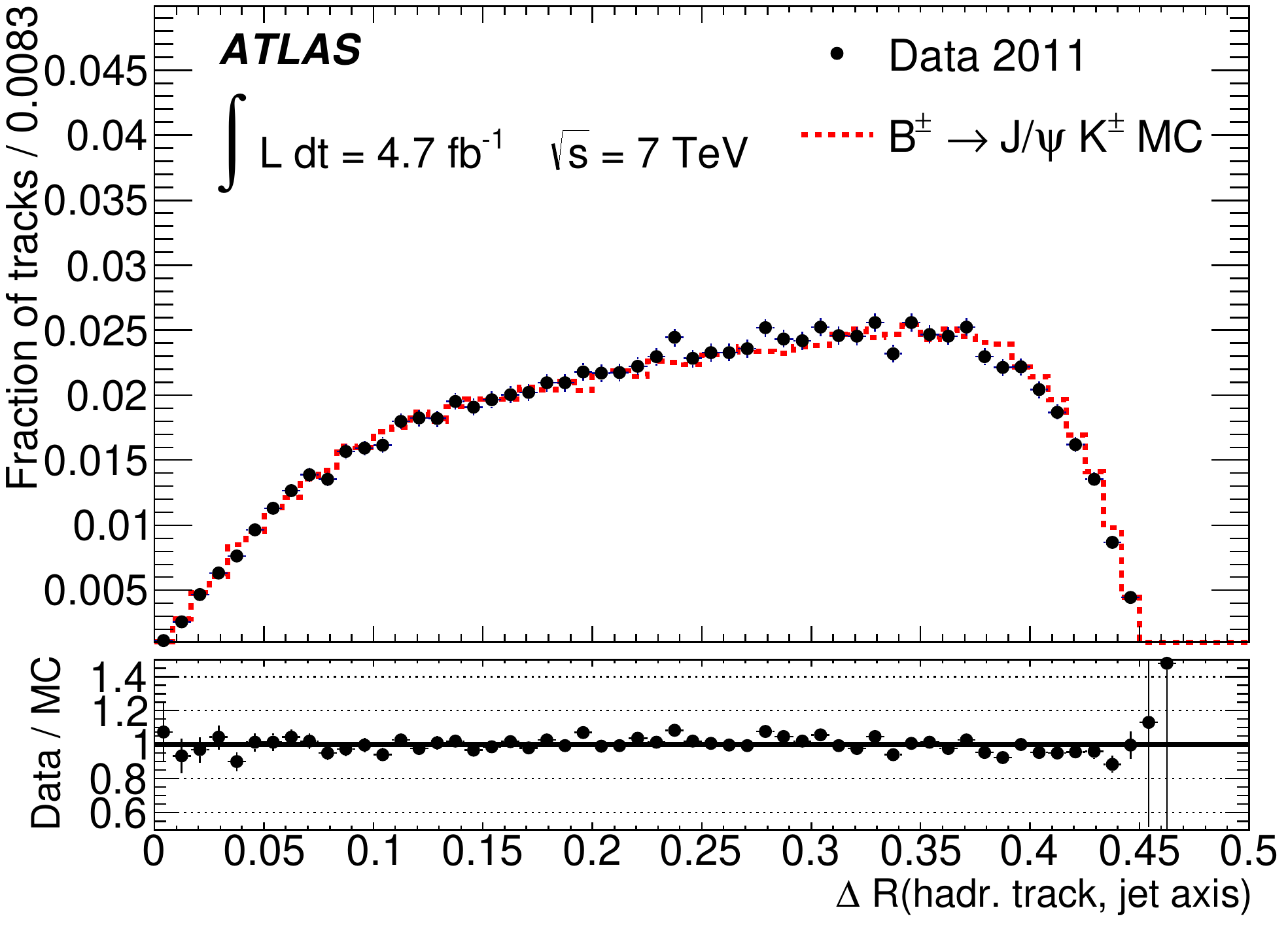}}
  \subfloat[]{\label{g:HaddRn_b}\includegraphics[width=0.49\textwidth]{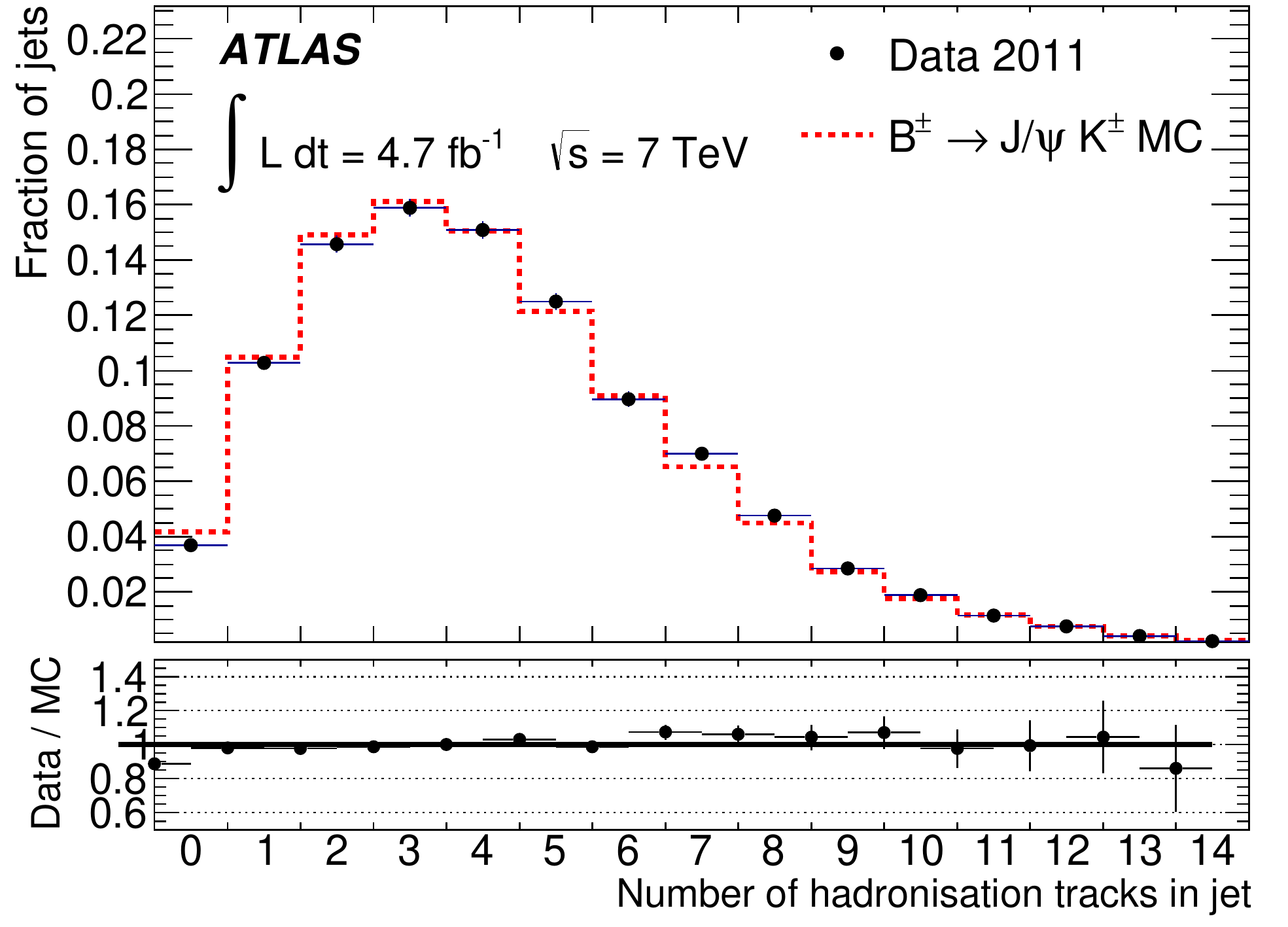}}
  \caption{Angular distance $\Delta R$ between the hadronisation tracks 
    and the jet axis (a) and the hadronisation track multiplicity (b).}
  \label{g:HaddRn}
\end{figure}

The reliability of the description of the hadronisation process in simulated events 
is verified by means of the second group of variables.
These include the angular distance of hadronisation tracks 
to the jet axis and the track multiplicities, as shown in Fig.~\ref{g:HaddRn}.
In both cases excellent agreement with data is found, reflecting the quality of the Monte Carlo generator tuning, 
with a small tendency of the simulation to underestimate the multiplicity.

\begin{figure}
  \subfloat[]{\label{g:nTag_a}\includegraphics[width=0.49\textwidth]{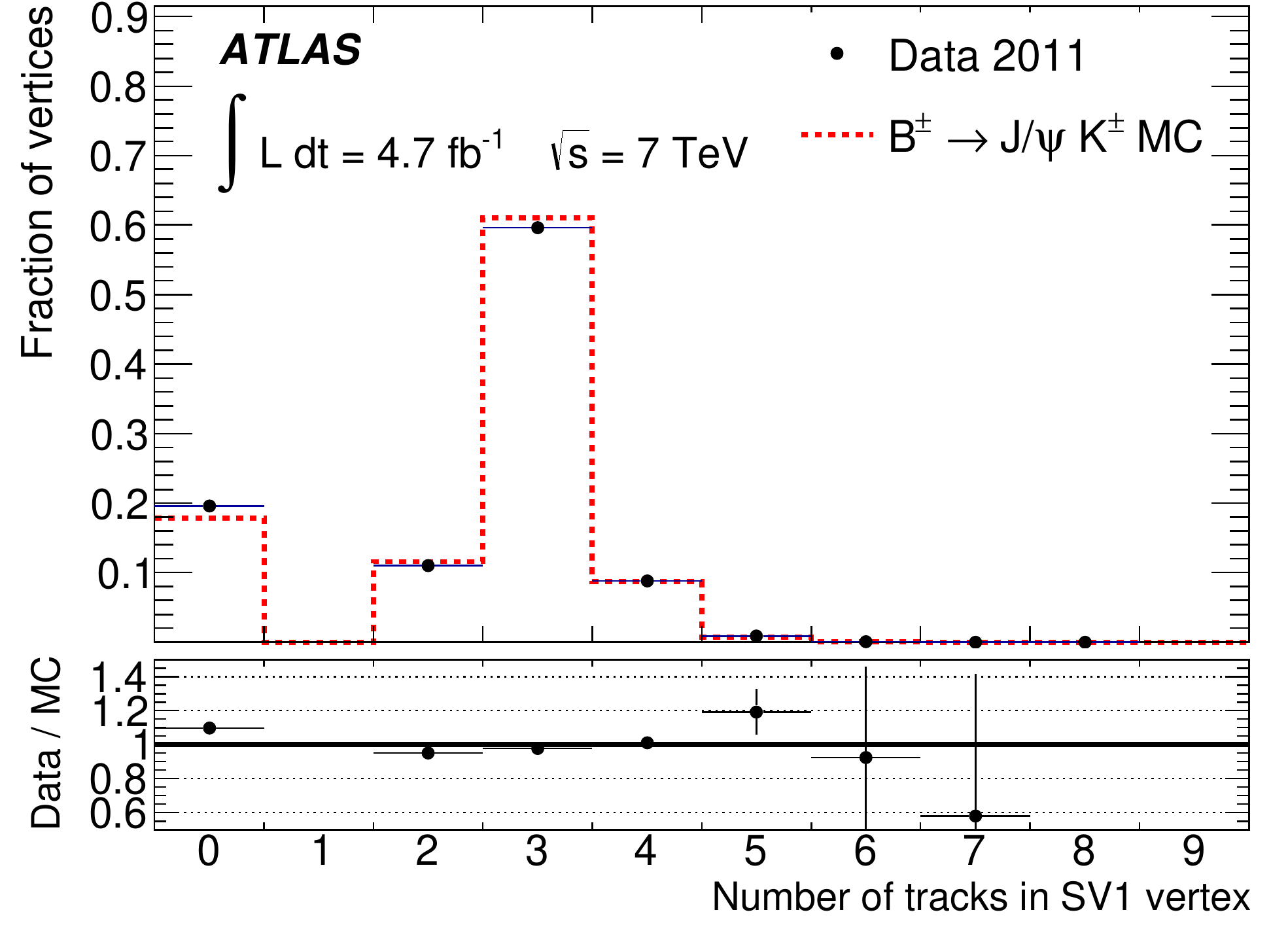}}
  \subfloat[]{\label{g:nTag_b}\includegraphics[width=0.49\textwidth]{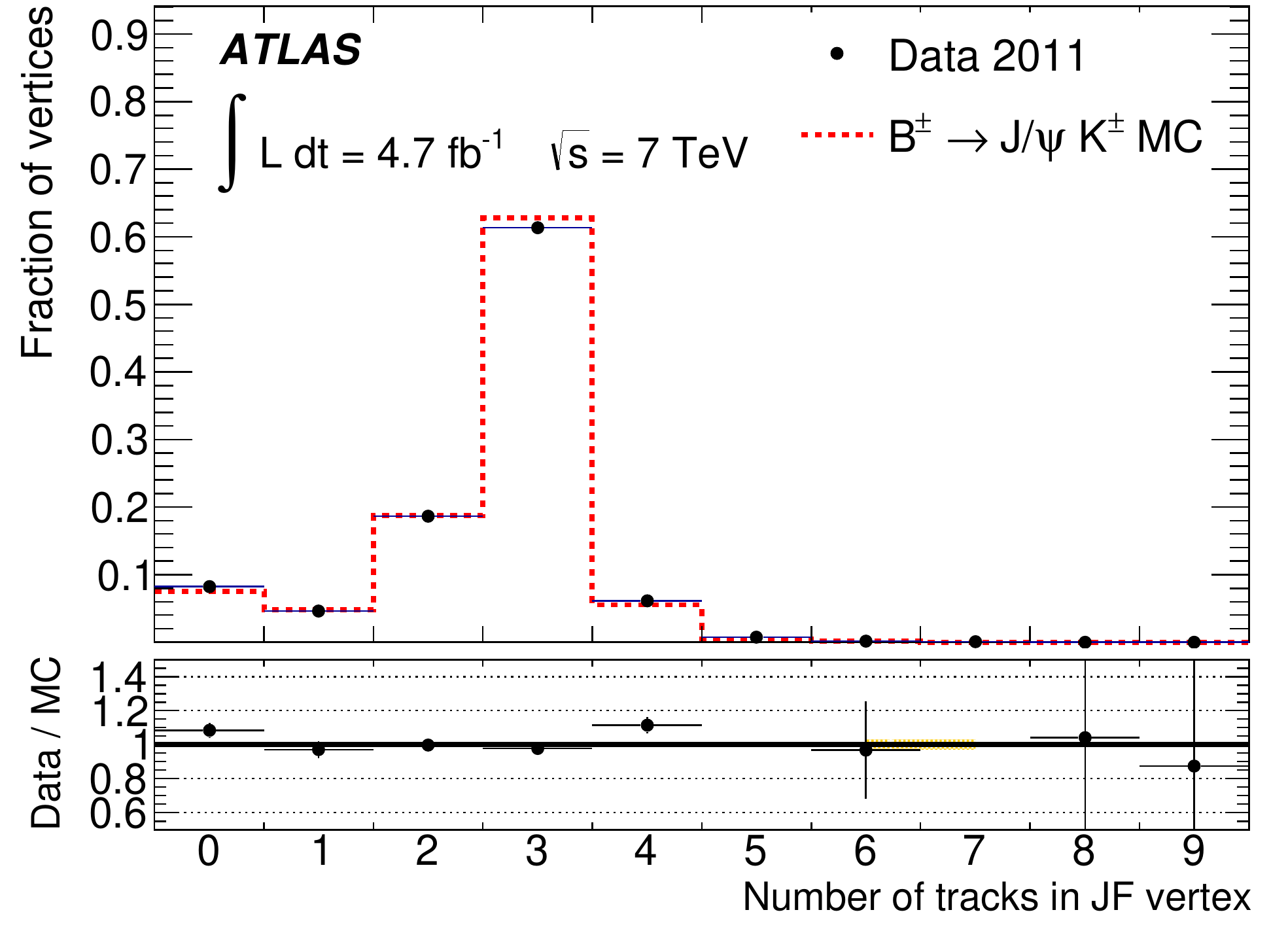}}
  \caption{Number of tracks associated with the displaced vertices reconstructed by the SV1 (a) and JetFitter (b) algorithms.
    Zero associated tracks means that no displaced vertex was reconstructed.}
  \label{g:nTag}
\end{figure}

\begin{figure}
  \subfloat[]{\label{g:TagCounter_a}\includegraphics[width=0.49\textwidth]{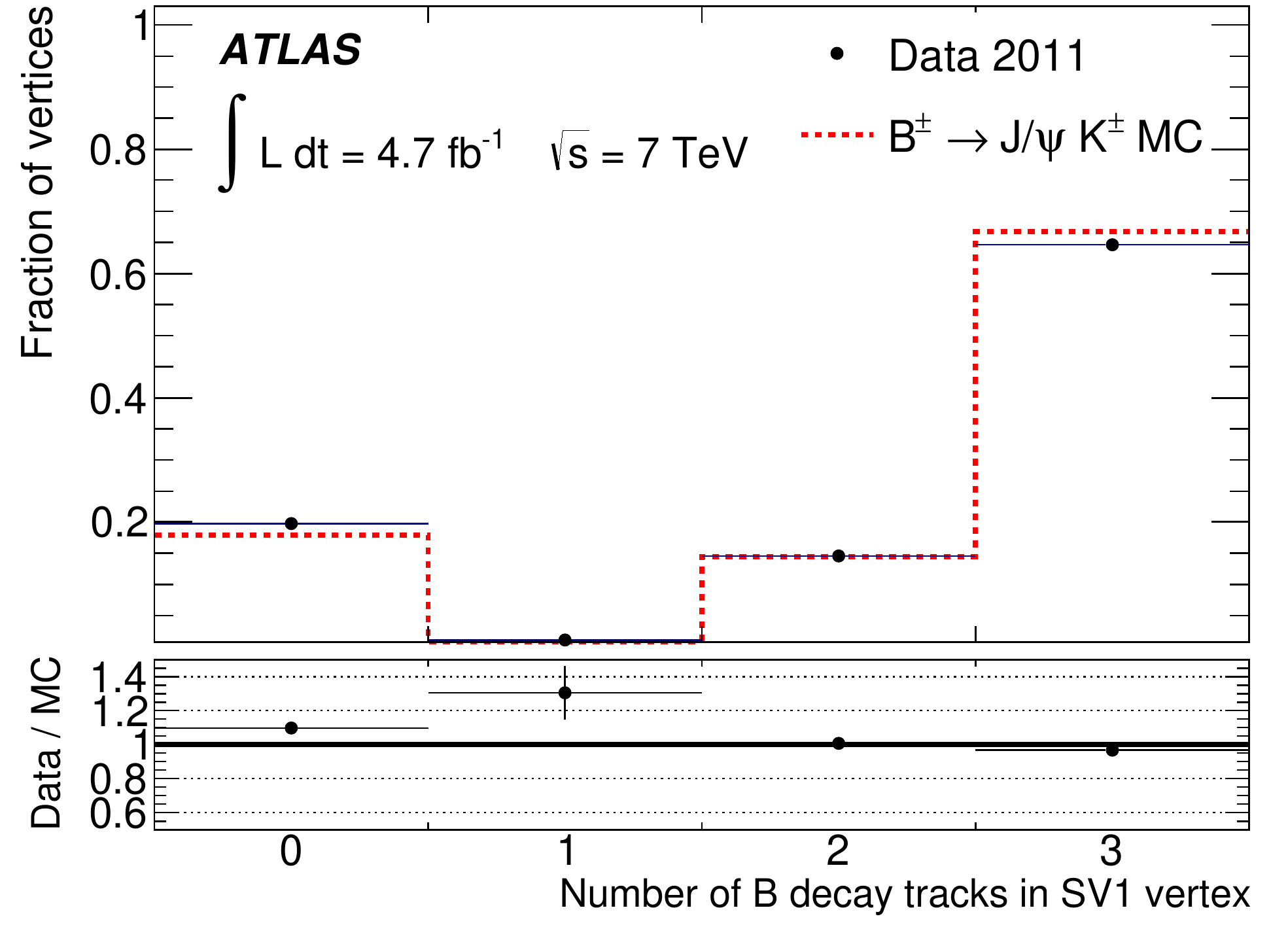}}
  \subfloat[]{\label{g:TagCounter_b}\includegraphics[width=0.49\textwidth]{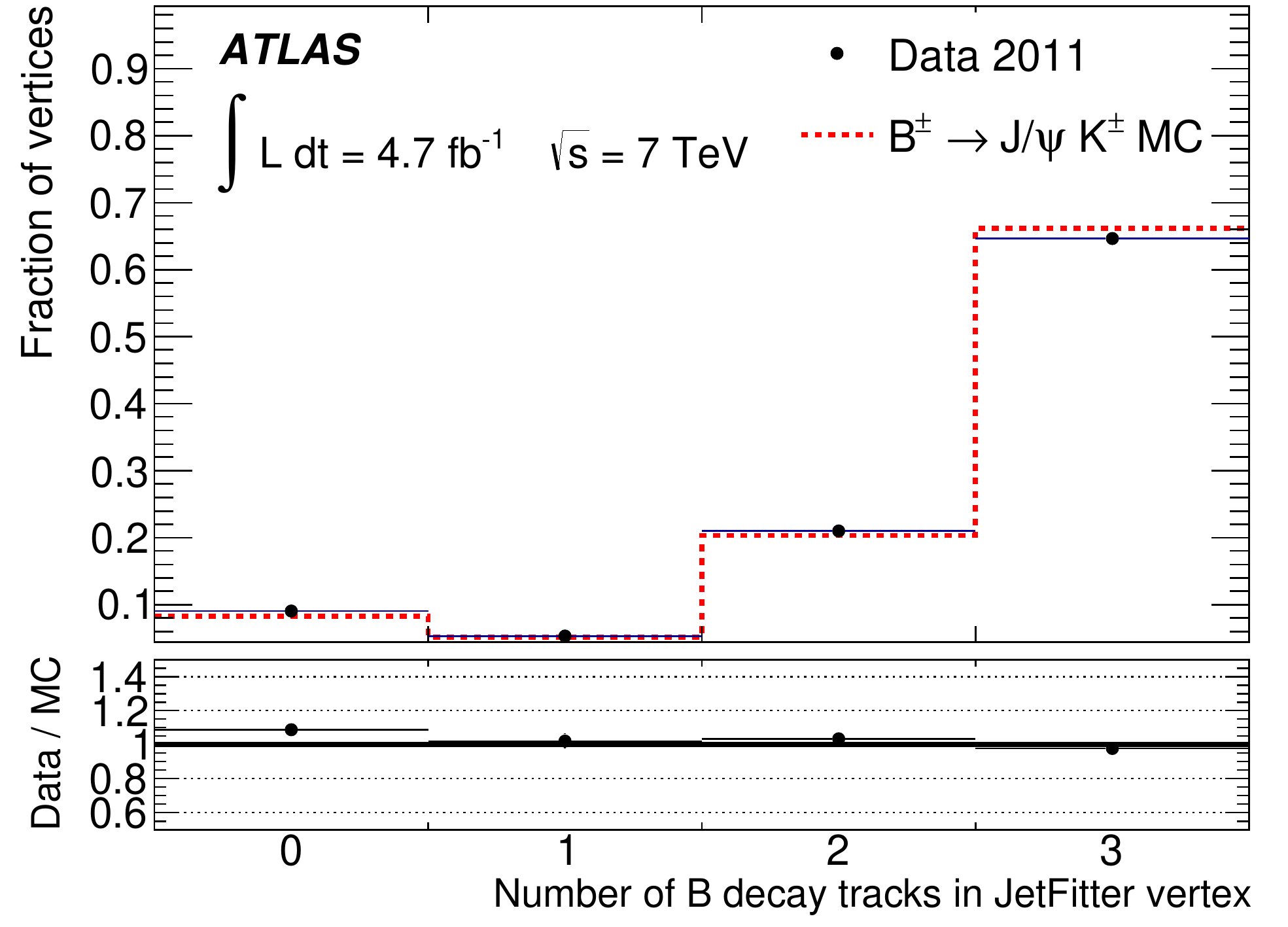}}
  \caption{Number of tracks from the reconstructed $B^{\pm}$ decay associated with the displaced vertices
    for the SV1 (a) and JetFitter (b) algorithms.}
  \label{g:TagCounter}
\end{figure}

The last comparison targets the performance of the $b$-tagging algorithms.
Given that the $B^{\pm}$ decay is completely reconstructed, a detailed
comparison of the secondary vertex-based algorithms is possible.
In Fig.~\ref{g:nTag} the comparison between data and simulation of the 
number of tracks associated with the displaced vertices reconstructed by the SV1 and JetFitter algorithms is shown. 
In addition, Fig.~\ref{g:TagCounter} compares the efficiencies in data and simulation of the SV1 and JetFitter algorithms in associating 
the $B^{\pm}$ decay products with the displaced vertices.
The agreement between data and simulation is good but shows, nevertheless,
slightly lower efficiency in data to select the tracks resulting from the $B^{\pm}$ decay.

In summary, the overall agreement between distributions in data and simulated events is
quite good, with only small differences in the impact parameter distributions
and hit and track multiplicities. The track \pt{} spectrum in the sample studied
here is soft, and therefore the discrepancies observed at high track \pt{} in
Section~\ref{sec:ipres} are not evident here.


\section{$b$-jet tagging efficiency calibration using muon-based methods}
\label{sec:beff_mubased}

The ideal sample for the calibration of flavour-tagging algorithms is composed of jets 
characterised by a strong predominance of a single flavour, 
whose fractional abundance can be measured from data.
For the $b$-jet tagging efficiency calibration, a good sample can be obtained by selecting jets containing a muon:
because of the semileptonic decay of the $b$ hadrons this sample is enriched in $b$ jets.

Two methods were used to measure the $b$-jet tagging efficiency in the inclusive
sample of jets containing a muon: \ptrel{} and system8.
The \ptrel{} method uses templates of the muon momentum transverse to
the jet axis to fit the fraction of $b$ jets before and after $b$-tagging to extract the $b$-jet tagging efficiency. 
The system8 method, developed within the D0 experiment~\cite{System8}, was designed to involve a minimal input 
from simulation and therefore to be less sensitive to the associated systematic uncertainties.
It applies three independent criteria to a data sample containing a muon associated 
with a jet to build a system of eight equations between observed and expected event counts.

\subsection{Data and simulation samples}
\label{sec:mujet_samples}

The events used in the analyses were collected with triggers that require a muon reconstructed from hits in the Muon 
Spectrometer and spatially matched to a calorimeter jet. In each jet \pt\ bin of the analyses,
the muon-jet trigger with the lowest jet threshold that has reached the efficiency plateau is used.

In the lower jet \pt\ region (up to 60~\GeV{}) jets with $E_{\rm T} > 10\GeV$ at 
the EF level are required. Starting from 60~\GeV{} up to 110~\GeV{} the analyses use
events with at least one jet with $E_{\rm T} > 10\GeV$ at the first trigger level, while 
for jet \pt\ above 110~\GeV{} the trigger threshold is increased to 30~\GeV{}.
During data taking each of the muon-jet triggers was prescaled to collect data at a fixed rate slightly below 1~Hz.

For quantities related to $b$ and $c$ jets, the analyses make use of a simulated muon-filtered inclusive jet sample, 
referred to below as the $\mu$-jet sample, where the events are required to have a muon with $\pt > 3\GeV$ at generator level.
The sample is generated with PYTHIA~\cite{pythia2}, utilising the ATLAS AUET2B LO** PYTHIA tune~\cite{ATL-PHYS-PUB-2011-014}.
A total of 25.5 million events have been simulated in four intervals of $\hat{p}_{\perp}$,
the momentum of the hard scatter process perpendicular 
to the beam line~\cite{pythia2}, starting from $\hat{p}_{\perp}=17\GeV$.
For estimates of inclusive flavour fractions, as well as quantities related to light-flavour jets, 
the analyses make use of an inclusive jet sample for which
the simulation has been carried out in six $\hat{p}_{\perp}$ intervals.
About 2.8 million events have been simulated per $\hat{p}_{\perp}$ interval.

To reduce the dependence on the modelling of \ptrel{} for muons in light-flavour jets,
the heavy-flavour content in the \ptrel{} sample is increased by requiring that there is 
at least one jet in each event, other than those used in the \ptrel{} measurement, 
with a reconstructed secondary vertex with a signed decay length significance $L/\sigma(L) > 1$. 
The same sample is used as a subsample, called $p$-sample in the system8 analysis.
This flavour-enhancement requirement is not applied in the sample used to derive the 
\ptrel{} template for light-flavour jets.

\subsection{Jet energy correction for semileptonic $b$ decays}
\label{sec:semilep_jes_corr}

The jet energy measurement in ATLAS is characterised using the calorimeter
response $R^{\rm calo}=\ptjet/\pt^{\rm truth}$, where $\pt^{\rm truth}$ is the
$\pt$ of a matched jet built of final state particles with a lifetime longer 
than 10~ps, except for muons and neutrinos~\cite{PERF-2012-01}. 
For $b$ jets containing semileptonic decays, however, a larger fraction of 
the momentum is carried by muons and neutrinos than for inclusive jets. 
Therefore an additional correction is applied based on the all-particle
response, $\mathcal{R}^{\rm all}=\pt^{\rm jet+\mu}/\pt^{\rm truth, all}$, 
where $\pt^{\rm jet+\mu}$ includes selected reconstructed muons in the jet cone
(while correcting for their mean energy loss in the calorimeters) and $\pt^{\rm truth, all}$ 
is the $\pt$ of a matched jet built of final state particles with a lifetime 
longer than 10~ps. This correction and its systematic uncertainties are described in detail
also in Ref.~\cite{PERF-2012-01}. 
The estimation of the effect of the systematic uncertainty on the calibrations is described in Section~\ref{sec:syst}.

\subsection[The \ptrel{} method]{The \ptrelBM{} method}
\label{sec:ptrel}

The number of $b$ jets before and after tagging can be obtained for a subset of all $b$ jets, 
namely those containing a reconstructed muon, using the variable \ptrel{} which is 
defined as the momentum of the muon transverse to the combined muon-plus-jet axis. 
Muons originating from $b$-hadron decays 
have a harder \ptrel{} spectrum than muons in $c$ and light-flavour jets. 
Templates of \ptrel{} in simulated events are constructed for $b$, $c$ and light-flavour jets separately, 
and these are fitted to the \ptrel{} spectrum of muons in jets in data
to obtain the fraction of $b$ jets before and after requiring a $b$-tag.

As the templates from $c$  and light-flavour jets have a rather similar shape, 
the fit can only reliably separate the $b$ jets from non-$b$ jets.
Therefore, the ratio of the $c$  and light-flavour fractions is
constrained in the fit to the value observed in simulated events, which 
in the pre-tagged sample ranges from 2 at low \pt\ to 0.7 at high \pt.
This ratio is then varied as a systematic uncertainty, as described in Section~\ref{sec:syst_clratio}.

\begin{figure}[htb]
  \begin{center}
    \subfloat[]{\label{fig:ptrel_fits_a}\includegraphics[width=0.49\textwidth]{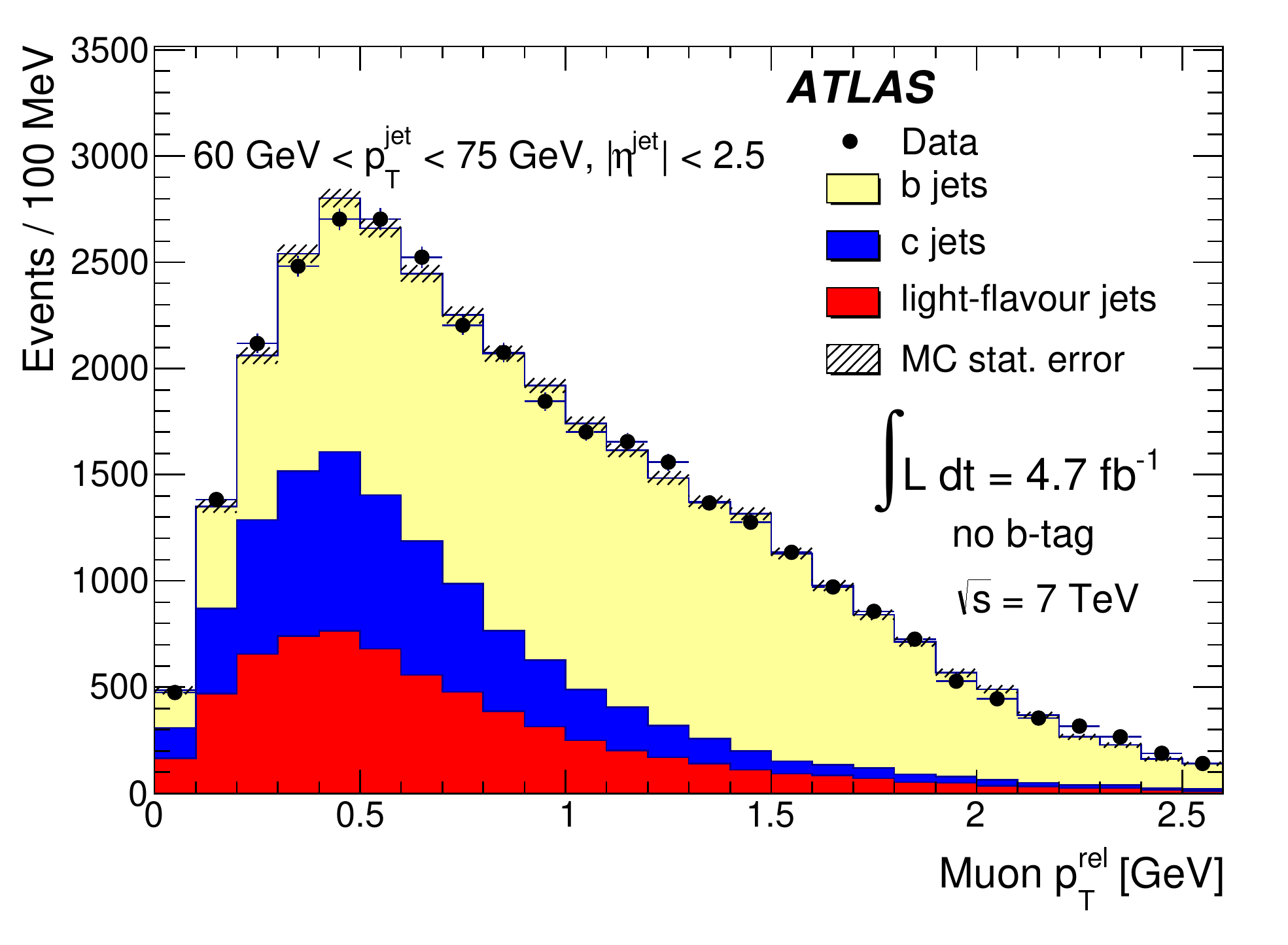}}
    \subfloat[]{\label{fig:ptrel_fits_b}\includegraphics[width=0.49\textwidth]{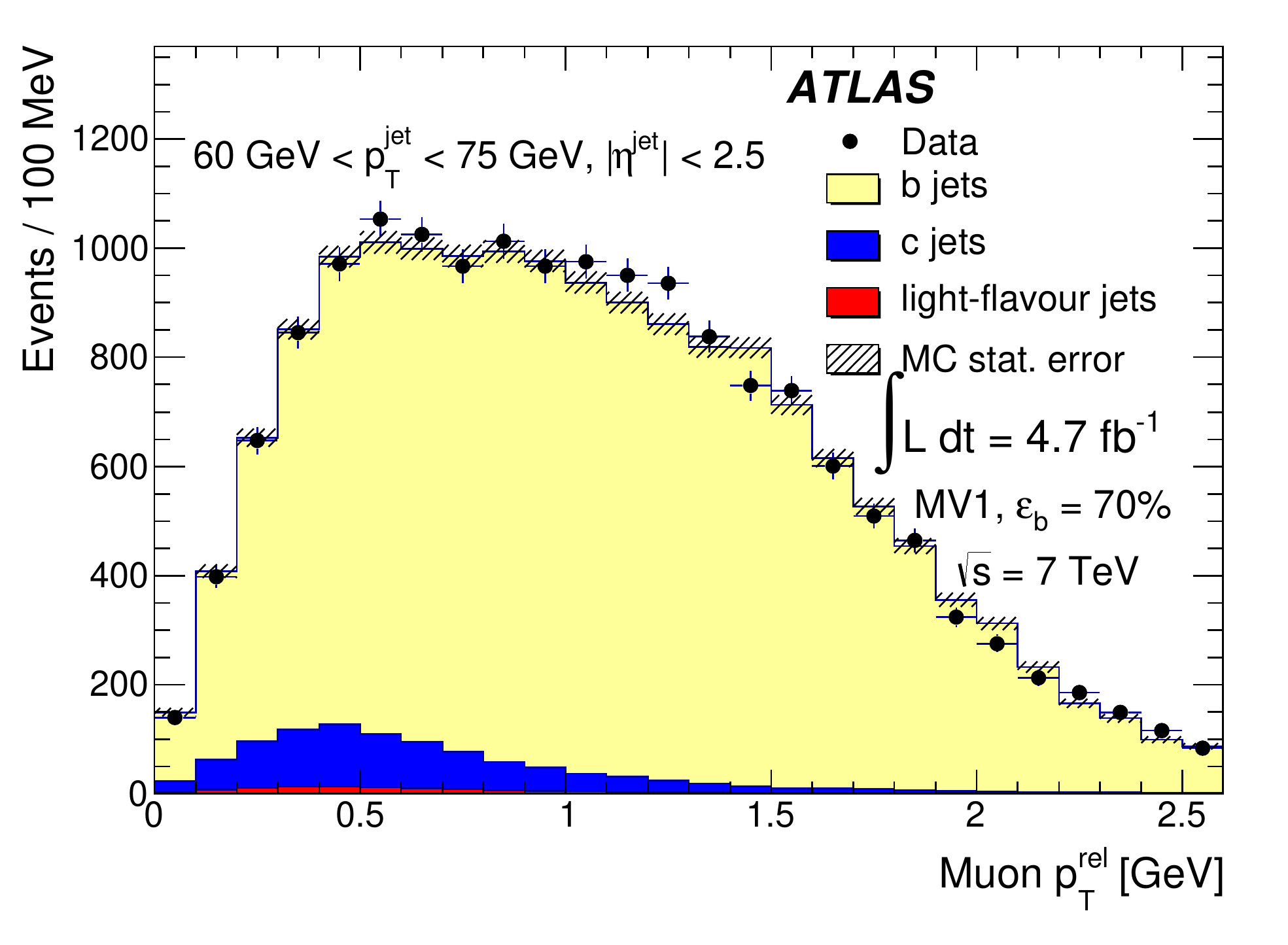}}
    \caption{Examples of template fits to the distribution of \ptrel, the momentum
      of the muon transverse to the combined muon-plus-jet axis, in data before (a) and 
      after (b) $b$-tagging by applying the MV1 tagging algorithm at 70\%
      efficiency, for jets with $60\GeV < \pt < 75\GeV$.
      \label{fig:ptrel_fits}}
  \end{center}
\end{figure}

Figure~\ref{fig:ptrel_fits} shows examples of template fits to the \ptrel{} distribution in data before (left) and 
after (right) $b$-tagging.
Having obtained the flavour composition of jets containing muons from the \ptrel{} fits, 
the $b$-jet tagging efficiency is defined as

\begin{equation}
\epsilon_b^{\rm data} = \frac{f_b^{\rm tag}\cdot N^{\rm tag}}{f_b \cdot N} \cdot C
= \frac{f_b^{\rm tag}\cdot N^{\rm tag}}{f_b^{\rm tag} \cdot N^{\rm tag} + f_b^{\rm untag} \cdot N^{\rm untag}} \cdot C
,
\label{eq:eff_b}
\end{equation}
where $f_b$ and $f_b^{\rm tag}$ are the fractions of $b$ jets in the pre-tagged and tagged samples of 
jets containing muons, and $N$ and $N^{\rm tag}$ are the total number
of jets in those two samples.
In practice, the second form of the expression is used (where $N^{\rm untag}$ and
$f_{b}^{\rm untag}$ denote the total number of events in the untagged sample and
the fitted $b$-jet fraction therein), since this explicit subdivision into
statistically independent samples allows for a proper computation of the
statistical uncertainty on $\epsilon_b^{\rm data}$.
The factor $C$ corrects the efficiency for the
biases introduced through differences between data and simulation in the modelling of the $b$-hadron direction 
and through heavy flavour contamination of the \ptrel{} template for light-flavour jets,
as described below.
The efficiency measured for $b$ jets with a semileptonically decaying $b$ hadron in data is compared to the 
efficiency for the same kind of jets in simulated events to compute the
corresponding data-to-simulation efficiency scale factor.

Both the pre-tagged and the tagged samples are fitted using templates derived from all jets passing the jet 
selection criteria defined in Section~\ref{sec:samples}.
The \ptrel{} templates for $b$  and $c$ jets are derived from the simulated $\mu$-jet sample, 
using muons associated with $b$ and $c$ jets, without requiring any $b$-tagging criteria.
It has been verified that the pre-tagged and tagged template shapes agree within statistical uncertainties. 
The template for light-quark jets is derived from muons in jets in a light-flavour dominated data sample.
The sample is constructed by requiring that no jet in the event is $b$-tagged by the 
IP3D+SV1 tagging algorithm,
using an operating point that yields a $b$-jet tagging efficiency of approximately 80\% in simulated \ttbar\ events.
This requirement rejects most events containing $b$ jets and yields a sample dominated by
$c$  and light-flavour jets. The $b$-jet contamination in this sample varies between 2\% and 6\% depending on the \pt\ bin.
The small bias introduced in the measurement from the $b$-jet contamination in the light-template is corrected for in the
final result.

As the \ptrel{} method 
is directly affected by how well the $b$-hadron direction and the calorimeter jet axis are modelled in the simulation,
a difference in the jet direction resolution between data and simulation, or an improper modelling
of the angle between the $b$ quark and the $b$ hadron in simulation would cause the
\ptrel{} spectra in simulation and data to disagree, introducing a bias in the measurement.
To study this effect, an independent jet axis was formed by the vector addition of the 
momenta of all tracks in the jet. The difference
between this track-based and the standard calorimeter-based 
jet axis in the azimuthal angle $\phi$ and the pseudorapidity $\eta$, 
$\Delta \phi ({\rm calo,track})$ and  $\Delta \eta ({\rm calo,track})$, was derived in both data and simulation.
The difference between the track-based jet axis direction and the calorimeter-based jet axis direction is observed 
to be larger in data than in simulation, and 
the $\phi$ and $\eta$ of the calorimeter-based jet axis in simulation 
were therefore smeared such that the $\Delta \phi ({\rm calo,track})$ and  $\Delta \eta ({\rm calo,track})$
distributions agreed better with those from data. 
No significant \pt\ dependence of the difference between the widhts in data and simulation was observed, and a 
smearing based on a Gaussian distribution with a width of 0.004 in $\phi$ and 0.008 in $\eta$
was found to give good agreement between data and simulation in all bins of jet \pt. 
The \ptrel{} templates for $b$  and $c$ jets were rederived from this smeared sample, and
the \ptrel{} distribution in data was fitted using these altered templates.  The
difference between using the unsmeared and smeared jet directions is then taken as a systematic uncertainty.

\subsection{The system8 method}
\label{sec:system8}

The system8 method~\cite{System8} uses three uncorrelated selection criteria to construct a system of
eight equations based on the number of events surviving any given subset of these criteria.
The system, which is fully constrained, is used to solve for eight unknowns:
the efficiencies for $b$ and non-$b$ jets to pass each of the
three selection criteria, and the number of $b$ and non-$b$ jets originally
present in the sample. As there are insufficient degrees of freedom to make a complete 
separation of the non-$b$ component into ($c$, $s$, $d$, $u$, $g$) jet flavours, these are combined into one category 
and denoted $cl$.
In simulated events, the flavour composition of the sample is relatively independent of jet \pt{}
in the range studied, while the efficiencies to pass each of the selection criteria have a strong \pt{} dependence. 

The three selection criteria chosen are:
\begin{itemize}
\item The lifetime-based tagging criterion under study.
\item The requirement $\ptrel> 700\MeV$.
\item The requirement of at least another jet in the event, other than the one containing the muon,
  with a reconstructed secondary vertex with a signed decay length significance $L/\sigma(L) > 1$. 
\end{itemize}
The resulting system of equations can be written as follows:
\begin{equation}
  \renewcommand{\baselinestretch}{1.2}\normalsize
  \begin{array}{lcrcr}
    n       & = & n_{b} & + & n_{cl}, \\
    p       & = & p_{b} & + & p_{cl}, \\
    n^{LT}   & = & \epsilon_{b}^{LT}n_{b} & + &  \epsilon_{cl}^{LT}n_{cl},\\
    p^{LT}   & = & \alpha_{6}\epsilon_{b}^{LT}p_{b} & + &
    \alpha_{4}\epsilon_{cl}^{LT}p_{cl},\\
    n^{MT}   & = & \epsilon_{b}^{MT}n_{b} & + &  \epsilon_{cl}^{MT}n_{cl},\\
    p^{MT}   & = & \alpha_{5}\epsilon_{b}^{MT}p_{b} & + &
    \alpha_{3}\epsilon_{cl}^{MT}p_{cl},\\
    n^{LT,MT} & = & \alpha_{1}\epsilon_{b}^{LT}\epsilon_{b}^{MT}n_{b} & + &
    \alpha_{2}\epsilon_{cl}^{LT}\epsilon_{cl}^{MT}n_{cl},\\
    p^{LT,MT} & = &
    \alpha_{7}\alpha_{6}\alpha_{5}\epsilon_{b}^{LT}\epsilon_{b}^{MT}p_{b} &
    + & \alpha_{8}\alpha_{4}\alpha_{3}\epsilon_{cl}^{LT}\epsilon_{cl}^{MT}p_{cl}.
    \label{sys8_eq}
  \end{array}
\end{equation}
In these equations, the superscripts $LT$ and $MT$ denote the lifetime tagging
criterion and soft muon tagging criterion, respectively. The $n$ and $p$ numbers
denote the size of the samples without ($n$) and with ($p$) the application of the
requirement of another jet; these samples are referred to as the ``$n$'' sample and the
``$p$'' sample, respectively.

Little correlation is expected between the variables used in the above criteria.
However, even if correlations between tagging algorithms are small in practice, they must be accounted for. 
This is accomplished through correction factors, $\alpha_{i}, i=1, \ldots, 8$, which are
defined as:

\begin{equation}
 \renewcommand{\baselinestretch}{1.2}\normalsize
 \setlength\arraycolsep{0.1em}
 \begin{array}{lclclcl}
   \alpha_{1} & = &
   \epsilon_{b}^{LT,MT,n}/(\epsilon_{b}^{LT,n}\epsilon_{b}^{MT,n}), &
   \hspace{20mm} &
   \alpha_{2} & = &
   \epsilon_{cl}^{LT,MT,n}/(\epsilon_{cl}^{LT,n}\epsilon_{cl}^{MT,n}), \\
   \alpha_{5} & = & \epsilon_{b}^{MT,p}/\epsilon_{b}^{MT,n}, &
   \hspace{20mm} &
   \alpha_{3} & = & \epsilon_{cl}^{MT,p}/\epsilon_{cl}^{MT,n}, \\
   \alpha_{6} & = & \epsilon_{b}^{LT,p}/\epsilon_{b}^{LT,n}, &
   \hspace{20mm} &
   \alpha_{4} & = & \epsilon_{cl}^{LT,p}/\epsilon_{cl}^{LT,n}, \\
   \alpha_{7} & = & \epsilon_{b}^{LT,MT,p}/(\epsilon_{b}^{LT,p}\epsilon_{b}^{MT,p}), &
   \hspace{20mm} &
   \alpha_{8} & = & \epsilon_{cl}^{LT,MT,p}/(\epsilon_{cl}^{LT,p}\epsilon_{cl}^{MT,p}). \\  
   \label{sys8_coeff}
 \end{array}
\end{equation}
A lack of correlation between two criteria thus implies that the related
correction factors are equal to unity.

As it is impossible to isolate independent corresponding samples in data,
these correlations are inferred from simulated samples.
The correction factors for $b$ jets, as well as the $c$-jet information used to
compute the $cl$ correction factors, are derived from the simulated
$\mu$-jet sample, while the light-flavour-jet information used to compute the
$cl$ correction factors is derived from the simulated inclusive jet sample. 
As light-flavour jets only rarely have reconstructed muons associated with them, 
the statistical uncertainty 
on the correction factors would be
unacceptably large if they were derived from muons matched to light-flavour jets in
simulation. Instead, a charged particle track, fulfilling the requirements made for the
Inner Detector track matched to reconstructed muons, is chosen at random and treated 
subsequently as if it were a muon.
To ensure that Inner Detector tracks model the kinematic properties of reconstructed muons in 
light-flavour jets, the tracks are weighted to account for the $\pt$- and $\eta$-dependent 
probability that a muon reconstructed as a track in the Inner Detector is also 
reconstructed in the Muon Spectrometer, as well as the sculpting of the muon kinematics by the 
muon trigger term. An additional correction factor is applied to account for the probability 
that a muon originating from an in-flight decay is associated with the jet.

As system8 only includes correction factors for $b$ and non-$b$ jets, 
the $c$ and light-flavour samples have to be combined to obtain the $cl$ correction factors.
The relative normalisation of the charm and light-flavour samples is inferred from the simulated 
inclusive jet sample, leading to a charm-to-light ratio in the $n$- and $p$-samples which ranges
from 0.6 to 1.5 depending on the sample and jet \pt{} bin.
The variation of the charm fraction in the combined sample is treated as a systematic uncertainty, 
as discussed in Section~\ref{sec:syst_clratio}.
The values of the correction factors depend on the tagging algorithm, operating
point and jet \pt{} bin.
For the MV1 tagging algorithm at 70\% efficiency, the correction factors for $b$
jets ($cl$ jets) range between 0.96 and 1.04 (between 0.93 and 1.15).

The system of equations is solved technically by minimising a $\chi^{2}$
function relating the observed event counts in the eight disjoint event
categories to the eight parameters $n_{b}$, $n_{cl}$, $p_{b}$, $p_{cl}$,
$\epsilon^{LT}_{b}$, $\epsilon^{LT}_{cl}$, $\epsilon^{MT}_{b}$, and
$\epsilon^{MT}_{cl}$. 
Since no degrees of freedom remain, the found minimum must have $\chi^{2}=0$.

\subsection{Systematic uncertainties}
\label{sec:syst}

The systematic uncertainties affecting the \ptrel{} and system8 methods are common to a large extent.
One important class of common systematic uncertainties are those addressing how well the simulation models 
heavy flavour production, decays and fragmentation. 
Other common systematic uncertainties are those arising from the imperfect knowledge
of the jet energy scale and resolution as well as the modelling of the additional pile-up interactions. 
A systematic uncertainty that applies only to the \ptrel{} analysis 
is the heavy-flavour contamination in the \ptrel{} light-flavour data control sample, while a 
systematic uncertainty that only applies to the system8 analysis arises from varying the muon 
\ptrel{} cut which is used as the soft muon tagging criterion.

The systematic uncertainties on the data-to-simulation scale factor 
$\sfb \equiv \epsilon_b^{\rm data}/\epsilon_b^{\rm sim}$ of the MV1 tagging algorithm 
at 70\% efficiency are shown in Tables~\ref{tab:systs_ptrel} and~\ref{tab:systs_system8} for the \ptrel{} and system8 methods respectively.
The estimates of the systematic uncertainties, especially in the
system8 analysis, suffer from the limited number of simulated events which leads to unphysical bin-to-bin variations in some cases. However, when the calibration results of several methods are combined (see Section~\ref{sec:combination}) 
these irregularities are smoothed out.

\begin{table}[htbp]
\begin{center}
   \scriptsize
   \setlength{\tabcolsep}{0.25pc}
   \begin{tabular}{l|ccccccccc}
\hline\hline
                                                 & \multicolumn{9}{c}{Jet $\pt\ \mathrm{[GeV]}$} \\
Source                                           & 20--30& 30--40& 40--50& 50--60& 60--75& 75--90&90--110&110--140&140--200 \\
\hline
Simulation statistics				 &  2.1	 &  1.8	 &  0.8	 &  1.4	 &  2.1	 &  2.3	 &  3.2	 &  4.0	 &  4.3\\
Simulation tagging efficiency			 &  0.8	 &  0.8	 &  0.4	 &  0.5	 &  0.8	 &  0.8	 &  1.2	 &  1.8	 &  0.9\\
Modelling of $g \to b\bar{b}$			 &  -	 &  -	 &  -	 &  -	 &  0.1	 &  0.2	 &  0.1	 &  0.1	 &  0.2\\
Modelling of $g \to c\bar{c}$                    &  0.6	 &  0.6	 &  1.7	 &  2.2  &  2.4	 &  4.5	 &  4.7	 &  6.4	 & 15 \\ 
$b$-hadron direction modelling		 	 &  0.3	 &  -	 &  0.6	 &  0.4	 &  0.6	 &  1.4	 &  1.3	 &  1.8	 &  6.0\\
$b$-fragmentation fraction			 &  0.1	 &  0.5	 &  0.2	 &  -	 &  0.2	 &  0.3	 & -	 &  0.4	 & 0.1\\
$b$-fragmentation function			 &  0.1	 &  -	 &  -	 &  -	 &  -	 &  -	 &  -	 &  0.2	 &  -\\
$b$-decay branching fractions			 &  -	 &  0.1	 &  0.1	 &  0.1	 &  0.1	 &  0.1	 &  -	 &  0.1	 &  -\\
$b$-decay \pstar\ spectrum			 & 0.9	 & 0.8	 & 1.0	 & 0.8	 & 0.9	 & 0.5	 &  0.1	 & 1.2	 & 1.0\\
Charm-light ratio				 &  1.8	 &  1.5	 &  1.1	 &  1.3	 &  0.9	 &  0.2	 &  0.2	 &  1.1	 &  6.6\\
Muon \pt{} spectrum				 & 0.2	 & 0.4	 & 0.4	 & 0.3	 & 0.4	 & 0.4	 & 0.1	 & 0.4	 & 0.8\\
Fake muons in $b$ jets				 &  -	 & -	 &  -	 &  -	 &  -	 &  -	 & -	 &  -	 &  -\\
\ptrel{} light-flavour template contamination		 & 0.2	 & 0.3	 & 0.6	 & 0.4	 & 0.6	 & 0.4	 & 0.3	 & 0.3	 & 0.3\\
Jet energy resolution				 &  0.4	 &  -	 &  0.1	 &  0.2	 &  0.2	 & 0.2	 &  0.8	 &  1.4	 &  1.6\\
Jet energy scale				 &  0.1	 &  0.1	 &  0.2	 &  0.4	 &  0.1	 &  0.4	 &  0.1	 &  0.1	 &  0.2\\
Semileptonic correction				 & 0.2	 &  0.1	 & 0.1	 &  -	 & 0.1	 & -	 & 0.1	 & 0.1	 & 0.3\\
Pile-up $\langle \mu\rangle$ reweighting			 &  0.4	 &  0.1	 &  -	 &  0.2	 &  0.1	 &  0.4	 &  -	 &  0.4	 &  -\\
Extrapolation to inclusive $b$ jets		 &  4.0	 &  4.0	 &  4.0	 &  4.0	 &  4.0	 &  4.0	 &  4.0	 &  4.0	 &  4.0\\
\hline
Total systematic uncertainty			 &  5.2	 &  5.0	 &  4.9	 &  5.2	 &  5.6	 &  6.9	 &  7.2	 &  9.3	 & 19 \\ 
Statistical uncertainty				 &  1.6	 &  1.5	 &  2.1	 &  3.1	 &  1.7	 &  2.7	 &  3.7	 &  2.9	 &  5.6\\
\hline
Total uncertainty                                &  5.4  &  5.2  &  5.3  &  6.1  &  5.9  &  7.4  &  8.1  &  9.7  & 20 \\ 
\hline\hline
\end{tabular}
\caption{
\label{tab:systs_ptrel}Relative statistical and systematic uncertainties, in \%, on the 
data-to-simulation scale factor \sfb\ from the \ptrel{} method for the MV1
tagging algorithm at 70\% efficiency.
Negligibly small uncertainties are indicated by dashes.
}
\end{center}
\end{table}

\begin{table}[htbp]
\begin{center}
   \scriptsize
   \setlength{\tabcolsep}{0.25pc}
   \begin{tabular}{l|ccccccccc}

\hline\hline
& \multicolumn{9}{c}{Jet $\pt\ \mathrm{[GeV]}$} \\
Source                                  &20--30& 30--40 & 40--50 & 50--60 & 60--75 & 75--90 & 90--110 & 110--140 & 140--200 \\
\hline
Simulation statistics                   &  2.1 &  1.5 &  0.6 &  0.8 &  1.2 &  1.3 &  2.1 &  3.6 &  2.9 \\
Simulation tagging efficiency           &    - &    - &    - &    - &    - &  0.2 &  0.2 &  0.4 &  0.5\\
Modelling of $g \to b\bar{b}$           &    - &    - &    - &    - &    - &  0.1 &    - &    - &  0.2 \\
Modelling of $g \to c\bar{c}$           &    - &    - &    - &    - &    - &    - &  0.2 &    - &  0.2 \\
$b$-hadron direction modelling          &  0.5 &    - &    - &  0.3 &  0.1 &  0.2 &    - &  0.4 &  1.2 \\
$b$-fragmentation fraction              &  2.1 &  1.8 &  1.8 &  2.1 &  1.2 &  1.8 &  2.5 &  0.9 &  0.8 \\
$b$-fragmentation function              &    - &  0.2 &    - &    - &    - &  0.3 &  0.4 &  0.5 &  0.8 \\
$b$-decay branching fractions           &    - &    - &    - &    - &    - &    - &    - &  0.1 &    - \\
$b$-decay \pstar\ spectrum              &  0.2 &  0.3 &    - &  0.1 &  0.3 &  0.1 &  0.3 &  0.1 &  0.1 \\
Charm-light ratio                       &  0.2 &    - &    - &  0.1 &  0.1 &  0.2 &    - &  0.3 &    - \\
Muon \pt{} spectrum                     &  3.1 &  2.0 &  1.4 &  0.9 &  0.4 &  0.4 &  0.2 &  0.6 &  2.7 \\
Fake muons in $b$ and $c$ jets          &  0.3 &  0.1 &    - &  0.2 &  0.1 &  0.4 &  0.2 &  0.5 &  0.3 \\
\ptrel{} cut variation                  &  0.7 &  1.3 &  1.0 &  1.4 &    - &  0.2 &  2.7 &  2.1 &  1.8 \\
Jet energy resolution                   &  1.7 &  0.9 &  2.8 &    - &    - &  0.2 &  0.8 &  2.3 &  1.3 \\
Jet energy scale                        &  0.1 &  0.2 &  0.6 &    - &    - &  0.6 &  0.4 &  1.0 &  0.4 \\
Semileptonic correction                 &  0.4 &  0.4 &  1.1 &  0.5 &  0.7 &  1.3 &  0.2 &  2.1 &  0.6 \\
Pile-up $\langle\mu\rangle$ reweighting & -    &  0.1 &    - &    - &  0.1 &    - &  0.4 &  0.4 &    - \\
Extrapolation to inclusive $b$ jets     &  4.0 &  4.0 &  4.0 &  4.0 &  4.0 &  4.0 &  4.0 &  4.0 &  4.0 \\
\hline
Total systematic uncertainty            &  6.2 &  5.3 &  5.7 &  4.9 &  4.4 &  4.9 &  5.9 &  6.8 &  6.4 \\
Statistical uncertainty                 &  2.0 &  1.4 &  1.8 &  2.6 &  1.5 &  2.2 &  3.4 &  2.7 &  4.1 \\
\hline
Total uncertainty                       &  6.5 &  5.5 &  5.9 &  5.6 &  4.7 &  5.4 &  6.9 &  7.3 &  7.6 \\
\hline \hline 
\end{tabular}
\caption{
\label{tab:systs_system8}Relative statistical and systematic uncertainties, in \%, 
on the data-to-simulation scale factor \sfb\ from the system8 method for the MV1
tagging algorithm at 70\% efficiency.
Negligibly small uncertainties are indicated by dashes.
}
\end{center}
\end{table}

\subsubsection*{Simulation statistics}

The limited size of the simulated event samples results in statistical fluctuations on the \ptrel{} templates in the case of the \ptrel{}
analysis, and in statistical uncertainties on the system8 correlation factors in the system8 analysis.

The effect from limited template statistics in the \ptrel{} analysis is assessed through pseudo-experiments.
In the system8 analysis, the limited statistics available for the samples used to estimate the correlation
factors is accounted for using an extra contribution to the fit $\chi^2$: 
\begin{equation}
  \chi^{2} \to \chi^{2\prime} = \chi^{2} + (\vec{\alpha}-\vec{\alpha}_{0})^{\mathrm{T}} V^{-1}(\vec{\alpha}-\vec{\alpha}_{0}).
  \label{eq:system8_mcstat}
\end{equation}
Here, $\vec{\alpha}$ represent eight additional fit parameters, $\vec{\alpha}_{0}$ are their estimates, and $V$ is the
corresponding covariance matrix.
To estimate the contribution to the uncertainty from this limited statistics, the uncertainty for the fit without this
addition is subtracted quadratically from the uncertainty for the fit including it.

In addition, the limited simulation statistics result in an uncertainty on the
denominator in the scale factor expression, denoted \emph{simulation tagging efficiency}
in Tables~\ref{tab:systs_ptrel} and~\ref{tab:systs_system8}. 

\subsubsection*{Modelling of gluon splitting to $b\bar{b}$ and $c\bar{c}$}

As the properties of jets with two $b$ or $c$ quarks inside (originating e.g. from gluon splitting)
are different from those containing only a single $b$  or $c$ quark, a possible mismodelling
of the fraction of double-$b$ or double-$c$ jets in simulation has to be taken into account.
Jets which have two associated $b$ quarks or $c$ quarks are either given a weight of zero or a weight of two 
(effectively removing or doubling the double-$b$ or double-$c$ contribution) when constructing the \ptrel{} templates
and system8 correlation factors. 
This uncertainty on $c$-jet production is the dominant one for the \ptrel{} analysis 
at high jet \pt{} because gluon splitting is a larger contribution to the total charm production in this regime.
Also, in this region the \ptrel{} variable has a reduced discriminating power and hence the fit 
becomes more sensitive to a change in the shape of the $c$-jet template.
The systematic effects tend to cancel in the ratios that define the system8 correction factors, leading to much reduced
systematic uncertainties for system8.

\subsubsection*{$b$-hadron direction modelling}

Both the \ptrel{} and the system8 analyses make use of the momentum of the
associated muon transverse to the combined muon-plus-jet axis, where the muon
plus jet axis is a measure of the $b$-hadron direction. A different jet
direction resolution in data and simulation would therefore affect both analyses. 
This is accounted for by smearing the calorimeter jet direction by 0.004 in $\phi$ 
and 0.008 in $\eta$, as discussed in Section~\ref{sec:ptrel}.
Both analyses use the result from the unsmeared template distributions as the central value and
treat the full difference with the result from the smeared distributions as a systematic uncertainty.

\subsubsection*{$b$-quark fragmentation}

An incorrect modelling of the $b$-quark fragmentation in simulation can affect the momentum spectrum of the muons from $b$ decays 
and thus alter which muons pass the selection criteria.
To investigate the impact of fragmentation on the data-to-simulation scale factor, the \ptrel{} templates and system8 correlation factors
have been rederived on a simulated sample 
where the $b$ fragmentation function was reweighted so that the average fraction 
of the $b$-quark energy given to the $b$ hadron was varied by 5\%.

The production fractions of the various $b$-flavoured hadrons have been measured both at LEP and the Tevatron~\cite{pdg2010,TeVbflavours}, and the results for $b$ baryon production are only compatible at the 2~$\sigma$ level. The production fractions in the simulated samples used in this paper are in reasonable agreement with the fractions as measured by LEP.
A systematic uncertainty is evaluated by considering the difference in the result obtained by reweighting all of 
the events so that the distribution of hadron species matches the measured Tevatron values.

\subsubsection*{$b$-hadron decay}

The spectrum of the muon momentum in the $b$-hadron
rest frame, denoted as \pstar, directly affects the shape of the \ptrel{}
distribution for $b$ jets. 
The \pstar\ spectrum has two components, direct $b \to \mu + X$ decays and 
cascade 
$b \to c/\overline{c} \to \mu + X$ decays. 
Their branching fractions are 
$\text{BF}(b\to \ell X) = (10.69 \pm 0.22)\%$ and 
$\text{BF}(b\to c/\overline{c} \to \ell X) = (9.62  \pm 0.53)\%$, 
respectively~\cite{pdg2010},
giving the ratio
$\text{BF}(b\to  \ell X) / \text{BF}(b\to c/\overline{c} \to  \ell X) = 1.11 \pm 0.07$, 
where $\ell$ denotes either
a muon or an electron.
This ratio of branching fractions has been varied within the quoted uncertainty.
To investigate the effect of variations of the \pstar\ spectra, a weighting 
function has been applied to the \pstar\ spectrum of muons from the direct $b \to \mu + X$ decay.
This weighting function has been derived by comparing the direct \pstar{} spectrum of $b \to \ell + X$ decays
in PYTHIA as used in the analysis with the corresponding spectrum measured in Ref.~\cite{delphipstare}.

\subsubsection*{Charm-to-light ratio \label{sec:syst_clratio}}

Both the \ptrel{} and system8 methods are sensitive to the relative fractions of $c$ and light-flavour jets in the simulation.
As the \ptrel{} templates for $c$ and light-flavour jets have a very similar shape, the \ptrel{} fits can become unstable
if both components are allowed to vary freely in the fit. Therefore the fits to the \ptrel{} templates are performed with the
ratio of the charm and light-flavour fractions fixed to the simulated value. In
the system8 method, the ratios of the $c$-jet
and light-flavour-jet fractions in the $n$ and $p$ samples are also fixed to their simulated values.
The relative fractions of $c$ and light-flavour jets in these samples affect the correction factors related to non-$b$ jets
($\alpha_2$, $\alpha_3$, $\alpha_4$ and $\alpha_8$).
In both analyses the impact of the constrained charm-to-light ratio has been addressed by varying the ratio up and down
by a factor of two. The charm-to-light ratio variation is one of the dominant systematic uncertainties in the \ptrel{} 
analysis, especially in the highest \pt\ bin where the $b$- and $c$-jet templates start to look very similar.
The system8 analysis, which only relies on the \ptrel{} cut to increase the fraction of $b$ jets in the sample, is less affected.

\subsubsection*{Fake muons in $b$ jets}

The \ptrel{} templates and system8 correlation factors are obtained from the simulated $\mu$-jet sample where a muon 
with $\pt>3\GeV$ is required at the generator level.
This filter suppresses $b$ jets containing a reconstructed muon from other sources compared to those where the muon originated from a $b$ decay
(about 30\% of the muons in this category, denoted as ``fake'' muons here, are indeed real muons produced by in-flight decays of light mesons;
the remainder are not true muons). 
The fraction of fake muons in this sample is therefore likely to be lower than in data.
As this could potentially impact the \ptrel{} $b$ template shapes, the \ptrel{} measurement 
has been repeated with the fake muon fraction in the $b$ template increased by a factor of three, which 
was found to have a negligible impact on the final result. In the system8 measurement, the
fake muon contribution is varied in both $b$ and $c$ jets, again with a negligible impact on the final result.

\subsubsection*{\ptrelBM\ light-flavour-template contamination}

In the \ptrel{} method, the templates for light-flavour jets are obtained from a light-flavour-enriched data sample.
A measurement bias can arise from $b$-jet contamination in the light-flavour template.
This $b$-jet contamination in the light-flavour template is estimated from
simulation to be between 4\% and 10\%, depending on the jet \pt\ bin. 
The bias introduced by this contamination is corrected for in the final result, and the result
of a 25\% relative variation of the $b$-jet contamination is taken as a systematic uncertainty.

\subsubsection*{\ptrelBM\ cut variation} 

The system8 analysis uses a cut on the \ptrel{} of the muon associated to the jet to arrive at a sample with enhanced
heavy flavour content. The \ptrel{} cut, which is nominally placed at 700~\MeV, was varied between 600~\MeV{} and 800~\MeV, and the
difference with respect to the nominal result was taken as a systematic uncertainty.

\subsubsection*{Jet energy scale and resolution}

A jet energy scale in simulation that is different from that in data would bias the $\pt$ spectrum of the 
simulated events used to extract the \ptrel{} templates and the system8 correlation factors.
The systematic uncertainty originating from the jet energy scale is obtained by scaling the $\pt$ of each jet in the 
simulation up and down by one standard deviation, according to the jet energy scale uncertainty~\cite{PERF-2012-01}.

To estimate the systematic uncertainty from the jet energy resolution, a smearing has been applied to the jet energies in simulation, 
corresponding to the jet energy resolution uncertainties as described in~\cite{PERF-2011-04}.

\subsubsection*{Semileptonic correction}

Both the \ptrel{} and system8 analyses measure the $b$-jet tagging efficiency of jets containing semileptonically decaying $b$ hadrons.
Both analyses therefore make an extra jet energy scale correction, described in Section~\ref{sec:semilep_jes_corr}, to correct the jet energy
to the inclusive $b$-jet scale. The uncertainty on this correction, which amounts to about 2\%, reflects how sensitive the correction 
is to the modelling of $b$ jets in the simulation, the correlation between the correction and the 
$b$-tag output weights and how well the correction agrees in data and simulation. The uncertainty on the semileptonic
correction is propagated through the \ptrel{} and system8 analyses as a systematic uncertainty. Systematic sources which affect both the semileptonic correction and the \ptrel{} templates or system8 correlation factors are varied in a correlated manner.

\subsubsection*{Pile-up $\langle \mu \rangle$ reweighting}

Simulation studies show that the impact on the $b$-tagging performance from the change
in pile-up conditions during 2011 is relatively small
compared to the precision of the \ptrel{} and system8 analyses. The change in light-flavour-jet rejection
at fixed $b$-jet tagging efficiency exceeds 5\% only for the tightest operating points.
With the $\langle \mu \rangle$ distribution in simulated events reweighted to
that in the 2011 data, as described in Section~\ref{sec:samples}, only the 3\%
relative uncertainty on the $\langle \mu \rangle$ scale factor applied prior to
the reweighting procedure affects the pile-up in simulated events.
It is accounted for by repeating the analyses after changing the scale factor
accordingly.

\subsubsection*{Extrapolation to inclusive $b$ jets}
\label{sec:syst_inclsf}

The \ptrel{} and system8 methods only measure the $b$-jet tagging efficiency in data for $b$ jets with a semileptonic $b$-hadron decay.
The $b$-jet tagging efficiency is different for these $b$ jets than for inclusive $b$ jets
(in simulated events their ratio decreases from about 1.1 for the lowest \pt{} jets,
to values agreeing with unity within about 1\% for jet $\pt \gtrsim 100\GeV$).
This is because the charged particle multiplicity is different in semileptonic and hadronic $b$-hadron decays,
and, most importantly, the jets used in the \ptrel{} and system8  analyses always contain a high-momentum and typically well-measured muon track
whereas the hadronic $b$ jets do not.
However, assuming that the simulation adequately models 
the relative differences in $b$-jet tagging efficiencies between semileptonic and hadronic $b$ jets, 
the same data-to-simulation scale factor is valid for both types of jets.

To investigate the validity of this assumption, the data-to-simulation scale factor was measured separately for jets with and
without muons using a high purity sample of $b$ jets in \ttbar\ dilepton events.
The ratio of the data-to-simulation scale factors for jets with and without muons was found to be consistent with unity for
all tagging algorithms and operating points. The uncertainty on the measurement, approximately 4\%, is assigned
as a systematic uncertainty on the data-to-simulation scale factors obtained with the muon-based methods.
However, due to the limited number of $b$ jets with a semileptonic $b$-hadron decay in the \ttbar\ dilepton events,
this analysis was not performed in bins of jet \pt{}.

In order to independently investigate effects that can potentially lead to different relative  
$b$-jet tagging efficiencies in semileptonic and hadronic $b$ jets in the simulation
-- and especially their jet \pt{} dependence -- properties of the $b$-hadron production process and
semileptonic and hadronic $b$-hadron decays have been studied. The procedure used is analogous to 
that used for the calibration of the $c$-jet tagging efficiency, which is described in 
a more formal way in Section~\ref{sec:inclSF}.
For the following quantities, the simulation has been adjusted to available measurements~\cite{PDG2014}: 
$b$-hadron production fractions, branching fractions of semileptonic $b$-hadron decays, relative branching fractions
of the dominant exclusive semileptonic $b$-hadron decays and topological branching fractions of $c$-hadron decays.
These adjustments result in very small shifts (well below 1\%) and, given the uncertainties of the 
measurements, negligible uncertainties on the ratio of tagging efficiencies for jets with semileptonic
and inclusive $b$-hadron decays. 
The effect of gluon splitting to $b\bar{b}$ also has a direct impact on the $b$-jet tagging efficiency
as predicted by the simulation. The variation mentioned before (assigning weights of zero or two)
results in a small change of the $b$-jet tagging efficiency in simulation, with some -- almost linear -- dependence
on jet \pt{}.
The charged particle multiplicity spectrum of $b$-hadron decays has a direct impact on the $b$-jet tagging efficiency,
and variations of the charged particle multiplicity spectrum of hadronic $b$-hadron decays 
do not cancel in the ratio of relative tagging efficiencies.
Since no dedicated measurements of the charged particle multiplicity spectrum of (hadronic) $b$-hadron decays
for an admixture of $b$ hadrons as present in high energy collisions
are available\footnote{Only the total number of charged particles in $b$-hadron decays has been measured~\cite{PDG2014}, but
not their spectrum.
Event-wise charged particle multiplicity spectra of $\Upsilon(4S)$ decays, corresponding to
an admixture of $B^+$-$B^-$ and $B^0$-$\bar{B^0}$ meson pairs, have been
measured in Ref.~\cite{CLEO:Multiplicities}.}, 
the potential effect has been studied by adjusting the charged particle spectrum to the one
predicted by \EvtGen~\cite{Lange:2001uf}.\footnote{Other scenarios have also been studied for cross-checks.}
This is the dominant effect on the extrapolation to inclusive $b$ jets; the impact from these variations on the ratio of 
the relative differences in $b$-jet tagging efficiencies between semileptonic and hadronic $b$ jets
is typically a few percent (depending on the scenario). The jet \pt{} dependence of this effect is negligible.
Since these studies do not show a significant jet \pt{} dependence, the uncertainty of 4\% from the measurements 
of data-to-simulation scale factors in jets with and without muons is applied
to all jet \pt{} bins.

\subsection{Results}
\label{sec:results}

\begin{figure}[htbp]
  \subfloat[]{\label{fig:ptrel_system8_MV1_a}\includegraphics[width=0.49\textwidth]{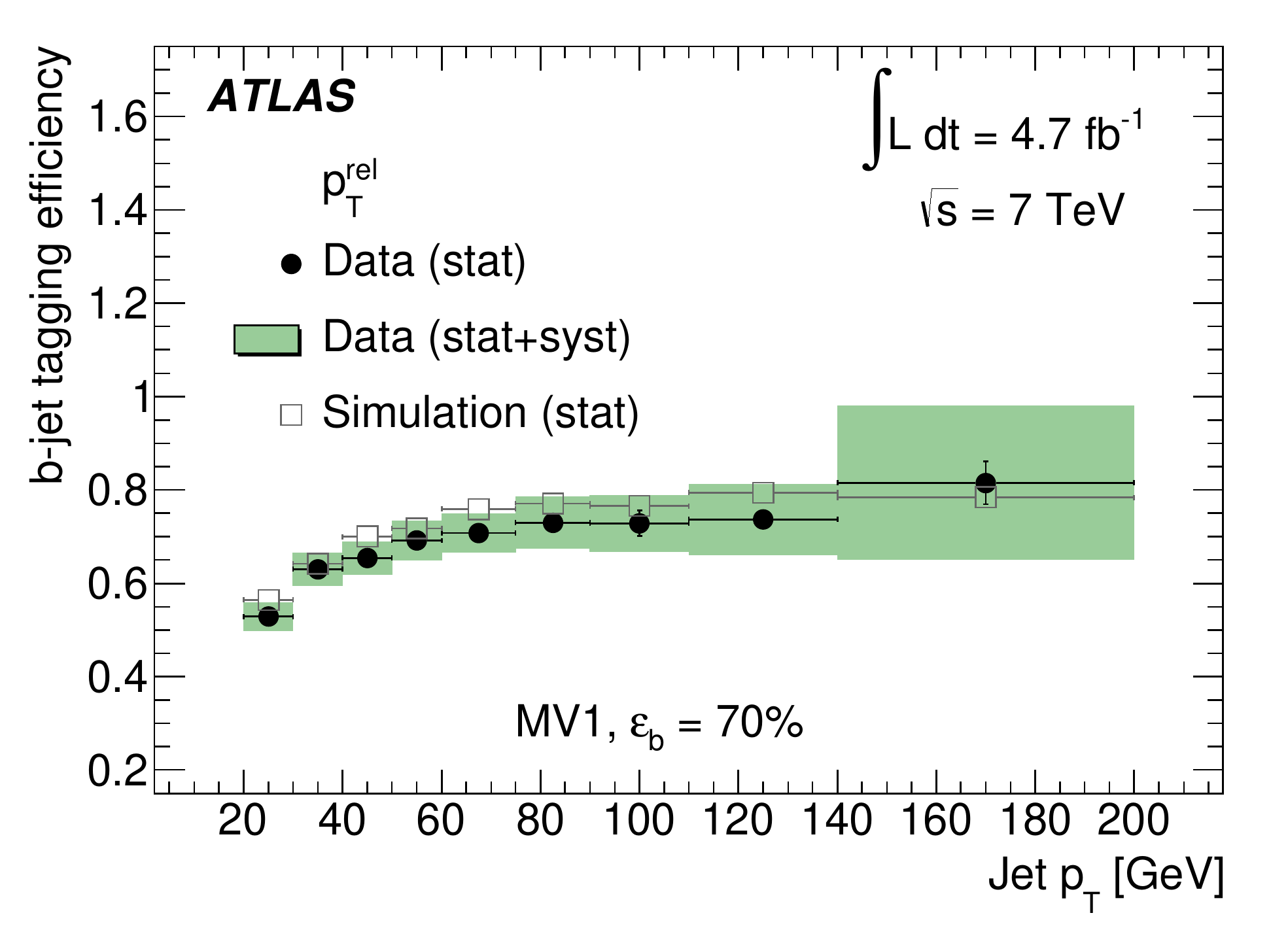}}
  \subfloat[]{\label{fig:ptrel_system8_MV1_b}\includegraphics[width=0.49\textwidth]{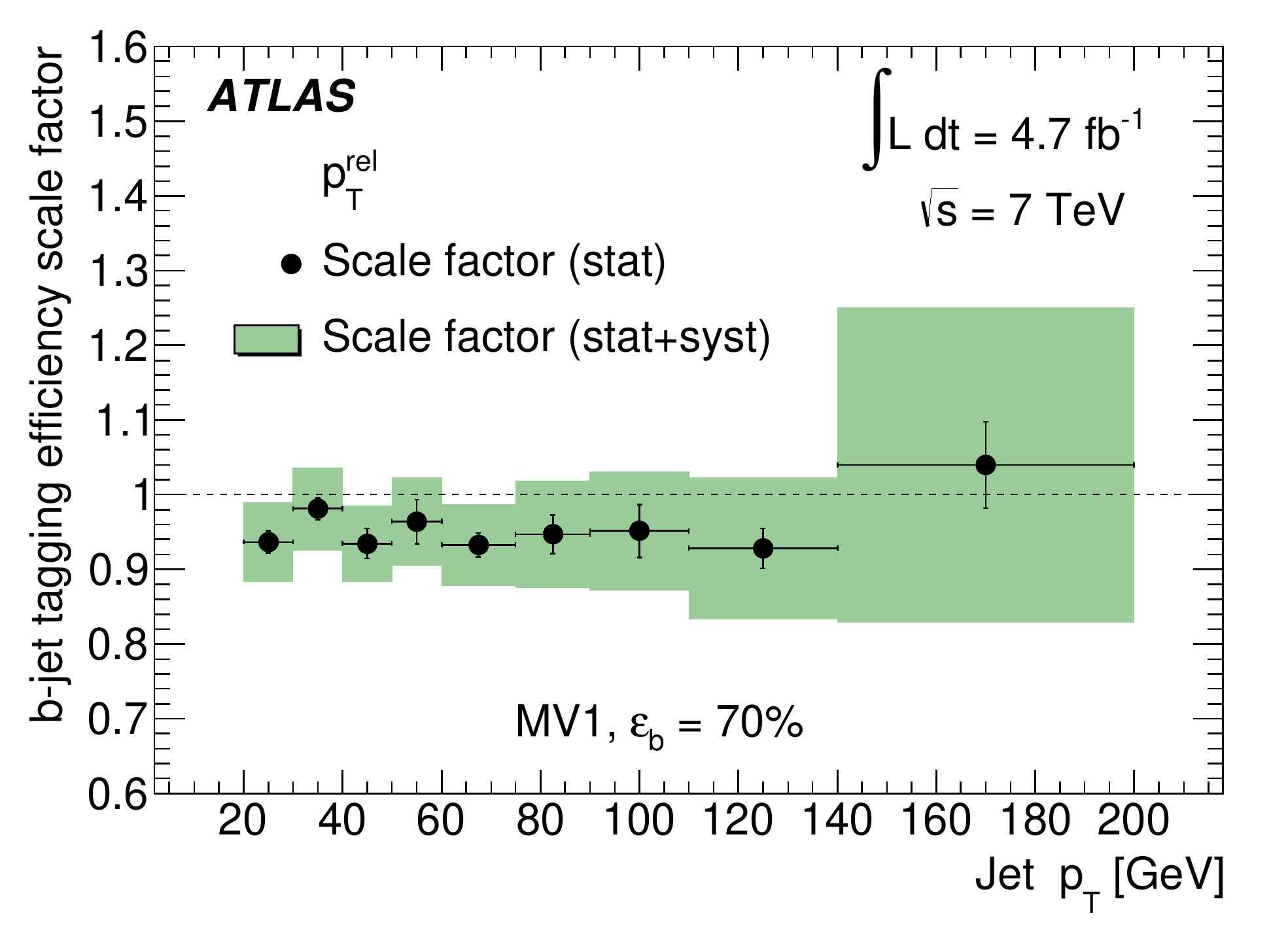}}\\
  \subfloat[]{\label{fig:ptrel_system8_MV1_c}\includegraphics[width=0.49\textwidth]{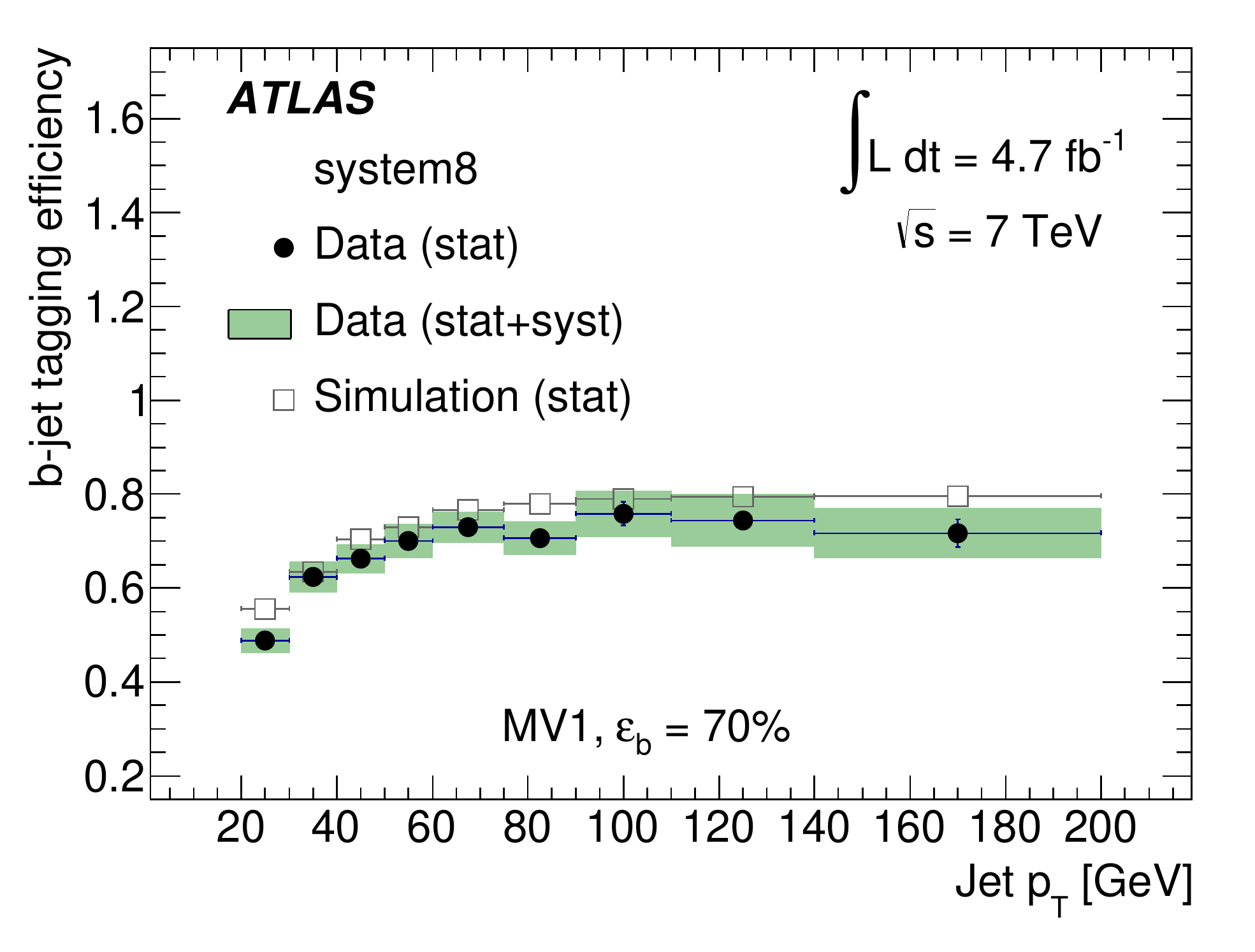}}
  \subfloat[]{\label{fig:ptrel_system8_MV1_d}\includegraphics[width=0.49\textwidth]{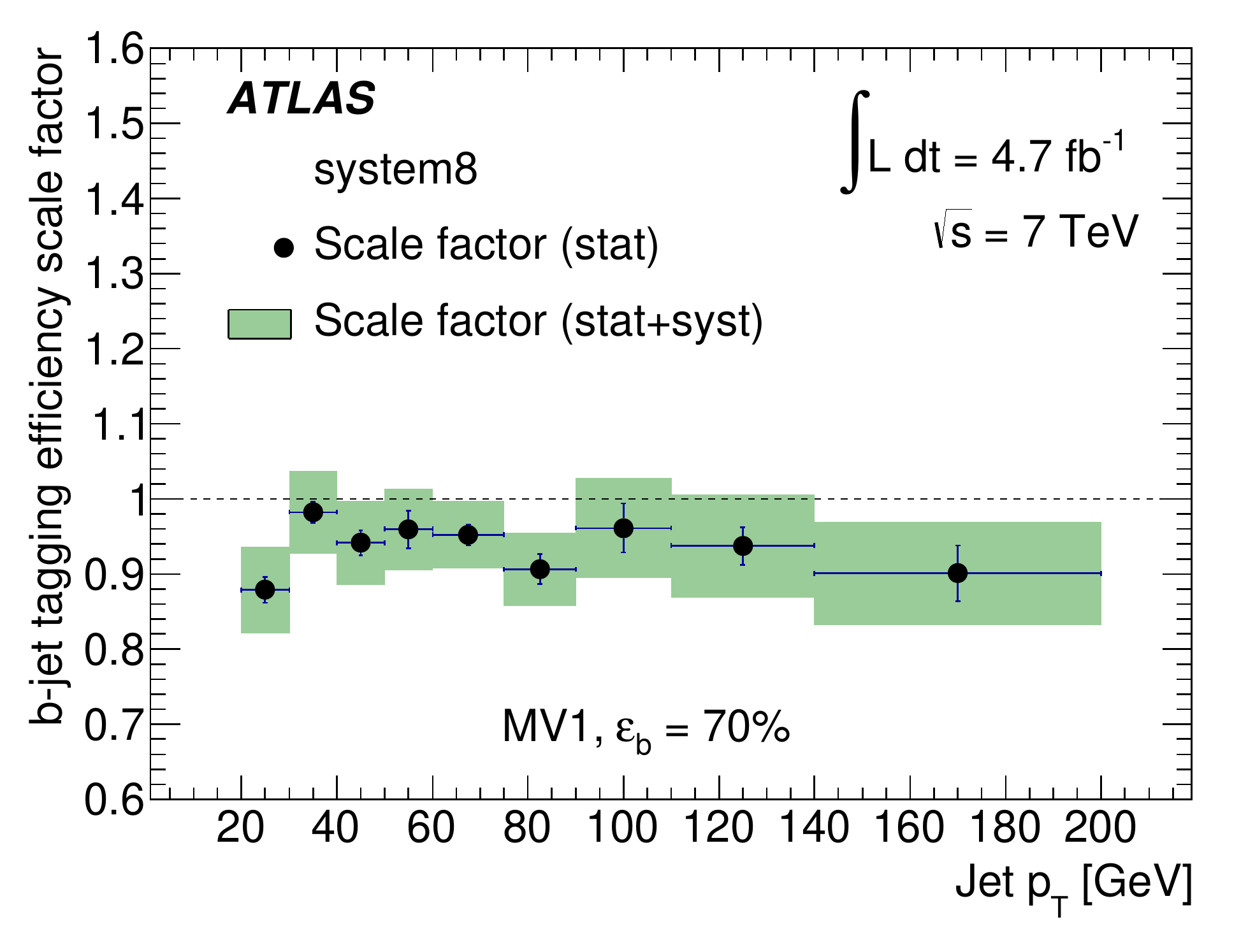}}
  \caption{The $b$-jet tagging efficiency in data and simulation (left) and data-to-simulation scale factor (right)
    for the MV1 tagging algorithm at 70\% efficiency obtained with the \ptrel{} (top) and system8 (bottom) methods.}
  \label{fig:ptrel_system8_MV1}
\end{figure}

The $b$-jet tagging efficiency measured in data using the \ptrel{} and system8 methods, the corresponding values from simulation and the resulting data-to-simulation
scale factors for the MV1 tagging algorithm at  70\% efficiency are shown in Fig.~\ref{fig:ptrel_system8_MV1}.
As the jets selected for the \ptrel{} and system8 measurements are different, 
the fraction of $b$-tagged jets are not necessarily equal.
On average the scale factor is about 0.95, which is typically one standard deviation lower than unity, and has no strong dependence on \pt.
It has to be kept in mind that the dominant systematic uncertainty from the extrapolation of the scale factors to an inclusive $b$-jet sample
is fully correlated between \pt{} bins.

\section{$b$-jet tagging efficiency calibration using $t\bar t$-based methods }
\label{sec:beff_ttbarbased}

The calibration methods described in the previous sections are  based on dijet events. 
At the LHC, the large $\ttbar$ production cross section provides an alternative source of events enriched in $b$ jets. 
With the large integrated luminosity of $\unit[4.7]{\ifb}$ collected during 2011,
the methods based on $\ttbar$ selections have become competitive
for the first time with the muon-based $b$-jet calibration methods described in Section~\ref{sec:beff_mubased}.
Compared to the muon-based methods, the $\ttbar$-based methods provide a $b$-tagging calibration measurement in an inclusive 
jet sample rather than a sample of semileptonic $b$ jets and cover a larger range in \pt.

In the following, four calibration methods are presented.
The tag counting method fits the multiplicity of $b$-tagged jets in $\ttbar$ candidate events 
while the {kinematic selection} method measures the $b$-tagging rate of the leading jets in the $\ttbar$ signal sample.
The kinematic fit method uses a fit of the $\ttbar$ event topology to extract a highly purified sample of $b$ jets
from which the $b$-jet tagging efficiency is obtained.
Finally, the combinatorial likelihood method improves the precision offered by
the kinematic selection method by exploiting the kinematic correlations between the jets in the event.

The kinematic selection method is applied to both the single-lepton and dilepton decay channels, 
whereas the tag counting method is presented only in the single-lepton channel.
By construction the kinematic fit method is restricted to the 
single-lepton channel. The combinatorial likelihood method is applied only to the dilepton channel.

\subsection{Simulation samples, event selections and background estimates}
\label{sec:selection}

The $\ttbar$ signal is simulated using MC@NLO interfaced to HERWIG
(\Powheg{}~\cite{bib:powheg,bib:powheg2} interfaced to \Pythia{}~\cite{pythia2}
in the case of the combinatorial likelihood analysis) with the mass of the 
top quark set to $\unit[172.5]{\GeV}$. The cross section is normalised to the approximate NNLO 
calculation from \textsc{Hathor} 1.2 \cite{bib:hator} using the MSTW 2008 $90\%$ parton
distribution function (PDF) sets \cite{bib:mstw2008},
incorporating PDF+$\alpha_{S}$ uncertainties according to the MSTW prescription
\cite{bib:mstw2009} cross checked with the approximate NNLO calculation of
Ref.~\cite{bib:nnlocacciari} as implemented in \textsc{Top++1.0} \cite{bib:topplusplus}.

For the main backgrounds, which consist of  $W/Z$ boson production in 
association with multiple jets, \Alpgen{} v2.13~\cite{bib:alpgen} is used, which implements the 
exact LO matrix elements for final states with up to six partons. Using the LO PDF set CTEQ6L1 
\cite{bib:cteq6l}, the following backgrounds are generated: $W$+jets events with up to five partons, 
$Z/\gamma^{\ast}$+jets events with up to five partons,
and diboson $WW$+jets, $WZ$+jets and $ZZ$+jets events. 
The MLM matching scheme~\cite{bib:alpgen,bib:MLM} of the \Alpgen{} generator is used to 
remove overlaps between the $n$ and $n + 1$ parton samples.
In the combinatorial likelihood analysis, \Sherpa{} v1.4.1~\cite{bib:sherpa} with
the CT10 PDF set~\cite{bib:ct10} is used instead for both $W+$jets and $Z+$jets
processes.

For all but the diboson processes, separate samples are generated that include $b\bar{b}$ 
quark pair production at the matrix element level. In addition, for the 
$W$+jets process,  separate samples containing $Wc$+jets and $Wc\bar{c}$+jets events are produced.
The same program employed for the generation of the $\ttbar$ signal (MC@NLO for
most analyses, \Powheg{} for the combinatorial likelihood analysis) is used for the
production of single-top $s$- and $Wt$-channel backgrounds.
\AcerMC{} is used for $t$-channel production.
The uncertainty due to the choice of $\ttbar$ generator is evaluated
by comparing the predictions of MC@NLO with those of \Powheg{} interfaced to HERWIG or PYTHIA.

\subsubsection{Event selection} \label{sec:sel_events}

Events in the single-lepton and dilepton $\ttbar$ channels are triggered using a high-\pt{} single-electron or single-muon trigger.
In addition to the objects described in Section~\ref{sec:intro} the $\ttbar$ analyses require isolated electrons and muons,
as well as missing transverse momentum.

In all $\ttbar$ analyses, both in the single-lepton and dilepton channels, the $b$-jet tagging 
efficiency measurement is performed in a sample comprising all electron and muon
combinations ($e$+jets and $\mu$+jets or $ee$, $\mu\mu$ and $e\mu$), including
muons and electrons resulting from $\tau$ lepton decays.

\subsubsection{Selection of the single-lepton sample} \label{sec:sel_ljets}

In the single-lepton channels ($e$+jets and $\mu$+jets), the following event selection 
is applied:
\begin{itemize}
  \item The appropriate single-electron (with trigger thresholds at 20 or 22
    \GeV, depending on the data taking period) 
    or single-muon trigger (with trigger threshold at $\unit[18]{\GeV}$) has fired.
  \item The event contains exactly one reconstructed lepton with
    $\pT > \unit[25]{\GeV}$ ($e$) or $\pT > \unit[20]{\GeV}$ ($\mu$), matching 
    the corresponding trigger object.
  \item The lepton must be isolated from any jet activity. Beyond requiring that
    the nearest jet must be separated by $\Delta R > 0.4$, the summed
    calorimeter transverse energies deposited in a cone of size $\Delta R < 0.2$
    around electron directions must be less than 3.5~\GeV; for muons this
    maximum value is 4~\GeV, in a cone of size $\Delta R < 0.2$, and in addition
    the summed track $\pt$ in a cone of size $\Delta R < 0.3$ must be less than
    4~\GeV. In all cases the contribution from the lepton itself is subtracted.
  \item The missing transverse momentum~\cite{PERF-2011-07} is required to be $\met >
    \unit[30]{\GeV}$ ($\met > \unit[20]{\GeV}$) in the $e$+jets ($\mu$+jets)
    channel and the transverse mass is required to be $m_{\mathrm{T}}(l\nu)>
    \unit[30]{\GeV}$ ($m_{\mathrm{T}}(l\nu) > \unit[60]{\GeV} - \met$) in the
    $e$+jets ($\mu$+jets) channel. The transverse
    mass is defined as $m_{\mathrm{T}}(l\nu) = \sqrt{2 \pT\met(1-\cos\Delta\phi(l\nu))}$,
    where $\Delta\phi(l\nu)$ is the azimuthal angular difference between the
    directions of the selected lepton and the missing transverse momentum. These cuts reduce the contribution 
    from the multijet background.
  \item The event is required to have at least four jets with $\pt > \unit[25]{\GeV}$, $|\eta |<2.5$
    and (if tracks are associated with the jet) $\mbox{JVF} > 0.75$.
\end{itemize}

\subsubsection{Background estimation in the single-lepton channel} \label{sec:ljets_bkg}

The background in the single-lepton channel is expected to be around 30\%.
The dominant background arises from $W$ boson production with 
associated jets ($W+$jets). Its estimate is based on the prediction from Monte Carlo 
simulation, corrected with scale factors derived directly from data. The correction of 
the overall normalisation is obtained with a charge asymmetry method 
\cite{TOPQ-2011-08}.
The flavour composition of the $W$+jets sample is measured with 
a tag counting method \cite{TOPQ-2011-14}, which provides 
scale factors for $Wb\bar{b}/c\bar{c}$+jets, $Wc$+jets and $W$ with light-flavour jets events
used to correct Monte Carlo simulation predictions.

The second most important contribution to the background comes from multijet production
and is measured directly in data using the matrix method. This method relies on finding 
a relationship between events with real and non-prompt or fake leptons, as described in Ref.~\cite{TOPQ-2010-01}.
Estimates of other backgrounds processes such as single top, diboson and $Z+$jets 
production are obtained from Monte Carlo simulation. 

\begin{figure}[h] 
  \centering
  \subfloat[]{\label{fig:sel_kinsel_lj_jets_a}\includegraphics[width=0.49\textwidth]{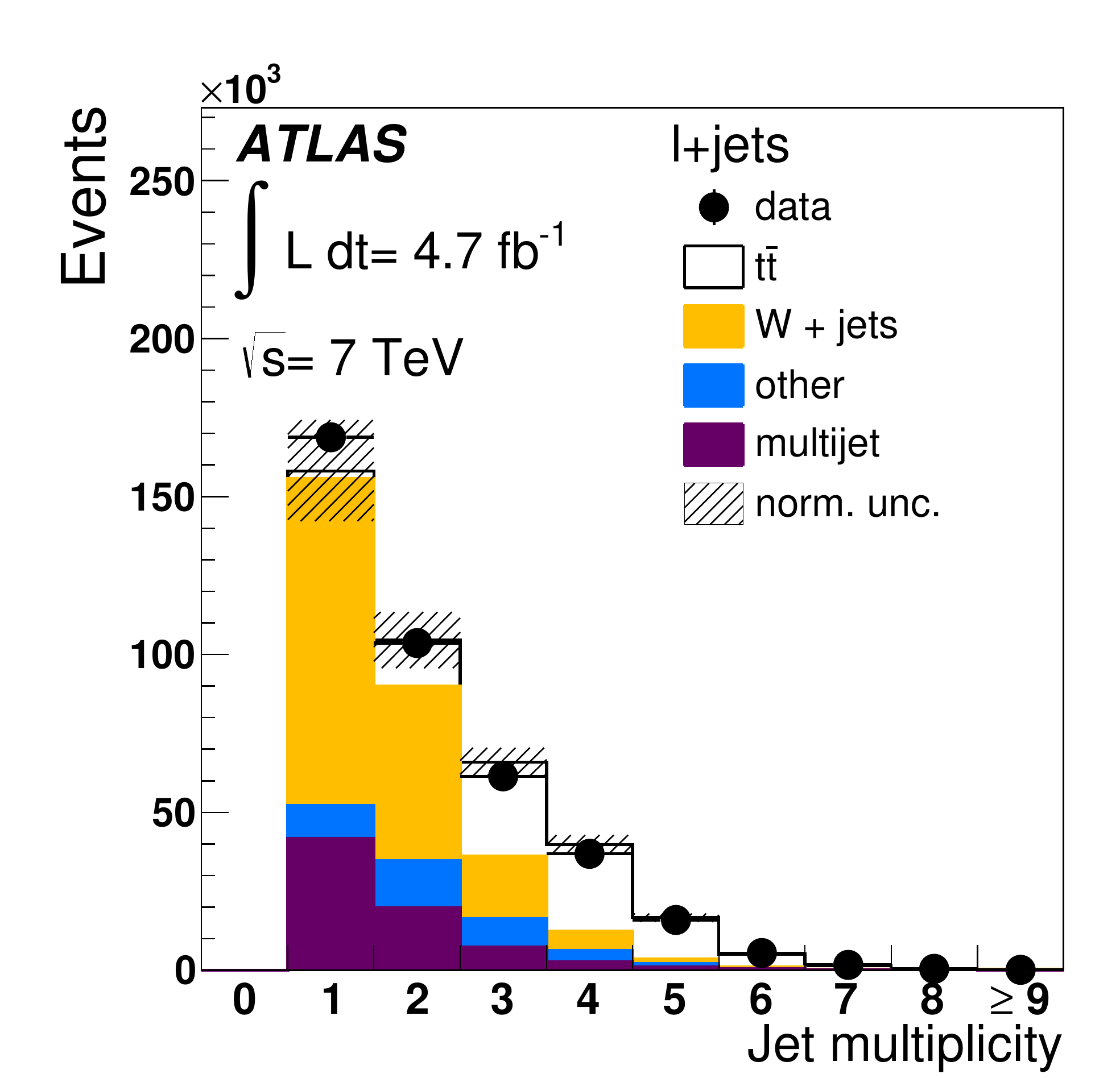}}
  \subfloat[]{\label{fig:sel_kinsel_lj_jets_b}\includegraphics[width=0.49\textwidth]{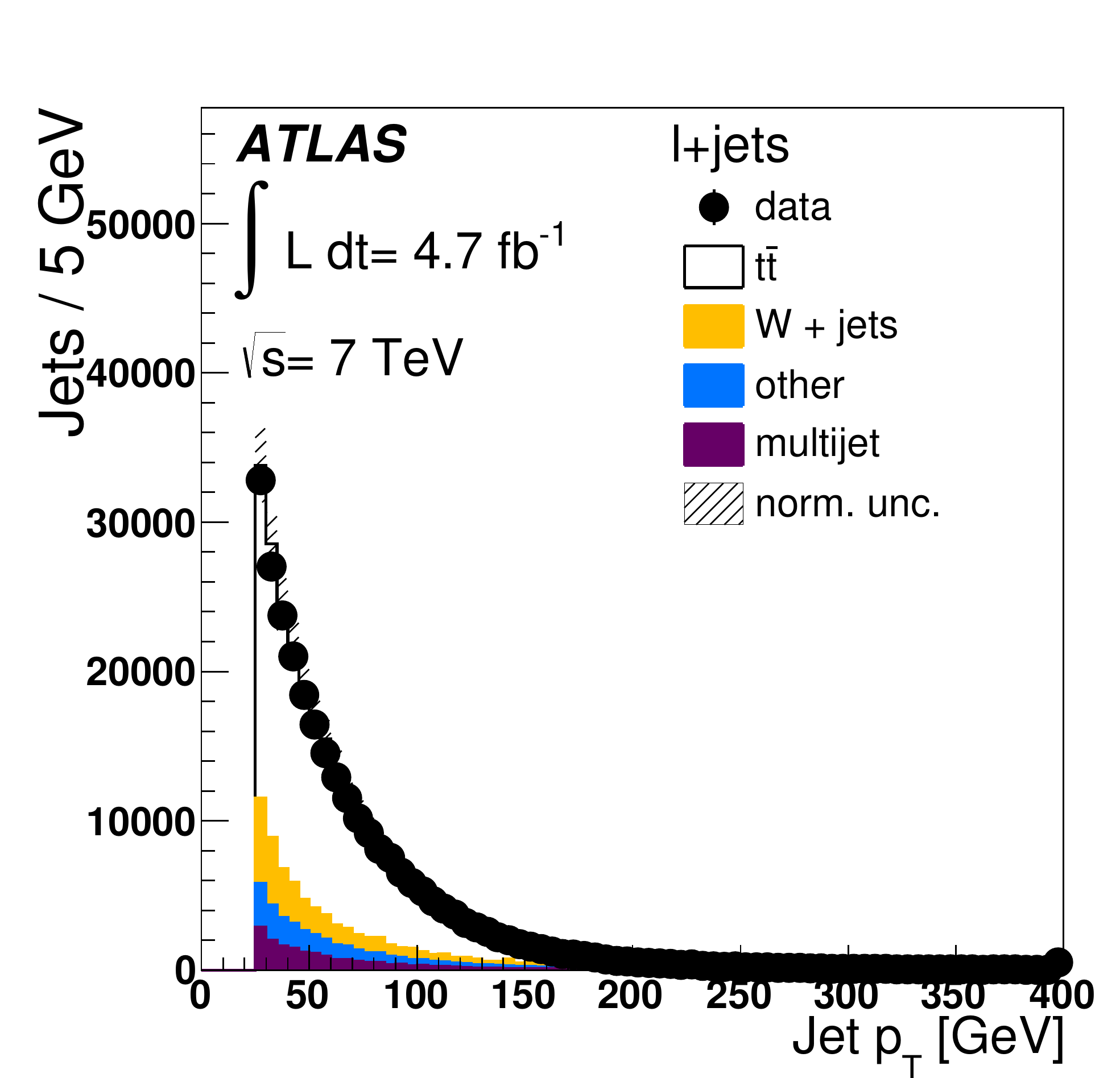}}
  \caption{Jet multiplicity in events passing the single-lepton selection (a) and the $\pT$ of all jets (b). 
    ``Other'' sums the contributions from $Z$+jets, single-top and diboson
    production. The normalisation uncertainties considered in the analysis are
    shown summed in quadrature; the last bin is an overflow bin.}
  \label{fig:sel_kinsel_lj_jets}
\end{figure}

Figure~\ref{fig:sel_kinsel_lj_jets} shows the jet multiplicity and jet $\pT$ distributions for data and simulated events. 
These distributions are sensitive to a correct description of the multijet and $W+$jets backgrounds, and they show a good 
agreement between the predicted background and signal contributions and data.

\subsubsection{Selection of the dilepton sample} \label{sec:sel_dilep}

A very pure sample of $\ttbar$ events with dileptonic decays ($ee$, $\mu\mu$ and $e\mu$)
can be obtained with the following event selection criteria:
\begin{itemize}
  \item A single-electron (trigger threshold at 20 or 22 \GeV{} depending on the data taking period) 
    or single-muon trigger (trigger threshold at $\unit[18]{\GeV}$) has fired.
  \item The event contains exactly two oppositely charged leptons ($ee$, $\mu\mu$ or
    $e\mu$), with the electron candidate satisfying $\pT > \unit[25]{\GeV}$, 
    and the muon candidate $\pT > \unit[20]{\GeV}$. At least 
    one of these must be associated with a lepton trigger object, and both
    leptons must satisfy the isolation requirements described in Section~\ref{sec:sel_ljets}.
  \item The event contains at least two jets with $\pT > \unit[25]{\GeV}$, $|\eta| < 2.5$ and 
    (if tracks are associated with the jet) $\mathrm{JVF} > 0.75$. The jet \pt{} threshold is lowered to 20~\GeV{} for
    the combinatorial likelihood method.
  \item In the $ee$ and $\mu\mu$ channels, to suppress backgrounds from
    $Z$+jets and multijet events the missing transverse momentum must satisfy
    $\met > \unit[60]{\GeV}$, and the invariant mass of the two leptons
    must differ by at least $\unit[10]{\GeV}$ from the $Z$ boson mass ($\Zboson$ mass veto): 
    $|m_{\ell\ell} - \mZ| > \unit[10]{\GeV}$.
    To suppress backgrounds from $\Upsilon$ and $\Jpsi$ decays, a low mass cut 
    of $m_{\ell\ell} > \unit[15]{\GeV}$ is applied.
  \item In the $e\mu$ channel, no $\met$ or $Z$ boson mass veto cuts are applied. However, 
    for all analyses except the combinatorial likelihood one,
    the scalar sum of the transverse momenta of the jets and of the charged leptons, $\HT(\ell,\mathrm{jets})$,
    must satisfy $\HT > \unit[130]{\GeV}$ to 
    suppress background from $Z (\rightarrow \tau\tau)$+jets production.
\end{itemize}

\subsubsection{Background estimation in the dilepton channel}
\label{sec:dilep_bkg}

The dilepton channel has a purity of 80\%--85\% depending on the lepton flavours. 
The dominant background originates from non-prompt or fake leptons 
from electron-like jets reconstructed as electrons or 
non-prompt leptons from a decay of a heavy-flavour hadron within a jet. 
This background receives contributions from $W$ boson production with associated jets,
$s$- and $t$-channel and $W t$ single top production,
the single-lepton decay of $\ttbar$ pairs and multijet events.

In most analyses, this background is estimated directly from data with a matrix 
method~\cite{TOPQ-2010-01} for each of the three channels separately; and
all background processes leading to two prompt leptons (diboson, $Z+$jets and single top in 
the $Wt$-channel) are directly taken from the simulation.
In the combinatorial likelihood analysis, the non-prompt and fake lepton
background is instead estimated from a sample where both leptons have the same
charge signs, and for which residual contributions predicted from simulation are
corrected.
The $Z+$jets background normalisation for each jet multiplicity bin in the
$e\mu$ channel is estimated using a data/MC normalisation factor obtained from
corresponding $ee$ and $\mu\mu$ data samples with $|m_{\ell\ell}-m_{Z}| < 10\GeV$;
in the combined $ee$ and $\mu\mu$ data
samples it is estimated from a sample where the suppression of the $Z$ resonance
is similarly replaced with a requirement $|m_{\ell\ell}-m_{Z}| < 10\GeV$.
  
\begin{figure}[!h]
  \centering
  \subfloat[]{\label{fig:sel_kinsel_dilep_jets_a}\includegraphics[width=0.48\textwidth]{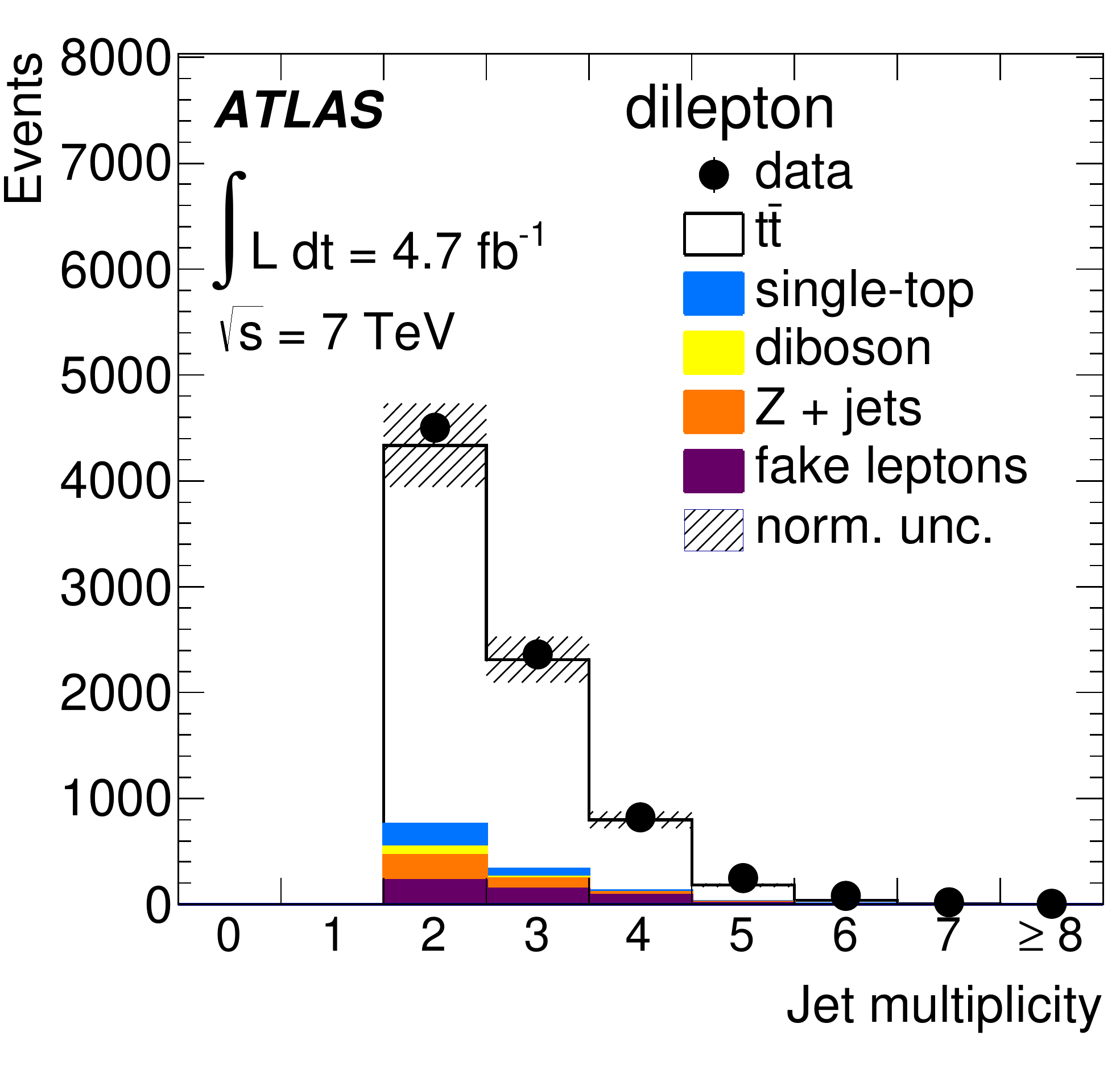}}
  \hfill
  \subfloat[]{\label{fig:sel_kinsel_dilep_jets_b}\includegraphics[width=0.48\textwidth]{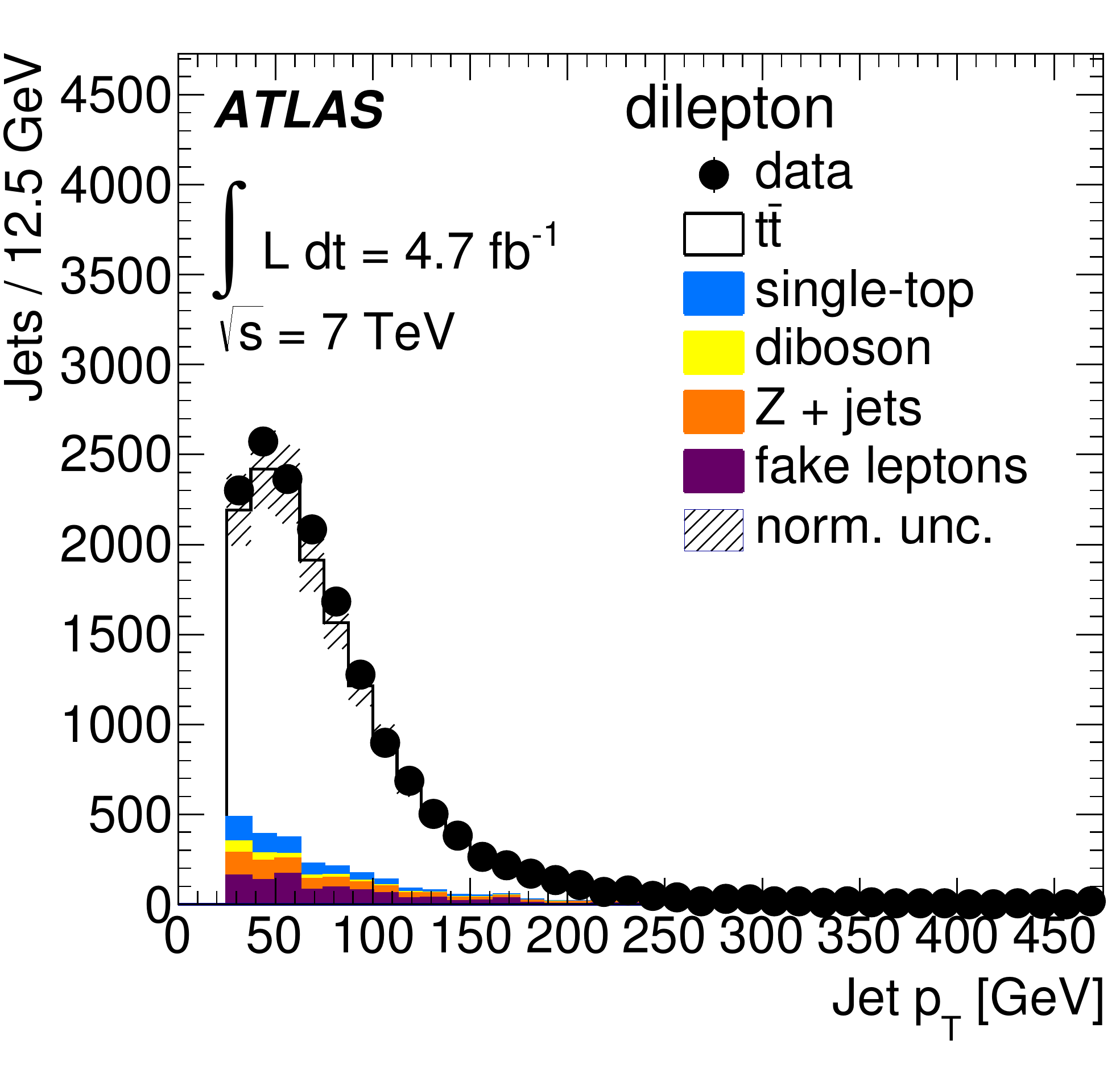}}
  \caption{Jet multiplicity (a) and the $\pT$ of all jets (b) in events passing the dilepton selection.
    The normalisation uncertainties considered in the analysis are
    shown summed in quadrature; the last bin is an overflow bin.}
  \label{fig:sel_kinsel_dilep_jets}
\end{figure}

Figure~\ref{fig:sel_kinsel_dilep_jets} shows the jet multiplicity and jet $\pT$ distributions for data
and simulated events. A good agreement between data and simulation can be seen.

\subsection{Tag counting method}
\label{sec:method_tagcount}

The tag counting method makes use of the fact that since the branching fraction of $t \to W b$ in the Standard 
Model is very close to unity, each $\ttbar$ event is expected to contain exactly two real $b$ jets. 
If there were no other sources of $b$ jets 
and if only $b$ jets were $b$-tagged, the expected number of events with two $b$-tagged jets 
would be $\varepsilon_b^2 N_{\mathrm{sig}} $ while the number of events with one $b$-tagged jet would be 
$\varepsilon_b \, (1-\varepsilon_b) \, 2N_{\mathrm{sig}}$, where $N_{\mathrm{sig}}$ is the number of 
$\ttbar$ signal events.

In reality, the number of reconstructed (or tagged) $b$ jets in a $\ttbar$ event will not necessarily be equal to 
two, since the $b$ jets from the top quark decays can be outside the detector acceptance, and additional 
$b$ jets can be produced through gluon splitting.
Moreover, $c$ jets and light-flavour jets, which come from the hadronic $W$-boson decay or initial or final state radiation,
can be tagged as $b$ jets. 
These effects are taken into account by evaluating the expected fractions, $F_{ijk}$, of events
containing $i$ $b$ jets, $j$ $c$ jets and $k$ light-flavour jets that pass the event selection.
The $F_{ijk}$ fractions are estimated from Monte Carlo simulation and are derived separately for the 
$\ttbar$ signal and the various background processes. The expected number of events with $n$ 
$b$-tagged jets is calculated as the sum of all these contributions. The $b$-jet tagging efficiency can be 
extracted by fitting the expected event counts to the observed counts.

The expected number of events with $n$ $b$-tagged jets, $<N_n>$, is calculated as
\begin{eqnarray}
  <N_n> & =  & \sum_{i,j,k} \bigg\{ ( \sigma_{\ttbar}\cdot \mathrm{BF} \cdot A_{\ttbar} \cdot \mathcal{L}
  \cdot F_{ijk}^{\ttbar} + N_{\rm bkg}\cdot F_{ijk}^{\rm bkg})
               \times \label{eqn:method_tagcount_lepjet} \\
  & & \sum_{i'+j'+k'=n}
  \binom{i}{ i'} \cdot{\varepsilon_b}^{i'} \cdot (1- \varepsilon_b)^{i-i'}\cdot
  \binom{j}{ j'} \cdot{\varepsilon_c}^{j'} \cdot (1- \varepsilon_c)^{j-j'}\cdot
  \binom{k}{ k'} \cdot{\varepsilon_l}^{k'} \cdot (1- \varepsilon_l)^{k-k'}
  \bigg\}, \nonumber
\end{eqnarray}
where $i$, $j$ and $k$ ($i'$, $j'$ and $k'$) represent the number of  pre-tagged (tagged) $b$, $c$, and light-flavour jets.
$F_{ijk}$ is the fraction of events containing $i$ $b$ jets, $j$ $c$ jets and $k$ 
light-flavour jets before any tagging requirement is applied in each $\pt$ bin.
BF is the branching fraction to each final state,
$A_{t\bar{t}}$ is the event selection efficiency for that particular final state,
and $\mathcal{L}$ is the integrated luminosity. The binomial coefficients account for 
the number of arrangements in which the $n$-tags can be distributed. The efficiencies to mistag a 
$c$ jet or light-flavour jet as a $b$ jet, $\varepsilon_c$ and $\varepsilon_l$ respectively, are
fixed to the values found in Monte Carlo simulation, but with data-driven
scale factors applied as obtained
with the methods described in Sections~\ref{sec:ceff_dstarbased} and~\ref{sec:mistag}.
$N_{\rm bkg}$ is the number of background events.

To apply the method as a function of $\pT$, the $F_{ijk}$ fractions are computed
in $\pT$ bins using only the jets in each event that fall in a given $\pt$ bin. 
For both signal and background the dominant fraction is 
$F_{000}$ which occurs when no jets fall in that particular $\pT$ bin. Since a single 
event can contribute to several $\pt$ bins, this method maximises the use of the 
available jets in the sample. 

The 0-tag bin is dominated in the single-lepton channel by 
inclusive jet and $W$+jets backgrounds and is therefore not included in the fit. 
The inclusive jet background is subtracted from the $n$-tag distribution prior to performing the fit since the
$F_{ijk}$ fractions cannot be estimated reliably from Monte Carlo simulation.
For the remaining background processes, dominated by $W$+jets, $F_{ijk}^{\rm bkg}$ values are calculated
from Monte Carlo simulations and included in the fit to extract the $b$-jet tagging efficiency.  

The extraction of parameters in Eq.~\ref{eqn:method_tagcount_lepjet} from the data is performed using a 
likelihood fit. The likelihood function used is 

\begin{equation}
  L = \mathrm{Gaus}(\sigma_{t\overline{t}}|\sigma_{t\overline{t},{\rm MC}},\delta_{\sigma_{t\overline{t},{\rm MC}}} )\
 \mathrm{Gaus}(N_{\rm bkg}|N_{\rm bkg,MC},\delta_{N_{\rm bkg}})\
 \prod_{n-{\rm tags}} \mathrm{Pois}(N_n|<N_n>).
\end{equation}
The number of events in each $n$-tag bin is described by a Poisson probability with an average value 
corresponding to the number of expected events. The $\ttbar$ cross section and $N_{\rm bkg}$ are floating 
parameters of the fit but are each constrained by a Gaussian distribution with a width of 
one standard deviation of the respective normalisation uncertainties. The uncertainty introduced by 
the Monte Carlo simulation statistics has been estimated from the uncertainties on the
$F_{ijk}$ fractions and is found to be negligible. 

\begin{figure}[!h] 
  \centering
  \subfloat[]{\label{fig:res_tagcount_lj_Ntag_a}\includegraphics[width=0.32\textwidth]{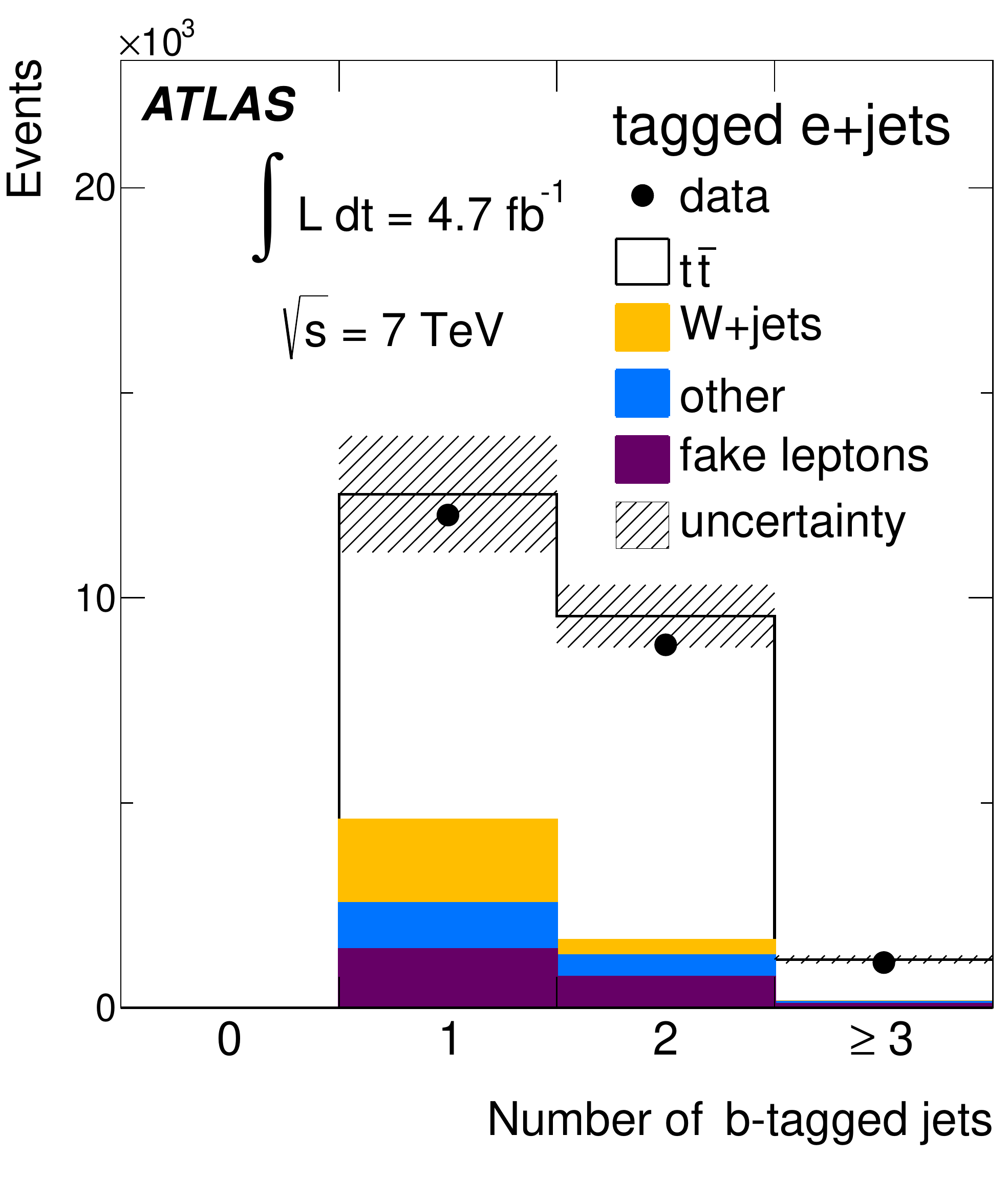}} 
  \subfloat[]{\label{fig:res_tagcount_lj_Ntag_b}\includegraphics[width=0.32\textwidth]{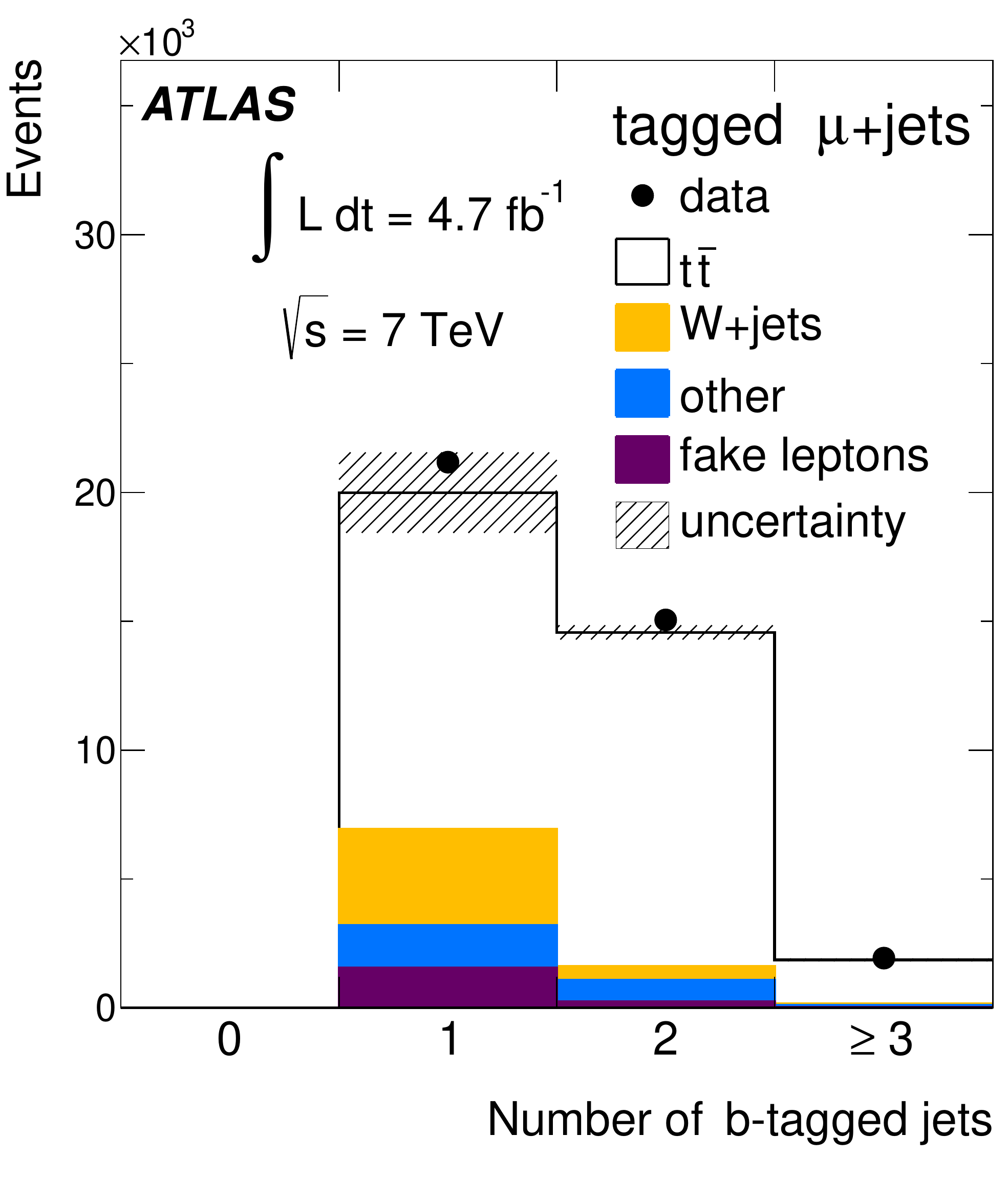}} \\
  \subfloat[]{\label{fig:res_tagcount_lj_Ntag_c}\includegraphics[width=0.32\textwidth]{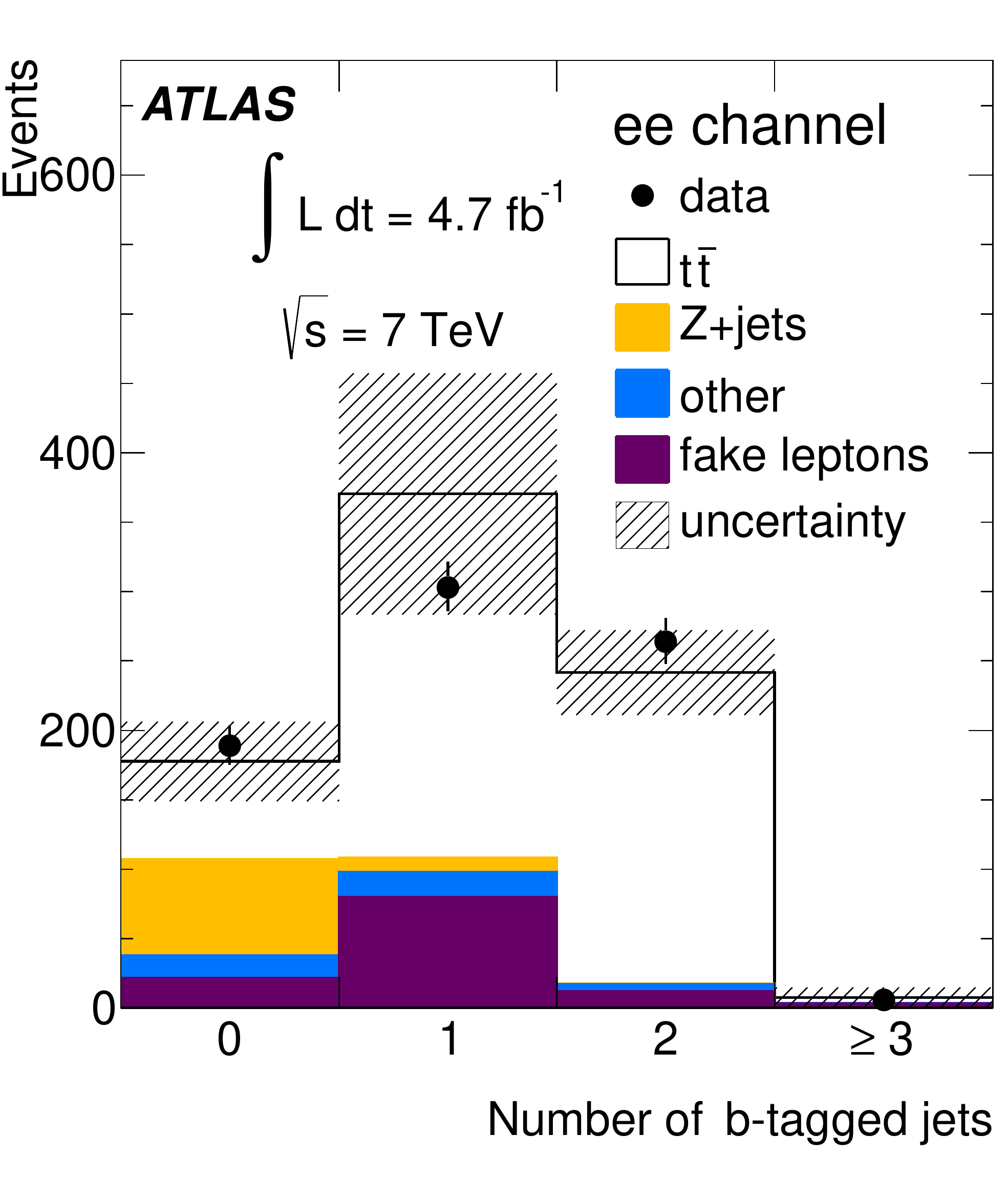}} 
  \subfloat[]{\label{fig:res_tagcount_lj_Ntag_d}\includegraphics[width=0.32\textwidth]{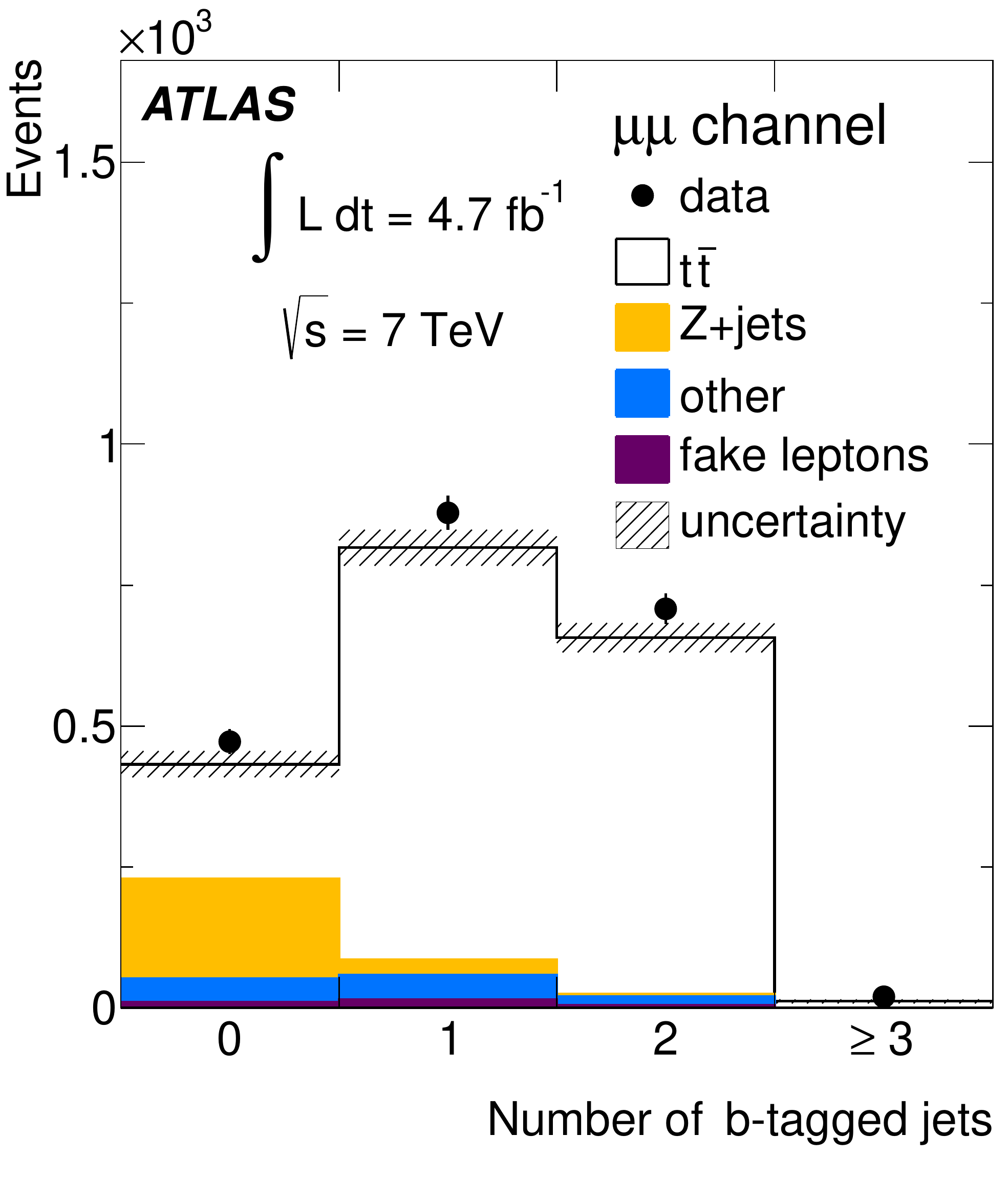}} 
  \subfloat[]{\label{fig:res_tagcount_lj_Ntag_e}\includegraphics[width=0.32\textwidth]{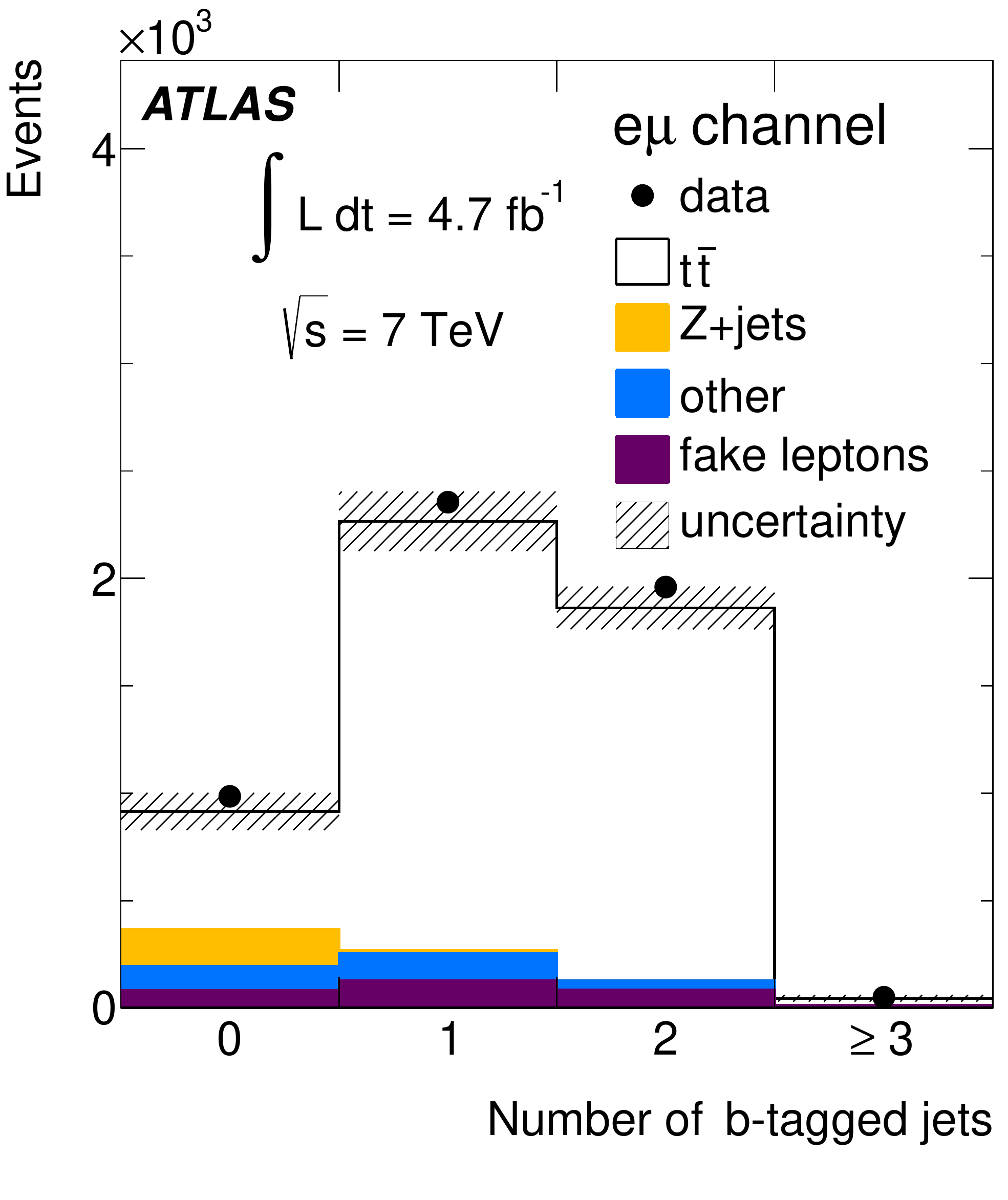}} 
  \caption{The $b$-tagged jet multiplicity distribution in Monte Carlo simulation superimposed on the
    distribution observed in data in the $e$+jets (a), $\mu$+jets (b), $ee$ (c), $\mu\mu$ (d) and $e\mu$ (e) channels.} 
  \label{fig:res_tagcount_lj_Ntag}
\end{figure}

Figure~\ref{fig:res_tagcount_lj_Ntag} shows the number of $b$-tagged jets in all of the channels
in comparison to the normalised Monte Carlo simulation using the fit result.
The measured $b$-tagged jet multiplicity distributions are well reproduced.

\subsection{Kinematic selection method}

The kinematic selection method relies on the knowledge of the flavour 
composition of the $\ttbar$ signal and background samples,
and extracts the $b$-jet tagging efficiency by measuring the fraction of $b$-tagged jets
in data.
Given an expected number of $b$, $c$ and light-flavour jets, as well as the $c$- and light-flavour-jet mistag efficiencies, 
the fraction of $b$-tagged jets in data is given by

\begin{equation} \label{eq:method_kinsel_dilep_ntag}
  f_{b-\mathrm{tag}} = \varepsilon_{b} f_{b} + \varepsilon_{c} f_{c} +
  \varepsilon_{l} f_{l} + \varepsilon_{\mathrm{fake/np}}f_{\mathrm{fake/np}},
\end{equation}
which can be rearranged to solve for the $b$-jet tagging efficiency, $\epsilon_b$:

\begin{equation}\label{eq:method_kinsel_bigeq}
  \varepsilon_{b} = \frac{1}{f_{b}} \cdot  \Big(f_{b-\mathrm{tag}} -
  \varepsilon_{c} f_{c} - \varepsilon_{l} f_{l} - 
  \varepsilon_{\mathrm{fake/np}}f_{\mathrm{fake/np}}
  \Big) .
\end{equation}

\begin{figure}[h!]
  \centering
  \subfloat[]{\label{fig:x_frac_a}\includegraphics[width=0.49\textwidth]{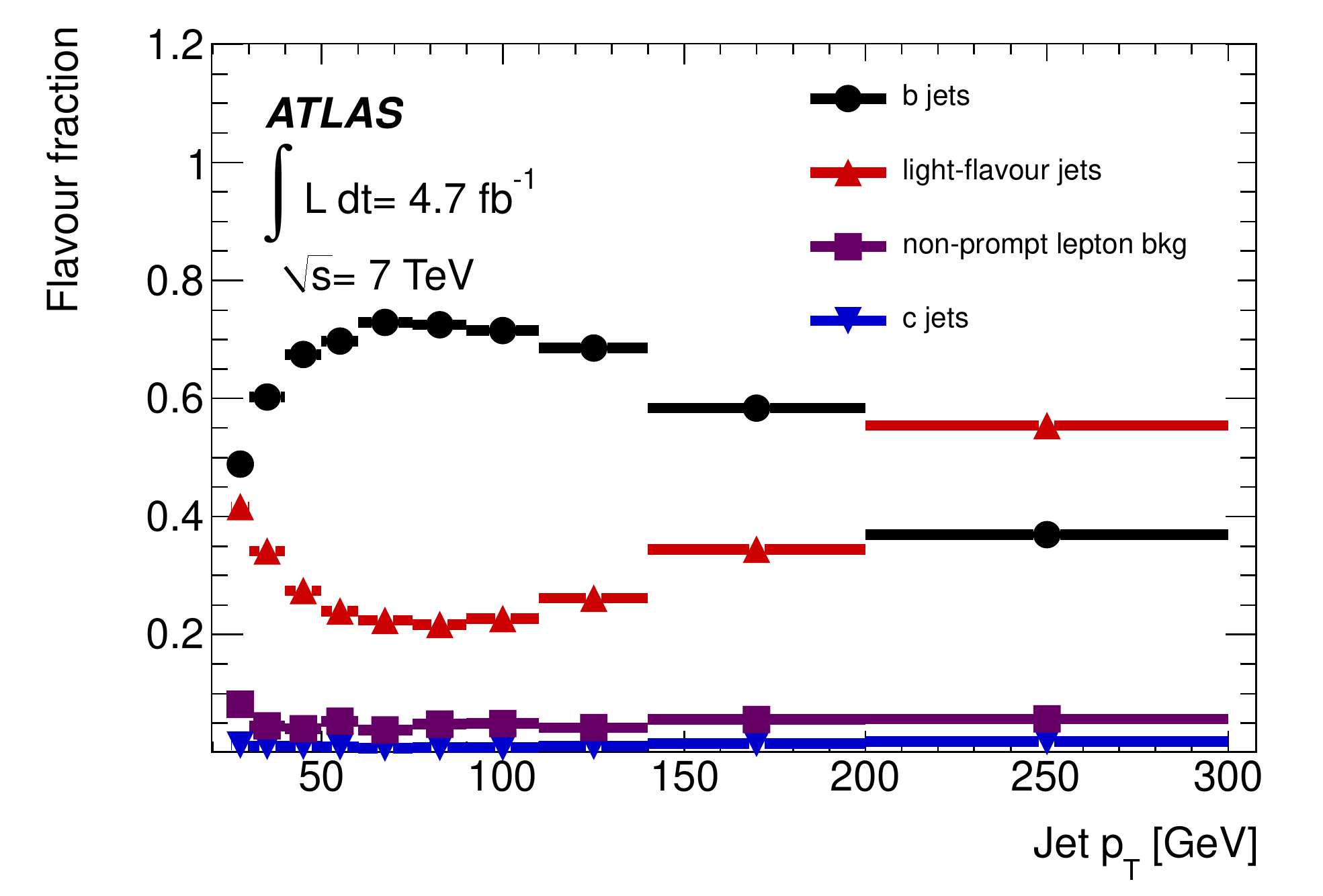}}
  \subfloat[]{\label{fig:x_frac_b}\includegraphics[width=0.49\textwidth]{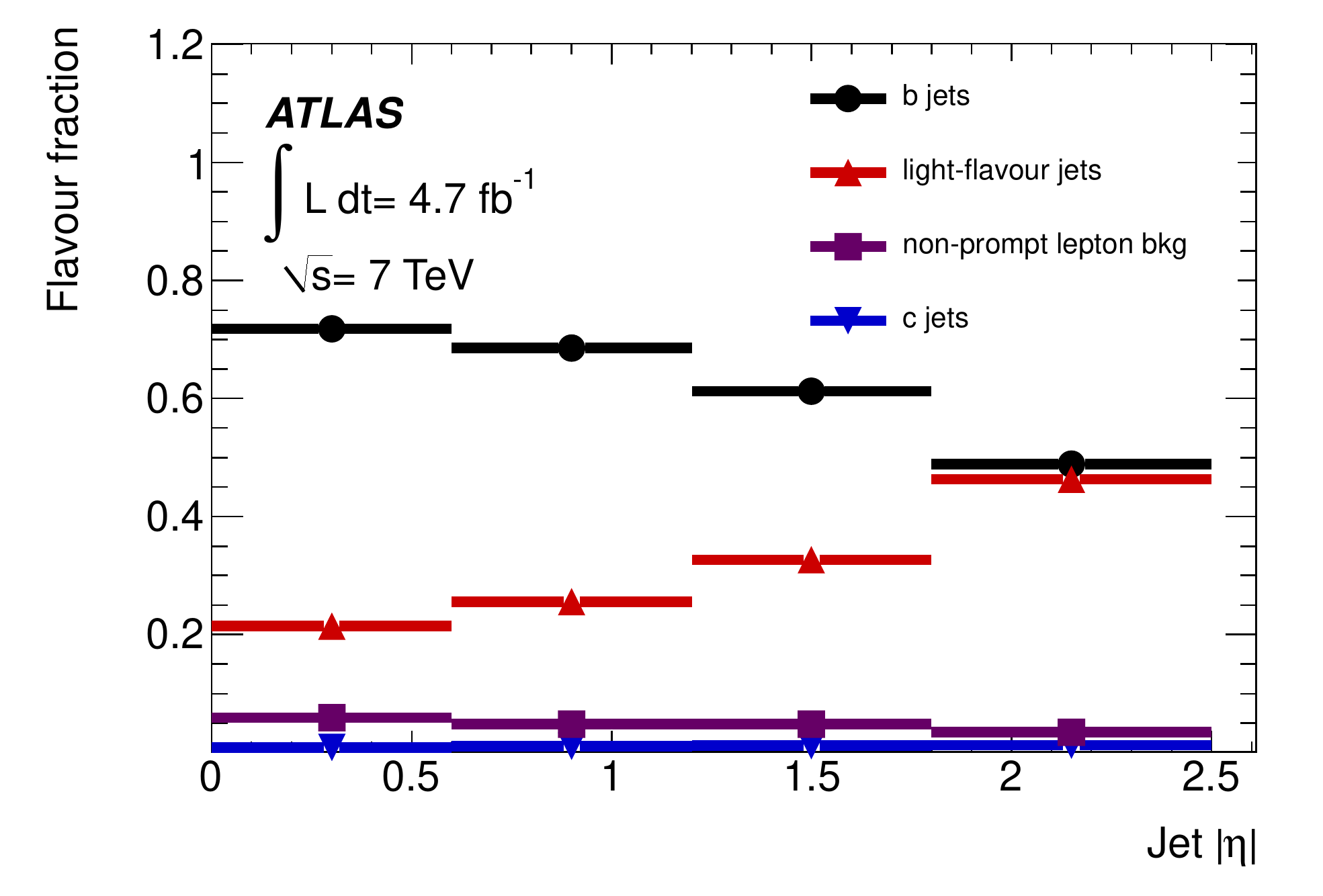}}
  \caption{Expected jet flavour composition of the two leading jets in the selected dilepton events
   as a function of jet $\pT$ (a) and $|\eta|$ (b).}
  \label{fig:x_frac}
\end{figure}

Here, $f_{b}$, $f_{c}$ and $f_{l}$ are the expected fractions of $b$, $c$, and light-flavour jets in data which are estimated from simulated events.
$\varepsilon_{c}$ and $\varepsilon_l$ are the mistag efficiencies for $c$  and light-flavour jets to be tagged as $b$ jets,
which are taken from Monte Carlo simulation but with data driven scale factors applied,
obtained with the methods described in Sections~\ref{sec:ceff_dstarbased} and~\ref{sec:mistag}.
$f_{\mathrm{fake/np}}$ is the fraction of jets from the non-prompt or fake lepton (in the dilepton channel)
or inclusive jet (in the single-lepton channel) background and is determined from data. The flavour fractions are 
calculated with respect to the sum of jets from Monte Carlo simulation
and obey the relation 
$f_{b} + f_{c} + f_{l} + f_{\mathrm{fake/np}} = 1$.
The flavour composition of the jet sample obtained after applying a dilepton selection is shown in Fig.~\ref{fig:x_frac} 
binned in both $\pT$ and $\eta$.
The expected fraction of $b$-tagged non-prompt lepton, fake lepton or inclusive jet events, $\varepsilon_{\mathrm{fake/np}}$, is
estimated from data, as detailed below.

The inclusive jet mistag rate in the single-lepton channel, $\varepsilon_{\mathrm{fake/np}}$, 
is measured in a data control region enriched in inclusive jet events.
The control region is obtained by reversing the $\met$ and $m_{\mathrm{T}}(l\nu)$ selection criteria:
\begin{itemize} 
\item $e$+jets channel: $\unit[5]{\GeV}< \met < \unit[30]{\GeV}$ and
  $m_{\mathrm{T}}(l\nu) < \unit[25]{\GeV}$.
\item $\mu$+jets channel: $\unit[5]{\GeV}< \met < \unit[15]{\GeV}$ or 
  $\met + m_{\mathrm{T}}(l\nu) < \unit[60]{\GeV}$.
\end{itemize}
Moreover, the leptons in the control region are only required to satisfy looser selection criteria 
(so-called loose leptons), following Ref.~\cite{TOPQ-2010-01}. Loose muons are not required to fulfil any isolation 
criteria, while the isolation criterion for loose electrons is less strict than that used in the baseline event selection. 
From the events measured in the control region in data the predicted contributions of the 
$\ttbar$, single top, diboson, $W$+jets and $Z$+jets processes obtained from Monte Carlo 
simulation are subtracted. 

In the dilepton channel, the fraction of $b$-tagged jets coming from the 
non-prompt or fake lepton background $\varepsilon_{\mathrm{fake/np}}$ is determined 
from events in which both leptons have the same charge.
The remaining event selection criteria are required to be satisfied. Since it is 
expected that neither the dileptonic $\ttbar$ decay nor the background processes $Z$+jets 
or single top produce same sign events (the contribution from events with a wrongly measured electron charge is expected to be small), 
a sample is obtained that is
dominated by events having at least one non-prompt or fake lepton. 

To increase the purity in the single-lepton channel, in addition to the selection described in Section~\ref{sec:sel_ljets},
the events  are also required to have at least one jet $b$-tagged 
with the MV1 tagging algorithm at an operating point that corresponds to an efficiency of $70\%$.
Based on which jet is $b$-tagged, the single-lepton sample is split into two sub-samples in the following way:
\begin{itemize}
  \item If the leading jet is $b$-tagged, the $b$-tagging rate of the next three
    jets is measured (L234 sample).
  \item If the next-to-leading jet is $b$-tagged, the $b$-tagging rate of the leading jet 
  is measured (L1 sample).
\end{itemize}
Subsequently, jets are divided in bins of $\pt$, in which the number of $b$-tagged jets  
from each selection is counted. To calculate the $b$-jet tagging efficiency, the
combined L1 and L234 sample is used.

In the dilepton channel, the $b$-jet fraction of the sample is increased by using only the 
two leading jets in each event, as this reduces the contamination of $c$  and light-flavour jets 
originating from initial and final state gluon radiation.

\subsection{Kinematic fit method}
\label{sec:method_kinfit}

The kinematic fit method is based on the selection of a high-purity $b$-jet sample by
applying a kinematic fit, similar to that described in Ref.~\cite{TOPQ-2011-15}, 
to the events passing the selection described in Section~\ref{sec:selection}.
The kinematic fit performed on the single-lepton $\ttbar$ event topology provides a mapping between the
reconstructed jets, the lepton, and the missing transverse momentum
onto the $b$ jets originating directly from the top quark decays and the 
jets (leptons) from the subsequent hadronic (leptonic) $W$-boson decay.
The kinematic fit exploits the masses of the two top quarks and $W$ bosons as constraints, 
leading to four constraints in total with one unmeasured parameter resulting in three degrees of freedom.
The fit, which is based on a $\chi^{2}$-minimisation method, is performed on all permutations of
the six highest-$\pt$ jets, and the permutation with the lowest $\chi^{2}$ is retained. 

The $b$-jet tagging efficiency is measured with the jet assigned by the fit to be the
$b$ jet on the leptonic side of the event (i.e., associated with the leptonic $W$-boson decay).
Choosing the lowest $\chi^{2}$ permutation of the kinematic fit selects the
correct jet association in about 60\% of the $\ttbar$ events.

In addition to the combinatorial background the sample still 
contains backgrounds from other processes, such as single top and $W$+jets events.
Nevertheless, the full $b$-jet weight distribution can be obtained
from data by using a statistical background subtraction.
This subtraction is done by dividing the sample into two orthogonal sub-samples 
based on the information about the jets associated to the hadronic side of the event
(i.e., associated with the hadronic $W$-boson decay): 
the first sub-sample (``signal sample'') is selected by applying additional 
cuts to increase fraction of correct mappings, while the second sub-sample 
(``background sample'') is enriched in incorrect mappings. The additional cuts applied 
to the signal sample are:
\begin{itemize}
  \item The jet identified by the kinematic fit to be the $b$ jet on the hadronic side of the event
  needs to be $b$-tagged by the MV1 tagging algorithm at the 70\% efficiency operating point.
  This is applied to suppress the $W$+jets events and incorrect permutations.
     \item The jets associated with the hadronic $W$-boson decay must not be $b$-tagged by 
    the MV1 $b$-tagging algorithm at the 70\% efficiency operating point.
  \item Only events with six or fewer jets with $\pt > \unit[25]{\GeV}$ are
    considered.
\end{itemize}

The background sample is instead defined by removing the $b$-tagging requirement on the hadronic-side 
$b$ jet and the jet multiplicity requirement and inverting the $b$-tagging veto on the jets 
associated to the $W$-boson decay:
\begin{itemize}
  \item At least one of the jets assigned by the kinematic fit to the hadronic $W$-boson decay is required
  to be $b$-tagged by the MV1 tagging algorithm at the 70\% efficiency operating point.
\end{itemize}

To verify that the signal sample is enriched in correct mappings at low values of
fit $\chi^2$, while the background sample is dominated by incorrect mappings at all values of fit
$\chi^2$, a truth-match based on a $\Delta R$ cut to the original
partons of the hard interactions is performed in Monte Carlo simulation. Here, groupings 
of partons, hadron-level jets and reconstructed jets are chosen in a way that minimises the 
sum of their respective distances in the $\eta-\phi$ plane. Such a triplet is considered to 
be matched if the respective sum of the three distances in the $\eta-\phi$-plane passes the 
requirement $\Delta R({\rm parton, hadron}) + \Delta R({\rm reco, hadron}) + \Delta R({\rm reco, parton}) <0.5$.
Due to e.g. unreconstructed jets, it will not always be possible to define
the above triplets, and thus a fraction of the events will remain unmatched.
The unmatched mappings remain in the analysis: in Monte Carlo simulations the truth
$b$ jets are taken into account independently of their matching status.

\begin{figure}[htb!]
  \centering
  \subfloat[]{\label{fig:figures_kinfit_chi2Signal_25-200_a}\includegraphics[width=0.49\textwidth]{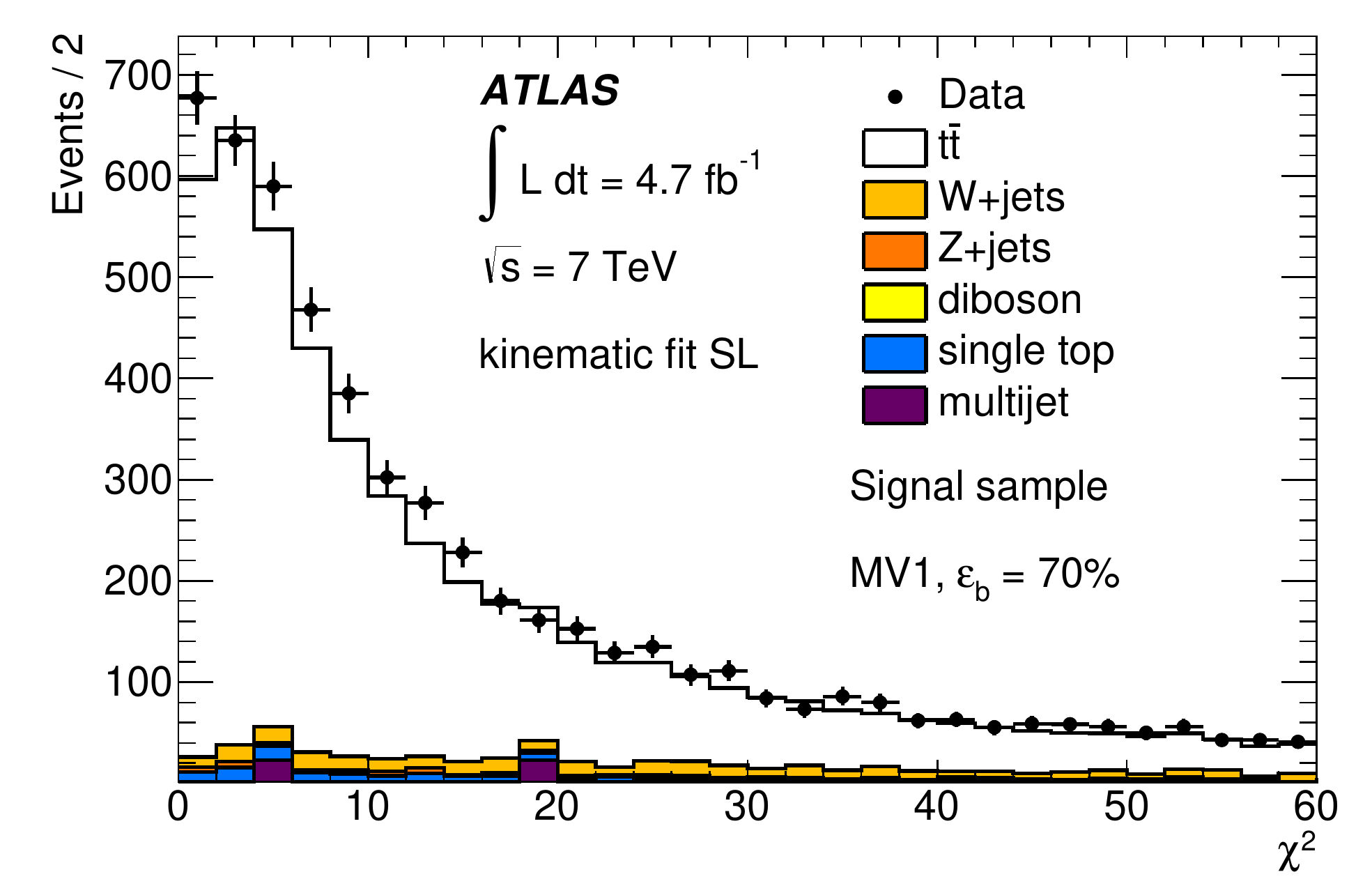}} 
  \subfloat[]{\label{fig:figures_kinfit_chi2Signal_25-200_b}\includegraphics[width=0.49\textwidth]{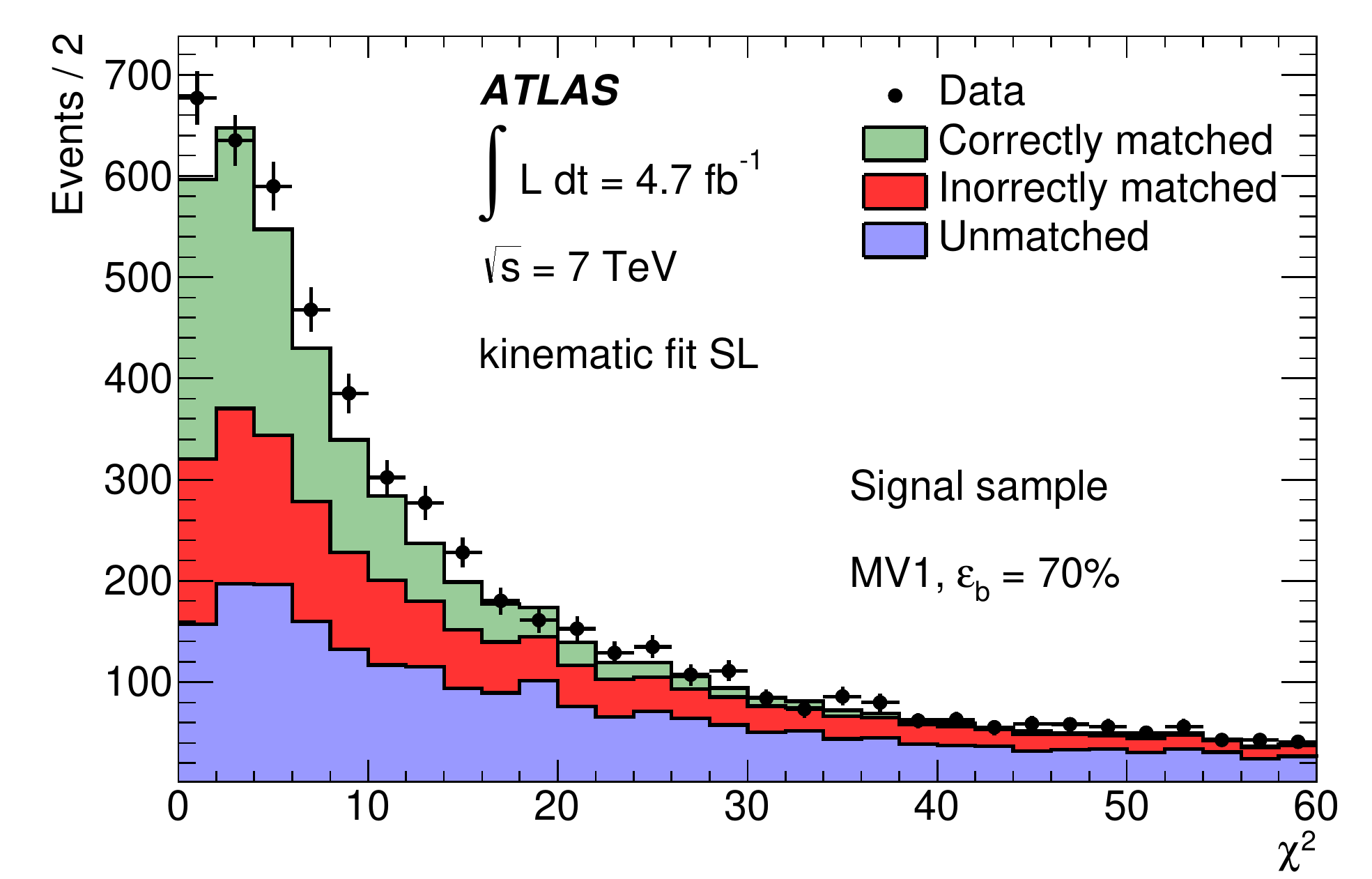}} \\
  \subfloat[]{\label{fig:figures_kinfit_chi2Signal_25-200_c}\includegraphics[width=0.49\textwidth]{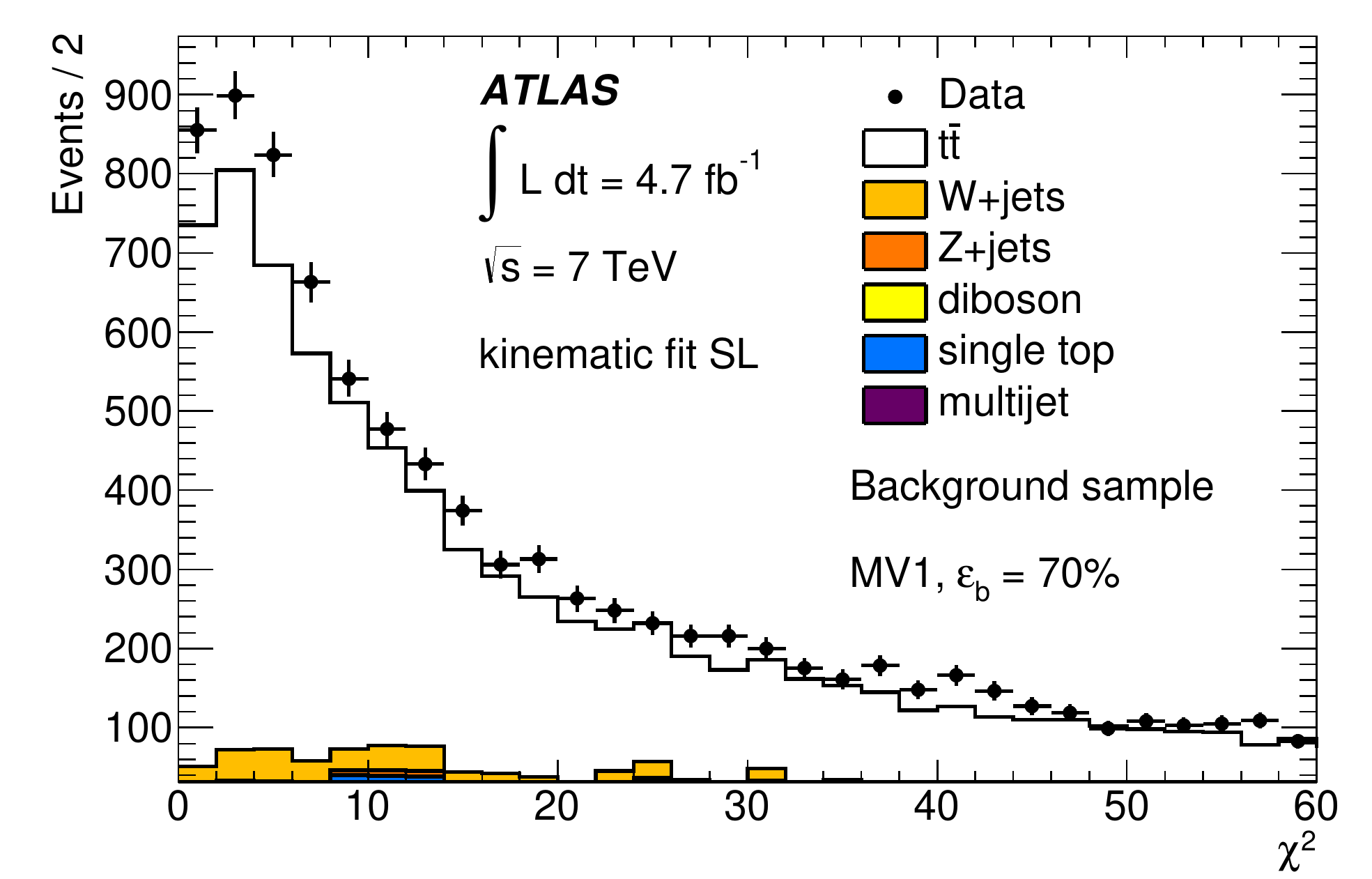}}
  \subfloat[]{\label{fig:figures_kinfit_chi2Signal_25-200_d}\includegraphics[width=0.49\textwidth]{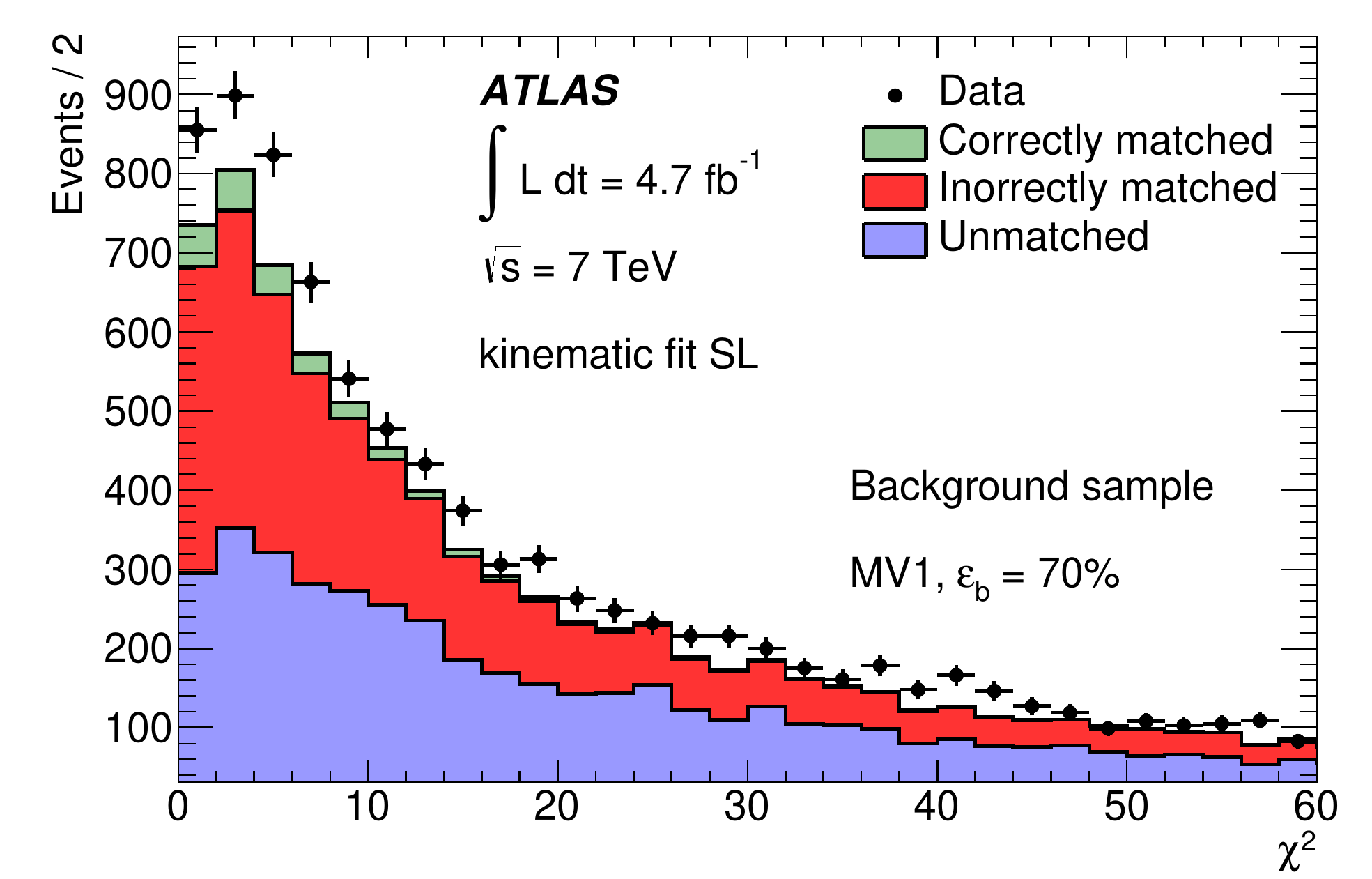}}
  \caption{The $\chi^2$ distributions of events in the signal (top) and background (bottom) samples. 
    The left plots show the various physics processes contributing to the sample
    while the right plots show the contribution from correctly matched,
    incorrectly matched and unmatched events, as obtained by the truth-match algorithm. 
    The simulated distributions are shown without uncertainties.}
  \label{fig:figures_kinfit_chi2Signal_25-200}
\end{figure}

The $\chi^2$ distributions of both the signal and background samples are shown in
Fig.~\ref{fig:figures_kinfit_chi2Signal_25-200}, together with the result of the truth-match.
As desired, the signal sample has a sizable fraction of correct mappings, while the 
background sample almost exclusively is made up of unmatched or incorrect mappings.
Furthermore, the correct mappings predominantly have low $\chi^2$ values, while the
high $\chi^2$-region is fully dominated by incorrect and unmatched mappings.

The remaining background of incorrect mappings in the signal sample selected from data can therefore
be estimated from the background sample. As events with high values of $\chi^{2}$ are 
predominantly incorrect mappings in both sub-samples, the background sample prediction can be
normalised at high $\chi^{2}$ values $(\chi^{2} > 25)$, by using the scale factor

\begin{equation}
  S_{\mathrm{BG}} = \frac{\int_{25}^{\infty}\mathrm{d}\chi^{2}_{S}}{\int_{25}^{\infty}\mathrm{d}\chi^{2}_{B}}.
  \label{eq:background_scale_kinfit}
\end{equation}

The background-subtracted $b$-tag weight distribution of the $b$ jet from the leptonic 
decay in the signal sample, from which the $b$-jet tagging efficiency is
eventually extracted, is subsequently derived by subtracting the $b$-tag weight distribution
in the background sample, scaled according to Eq~.\ref{eq:background_scale_kinfit}.

For the background subtraction method to work correctly it is imperative that the 
shape of the $\chi^2$ distribution of the non-$b$ portion of the background sample 
agrees with that of the non-$b$ portion of the signal sample.
This has been found to be the case for all the $b$-tagging algorithms tested.

The measurement of the $b$-jet tagging efficiency is based on the $b$-tag weight distribution 
of the sample of $b$ jets on the leptonic side of the event.
An important advantage of this method is that a continuous calibration of the $b$-tag weight distribution 
is feasible, as the full distribution is reconstructed.
The $b$-jet tagging efficiency for a given operating point, corresponding to a certain weight cut $w_{\rm cut}$,
can be calculated using the (normalised to unity) weight distribution $T(w)$ of the selected 
$b$-jet sample after the background subtraction by integration above the threshold $w_{\rm cut}$:

\begin{equation}
  \varepsilon(w_{\rm cut}) = \int_{w_{\rm cut}}^{\infty} T(w) \, \mathrm{d}w. \label{eq:kinfit_weight}
\end{equation}
Depending on the available statistics the measurement of the $b$-jet tagging 
efficiency can be binned in any variable, for example $\pt$ or $\eta$.

\begin{figure}[htb!]
  \centering
  \subfloat[]{\label{fig:figures_kinfit_MV1_25-200_calibration_chain_a}\includegraphics[width=0.49\textwidth]{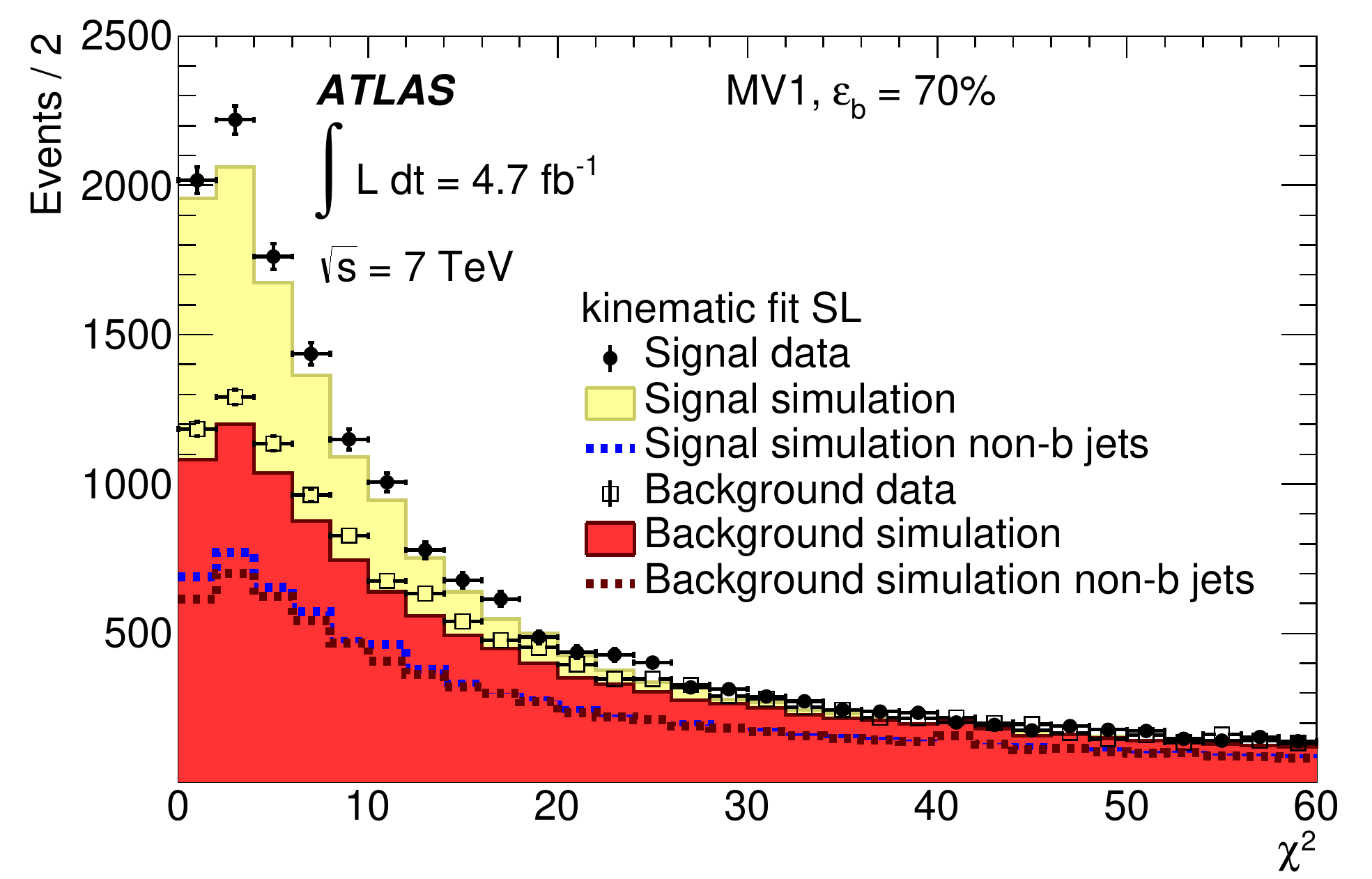}}
  \subfloat[]{\label{fig:figures_kinfit_MV1_25-200_calibration_chain_b}\includegraphics[width=0.49\textwidth]{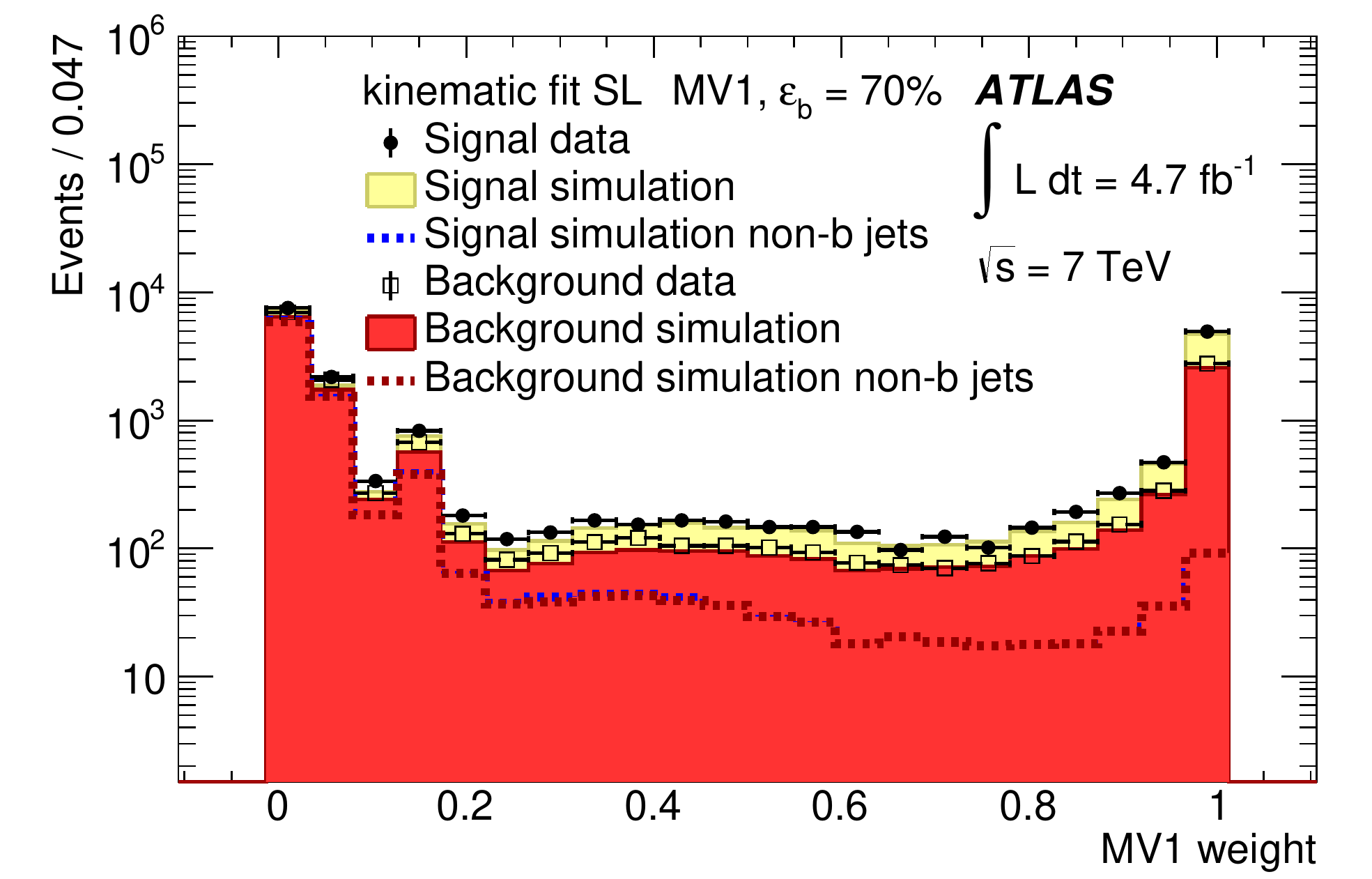}}
  \caption{The $\chi^2$ (a) and MV1 weight (b) distributions in the signal and scaled background samples with 
    the respective non-$b$ contributions outlined. The simulated distributions are shown without uncertainties.}
  \label{fig:figures_kinfit_MV1_25-200_calibration_chain}
\end{figure}

\begin{figure}[htb!]
  \centering
  \includegraphics[width=0.6\textwidth]{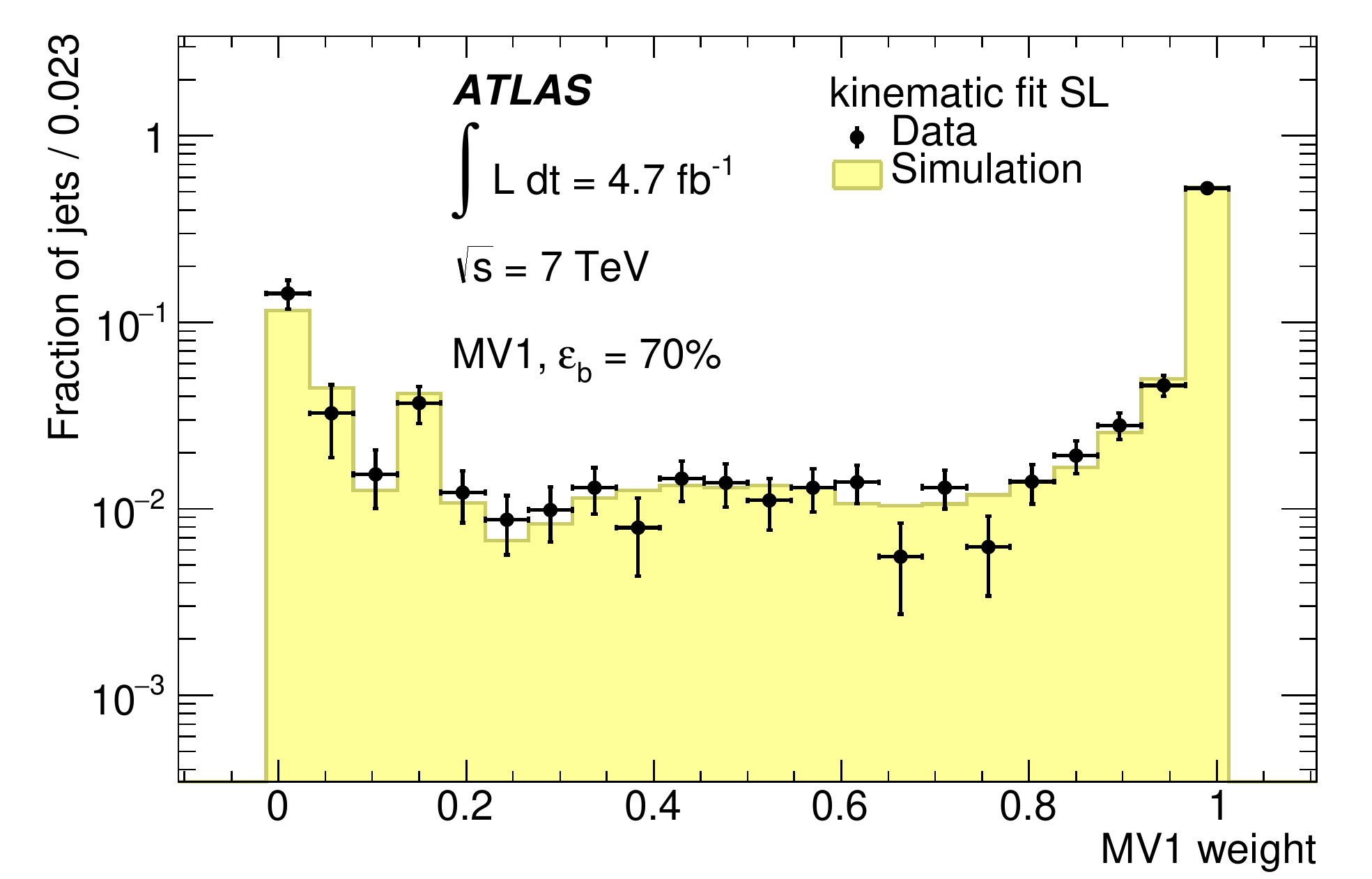}
  \caption{The background-subtracted MV1 weight distribution in data, compared to the
    MV1 distribution for $b$ jets in the simulated signal sample.}
  \label{fig:figures_kinfit_MV1_subtracted}
\end{figure}

The complete sequence of calibration steps for the MV1 $b$-tagging algorithm for
jets with $\unit[25]{\GeV} < \pt < \unit[200]{\GeV}$ is presented in 
Figs.~\ref{fig:figures_kinfit_MV1_25-200_calibration_chain} and~\ref{fig:figures_kinfit_MV1_subtracted}.
After scaling the background sample (Fig.~\ref{fig:figures_kinfit_MV1_25-200_calibration_chain})
the prescription results in a background-subtracted distribution of the MV1 weight (Fig.~\ref{fig:figures_kinfit_MV1_subtracted}).
Using Eq.~\ref{eq:kinfit_weight} the efficiency is then derived.
It is shown that the method applied to simulated events (``expected'') describes the distribution
obtained from the sample of true $b$ jets in Monte Carlo simulated events.
The ratio displayed is the efficiency measured in data divided by the efficiency calculated from true $b$ jets in simulated events.

\subsection{Combinatorial likelihood method}
\label{sec:method_likelihood}

The combinatorial likelihood method is intended to increase the precision by
exploiting the kinematic correlations between the jets in the dilepton sample.
Like the kinematic selection method, this method relies on an a priori
knowledge of the flavour composition of the $\ttbar$ signal and background
samples.

Events with either two or three jets selected using the criteria detailed in
Section~\ref{sec:sel_dilep} are used in the analysis.
The analysis is carried out separately for two- and three-jet events, and for
the $e\mu$ and combined $ee$ and $\mu\mu$ channels. In detail, the $b$-jet
tagging efficiency determination in each of the resulting four channels uses an
unbinned maximum likelihood fit.

In the two-jet case the following per-event likelihood function is adopted:
\begin{eqnarray}
  {\cal L}\left(p_{\mathrm{T},1},p_{\mathrm{T},2},w_{1},w_{2}\right) 
  = \frac{1}{2}  \sum_{(i,k)} &[& f_{bb} {\mathcal{P}}_{bb}\left(p_{\mathrm{T},i},p_{\mathrm{T},k}\right) 
    {\mathcal{P}}_{b}\left(w_{i}\vert p_{\mathrm{T},i}\right) 
    {\mathcal{P}}_{b}\left(w_{k}\vert p_{\mathrm{T},k}\right) \label{eq:likelihood-2j}\\
    & + &   f_{bj} {\mathcal{P}}_{bj}\left(p_{\mathrm{T},i},p_{\mathrm{T},k}\right) 
    {\mathcal{P}}_{b}\left(w_{i}\vert p_{\mathrm{T},i}\right) 
    {\mathcal{P}}_{j}\left(w_{k}\vert p_{\mathrm{T},k}\right)  \nonumber \\
    & + &   f_{jj} {\mathcal{P}}_{jj}\left(p_{\mathrm{T},i},p_{\mathrm{T},k}\right) 
    {\mathcal{P}}_{j}\left(w_{i}\vert p_{\mathrm{T},i}\right) 
    {\mathcal{P}}_{j}\left(w_{k}\vert p_{\mathrm{T},k}\right) ~ ], \nonumber
\end{eqnarray}
where:
\begin{itemize}
  \item the indices $(i,k)$ run over $(1,2)$ and $(2,1)$;
  \item $f_{bb}$, $f_{bj}$ are the two independent jet flavour fractions, and $f_{jj} = 1 - f_{bb} - f_{bj}$;
  \item ${\mathcal{P}}_{f} \left(w \vert p_{\mathrm{T}} \right)$ is the PDF (probability density 
    function) for the $b$-tagging discriminant or \emph{weight} for a jet of flavour $f$, for 
    a given transverse momentum;\footnote{This means that, regardless of the jet $\pt$,
      the integral of the PDF over the $b$-tagging weight variable is one.}
  \item and ${\mathcal{P}}_{f_{1}f_{2}} \left(p_{\mathrm{T},1},p_{\mathrm{T},2}\right)$ is the two-dimensional 
    PDF for $[p_{\mathrm{T},1},p_{\mathrm{T},2}]$ for the flavour combination $[f_{1},f_{2}]$.
\end{itemize}

All PDFs are implemented as binned histograms. For example, for $N$ $\pt$ bins,
${\mathcal{P}}_{f_{1}f_{2}} \left(p_{\mathrm{T},1},p_{\mathrm{T},2}\right)$ is expressed 
as an $N \times N$ binned histogram. For the symmetric $bb$ and $ll$ combinations, 
the PDF is symmetrised, reducing the number of independent bins to determine from
$N^{2} - 1$ to $N \times \left( N + 1 \right)/2 - 1$ which 
reduces the statistical fluctuations from the Monte Carlo simulation; as a consequence, the
explicit symmetrisation expressed by Eq.~\ref{eq:likelihood-2j} for these
combinations is for notational convenience only. The flavour PDFs ${\mathcal{P}}_{f} 
\left(w \vert \pt\right)$ are defined in a similar way, with one binned histogram 
for each $\pt$ bin. All PDFs are determined from simulation, except for the 
$b$-jet weight PDF, which contains the information to be extracted from the data. 

A histogram with only two bins is used to describe the $b$-weight PDF for each $\pt$ bin,
with the bin above the cut value corresponding to the $b$-jet tagging efficiency.
The $b$-jet tagging efficiency then corresponds to

\begin{equation}
  \epsilon_{b}\left(\pt \right) = 
  \int_{w_{\rm cut}}^{\infty} dw' {\mathcal{P}}_{b}\left(w',\pt \right). \nonumber
\end{equation}

The likelihood function distinguishes between the different flavour fractions, but not
between signal and background processes.
To extract ${\mathcal{P}}_{b}\left(w \vert \pt\right)$ in bins of $\pt$, the flavour 
fractions $f_{f_{1},f_{2}}$, the ${\mathcal{P}}_{f_{1}f_{2}}\left(p_{\mathrm{T},1},p_{\mathrm{T},2}\right)$ 
and the non-$b$-jet $b$-weight PDFs are determined from simulation.

A slightly more complex likelihood function is 
defined for the three-jet case, which is conceptually analogous but needs 
to consider that the jet flavour combinations are increased to four. Accordingly, 
there are up to $3!=6$ equivalent jet combinations the likelihood 
needs to be summed over, as they are a priori indistinguishable in data. The formalism in this case is:

\small
\begin{eqnarray}
  {\cal L}\left(p_{\mathrm{T},1},p_{\mathrm{T},2},p_{\mathrm{T},3},w_{1},w_{2},w_{3}\right) 
  = \frac{1}{6} \sum_{(i,k,l)} &[& f_{bbb} {\mathcal{P}}_{bbb}\left(p_{\mathrm{T},i},p_{\mathrm{T},k},p_{\mathrm{T},l}\right) 
    {\mathcal{P}}_{b}\left(w_{i}\vert p_{\mathrm{T},i}\right) 
    {\mathcal{P}}_{b}\left(w_{k}\vert p_{\mathrm{T},k}\right) 
    {\mathcal{P}}_{b}\left(w_{l}\vert p_{\mathrm{T},l}\right) \nonumber \\
    & + &   f_{bbj} {\mathcal{P}}_{bbj}\left(p_{\mathrm{T},i},p_{\mathrm{T},k},p_{\mathrm{T},l}\right) 
    {\mathcal{P}}_{b}\left(w_{i}\vert p_{\mathrm{T},i}\right) 
    {\mathcal{P}}_{b}\left(w_{k}\vert p_{\mathrm{T},k}\right)  
    {\mathcal{P}}_{j}\left(w_{l}\vert p_{\mathrm{T},l}\right) \nonumber \\
    & + &   f_{bjj} {\mathcal{P}}_{bjj}\left(p_{\mathrm{T},i},p_{\mathrm{T},k},p_{\mathrm{T},l}\right) 
    {\mathcal{P}}_{b}\left(w_{i}\vert p_{\mathrm{T},i}\right) 
    {\mathcal{P}}_{j}\left(w_{k}\vert p_{\mathrm{T},k}\right) 
    {\mathcal{P}}_{j}\left(w_{l}\vert p_{\mathrm{T},l}\right)  \nonumber \\ 
    & + &   f_{jjj} {\mathcal{P}}_{jjj}\left(p_{\mathrm{T},i},p_{\mathrm{T},k},p_{\mathrm{T},l}\right) 
    {\mathcal{P}}_{j}\left(w_{i}\vert p_{\mathrm{T},i}\right) 
    {\mathcal{P}}_{j}\left(w_{k}\vert p_{\mathrm{T},k}\right) 
    {\mathcal{P}}_{j}\left(w_{l}\vert p_{\mathrm{T},l}\right) ~], \nonumber 
\end{eqnarray}
\normalsize
where the indices $(i,k,l)$ run over all possible permutations of the three jets,
and the various PDFs are defined in a similar way to the two-jet case. 

In order to simplify the determination of ${\mathcal{P}}_{f_{1}f_{2}f_{3}}
\left(p_{\mathrm{T},1},p_{\mathrm{T},2},p_{\mathrm{T},3}\right)$ from simulations, which otherwise requires 
prohibitive simulation statistics, the following factorisation assumption is made:
\begin{equation}
  {\mathcal{P}}_{f_{1}f_{2}f_{3}}\left(p_{\mathrm{T},1},p_{\mathrm{T},2},p_{\mathrm{T},3}\right) = 
  {\mathcal{P}}_{f_{1}}\left(p_{\mathrm{T},1}\right)
  {\mathcal{P}}_{f_{2}}\left(p_{\mathrm{T},2}\right)
  {\mathcal{P}}_{f_{3}}\left(p_{\mathrm{T},3}\right) .
\end{equation}

The effect of this approximation was tested, along with the entire fitting
method, using closure tests based on simulated events. In these tests 
the fit procedure is applied to the Monte Carlo simulation itself in order to check for possible biases.
The tests have been performed for all four channels ($e\mu$ and $ee+\mu\mu$, 2
and 3 jets) and for several different $b$-jet tagging efficiency points. These
tests all yield efficiencies compatible with their known inputs within
the statistical uncertainties, with systematic effects from non-closure found to
be much less than $1\%$.

\subsection{Systematic uncertainties}
\label{sec:ttbar_syst}

The tag counting and kinematic selection analyses share most systematic uncertainties,
and many sources of uncertainty are also in common with the combinatorial likelihood analysis.
One important class of common systematic uncertainties are those addressing how well the 
simulation models the $\ttbar$ production process, including generator dependence, initial state radiation, and
heavy flavour fragmentation.
The estimation of the contamination from the main background processes is another source of systematic uncertainties.
Furthermore, common systematic uncertainties are those arising from the imperfect knowledge of the jet 
energy scale and resolution.
The dominant systematic uncertainty in the kinematic fit method arises from uncertainties in the background subtraction.

All individual contributions to the systematic uncertainties are summarised in 
Tables~\ref{tab:tagcount_MV1_70} through \ref{pdf-tab-sys-MV1-70}, in bins of
jet $\pt$ and for the MV1 $b$-tagging algorithm 
at an operating point corresponding to a nominal efficiency of 70\%. 
While the total uncertainties on the results of the tag counting and the
kinematic selection methods are dominated by systematic uncertainties, the kinematic fit
method is limited by data statistics. In the combinatorial likelihood method the statistical and 
systematic uncertainties are approximately equal in size.
The estimates of some systematic uncertainties include a sizeable statistical component 
which leads to unphysical bin-to-bin variations in some cases. 
However, when the calibration results 
of several methods are combined (see Section~\ref{sec:combination}) these irregularities are
smoothed out.

\begin{table}[tbp]
  \begin{center}
    \scriptsize
    \setlength{\tabcolsep}{0.4pc}
    \begin{tabular}[t]{l|ccccccccccc}
      \hline
      \hline
                            & \multicolumn{10}{c}{Jet $\pt\ [\GeV]$} \\ 
      Source                & 25--30 &30--40 &40--50 &50--60 & 60--75 & 75--90 & 90--110& 110--140& 140--200 &200--300\\
      \hline
      IFSR                  & 4.0  & 2.5  &  2.8 & 3.1 & 2.9 &  2.1 &  3.5 &  2.1 &  4.2 &  7.9 \\
      Generator             & 3.7  & 2.7  &    - & 0.6 & 5.9 &  0.6 &  6.4 &  1.6 &  0.6 &  1.0 \\
      Fragmentation         & 2.3  & 8.0  &  2.9 & 3.7 & 2.9 &  3.4 &  2.4 &  4.1 &  3.5 &  4.4 \\
      $W$ + jets            & 2.6  & 0.5  &  0.3 & 0.3 & 0.2 &  0.5 &  0.3 &  0.7 &  0.5 &  1.1 \\
      Single top            & 0.7  & 0.5  &  0.1 & 0.3 & 0.1 &  0.2 &  0.3 &  0.8 &  0.9 &  0.4 \\
      $Z$ + jets            & 1.3  & 0.3  &  0.3 & 0.2 & 0.3 &  0.2 &  0.2 &  0.6 &  0.6 &  0.6 \\
      Diboson               & 0.7  & 0.4  &    - & 0.1 & 0.2 &  0.3 &  0.3 &  0.7 &  0.7 &  0.1 \\
      Multijet              & 9.9  & 1.4  &  4.3 & 2.9 & 1.2 &  6.7 &  3.3 &  4.0 &  1.3 &  4.1 \\
      Jet energy scale      &   21 & 6.2  &  3.3 & 1.0 & 2.3 &  6.3 &  4.8 &  7.0 &  7.4 &   17 \\
      Jet energy resolution &   14 & 9.8  &   11 & 3.8 & 0.8 &  1.4 &  3.2 &  4.0 &  4.8 &  4.2 \\
      Jet reconstruction eff. & 0.8& 0.5  &  0.1 & 0.2 & 0.2 &    - &  0.2 &  0.3 &  0.2 &    - \\
      Jet vertex fraction   &   17 & 0.4  &  0.2 &   - & 0.3 &  0.3 &  0.3 &  0.8 &  0.7 &  2.2 \\
      $\varepsilon_{c}$      & 3.5  & 1.8  &  1.9 & 1.7 & 1.6 &  1.4 &  2.0 &  1.2 &  1.3 &  2.1 \\
      $\varepsilon_{l}$      &  14  & 1.1  &  0.6 & 0.5 & 0.8 &  0.6 &  0.3 &  0.8 &  0.6 &  1.1 \\
      $\met$                & 0.8  & 0.6  &  0.2 & 0.3 & 0.3 &  0.4 &  0.2 &  0.9 &  0.9 &  0.1 \\
      Lepton eff./res.      & 2.4  & 1.5  &  0.5 & 0.7 & 0.4 &  0.8 &  0.6 &  2.0 &  2.2 &  1.3 \\
      Luminosity            & 2.9  & 0.8  &  0.7 & 0.6 & 0.4 &  0.7 &  0.9 &  1.0 &  1.4 &  3.4 \\
      \hline
      Total systematic uncertainty &  36  &   15 &   13 & 7.2 & 8.0 &   10 &   10 &   11 &   11 &   21 \\
      Statistical uncertainty & 5.2  & 2.6  &  2.9 & 3.0 & 2.7 &  3.6 &  3.4 &  3.4 &  3.8 &  4.0 \\
      \hline
      Total uncertainty       &  36  &  15  &   14 & 7.8 & 8.5 &   11 &   11 &   11 &   12 &   21 \\
      \hline
      \hline
    \end{tabular}
    \caption{Relative uncertainties (in \%) for the tag counting method in the single lepton $\ttbar$ channel
      ($e$+jets and $\mu$+jets combined), for the
      MV1 algorithm at an operating point corresponding to a 70\% tagging efficiency.
      Negligibly small uncertainties are indicated by dashes.
    }
    \label{tab:tagcount_MV1_70}
  \end{center}
\end{table}

\begin{table}[tbp]
  \begin{center}
    \scriptsize
    \setlength{\tabcolsep}{0.4pc}
    \begin{tabular}{ l|c c c c c c c c c c c}
      \hline
      \hline
                                   & \multicolumn{10}{c}{Jet $\pt\ [\GeV]$} \\ 
      Source                       & 25--30 &30--40 &40--50 &50--60 & 60--75 & 75--90 & 90--110& 110--140& 140--200 &200--300\\ 
      \hline
      IFSR                         & 4.3 & 3.2 & 2.7 & 3.1 & 3.6 & 4.0 & 3.6 & 4.0 & 5.2 & 8.0 \\
      Generator                    & 0.5 & 0.2 & - & 0.2 & 0.3 & 0.1 & 0.6 & 1.2 & 1.9 & 3.9 \\
      Fragmentation                & 0.1 & 1.0 & 1.9 & 1.1 & 2.0 & 1.1 & 0.8 & 0.7 & 0.3 & 2.9 \\
      $W$ + jets                   & 1.8 & 1.5 & 1.0 & 0.8 & 0.9 & 0.8 & 0.8 & 0.9 & 1.2 & 1.6 \\
      $t \bar{t}$                  & 1.0 & 0.9 & 0.8 & 0.9 & 0.9 & 0.9 & 1.0 & 1.1 & 1.4 & 1.5 \\
      Single top                   & - & - & - & - & - & - & - & - & 0.1 & 0.2 \\
      $Z$ + jets                   & - & 0.1 & 0.1 & 0.1 & 0.2 & 0.2 & 0.2 & 0.3 & 0.4 & 0.5 \\
      Diboson                      & - & - & - & - & - & - & - & - & - & - \\
      Multijet                     & 1.5 & 1.3 & 1.4 & 1.7 & 1.6 & 1.4 & 1.6 & 1.6 & 1.9 & 1.9 \\
      Jet energy scale             & 5.7 & 3.2 & 2.2 & 1.1 & 0.8 & 0.4 & 0.6 & 1.2 & 1.2 & 2.8 \\
      Jet energy resolution        & 5.0 & 0.2 & 0.5 & 1.4 & 0.3 & 0.8 & 0.6 & 0.6 & 0.6 & 0.7 \\
      Jet reconstruction eff.      & - & - & - & - & - & - & - & - & 0.1 & - \\
      Jet vertex fraction          & 0.2 & 0.3 & - & - & 0.2 & 0.1 & - & - & - & 0.4 \\
      $\varepsilon_{c}$             & 0.6 & 0.4 & 0.3 & 0.3 & 0.3 & 0.4 & 0.4 & 0.4 & 0.5 & 0.7 \\
      $\varepsilon_{l}$             & 0.3 & 0.3 & 0.4 & 0.4 & 0.6 & 0.7 & 0.9 & 1.1 & 1.4 & 2.0 \\
      $\varepsilon_\text{fake/np}$  & 1.7 & 0.4 & 0.2 & 0.4 & 0.4 & 0.4 & 0.4 & 0.5 & 0.5 & 0.4 \\
      $\met$                       & 0.1 & - & - & - & - & - & - & - & - & - \\
      Lepton eff./res.             & - & - & - & - & - & - & - & - & 0.1 & 0.1 \\
      Luminosity                   & 0.2 & 0.2 & - & - & - & - & - & - & 0.1 & 0.1 \\
      \hline
      Total systematic uncertainty & 9.3 & 5.2 & 4.4 & 4.3 & 4.7 & 4.7 & 4.5 & 5.1 & 6.5 & 10 \\
      Statistical uncertainty      & 5.1 & 3.1 & 2.7 & 2.5 & 1.9 & 2.0 & 2.0 & 2.2 & 2.7 & 5.4 \\
      \hline
      Total uncertainty            &  11 & 6.0 & 5.2 & 5.0 & 5.1 & 5.1 & 4.9 & 5.6 & 7.0 & 12 \\
      \hline
      \hline
    \end{tabular} 
    \caption{Relative uncertainties (in \%) for the kinematic selection method in
      the single lepton \ttbar\ channel ($e$+jets and $\mu$+jets combined), for the
      MV1 algorithm at an operating point corresponding to a 70\% tagging
      efficiency.
      Negligibly small uncertainties are indicated by dashes.}
    \label{tab-sys-mv1_eff70}
  \end{center} 
\end{table}

\begin{table}[tbp]
  \begin{center}
    \scriptsize
    \setlength{\tabcolsep}{0.4pc}
    \begin{tabular}{l|cccccccccc}
      \hline 
      \hline 
                                  & \multicolumn{10}{c}{Jet $\pt\ [\GeV]$} \\ 
      Source                      & 25--30 &30--40 &40--50 &50--60 & 60--75 & 75--90 & 90--110& 110--140& 140--200 &200--300\\ 
      \hline
      IFSR                        & 5.0 & 4.0 & 4.0 & 4.2 & 3.5 & 3.8 & 4.7 & 4.7 & 6.0 & 9.2 \\
      Generator                   & 1.1 & 0.7 & 1.3 & 1.0 & 0.6 & 2.1 & 1.2 & 0.5 & 2.9 & 8.1 \\
      Fragmentation               & 2.7 & 1.4 & 1.2 & 0.9 & 1.1 & 0.9 & 0.3 & 1.2 & 1.2 & 3.9 \\
      $Z$ + jets                  & 2.2 & 1.8 & 1.4 & 1.3 & 1.2 & 1.2 & 1.3 & 1.7 & 2.5 & 5.3 \\
      Fake/non-prompt lepton      & 3.2 & 1.1 & 1.3 & 1.3 & 1.1 & 1.2 & 1.5 & 1.0 & 2.0 & 0.4 \\
      $t \bar{t}$                 & 1.8 & 1.1 & 0.9 & 0.7 & 0.6 & 0.7 & 0.7 & 0.7 & 1.1 & 1.2 \\
      Single top                  & - & - & - & - & - & - & - & - & - & 0.1 \\
      Diboson                     & 0.2 & 0.1 & - & - & - & - & - & - & - & - \\
      Jet energy scale            & 8.3 & 3.1 & 2.0 & 1.3 & 0.3 & - & 0.3 & 1.0 & 1.3 & 2.9 \\
      Jet energy resolution       & 0.9 & 1.1 & 1.3 & 0.3 & 0.6 & - & 0.5 & 0.1 & 0.5 & 0.5 \\
      Jet reconstruction eff.     & 0.2 & - & 0.2 & - & - & - & - & 0.1 & - & 0.2 \\
      Jet vertex fraction         & 0.3 & 0.3 & 0.1 & - & - & - & - & - & 0.2 & - \\
      $\varepsilon_{c}$            & 0.2 & - & - & - & - & - & - & 0.1 & 0.2 & 0.5 \\
      $\varepsilon_{l}$            & 0.4 & 0.2 & 0.1 & 0.1 & - & 0.1 & - & 0.1 & 0.3 & 1.1 \\
      $\varepsilon_{\text{fake/np}}$ & 1.2 & 1.3 & 0.8 & 1.5 & 0.9 & 1.3 & 1.1 & 1.1 & 0.9 & 3.8 \\
      $\met$                      & - & - & - & - & - & - & - & - & 0.1 & 0.2 \\
      Lepton eff./res.            & 0.4 & 0.2 & 0.2 & 0.2 & - & 0.2 & - & 0.2 & 0.4 & 1.4 \\
      Luminosity                  & 0.3 & - & - & - & - & - & 0.1 & - & 0.2 & - \\
      \hline
      Total systematic uncertainty&  11 & 6.1 & 5.5 & 5.2 & 4.2 & 5.0 & 5.4 & 5.6 & 7.7 & 15 \\
      Statistical uncertainty     & 5.4 & 3.2 & 2.6 & 2.6 & 2.1 & 2.4 & 2.5 & 2.9 & 3.7 & 11 \\
      \hline
      Total uncertainty           &  12 & 6.9 & 6.1 & 5.9 & 4.7 & 5.5 & 5.9 & 6.3 & 8.6 & 18 \\
      \hline
      \hline
    \end{tabular}  
    \caption{Relative uncertainties (in \%) for the kinematic selection method in
      the dilepton \ttbar\ channel ($ee$, $\mu\mu$ and $e\mu$ combined), for the MV1
      algorithm at an operating point corresponding to a 70\% tagging efficiency.
      Negligibly small uncertainties are indicated by dashes.}
    \label{tab:kinsel_dilep_MV1_70_sys} 
  \end{center}
\end{table}

\begin{table}[tbp]
  \begin{center}
    \scriptsize
    \setlength{\tabcolsep}{0.4pc}
    \begin{tabular}{l|ccccccccc}
      \hline\hline
                              & \multicolumn{9}{c}{Jet $\pt\ [\GeV]$} \\ 
      Source                  & 25--30 &30--40 &40--50 &50--60 & 60--75 & 75--90 & 90--110&110--140&140--200 \\ 
      \hline
      IFSR                    & 1.8 & 2.1 & 2.1 & 1.9 & 2.5 & 2.6 & 2.2 & 2.0 & 1.8 \\
      Generator               & 0.8 & 0.3 & 0.6 & 1.7 & 0.2 & 0.4 & 0.4 & 0.9 & 0.9 \\
      Fragmentation           & 0.2 & 0.4 &   - & 0.4 & 1.5 & 0.3 & 0.4 & 2.0 & 1.7 \\
      Jet energy scale        & 0.4 & 0.7 & 0.4 & 1.2 & 0.9 & 0.4 & 0.7 & 2.1 & 0.8 \\
      Jet energy resolution   &   - &   - &   - &   - &   - &   - &   - &   - & - \\
      Jet reconstruction eff. & 0.2 &   - &   - & -   &   - &   - &   - &   - & - \\
      $\met$                  & 1.3 & 1.8 & 1.2 & 0.3 & 0.6 & 0.4 & 2.2 & 2.8 & 0.7 \\
      Lepton eff./res.        & 0.3 & 0.2 & 0.2 & 0.3 &   - &   - &   - &   - &   - \\
      Top quark mass          & 1.4 & 1.0 & 1.2 & 1.1 & 0.4 & 0.4 & 0.6 & 0.1 & 1.1 \\
      $\chi^2$ cut            & 8.8 &  17 & 6.7 & 9.8 & 4.1 &  11 & 5.0 & 6.4 & 5.3 \\
      Pretag cut              & 2.5 & 2.9 & 0.3 & 1.3 & 5.3 & 4.9 & 3.4 & 2.6 & 0.4 \\
      \hline
      Total systematic uncertainty& 9.5 & 18 & 7.2 & 10 & 7.4 & 13 & 6.9 & 8.3 & 6.1 \\
      Statistical uncertainty &  18 &  10 & 8.9 & 9.3 & 5.5 & 7.7 & 6.2 & 6.3 & 6.7 \\
      \hline
      Total uncertainty       &  20 &  21 &  12 &  14 & 9.3 &  15 & 9.2 &  10 & 9.1 \\
      \hline\hline
    \end{tabular} 
    \caption{Relative uncertainties (in \%) for the kinematic fit method in the
      single-lepton \ttbar\ channel ($e$+jets and $\mu$+jets combined), for the MV1
      algorithm at an operating point corresponding to a 70\% tagging efficiency.
      Negligibly small uncertainties are indicated by dashes.}
    \label{kinfit-tab-sys-MV1-70}
  \end{center}
\end{table}

\begin{table}[tbp]
  \begin{center}
    \scriptsize
    \setlength{\tabcolsep}{0.4pc}
 \begin{tabular}{l|cccccccccc}
\hline\hline
& \multicolumn{10}{c}{Jet $\pt\ [\GeV]$} \\ 
Source & 20--30  & 30--40 & 40--50  & 50--60 & 60--75 & 75--90 & 90--110  & 110--140 & 140--200 & 200--300 \\
\hline
IFSR                       & 1.1 & 1.0 & 1.3 & 0.6 & 0.9 & 0.9 & 1.0 & 0.9 & 1.2 & 2.0 \\
Generator                  & 1.2 & 1.4 & 0.9 & 2.1 & 0.9 & 1.4 & 0.9 & 2.0 & 1.1 & 0.8 \\
Fragmentation              & 0.5 & 0.7 & 1.5 & 1.2 & 1.1 & 0.7 & 0.9 & 0.4 & 1.2 & 1.9 \\
Top \pt{} reweighting      & 0.1 & 0.2 & 0.2 & 0.3 & 0.1 & - & 0.1 & 0.4 & 1.4 & 4.6 \\
$Z$+jets                   & 1.1 & 0.9 & 0.6 & 1.2 & 0.4 & 0.5 & 0.6 & 0.8 & 1.4 & 2.9 \\
Fake/non-prompt lepton     & 0.3 & 0.2 & 0.2 & 0.2 & 0.2 & 0.4 & 0.2 & 0.2 & 0.3 & 0.7 \\
Single top                 & 0.8 & 0.6 & 0.3 & 0.3 & 0.3 & 0.2 & 0.2 & 0.1 & 0.2 & 0.2 \\
Diboson                    & 0.8 & 0.4 & 0.4 & 0.4 & 0.5 & 0.3 & 0.7 & 0.8 & 1.3 & 3.9 \\
Jet energy scale           & 3.5 & 1.4 & 0.6 & 0.8 & 0.6 & 0.4 & 0.7 & 0.9 & 1.4 & 2.9 \\
Jet energy resolution      & 1.7 & - & 0.8 & 0.4 & 0.2 & 0.2 & 0.2 & 0.1 & 0.6 & 1.1 \\
Jet vertex fraction        & 0.1 & - & 0.1 & 0.2 & 0.2 & 0.2 & 0.2 & 0.3 & 0.4 & 0.5 \\
$\varepsilon_{l}$           & 1.2 & 0.5 & 0.2 & 0.3 & 0.2 & 0.1 & 0.1 & 0.2 & 0.3 & 1.0 \\
\met                       &  0.1 & 0.2 & 0.1 & 0.1 & 0.1 & 0.1 & 0.2 & 0.2 & 0.4 & 0.8 \\
Lepton eff./res.           & 0.1 & 0.1 & 0.1 & 0.1 & 0.1 & -   & -   & -   & 0.1 & 0.1 \\
Pile-up $\langle \mu \rangle$ reweighting & 0.2 & - & 0.2 & 0.1 & 0.1 & - & 0.1 & - & 0.1 & 0.2 \\
\hline
Total systematic uncertainty & 4.7 & 2.6 & 2.6 & 3.0 & 2.0 & 2.0 & 2.0 & 2.7 & 3.6 & 8.0\\
Statistical uncertainty    & 3.7 & 2.7 & 2.3 & 2.3 & 1.9 & 2.1 & 2.1 & 2.5 & 3.4 & 8.3\\
\hline
Total uncertainty          & 6.0 & 3.8 & 3.5 & 3.8 & 2.8 & 2.9 & 3.0 & 3.7 & 4.9 & 12 \\
\hline\hline
\end{tabular} 
\caption{Relative uncertainties (in \%) for the combinatorial likelihood method in the 
  dilepton \ttbar\ channel ($ee$, $\mu\mu$ and $e\mu$ combined), 
  for the MV1 algorithm at an operating point corresponding to a 70\% 
  tagging efficiency. Negligibly small uncertainties are indicated by dashes.}
\label{pdf-tab-sys-MV1-70}
\end{center}
\end{table}

\subsubsection*{Initial and final state radiation}

Initial and final state radiation (IFSR) directly affects the flavour composition of the
$\ttbar$ events. The associated systematic uncertainty due to IFSR is estimated by studies 
using samples generated with \AcerMC~\cite{bib:acer} interfaced to \Pythia, and by varying the 
parameters controlling IFSR in a range consistent with experimental 
data~\cite{STDM-2011-03,TOPQ-2011-21}. 
In the combinatorial likelihood method, the IFSR parameters are also varied in single top $Wt$-channel events.

\subsubsection*{Generator and fragmentation dependence}

The baseline generator MC@NLO+\Herwig{} may not correctly predict the kinematic distribution 
of $\ttbar$ events, which may result in differences in the acceptance and flavour 
composition of selected events. A systematic uncertainty is assigned to the choice 
of Monte Carlo generator (Generator) by comparing the results produced
with the baseline $\ttbar$ generator with those produced with events simulated with \Powheg+\Herwig{}.
In the combinatorial likelihood method, which uses \Powheg+\Herwig{} as the baseline generator, 
a comparison between \Powheg+\Herwig{} and \Alpgen{}+\Herwig{} is done instead.
Uncertainties in the fragmentation modelling (Fragmentation) are estimated by comparing results 
between events generated with \Powheg+\Herwig{} and those generated using \Powheg+\Pythia{}.

\subsubsection*{Background normalisation}

In all analyses the dominant backgrounds are estimated using data-driven techniques.
In the single-lepton final state, the dominant background comes from $W$+jets production,
and the normalisation of this background is varied by 13\% based on the consideration
of the various scale factors to correct the expectations derived from simulated events. In the dilepton 
final states the $Z$+jets normalisation uncertainty depends on the number of jets in 
the final state. An inclusive normalisation uncertainty of 4\% is assumed, and
following Ref.~\cite{Alwall:2007fs}, an additional term of 
24\% per jet is added in quadrature.
In the combinatorial likelihood method the normalisation of the $Z$+jets background is varied by $\unit[20]{\%}$.
In the single-lepton analyses, where the $Z$+jets 
background is substantially smaller, it is normalised to the theoretical cross section and 
varied by 60\% \cite{TOPQ-2010-01}.

The multijet background in the kinematic selection measurement is
varied by 50\% in the $e$+jets channel, 
covering any differences in kinematic
distributions arising from mismodelling of the multijet background. In
the $\mu$+jets channel, by comparing estimates based on two different control
regions, the uncertainty on the multijet background normalisation can be
reduced to 20\%.
The non-prompt or fake lepton background in the kinematic selection dilepton 
and combinatorial likelihood analyses is varied by 50\%.

The single top and diboson backgrounds are normalised to their theoretical cross
sections, and the corresponding uncertainties (8\% for the single top $Wt$
channel~\cite{bib:WtXsec}, 4\% for the $s$- and $t$-channel single top
production processes~\cite{bib:tXsec,bib:sXsec}, and 5\% for the diboson
background) are accounted for in the analysis.
In the combinatorial likelihood method, the relative \ttbar\ to single top normalisation is varied by $\unit[25]{\%}$ 
in the two-jet bin and $\unit[35]{\%}$ in the three-jet bin, motivated by scale
uncertainties and parton distribution function systematic variations.

\subsubsection*{Background flavour composition}

The flavour composition of all background samples except $W$+jets and in some cases 
$Z$+jets is taken from simulation.
No systematic uncertainty on the flavour composition for these samples is assigned.
For the $W$+jets background the
normalisations of heavy flavour (HF) events ($W\bbar$+jets, $W\ccbar$+jets 
and $Wc$+jets) are varied within their uncertainties. 
Sources of systematic uncertainty that affect the HF scale factors in $W$+jets 
events often also affect the calibration methods described in this
paper directly. Examples of such systematic uncertainties are the uncertainties 
on the $\ttbar$ cross section and $W$+jets normalisation. To account
for such correlations, these uncertainties are evaluated by coherently evaluating their 
impact on all components of the analysis.
In the combinatorial likelihood method the heavy flavour component of the $Z$+jets 
background is varied up and down by 100\% and 50\% respectively,
which is conservative compared to the variations between data and simulated 
events observed in $Z+b$ measurements~\cite{STDM-2011-22}.

\subsubsection*{Background modelling}

In the combinatorial likelihood method, the uncertainty from the modelling of $Z$+jets is estimated by comparing events 
generated with \Sherpa{} to those generated with \Alpgen.
The impact of the modelling of diboson events is investigated by comparing \Alpgen{} to \Herwig.
In the other analyses, this uncertainty is estimated to be negligible compared
to the corresponding background normalisation and background flavour composition
uncertainties, and is neglected.

\subsubsection*{Jet reconstruction efficiency, energy scale and resolution}

The systematic uncertainty originating from the jet energy scale~\cite{PERF-2012-01} is obtained
by scaling the $\pT$ of each jet in the simulation up and down by the estimated uncertainty on the jet energy scale. 
In the combinatorial likelihood method, the independent components of the jet energy scale uncertainty
are applied separately, as discussed in Ref.~\cite{PERF-2012-01}. 
The nominal jet energy resolution in Monte Carlo simulation and data are found to be compatible, 
but a systematic uncertainty is assigned to cover the effect of possible residual 
differences by smearing the jet energy in simulated events.
The full difference from the nominal result is taken as the uncertainty.
The jet reconstruction efficiency was derived using a tag-and-probe method in dijet events
and found to be compatible with a measurement using simulated $\ttbar$ events. 
A systematic uncertainty is assigned to cover the effect of possible residual differences
by randomly rejecting jets based on the measured jet reconstruction efficiency. 

\subsubsection*{Jet vertex fraction}

The modelling of the jet vertex fraction in simulated events has been studied in
$Z \to \ell \ell$+jets events. Jets from the hard scatter interaction are selected from 
events where one jet is produced back-to-back with a high-$p_T$ $Z$ boson, while jets
from pile-up interactions are selected from events where the $Z$ boson is produced almost
at rest.
Correction factors, bringing the efficiency of a jet either from the
hard scatter vertex or from a pile-up vertex to pass the jet vertex fraction cut in simulated 
events to agree with that measured in data, are applied. 
These correction factors are then varied within their uncertainties.

\subsubsection*{Mistag efficiencies}

In both the tag counting and the kinematic selection methods, the mistag efficiencies for $c$ 
and light-flavour jets, $\epsilon_c$ and $\epsilon_l$, directly enter the expression used to obtain 
the $b$-jet tagging efficiency. The efficiencies in simulated events are adjusted by the data-to-simulation 
scale factors obtained with the methods descried in Sections~\ref{sec:ceff_dstarbased} and~\ref{sec:mistag}.
The efficiencies are then varied within the 
uncertainties on these correction factors, which range from approximately 12\% to 50\%.

In the kinematic selection methods the $b$-jet tagging efficiency 
$\varepsilon_{\rm fake/np}$  for jets from the multijet background in the single-lepton 
analysis and the non-prompt or fake lepton background  in the dilepton analysis  is 
measured in a control region in data. In the dilepton analysis an uncertainty 
of 50\% is assumed, while in the single-lepton analysis the uncertainty 
is obtained by comparing the baseline result with the $b$-jet tagging efficiencies 
measured in events in which the requirement of an isolated electron is replaced by that of a jet 
with a large electromagnetic energy fraction, the so called
jet-electron model~\cite{TOPQ-2011-14}.

\subsubsection*{Missing transverse momentum}

In the combinatorial likelihood method, the uncertainty due to the modelling of the soft-terms used in the $\met$ calculation is accounted for. 
In addition the variation in the $\met$ due to the jet, electron and muon uncertainties are taken into account
when the corresponding systematics are varied.

\subsubsection*{Lepton trigger and identification efficiency, energy scale and resolution}

The modelling in simulation of the lepton trigger, reconstruction and selection efficiencies as well as the energy resolution and scaling 
have been assessed using tag-and-probe techniques in $\Zee$ and $\Zmm$ events, as described in Ref.~\cite{TOPQ-2011-02}. 
The correction factors obtained are further varied within their uncertainties.

\subsubsection*{Luminosity}

The uncertainty on the integrated luminosity affects the measurement of the $b$-jet tagging
efficiency due to the change in the overall normalisation of the backgrounds estimated
from simulation. The integrated luminosity has been measured with a 
precision of 3.9\% following the methods described in Ref.~\cite{DAPR-2010-01}.

\subsubsection*{Pile-up $\langle\mu\rangle$ reweighting}

In all analyses, the Monte Carlo simulation is reweighted on an event-by-event basis to reproduce the 
distribution of the average number of proton-proton interactions ($\mu$) measured in data, after scaling 
$\mu$ in the Monte Carlo simulation as described in Sec.~\ref{sec:samples}. In the combinatorial likelihood method,
the associated systematic uncertainty is evaluated as described in Section~\ref{sec:syst},
while the other analyses probe the indirect effect of
pile-up on the results through pile-up related uncertainties in object 
modelling such as the jet energy scale and missing transverse momentum corrections ($\met$ pile-up).

\subsubsection*{Top momentum reweighting}

In the combinatorial likelihood analysis, the jet \pt\ spectrum in data is found to be softer than the 
prediction from the \Powheg+\Pythia{} \ttbar\ sample. 
The distribution of the average \pt\ of the top and anti-top quark is therefore reweighted at truth level
according to the unfolded measurement performed on 2011 $\sqrt{s} = 7\TeV$ data~\cite{TOPQ-2013-07}. 
The corresponding systematic uncertainty is taken to be 100\% of the correction. 

\subsubsection*{Top pair production cross section}

In the kinematic selection method, the $\ttbar$ cross section is used to normalise the expected $\ttbar$ 
signal relative to the backgrounds. The uncertainty on the predicted $\ttbar$ cross section is $10\%$ \cite{bib:TtbarXsec}.

\subsubsection*{Top quark mass}

The kinematic fit method involves a top quark mass constraint.
To estimate the uncertainty from this source, the measurement has been repeated with simulated 
events with a top mass of 170 and 175~\GeV{} and the change in the results is taken as a systematic uncertainty.

\subsubsection*{$\chi^2$ cut}

The kinematic fit method normalises the background sample in the high $\chi^2$ region.
The $\chi^2$ value used to define the high $\chi^2$ region has been varied from 20 to 50, and
the effect on the final result is taken as a systematic uncertainty. 

\subsubsection*{Pretag cut}

In the kinematic fit method, the jets originating from the decay of the $W$ boson are distinguished from
those originating from bottom quarks by means of a $b$-tagging requirement. Nominally jets are $b$-tagged 
by the MV1 algorithm at the 70\% operating point. The measurement has been repeated using the 75\%
operating point, and the full difference to the nominal result is taken as a systematic uncertainty.

\subsection{Results}
\label{sec:ttbar_results}

\begin{figure}[ht!] 
  \centering
  \subfloat[]{\label{fig:results_SFsMV1_ljet_a}\includegraphics[width=0.49\textwidth]{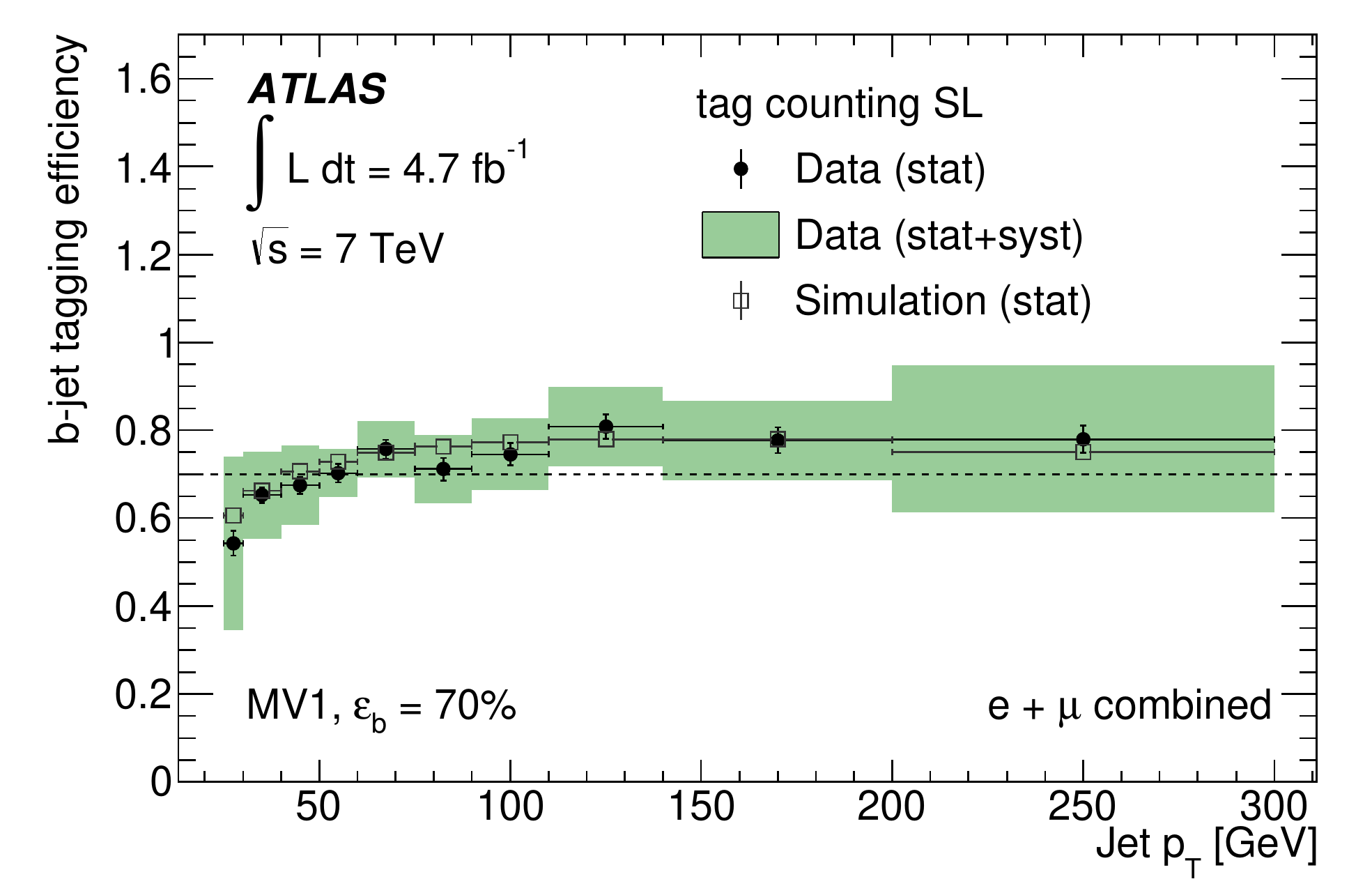}}
  \subfloat[]{\label{fig:results_SFsMV1_ljet_b}\includegraphics[width=0.49\textwidth]{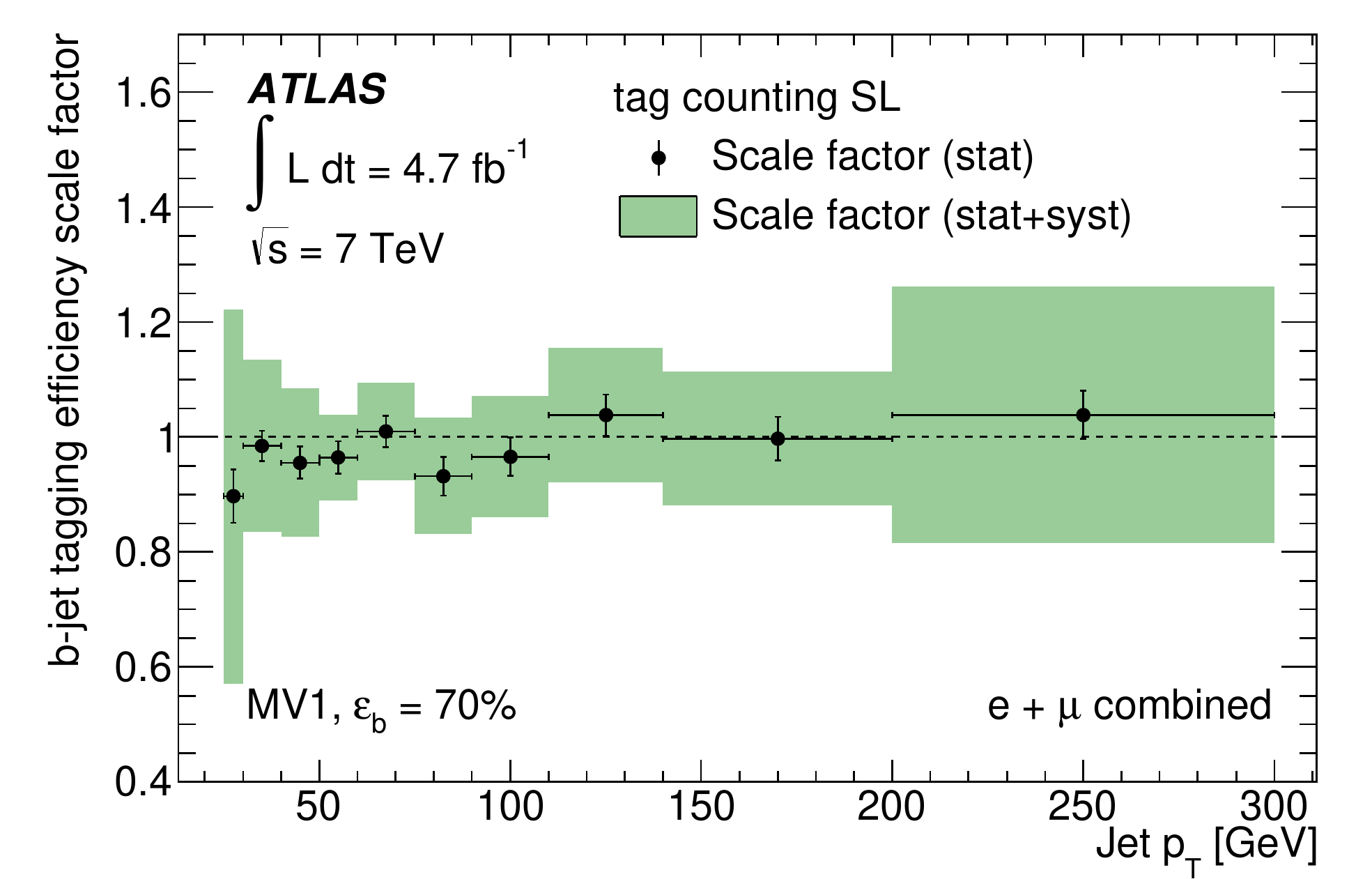}}  \\
  \subfloat[]{\label{fig:results_SFsMV1_ljet_c}\includegraphics[width=0.49\textwidth]{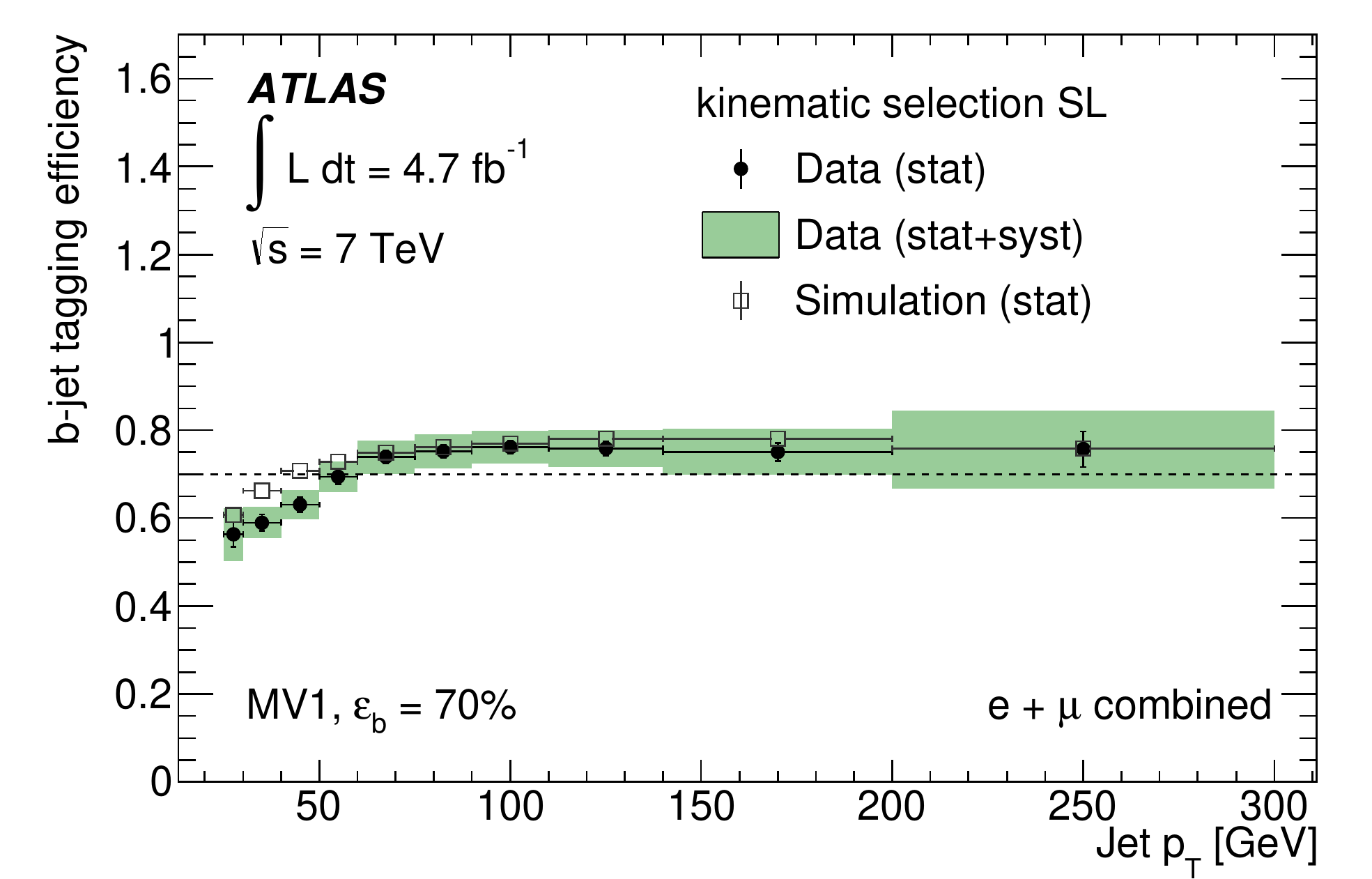}}
  \subfloat[]{\label{fig:results_SFsMV1_ljet_d}\includegraphics[width=0.49\textwidth]{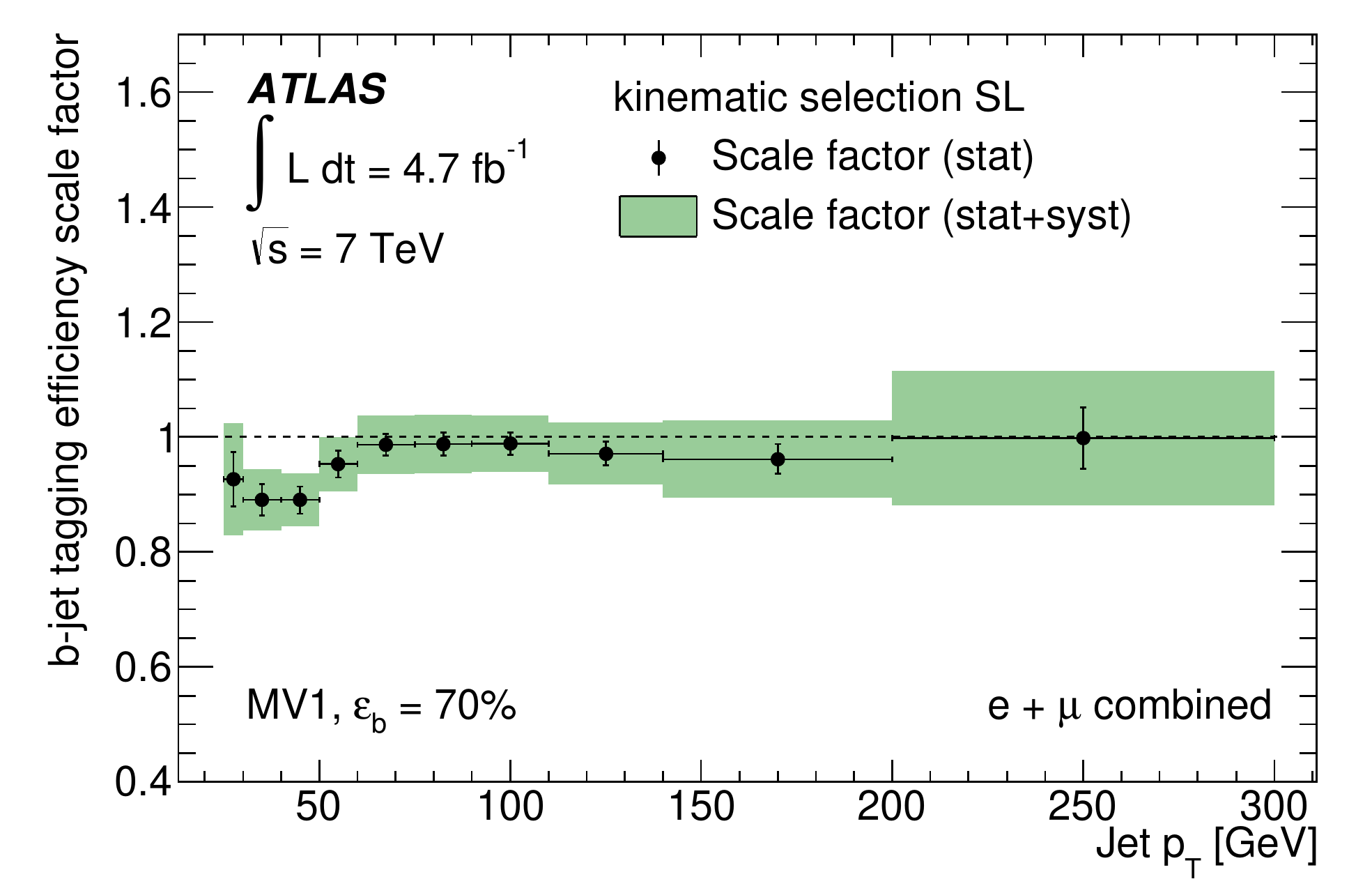}} \\
  \subfloat[]{\label{fig:results_SFsMV1_ljet_e}\includegraphics[width=0.49\textwidth]{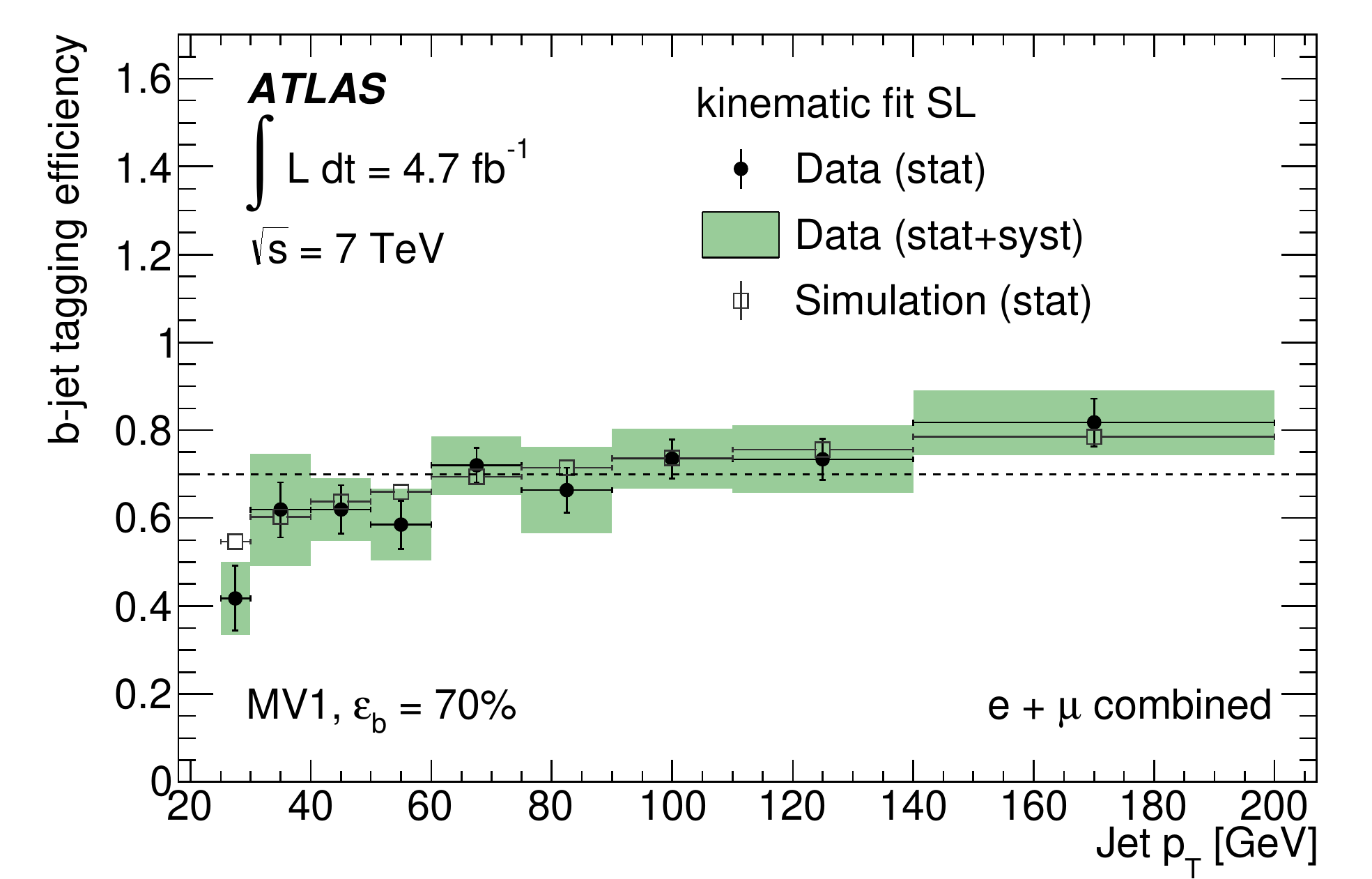}}
  \subfloat[]{\label{fig:results_SFsMV1_ljet_f}\includegraphics[width=0.49\textwidth]{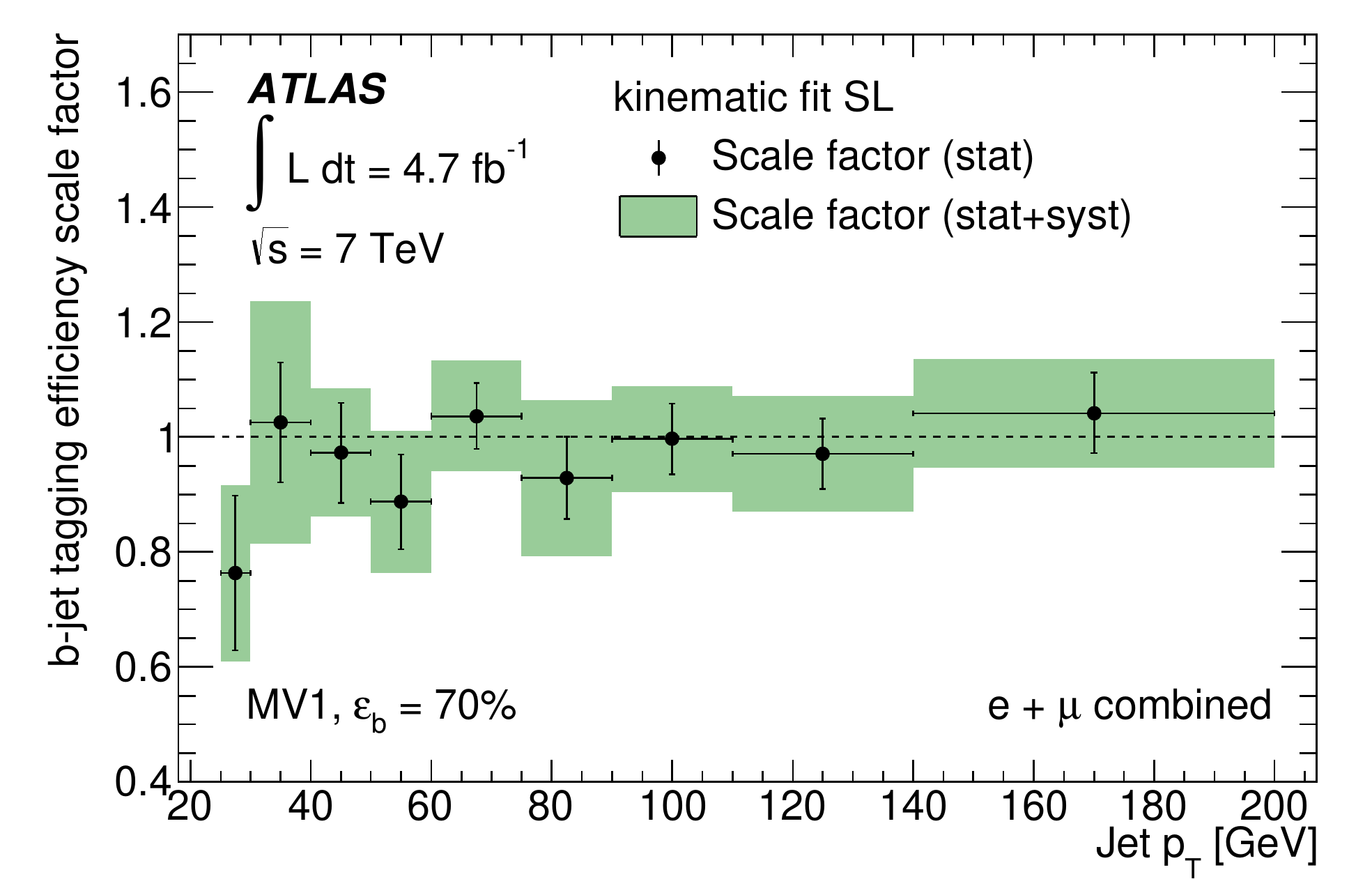}}
  \caption{The measured $b$-jet tagging efficiency in data compared to that in simulation (left) and 
    the resulting scale factors (right) for the MV1 algorithm at 70\% $b$-jet tagging efficiency
    for the single-lepton $\ttbar$ tag counting (top), kinematic selection (middle) and kinematic fit (bottom) methods. 
    The error bars show the statistical uncertainties while the green band indicates the total uncertainty.}
  \label{fig:results_SFsMV1_ljet}
\end{figure}

\begin{figure}[ht!] 
  \centering
  \subfloat[]{\label{fig:results_SFsMV1_dilep_c}\includegraphics[width=0.49\textwidth]{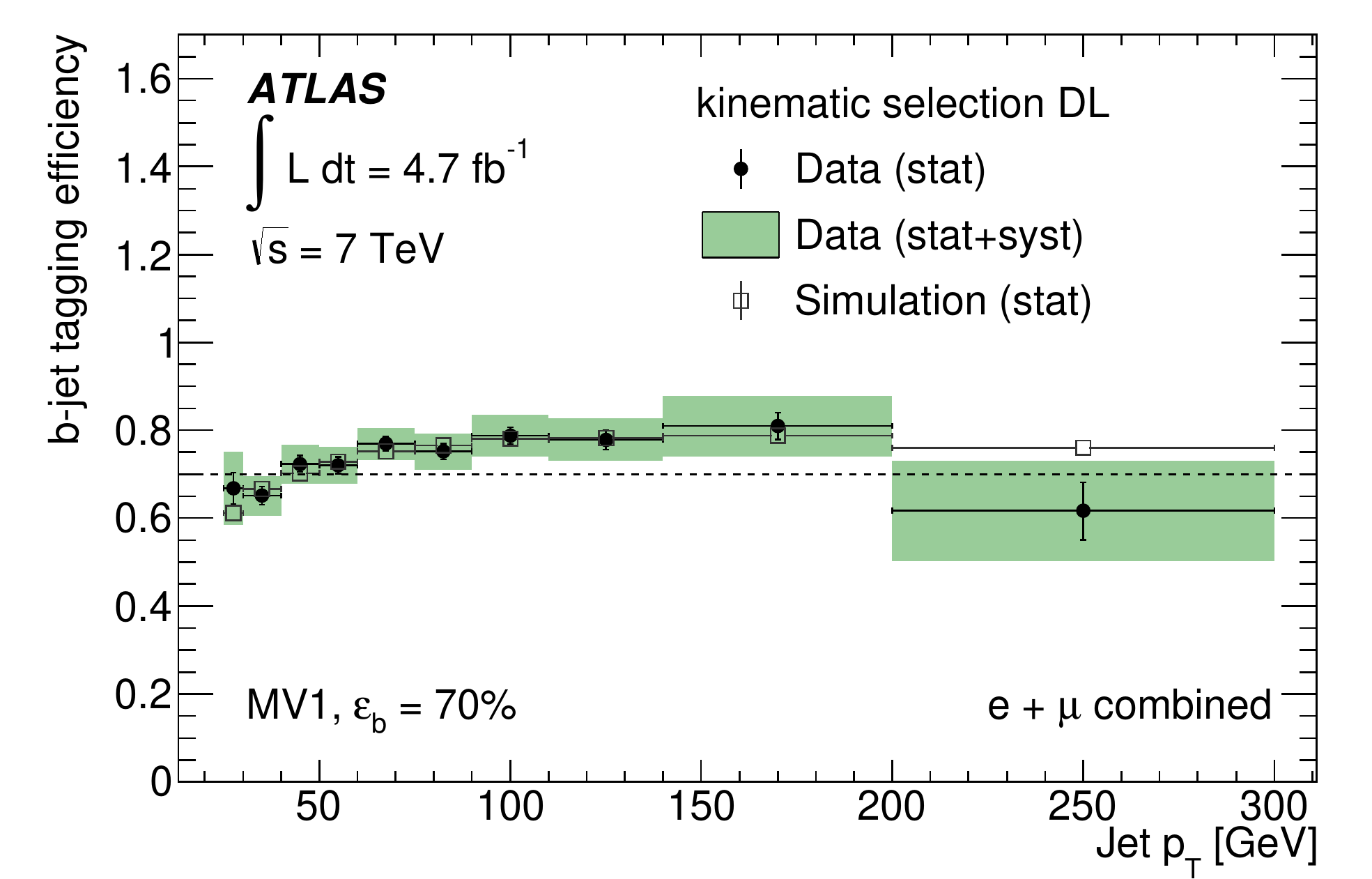}} 
  \subfloat[]{\label{fig:results_SFsMV1_dilep_d}\includegraphics[width=0.49\textwidth]{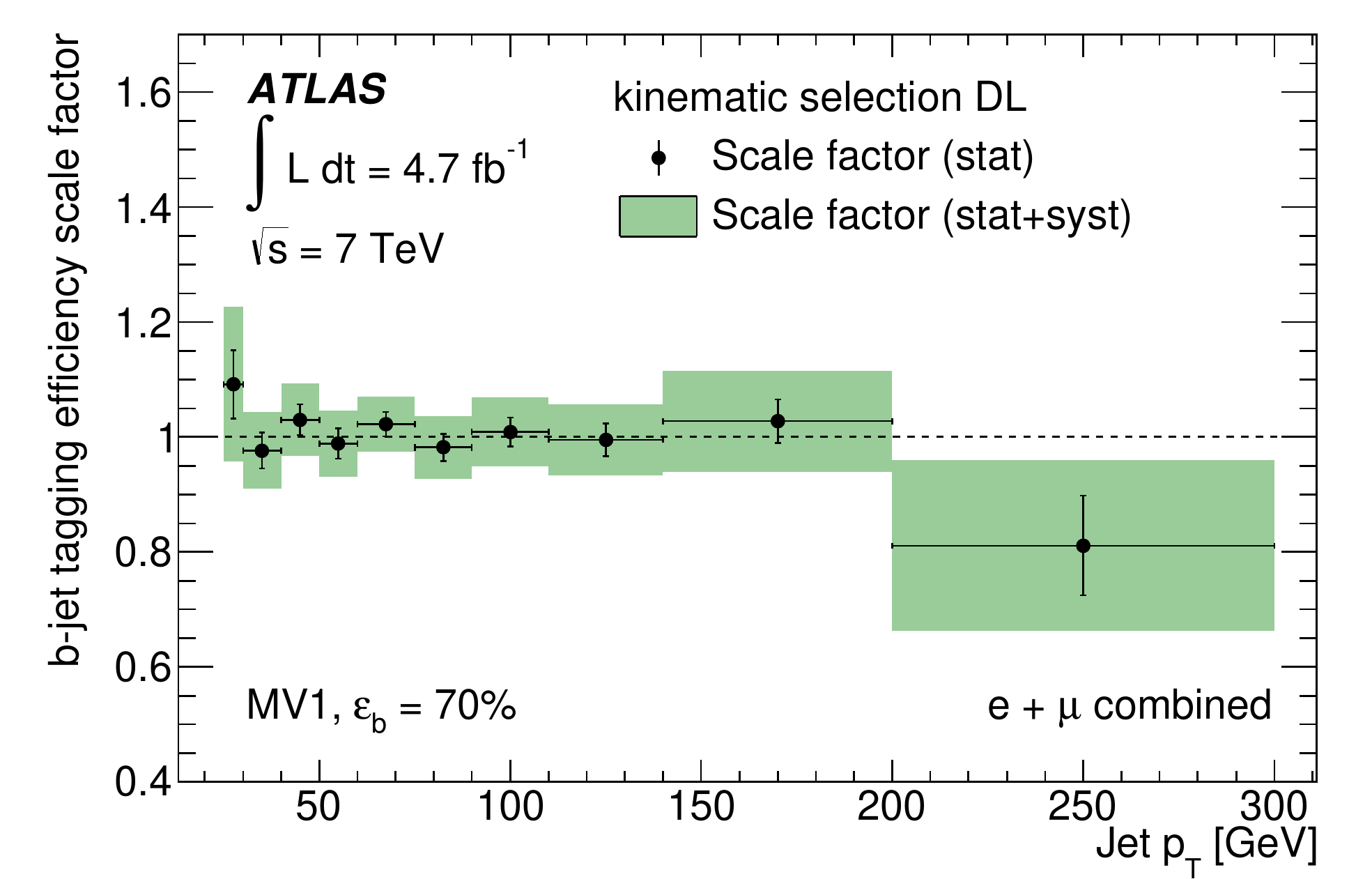}} \\ 
  \subfloat[]{\label{fig:results_SFsMV1_dilep_e}\includegraphics[width=0.49\textwidth]{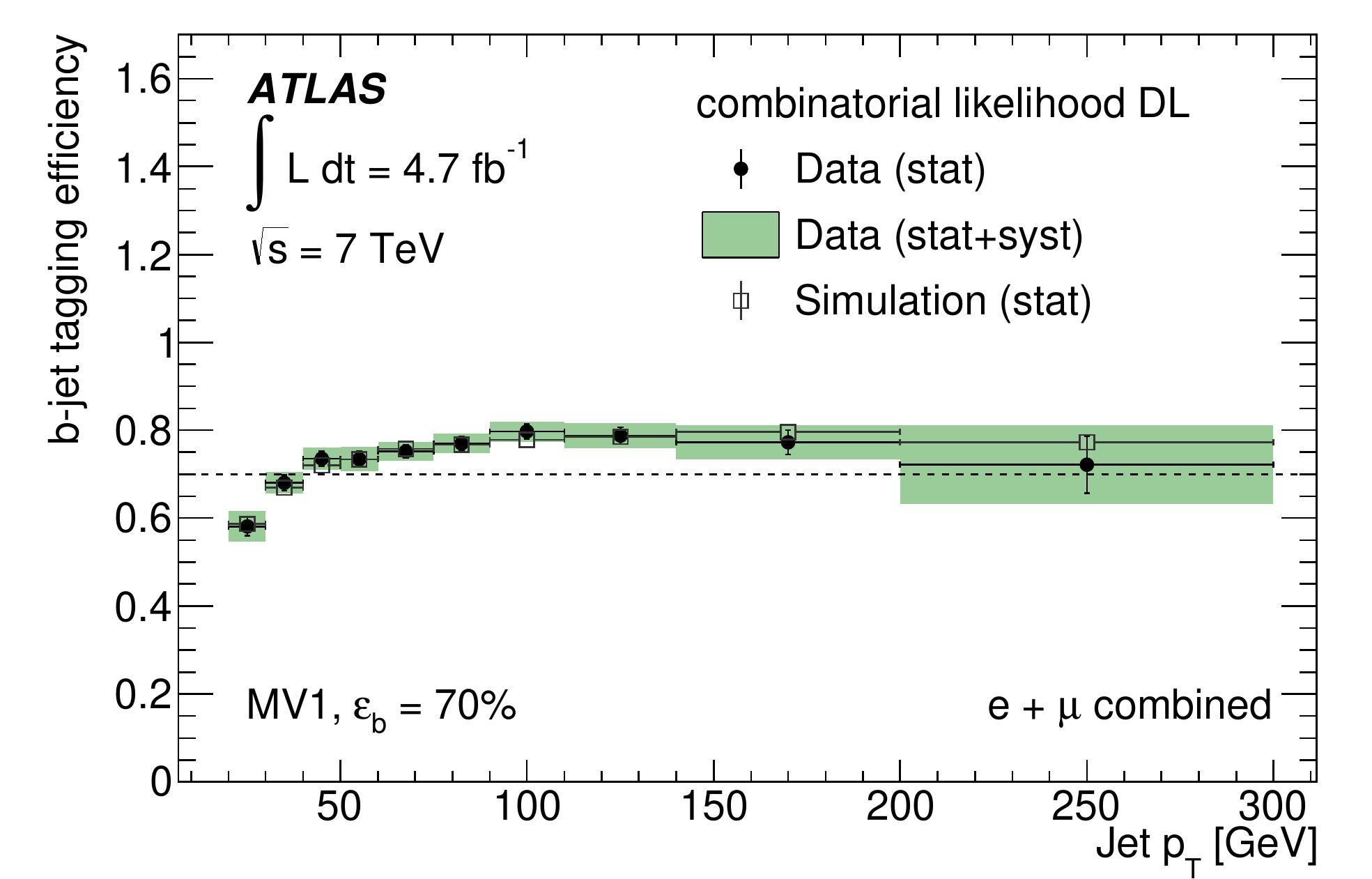}} 
  \subfloat[]{\label{fig:results_SFsMV1_dilep_f}\includegraphics[width=0.49\textwidth]{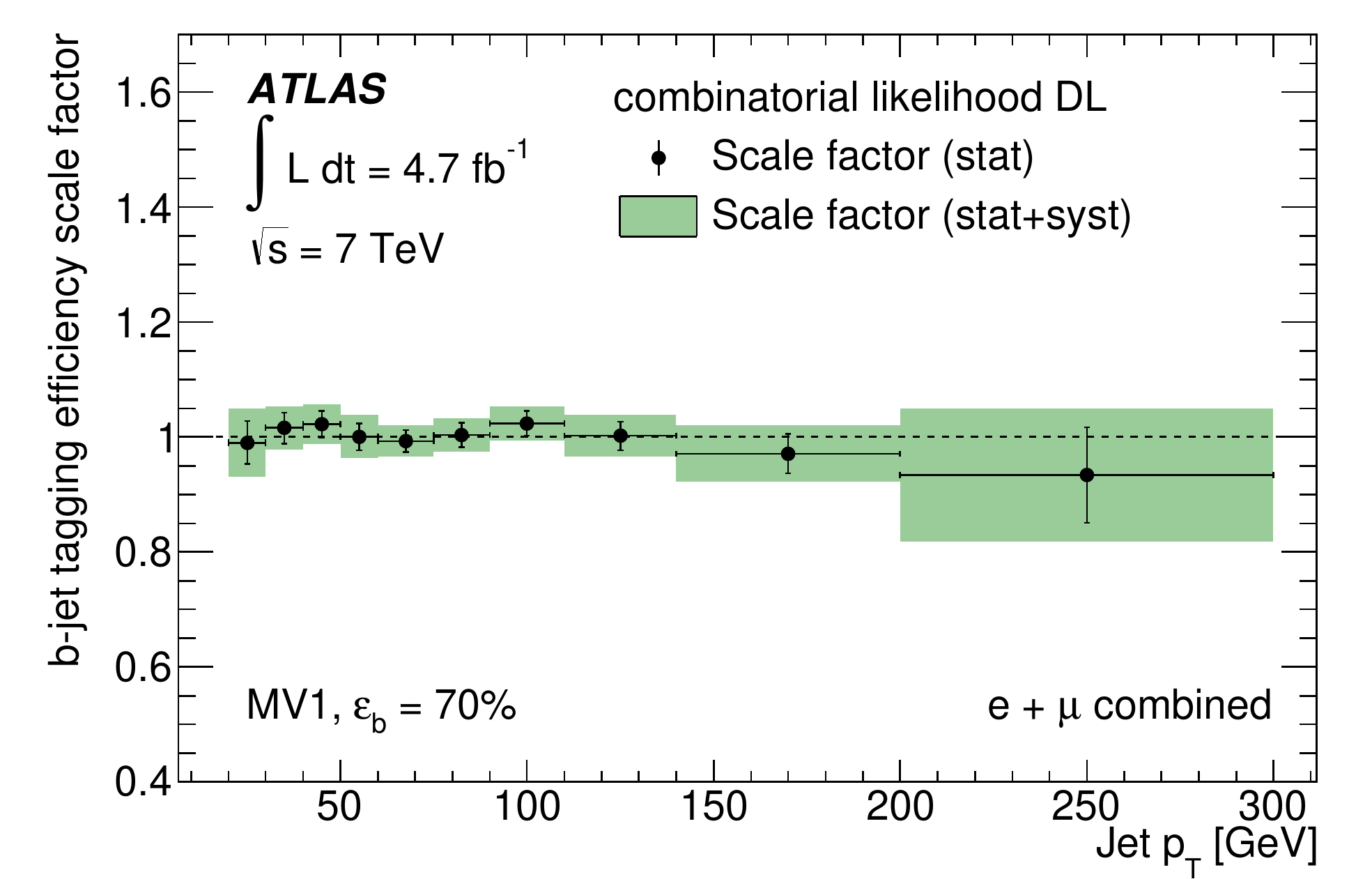}} 
 \caption{The measured $b$-jet tagging efficiency in data compared to that in simulation (left) and 
    the resulting scale factors (right) for the MV1 algorithm at 70\% $b$-jet tagging efficiency
    for the dilepton $\ttbar$ kinematic selection (top) and combinatorial likelihood (bottom) methods.
    The error bars show the statistical uncertainties while the green band indicates the total uncertainty.}
   \label{fig:results_SFsMV1_dilep}
\end{figure}

\begin{figure}[ht!] 
  \centering
  \subfloat[]{\label{fig:results_SFsMV1_dilep_eta_a}\includegraphics[width=0.49\textwidth]{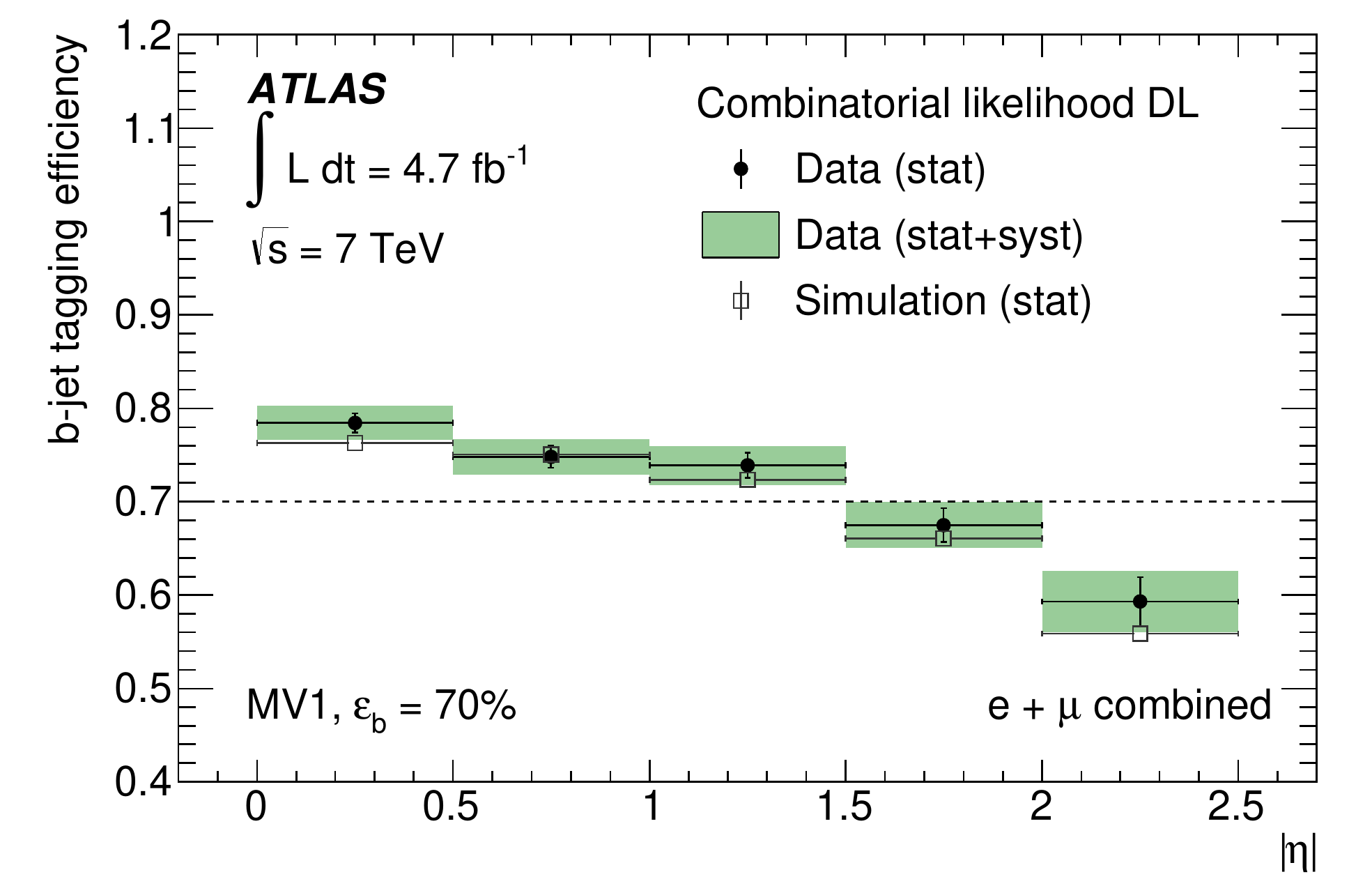}} 
  \subfloat[]{\label{fig:results_SFsMV1_dilep_eta_b}\includegraphics[width=0.49\textwidth]{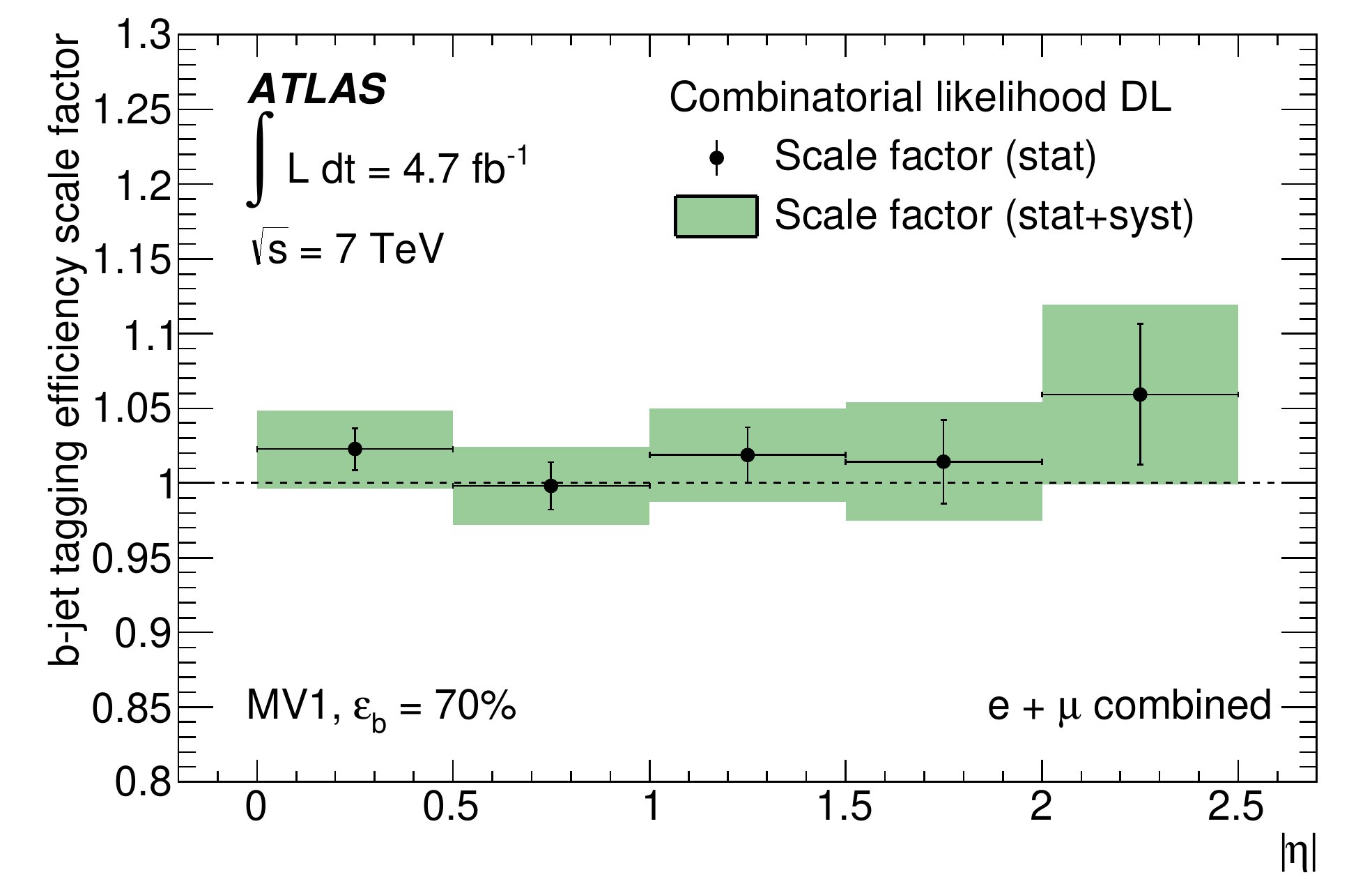}} 
  \caption{The measured $b$-jet tagging efficiency in data compared to that in simulation (left) and 
    the resulting scale factors (right) as a function of $|\eta|$ for the MV1 algorithm 
    at 70\% $b$-jet tagging efficiency for the dilepton $\ttbar$ combinatorial likelihood method.
    The error bars show the statistical uncertainties while the green band indicates the total uncertainty.}
   \label{fig:results_SFsMV1_dilep_eta}
\end{figure}

The $b$-jet tagging efficiency measured in data, the corresponding values 
from simulation and the resulting data-to-simulation scale factors for the MV1 tagging algorithm at 70\% efficiency 
are shown as a function of the jet \pT{} in Figs.~\ref{fig:results_SFsMV1_ljet} and~\ref{fig:results_SFsMV1_dilep}
for the single-lepton and dilepton analyses, respectively. 
The agreement in the scale factors among all the methods is very good. 
The scale factors are close to unity, with an uncertainty ranging from 4\% to about 40\%, depending on the jet \pT{} and on the calibration method.
The efficiencies and scale factors as a function of $|\eta|$ are shown in Fig.~\ref{fig:results_SFsMV1_dilep_eta} for the combinatorial likelihood method. 
No significant scale factor dependence on $|\eta|$ is observed.

\subsection{The \ptrel{} method in $t\bar t$ events}

To further the understanding of the results obtained from muon-based calibration
methods in dijet events, a \ptrel{} analysis (as discussed already in
Section~\ref{sec:beff_mubased}) has been performed on the single-lepton $t\bar t$
sample.
As the $t\bar t$ sample has a high $b$-jet purity and no trigger bias on $b$
jets, this $t\bar t$-based measurement provides a useful cross check to the dijet-based
\ptrel{} measurement. 
Furthermore, this allows the application of different calibration methods
using a very similar physics process, helping to understand potential biases originating from
either different calibration analyses or the use of event samples from different physics processes. 

The analysis takes the muon \ptrel{} templates of $b$, $c$, and light-flavour
jets from simulation.
It has been verified through studies of simulated events that there are
several sources of contamination (geometric overlaps from the close-by muons)
to the muons in light-flavour jets.
The dominant source is the isolated lepton that the other $W$ boson in $t\bar t$
dilepton events decays to.
This geometric overlap can be reduced by requiring the muons associated with jets 
that have a charge opposite to that of the isolated lepton in single-lepton \ttbar\ events 
to have $\pT < 20 \GeV$.
To reduce the heavy-flavour contamination to the light-flavour \ptrel{}
template, specific jet isolation 
($\Delta R (\mu, \mbox{other jets})> 0.8$) and muon selection
($\Delta R(\mu, {\rm jet})< 0.3$) criteria are imposed.

The two major sources of background in this analysis are $W$+jets and multijet events.
With some minor exceptions, the backgrounds are estimated as described in Section~\ref{sec:ljets_bkg}.
The \ptrel{} distributions in $W$+jets, $Z$+jets, single top and diboson events are obtained from simulation.
The $b$, $c$, and light-flavour \ptrel{} templates from $W$+jets samples
(together with templates from $Z$+jets, single top and diboson) are added to the
\ptrel{} templates obtained from the simulated $t\bar t$ sample when performing the template fit.
For the multijet background, both in $\mu$+jets and $e$+jets channels,
the muon \ptrel{} shape and normalisation are obtained using the data-driven matrix methods described in Section~\ref{sec:ljets_bkg}.
Both in the pre-tag event sample (with muon-in-jet requirements) and in the tagged
event sample (further requiring at least one jet being tagged by the $b$-tagging algorithm to be calibrated), 
the multijet contribution is estimated by using the fake rate $\epsilon_{\rm fake}$ of the tagged sample.

\begin{figure}[!h] 
  \begin{center}
    \begin{tabular}{c}
      \includegraphics[width=0.6\linewidth]{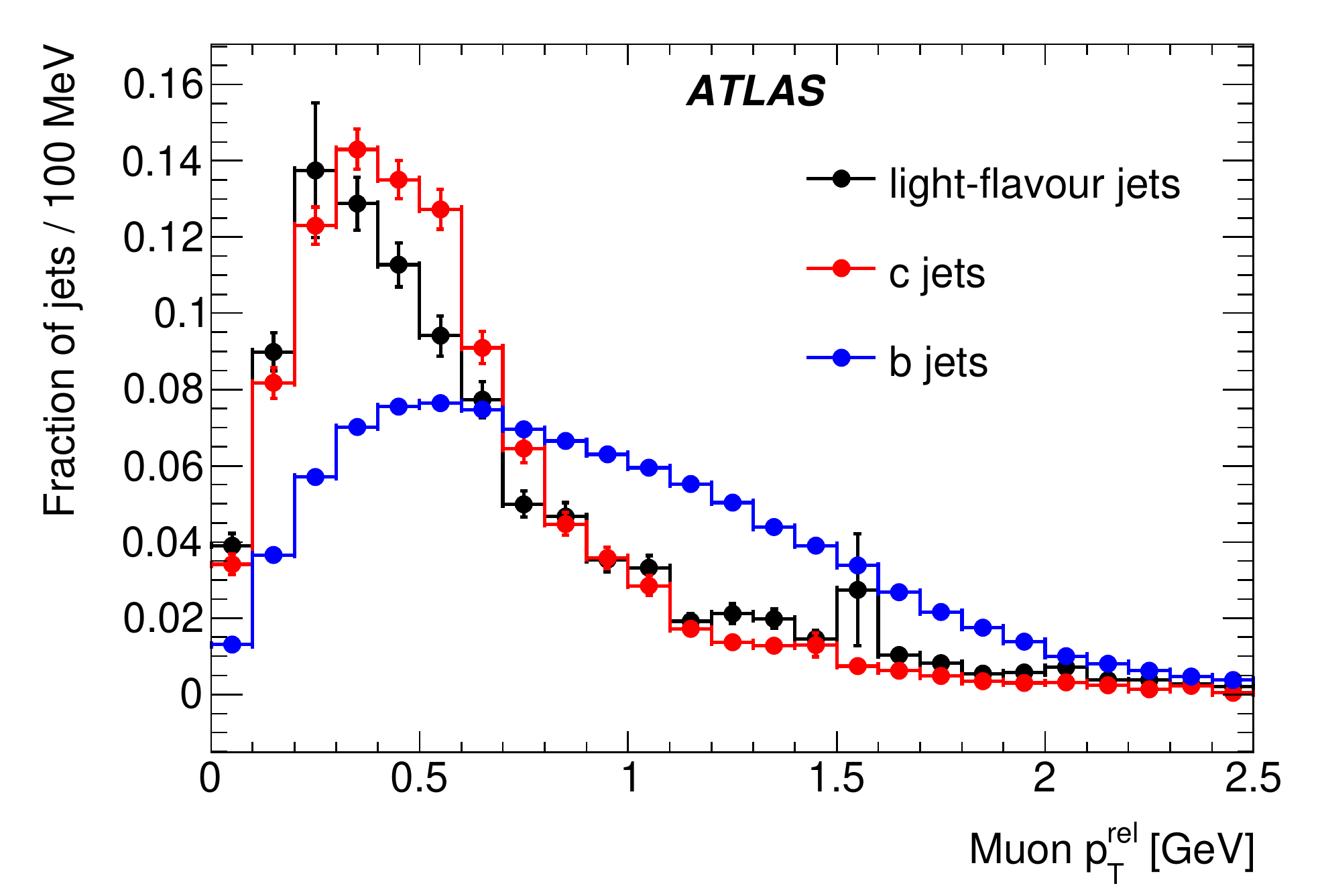}
    \end{tabular}
    \caption{Examples of $b$, $c$, light-flavour jet \ptrel{} templates in simulated $t\bar t$ events.}
    \label{fig:pTrel_ttbar_template}
  \end{center}
\end{figure}

\begin{figure}[!h] 
  \begin{center}
    \subfloat[]{\label{fig:pTrel_ttbar_fit_a}\includegraphics[width=0.49\textwidth]{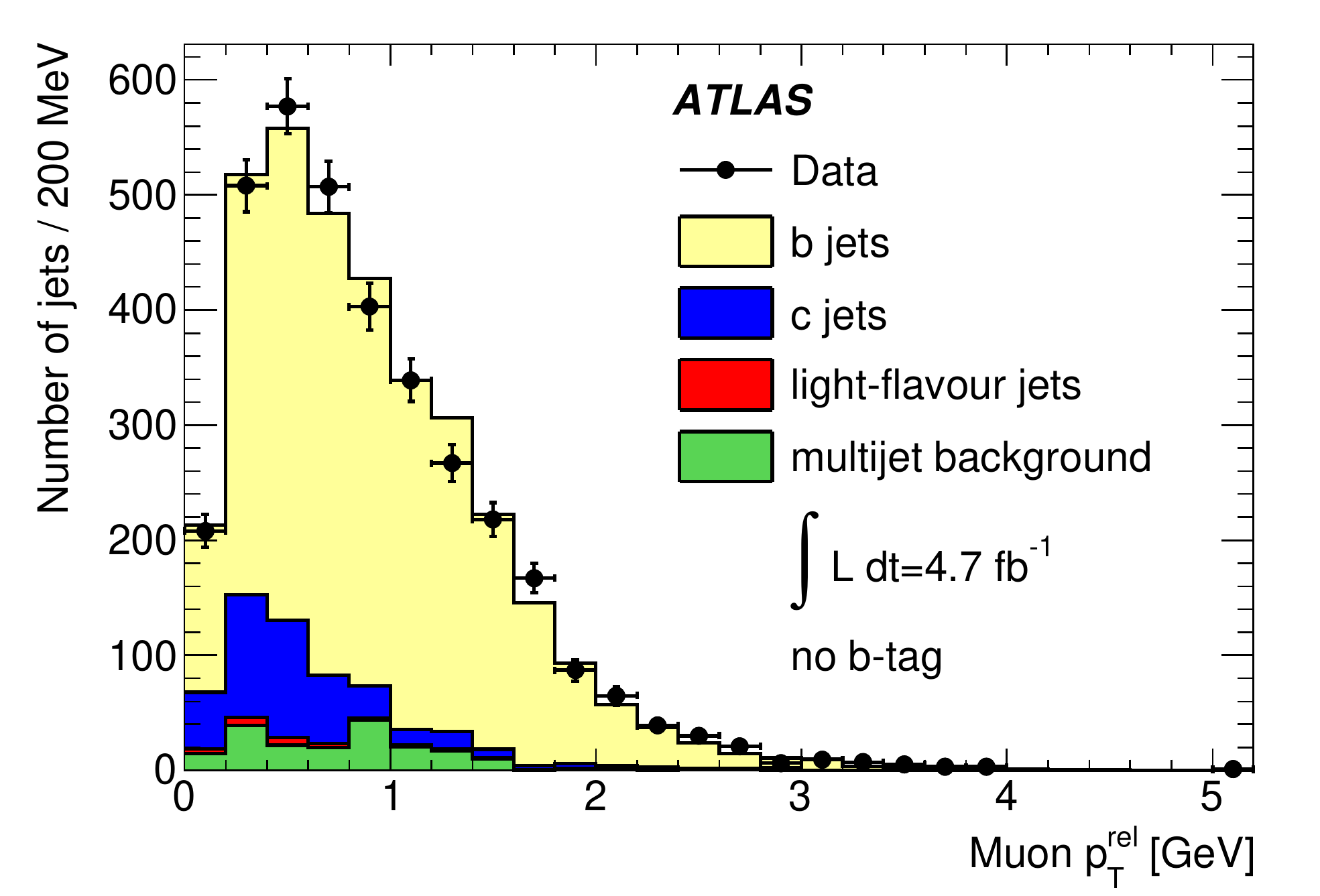}}
    \subfloat[]{\label{fig:pTrel_ttbar_fit_b}\includegraphics[width=0.49\textwidth]{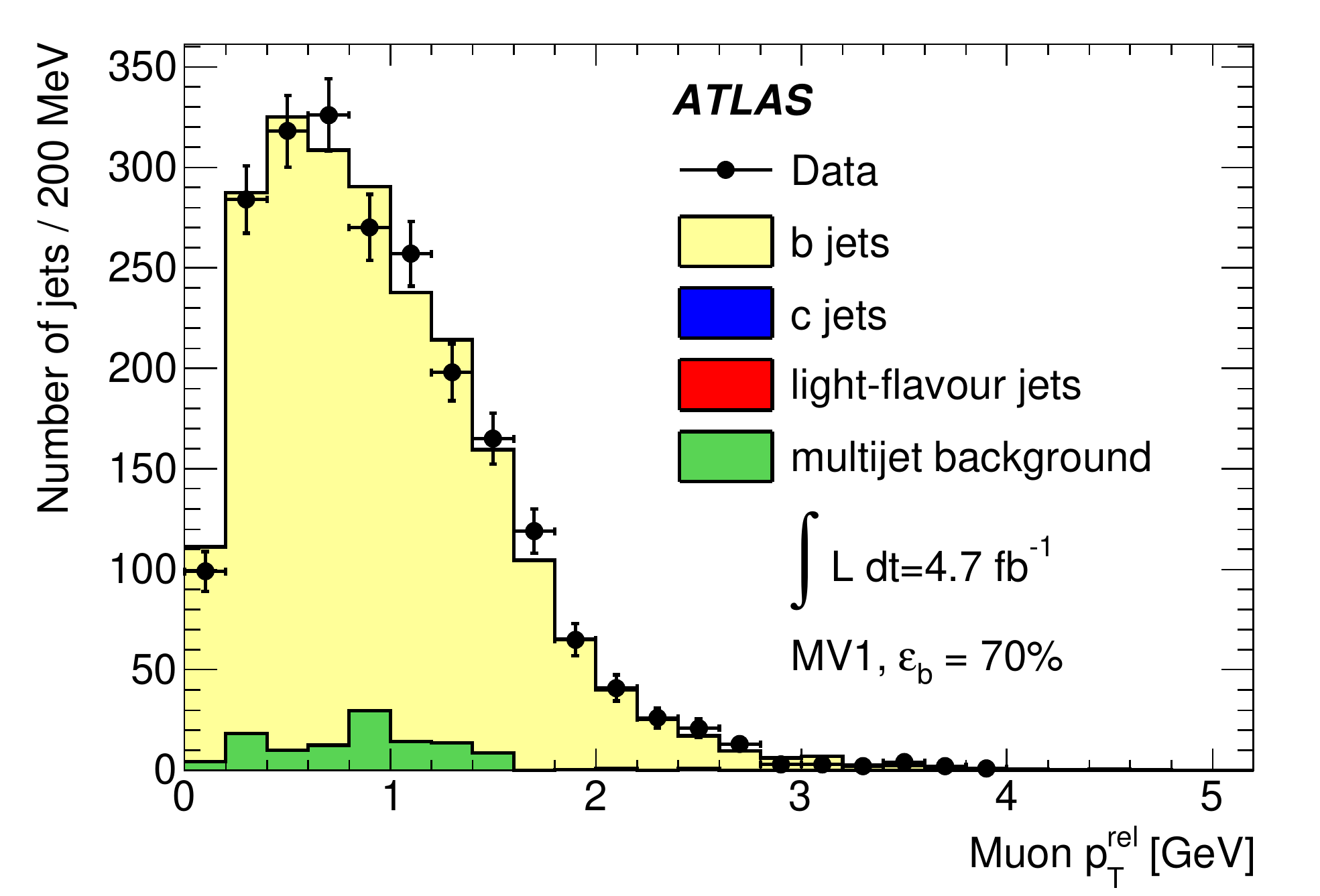}}
    \caption{Examples of template fits to the \ptrel{} distribution in data before (a) and after (b) $b$-tagging 
      with the MV1 algorithm at 70\% $b$-jet tagging efficiency, for jets with $60\GeV < \pt < 90\GeV$.}
    \label{fig:pTrel_ttbar_fit}
  \end{center}
\end{figure}

Examples of the $b$, $c$, and light-flavour jet \ptrel{} templates obtained from
simulation are shown in Fig.~\ref{fig:pTrel_ttbar_template}.
The multijet \ptrel{} template is treated as a fourth template besides the
$b$, $c$, and light-flavour jet \ptrel{} templates and is kept fixed
at the estimated normalisation when performing the fit, while the normalisations
of the $b$, $c$, and light-flavour jets are adjusted in the fit to the data.
Figure~\ref{fig:pTrel_ttbar_fit} shows an example of \ptrel{} template fit
results before and after tagging requirements.

\begin{figure}[!h]
  \begin{center}
    \subfloat[]{\label{fig:Effi_SF_pTrel_ttbar_a}\includegraphics[width=0.49\textwidth]{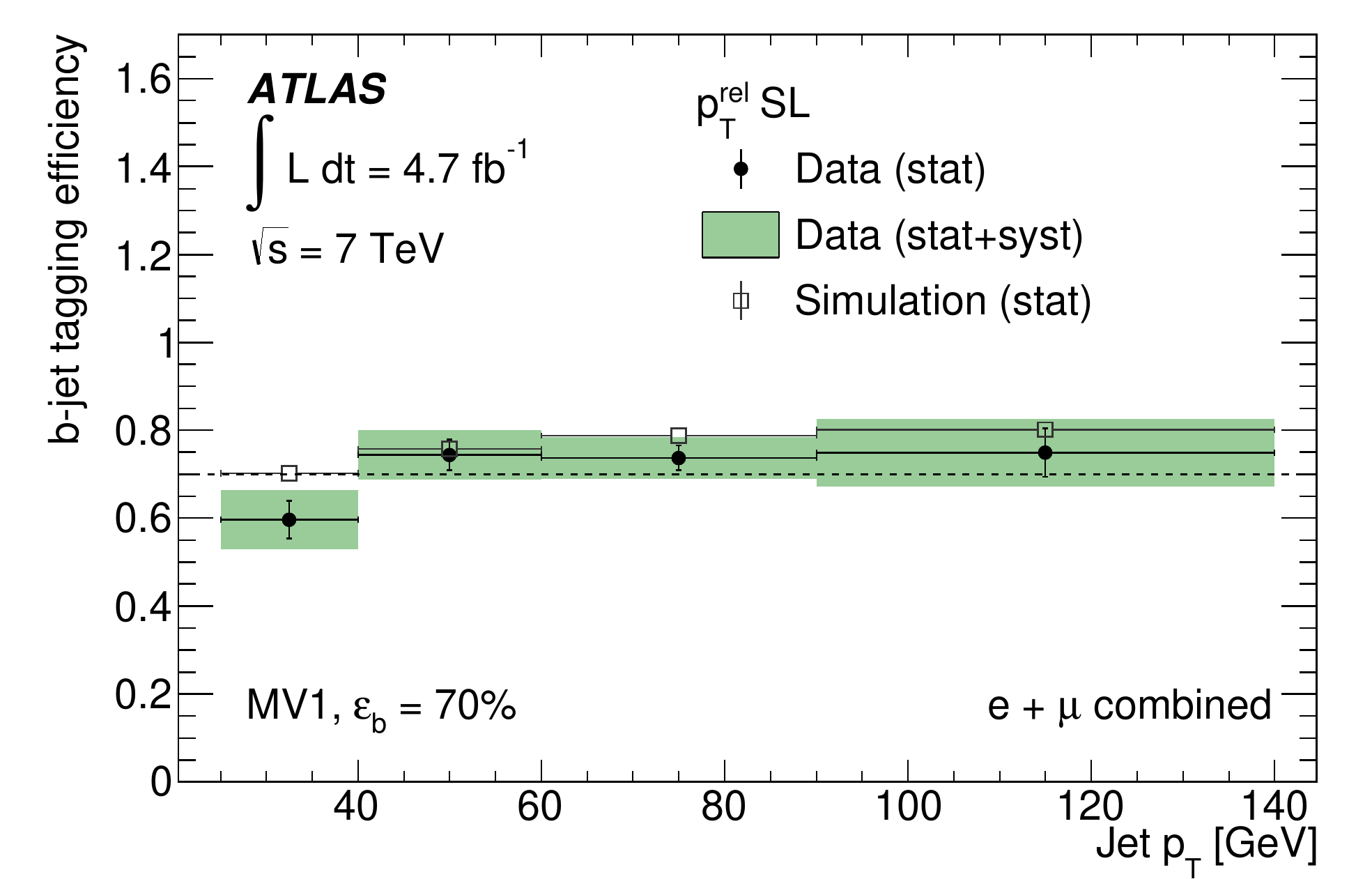}}
    \subfloat[]{\label{fig:Effi_SF_pTrel_ttbar_b}\includegraphics[width=0.49\textwidth]{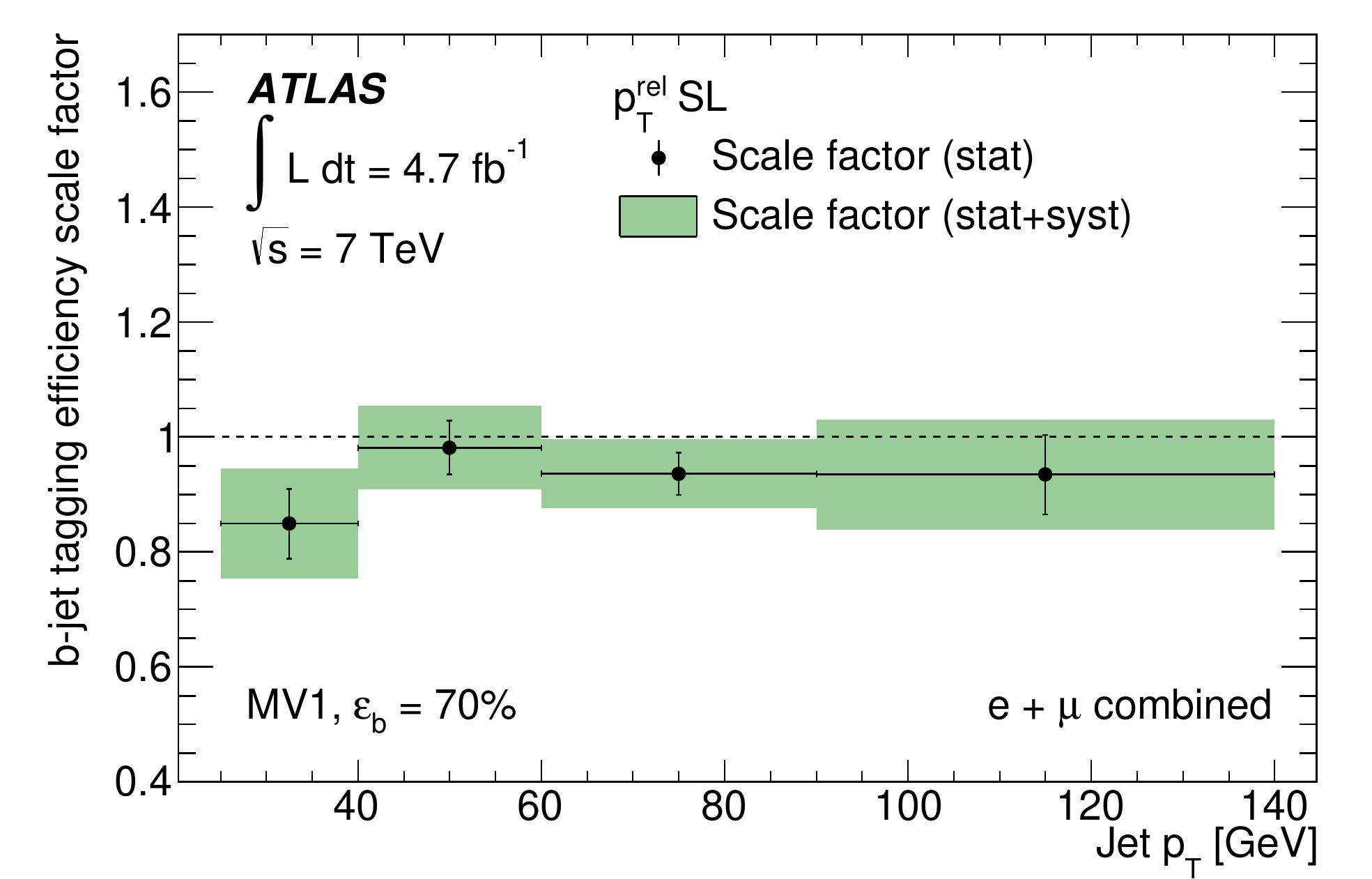}}
    \caption
    {The $b$-jet tagging efficiency in data and simulation (a) and the data-to-simulation scale factor (b) for the MV1 algorithm at 
      70\% $b$-jet tagging efficiency for the \ptrel\ method in the single-lepton \ttbar\ sample ($e$+jets and $\mu$+jets combined).}
    \label{fig:Effi_SF_pTrel_ttbar}
  \end{center}
\end{figure}

The $b$-jet tagging efficiencies measured in data, the true $b$-jet tagging efficiencies in
simulation, and the $b$-jet tagging efficiency data-to-simulation scale factors
for the MV1 tagging algorithm at 70\% efficiency are shown in Fig.~\ref{fig:Effi_SF_pTrel_ttbar}
as a function of jet \pT.
Within sizeable statistical and systematic uncertainties, the data-to-simulation
scale factors are consistent both with unity and with the scale factors obtained
using the \ptrel{} method in the dijet sample.
Therefore, no evidence exists for any biases in the results originating from
trigger biases and the lower $b$-jet fraction in the dijet analysis.
The results are consistent also with the other calibration results described in
the present section.
However, the uncertainties are too large for this method to address
possible differences observed between the dijet-based and $t\bar{t}$-based calibration
results. The compatibility between these two results is discussed in more detail
in the following section.

\begin{table}[htbp]
  \begin{center}
    \scriptsize
    \setlength{\tabcolsep}{0.4pc}
    \begin{tabular}{l|cccc}
      \hline
      \hline
      & \multicolumn{4}{c}{Jet $\pt\ \mathrm{[GeV]}$} \\ 
      Source                                  & 25--40   & 40--60   & 60--90   & 90--140  \\ 
      \hline
      IFSR                                    &  4.0     & 3.1      & 3.5      & 2.8 \\
      Generator                               &  0.4     & 3.6      & 3.5      & 2.8 \\
      Fragmentation                           & 4.1      & 2.9      & 2.4      & 1.1 \\
      $W$+jets                                & 0.3      &  0.2     & 0.1      & 0.2 \\
      Multijet                                & 4.1      &  1.0     & 0.6      & 1.0 \\
      Other backgrounds                       & 0.6      & -  & 0.1      & - \\
      Simulation statistics                   & 2.5      &  2.9     & 1.6      & 2.9 \\
      Modelling of $b$ and $c$ production     & 0.4      &  0.3     & 1.6      & 1.2 \\
      $b$-hadron direction modelling          & 1.0      &  0.7     & 1.2      & 1.7 \\
      Semileptonic correction                 & 0.5      &  0.6     & 1.4      & 0.4 \\
      Dilepton contamination in light-flavour template &  0.8    &  0.4     & 0.1      & 0.1 \\
      $b$-, $c$-jet muon contamination in light-flavour template &  0.2& 0.4& -  & 0.3 \\
      Jet energy scale                        & 0.7      &  2.8     & 3.0      & 1.7 \\
      Jet energy resolution                   & 5.2      &  2.4     & -  & 2.0 \\
      Jet reconstruction efficiency           & 0.6      &  0.2     & 0.1      & 0.1 \\
      Jet vertex fraction                     & 1.2      &  1.2     & 0.4      & -  \\
      $W$+HF normalisation                    & 0.8      &  0.4     & 0.5      & 0.2 \\
      $\met$                                  & 0.3      &  0.3     & -  & 0.3 \\
      Lepton efficiency/resolution            & 0.1      & 0.4      &  0.1     & 0.3 \\  
      Pile-up $\langle\mu\rangle$ reweighting & 0.3      &  0.7     & -  & - \\
      Extrapolation to inclusive $b$ jets     & 4.0      &  4.0     & 4.0      & 4.0 \\
      \hline
      Total systematic uncertainty            &   10     &  8.6     & 8.0      & 7.4 \\
      Statistical uncertainty                 & 7.2      &  4.8     & 4.0      & 7.5   \\
      \hline
      Total uncertainty                       &  13      &  9.8     & 8.9      &  11 \\
      \hline
      \hline
    \end{tabular}
    \caption{
      Relative uncertainties (in \%) on the
      data-to-simulation scale factor \sfb\ from the \ptrel{} analysis performed on
      the $t\bar t$ sample, for the MV1 tagging algorithm at 70\% efficiency. 
      Negligibly small uncertainties are indicated by dashes.
    }
    \label{tab:Uncertainty_pTrel_ttbar}
  \end{center}
\end{table}

The analysis shares many systematic uncertainties with the \ptrel{} analysis performed in the dijet sample, 
in particular those that affect the \ptrel{} template shapes, as discussed in Section~\ref{sec:syst}.
Besides these, most of the systematic uncertainties affecting other
$t\bar t$-based $b$-jet tagging efficiency calibrations have also been taken into
consideration; details can be found in Section~\ref{sec:ttbar_syst}.
The systematic uncertainties on the light-flavour \ptrel{} template
contamination are calculated through independent variations by 100\% of the
estimated dilepton, $b$-jet, and $c$-jet contaminations.
A 13\% uncertainty on the $W$+jets normalisation is assumed.
The normalisation of the multijet background is varied by 50\%,
based on the impact of either using the $\epsilon_\mathrm{fake}$ obtained in the pre-tag or tagged multijet samples.
The $b$-jet tagging efficiency scale factor uncertainties for the MV1 tagging algorithm at the 70\%
efficiency working point are summarised in Table~\ref{tab:Uncertainty_pTrel_ttbar}.
The table includes the complete set of systematic uncertainties, required if the scale factors were to be applied to an inclusive $b$-jet sample.
However, when comparing the results from this method to those obtained from muon-based calibration methods in dijet events,
the uncertainty associated with the extrapolation to inclusive $b$ jets should not be considered.
Analogously, when comparing to results from other \ttbar-based measurements, many \ttbar-related uncertainties will partially cancel.

\section{Combination of $b$-jet efficiency calibration measurements}
\label{sec:combination}

To obtain the best overall precision of the $b$-jet tagging efficiency calibration measurements, a combination of the results
is performed. In each jet \pt{} bin, the best estimate of the true data-to-simulation scale factor 
$\hat{\kappa}_{i}$ is extracted by maximising the likelihood that each measurement $\kappa_i$,
associated with a statistical uncertainty $\delta \kappa_i^{{\rm stat}}$ and a set of systematic uncertainties 
$\delta \kappa_{il}^{\rm syst}$, originates from a Gaussian probability density function $\mathcal{P}_i$ with mean value $\hat{\kappa}_{i}$.
The combination of $N$ measurements is performed by maximising the likelihood

\begin{equation}
  \mathcal{L} = \prod_{i=1}^N G(\kappa_{i} | \hat{\kappa}_{i}
  (1+\sum_{l=1}^{m} \delta \kappa_{il}^{\rm syst}\lambda^{\rm syst}_{l}), \delta \kappa_{i}^{\rm stat})
  \,\prod_{l=1}^m G(\lambda^{\rm syst}_{l}|0,1).
\end{equation}
Here, $G(\kappa_i | \hat{\kappa}_{i}(1+\sum_{l=1}^m{\delta \kappa_{il}^{\rm syst}}\lambda^{\rm syst}_{l}),\delta \kappa_i^{\rm stat})$
are Gaussian distributions centred at 
$\hat{\kappa}_{i}(1+\sum_{l=1}^m \delta \kappa_{il}^{\rm syst}\lambda^{\rm syst}_{l})$
with a width of $\delta \kappa_i^{{\rm stat}}$, evaluated at point $\kappa_i$.
The parameters $\lambda^{\rm syst}_{l}$ control the shifts of the central values of the Gaussian distributions, and 
in turn are constrained by Gaussian probability terms $G(\lambda^{\rm syst}_{l}|0,1)$.

The combination is performed separately for each \pt{} bin, 
assuming that 
all sources of uncertainty are 
correlated between different measurements within a single bin.
An alternative likelihood
that includes all the \pt{} bins in a global fit 
is used to estimate the impact on the final result
of the assumption of bin-to-bin correlations. 
In this alternative likelihood, systematic uncertainties that are
correlated among different bins are expressed through a single variable (and only one constraint),
while systematic uncertainties
specific to each individual bin (i.e. MC simulation statistical uncertainties) are implemented adding a different
variable (and independent constraint) for each bin.
In the case of correlated systematic uncertainties, the relative sign of the uncertainty in each individual measurement is taken into account.
Each systematic uncertainty on the final scale factor has a positive (negative) sign if the difference between the shift in the
scale factor when applying a positive and a negative variation of the underlying parameter is positive 
(negative), i.e., if $\sfb({\rm up}) - \sfb({\rm down}) > 0 \ (< 0)$.

The final $b$-jet tagging efficiency scale factors are combinations of three individual measurements: the results of the dijet-based \ptrel{} and
system8 analyses shown in Section~\ref{sec:results}, and those of the \ttbar-based combinatorial likelihood method shown in Section~\ref{sec:ttbar_results}.
The latter enters the combination with four individual channels,
corresponding to the $e\mu$ and combined $ee$ and $\mu\mu$ channels,
each separately for two- and three-jet events.
The fit quality and hence the overall compatibility between the different measurements is evaluated by computing the global $\chi^2$

\begin{equation}
  \chi^2 = \sum_{i=1}^{N_{\rm all}} \sum_{j=1}^{N_{\rm all}} \left( \kappa_i - \hat{\kappa}_i \right) C_{ij}^{-1} \left( \kappa_j - \hat{\kappa}_j \right),
  \label{eq:combination_ch2}
\end{equation}
where $N_{\rm all}$ refers to the total number of measurements in all $\pt$ bins and the covariance matrix $C_{ij}$ accounts
for correlations within and between different \pt{} bins.

As the \ptrel{} and system8 analyses use partly overlapping samples, the statistical uncertainty
is partially but not fully correlated.
The correlation coefficients have been derived using toy experiments in which somewhat
simplified versions of the \ptrel\ and system8 analyses were performed.
The statistical correlation of the two analyses was found to be moderate, e.g.,
below 50\% for the MV1 algorithm at the 70\% efficiency operating point.
The smallest correlations are observed in the \pt{} bins that
suffer from large statistical uncertainties, while the largest correlations are found for bins in the lower \pt{}
range where the statistical uncertainties are smaller. There are also \pt{} bins in which the two analyses use
different but highly prescaled triggers, leading to a negligible correlation.
The correlation of the statistical uncertainty is accounted for in the combination by dividing it into two components, 
one which is treated as fully correlated and one which is treated as uncorrelated.
The systematic uncertainty arising from limited simulation statistics is treated as fully uncorrelated between \pt\ bins but fully correlated between the \ptrel\ and system8 analyses for a given bin.
The main effect of this combination is a slightly reduced uncertainty at low
$\pt$ compared to the uncertainty resulting from the system8 calibration alone.

\begin{figure}[hbtp]
  \begin{center}
    \includegraphics*[width=1.0\textwidth]{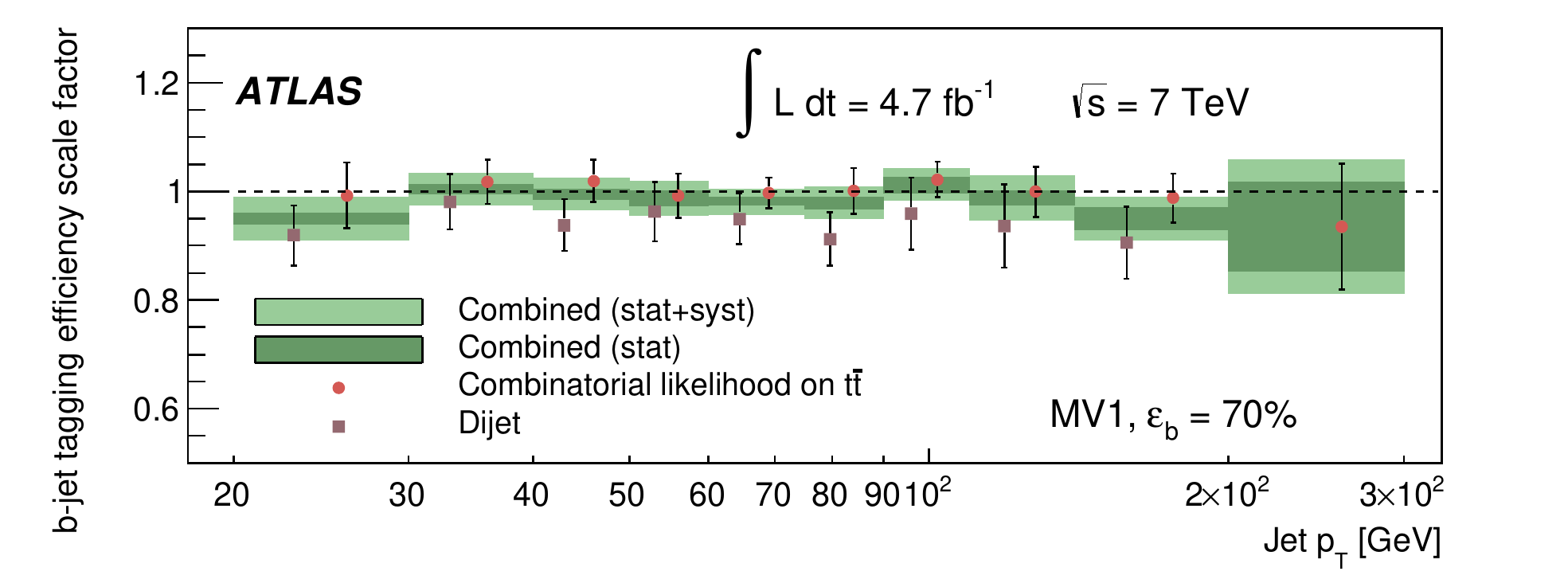}
  \end{center}
  \caption{The $b$-jet tagging efficiency data-to-simulation scale factor \sfb\ for the MV1 tagging algorithm at 70\% efficiency,
    obtained by combining the dijet-based \ptrel{} and system8 results with
    the \ttbar-based combinatorial likelihood results.}
  \label{fig:combeff}
\end{figure}

The agreement in the data-to-simulation scale factors between the \ptrel{} and system8 methods is very good.
Their combination is shown, for the MV1 tagging algorithm at the 70\% operating point, as the dijet result in Fig.~\ref{fig:combeff},
along with the results of the combinatorial likelihood analysis and their combination.
It is seen that dijet scale factors tend to be systematically lower, by about 5\%, than those resulting from the combinatorial likelihood analysis.
However, both bin-by-bin and globally they are consistent with each other; the
value of $\chi^2/N_{\text{dof}}$ in this combination turns out to be $0.95$, with $N_{\text{dof}}=48$, 
corresponding to a probability to obtain a $\chi^2$ larger than the observed one of 57\%,
providing evidence for a spread between measurements commensurate with the assigned uncertainties.
The combination results have also been compared with those obtained with the
``Best Linear Unbiased Estimates'' (BLUE) method ~\cite{BLUE}.
The  two sets of results show good agreement, as well as similar $\chi^{2}$ compatibility estimates.
For this particular tagging algorithm and operating point, the efficiency scale factors are consistent with
unity in the kinematic range covered by the analyses.

\section{$b$-jet efficiency calibration of the soft muon tagging algorithm}
\label{tandp}

A jet is considered tagged by the soft muon tagging (SMT) algorithm if it contains a reconstructed 
muon fulfilling the criteria listed in Section~\ref{sec:smt_tagger}.
The efficiency with which a $b$ jet in data passes such a tagging requirement is determined in a three-step approach.
Data-to-simulation scale factors for the efficiency to reconstruct a muon are obtained from a tag-and-probe
method, as described in Refs.~\cite{ATLAS-CONF-2011-063,ATLAS-CONF-2012-125}.
The efficiency with which a reconstructed muon passes the SMT selection criteria is measured using a 
tag-and-probe method in samples of isolated muons produced in $\Jpsi \to \mu\mu$ 
and $Z \to \mu\mu$ decays as described in Section~\ref{sec:SMT_TP}.
Finally, the probability that a $b$ jet of a certain \pt{} and $\eta$ contains a reconstructed
muon is derived from the same simulated \ttbar{} sample used also in the
\ttbar-based $b$-jet efficiency measurements; systematic uncertainties are assigned to account for 
possible mismodelling of this probability. These three parts are then combined into a measurement of the SMT
$b$-jet tagging efficiency.

\subsection{Data and simulation samples}

For the tag-and-probe method, events in the $\Jpsi \to \mu\mu$ sample are collected using a set of  
muon triggers optimised to be 
efficient at low~\pt~\cite{ATLAS-CONF-2011-021}, while events in the $Z \to \mu\mu$ sample are required to have a tag muon that
is accepted by the lowest-\pt{} unprescaled trigger available in a given data taking period (which imposes a muon 
\pt{} requirement of 18~\GeV)~\cite{ATLAS-CONF-2012-099}.

The $\Jpsi \to \mu\mu$ and $Z \to \mu\mu$ events in data are compared to simulated dimuon events, generated with {\sc PYTHIA}.
A simulated sample of \ttbar{} events, generated with {\sc MC@NLO} interfaced to {\sc HERWIG},
is used to derive the $b$-jet tagging efficiency of the SMT algorithm.

\subsection{Tag-and-probe based SMT muon efficiency measurement}
\label{sec:SMT_TP}

The tag muon is required to pass stringent selection criteria to ensure that it is of good quality.
A probe muon is then selected as a reconstructed muon that passes looser quality criteria but fulfils the 
requirement that the invariant mass of the tag-probe pair is consistent with that 
of the \Jpsi{} meson ($3096.916 \pm 0.011 \mev$~\cite{pdg2010})
or the $Z$ boson ($91.1876 \pm 0.0021\GeV$~\cite{pdg2010}).

All tag and probe muons are required to be measured both in the tracking
detector and in the muon system, and to satisfy the selection criteria mentioned in 
Section~\ref{sec:smt_tagger}.
The $\Jpsi \to \mu\mu$ events are further required to satisfy the following selection criteria, in line with 
the requirements for the study of the muon reconstruction efficiency, motivated in Ref.~\cite{ATLAS-CONF-2012-125}:
\begin{itemize}
\item The tag muon is required to have $|\eta | <2.5$ and $\pt > 4 \GeV$. The transverse and longitudinal
  impact parameters must fulfil $|d_{0}| <0.3$~mm, $|z_{0}| <1.5$~mm, $|S_{d_{0}}| < 3$ and $|S_{z_{0}}| < 3$.
\item The probe muon is required to have $|\eta | <2.5$ and $p > 3 \GeV$.
\item The tag and probe muons must be of opposite charge, have a common vertex
  with vertex fit $\chi ^{2} < 6$, and must satisfy
  $\Delta R ({\rm tag,probe}) <3.5$ and $2 \GeV < m_{\mu\mu}({\rm tag,probe}) < 4 \GeV$.
\end{itemize}
The $Z \to \mu\mu$ events are instead required to pass the following selection criteria, 
coherently with those adopted in the study of the muon reconstruction efficiency, described in Ref.~\cite{ATLAS-CONF-2011-063}:
\begin{itemize}
\item The tag muon is required to have $|\eta | <2.4$ and $\pt > 20 \GeV$.
\item The probe muon is required to have $|\eta | <2.5$ and $\pt > 7 \GeV$.
\item Both muons are required to have longitudinal impact parameter 
  $|z_{0}| <10$~mm and to be isolated from other tracks according to 
  $\sum \pt^{0.4}<0.2 \pt^{\rm muon}$, where the variable
  $\sum \pt^{0.4}$ is the sum of the~\pt{} of ID tracks in a $\Delta R=0.4$ cone around the muon.
\item The tag and probe muons must be of opposite electric charge, have 
  $\Delta \phi ({\rm tag,probe}) > 2.0$ and $|m_{\mu\mu}({\rm tag,probe}) - m_{Z}| < 10 \GeV$.
\end{itemize}

The efficiency for a muon to satisfy the SMT selection criteria, $\varepsilon_{\mu-\rm{SMT}}$, is then measured as

\begin{equation}
  \varepsilon_{\mu-\rm{SMT}} = \frac{N_{\mu}^{\rm SMT}}{N_{\mu}},
  \label{eq:x2eff}
\end{equation}
where $N_{\mu}$ refers to the number of probe muons either in the $\Jpsi \to \mu\mu{}$ or the $Z \to \mu\mu{}$ sample
and $N_{\mu}^{\rm SMT}$ refers to the number of probe muons passing the SMT selection criteria.

The background in the $\Jpsi \to \mu\mu{}$ sample before and after applying the SMT algorithm 
is estimated using fits to the dimuon invariant mass distribution.
In the invariant mass range between 2.5 and 3.6 \GeV, a second order polynomial
and a Gaussian are used to describe the background and signal, respectively.
The fitted background within $3\sigma$ of the mean of the signal peak is then
subtracted from the observed event counts in this range. The remainder is
assumed to be the number of $\Jpsi$ mesons,
and used to estimate the efficiency through Eq.~\ref{eq:x2eff}.
The mean of the Gaussian obtained in the fit to the pre-tag sample is used in the
fit to the tagged sample.

In the $Z \to \mu\mu$ sample, the backgrounds considered are 
$Z\rightarrow\tau\tau$, $W\rightarrow\mu\nu$, $W\rightarrow\tau\nu$,
$t\overline{t}$, $b\overline{b}$ and $c\overline{c}$ processes.
The background contribution in the $Z \to \mu\mu$ sample, estimated with simulated events, is found to be negligible
(less than one per mille). Hence, no background subtraction is applied.

\subsection{$b$-jet tagging efficiency measurement}

The efficiency with which a $b$ jet is tagged by the SMT algorithm, $\varepsilon_{b}$, is defined as

\begin{equation}
  \varepsilon_{b}^{\rm data} = \varepsilon_{b}^{\rm sim} \cdot
  \frac{\varepsilon_{\mu-\rm{reco}}^{\rm data}}{\varepsilon_{\mu-\rm{reco}}^{\rm sim}}
  \cdot
  \frac{\varepsilon_{\mu-\rm{SMT}}^{\rm data}}{\varepsilon_{\mu-\rm{SMT}}^{\rm sim}},
  \label{eq:jeteff}
\end{equation}
where $\varepsilon_{b}^{\rm sim}$ is later referred to as the uncalibrated $b$-jet tagging 
efficiency and $\varepsilon_{b}^{\rm data}$ as the data-calibrated $b$-jet tagging 
efficiency.
The correction factor ${\varepsilon_{\mu-\rm{SMT}}^{\rm data}}/{\varepsilon_{\mu-\rm{SMT}}^{\rm sim}}$, which
corrects the efficiency with which a muon passes the SMT selection criteria, is obtained with the
tag-and-probe method as described in Section~\ref{sec:SMT_TP}. The correction factor
${\varepsilon_{\mu-\rm{reco}}^{\rm data}}/{\varepsilon_{\mu-\rm{reco}}^{\rm sim}}$, which corrects the
muon reconstruction efficiency, is obtained from tag-and-probe studies of $\Jpsi \to \mu\mu$ and 
$Z \to \mu\mu$ decays as described in Refs.~\cite{ATLAS-CONF-2011-063,ATLAS-CONF-2012-125}. The muon reconstruction efficiency is generally well modelled
by the simulation, with scale factors compatible with unity for most of the detector region.
The simulated sample from which $\varepsilon_{b}^{\rm sim}$ is derived, is corrected for known differences between data and 
simulation in the $b$-jet modelling. The simulated branching fractions of the various $b$-hadron decay modes 
giving rise to a muon are scaled to the world average values \cite{pdg2010}, both for direct decays and 
sequential decays via charm quarks and $\tau$ leptons. The jet energy in the simulated sample is corrected to 
match the scale and resolution observed in data, extracted from inclusive dijet samples and 
$\gamma/Z$+jet samples~\cite{PERF-2011-03}. A correction of the jet energy to account for 
the momentum of the neutrino and the muon in semileptonic $b$-hadron decays is also applied, using the same 
procedure and corrections as in Section~\ref{sec:syst}.
Residual differences in the modelling of $b$ decays ($b$-quark fragmentation and hadronisation),
together with the kinematics of the hard scatter process, are accounted for by
assigning systematic uncertainties derived from the comparison between the
nominal simulated samples and alternative models.

\subsection{Systematic uncertainties}

The systematic uncertainties associated with the $b$-jet tagging efficiency measurement
are summarised in Table~\ref{tab:smt_bjet_eff_uncalib_syst}.

\begin{table}[!htbp]
  {\scriptsize
    \begin{center}
      \begin{tabular}{l|ccccccccc}
        \hline \hline
        & \multicolumn{9}{c}{Jet \pt{} [\GeV]} \\
        Source                      & 30--40  & 40--50  & 50--60  & 60--70  & 70--80  & 80--90  & 90--100 & 100--130& 130--170 \\
        \hline
        SMT muon efficiency         &     0.4 &     0.6 &     0.6 &     0.7 &     0.6 &     0.6 &     0.6 &     0.6 &     0.6  \\
        Muon reco efficiency        &     1.8 &     1.3 &     1.1 &     1.0 &     1.0 &     1.0 &     0.9 &     0.9 &     0.9  \\
        MC generator                &     2.3 &     2.0 &     1.0 &     0.8 &     1.3 &     1.3 &     1.4 &     0.3 &     2.8  \\
        Parton shower/fragm. model  &     3.0 &     5.2 &     5.6 &     6.4 &     6.3 &     6.0 &     5.5 &     4.1 &     3.9  \\
        IFSR                        &     1.4 &     0.6 &     1.2 &     1.0 &     0.4 &     1.4 &     1.6 &     1.0 &     1.8  \\
        Parton distribution function&     0.3 &     0.3 &     0.1 &     0.1 &     0.3 &     0.2 &     0.5 &     0.4 &     0.9  \\
        Branching fraction rescaling&     3.1 &     3.2 &     3.2 &     3.3 &     3.3 &     3.4 &     3.6 &     3.6 &     3.6  \\
        \hline
        Total systematic uncertainty &     4.5 &     6.6 &     6.8 &     7.5 &     7.4 &     7.5 &     7.1 &     5.8 &     6.6  \\
        \hline\hline
      \end{tabular}
    \end{center}
    \caption{Relative systematic uncertainties, in \%, for the $b$-jet tagging efficiency of the
      SMT algorithm as a function of jet \pt.}
    \label{tab:smt_bjet_eff_uncalib_syst}
  }
\end{table}

\subsubsection{SMT muon efficiency uncertainties} \label{sec:mubase_smt_eff}

The uncertainties on the efficiency of a probe muon passing the SMT selection criteria are described below.
The quadratic sum of these uncertainties constitute the SMT muon efficiency uncertainty.

\subsubsection*{$\Jpsi{}$ fit uncertainties}

In the $\Jpsi \to \mu\mu{}$ sample, the difference in the efficiency estimates obtained
using $3\sigma$ and $5\sigma$ $\Jpsi$ signal mass ranges 
($\varepsilon_{5\sigma}-\varepsilon_{3\sigma}$) is assigned as an uncertainty.
Moreover, for a fixed fit range, the values of the three fit parameters are varied within their uncertainties and the
difference in the efficiency obtained with the resulting minimum and maximum background estimates
($\varepsilon_{\textrm{bkgmax}}-\varepsilon_{\textrm{bkgmin}}$) is assigned as a systematic uncertainty.

\subsubsection*{$Z \to \mu\mu$ selection}

In the $Z \to \mu \mu$ sample, systematic uncertainties are estimated by varying the selection cuts, 
tightening the $|m_{\mu\mu} - m_{Z}|$ cut to 6 \GeV~and loosening it to 
$m_{Z} - 20 \GeV \le m_{\mu\mu} \le m_{Z} + 10 \GeV$, and varying the isolation cut on
$\frac{\sum p^{0.4}_{\rm T}}{ p^{\rm muon}_{\rm T} }$ to 0.1 and 0.3. 

\subsubsection*{Isolation dependence}

As the isolation profile of inclusive muons from $b$-hadron decays is very
different from that for muons produced from a~\Jpsi{} or a $Z$ boson,
the dependence of the SMT muon efficiency scale factors on the muon isolation has been investigated. 
Three isolation variables were chosen for this investigation. These were the transverse energy deposited in the 
calorimeter in cones of three different sizes (corresponding to opening angles of 0.2, 0.3 and 0.4 radians) 
around the central
muon.\footnote{To exclude the signal from the central muon, the transverse energy deposited in a cone of 
opening angle 0.1 in the centre of these larger cones is not counted.}
The other isolation variables are the number of tracks and the summed \pt\ of the tracks found 
in cones of the same size.

While the statistics in the data sample is enough to cover a wide isolation range, the more limited
statistics in the simulated sample does not allow to derive data-to-simulation scale factors
for muons surrounded by a large amount of transverse energy or a large number of tracks. 
Within the range studied however, the data-to-simulation scale factor 
found is consistent with unity, meaning there is no dependence of the muon
efficiency on the isolation of the probe muon. Thus no systematic uncertainty from this 
source has been assigned.

\subsubsection{$b$-jet tagging efficiency uncertainties}
\label{sec:syst_smt_jet_extrapolation}

\subsubsection*{Muon reconstruction efficiency}

The muon reconstruction efficiency in simulated events is corrected with data-to-simulation scale factors
obtained from tag-and-probe studies of $\Jpsi \to \mu\mu$ and $Z \to \mu\mu$ decays~\cite{ATLAS-CONF-2011-063,ATLAS-CONF-2012-125}.
The uncertainties on these scale factors are propagated to the $b$-jet tagging efficiency. 

\subsubsection*{Generator dependence}

The MC generator uncertainty covers the modelling of the $b$-quark kinematics as
a result of the hard interaction in different NLO generators.
This is evaluated by comparing the efficiencies in \Alpgen+\Herwig{}~\cite{bib:alpgen}
and \Powheg+\Herwig{} samples to that obtained using the baseline MC@NLO+\Herwig{} sample.

\subsubsection*{Parton shower and fragmentation model}

Different showering and fragmentation models may result in different 
kinematics of the soft muon from the $b$-hadron decay.
This is taken into account by comparing the efficiencies obtained using two samples of $t\bar{t}$ events,
generated with \Powheg, one of which has been showered by \Herwig{} and
the other by \Pythia.
To ensure that only the effects of the parton showering models are compared, rather
than the difference in decay tables used by each hadronisation routine, the
branching fraction of $b \rightarrow \mu X$ decays (both direct and sequential) in
both simulated samples is re-weighted to match that in Ref.~\cite{pdg2010}.

\subsubsection*{Initial and final state radiation}

The systematic uncertainty due to initial and final state radiation (IFSR) in
the $t\bar{t}$ events is estimated by studies using samples generated with
\AcerMC{}~\cite{bib:acer} interfaced to \Pythia{}, and by varying the parameters
controlling IFSR in a range consistent with experimental data
\cite{STDM-2011-03,TOPQ-2011-21}.

\subsubsection*{Parton distribution function}

The uncertainty on the parton distribution functions (PDFs) translates into an
uncertainty on the $b$-quark kinematics. This is accounted for by using three different PDF sets, the nominal CT10~\cite{bib:ct10} as well as MSTW~\cite{bib:mstw2008} and NNPDF~\cite{NNPDF}. The PDFs are varied based on the uncertainty along each of the PDF eigenvectors. Each variation is evaluated via an event-by-event re-weighting of the simulated $t\bar{t}$ sample. The total uncertainty assigned is the envelope of all PDF uncertainties.

\subsubsection*{Branching fraction rescaling}

The rates of the direct and sequential decays of $b$ quarks to muons are rescaled in the simulated samples to match the world average values. The experimental uncertainty on each of these values \cite{pdg2010} is propagated to the final result. The final uncertainty on the BF rescaling is larger at higher $\ptjet$, due to the larger relative weight of decays through double charm creation (like $b\rightarrow c\bar{c}\rightarrow \mu$), whose BF is known with a 30$\%$ accuracy.

\subsection{Results}

\begin{figure}[htbp]
  \begin{center}
    \subfloat[]{
      \label{fig:AllJpsiEffSfEta}
      \includegraphics[width=0.49\textwidth]{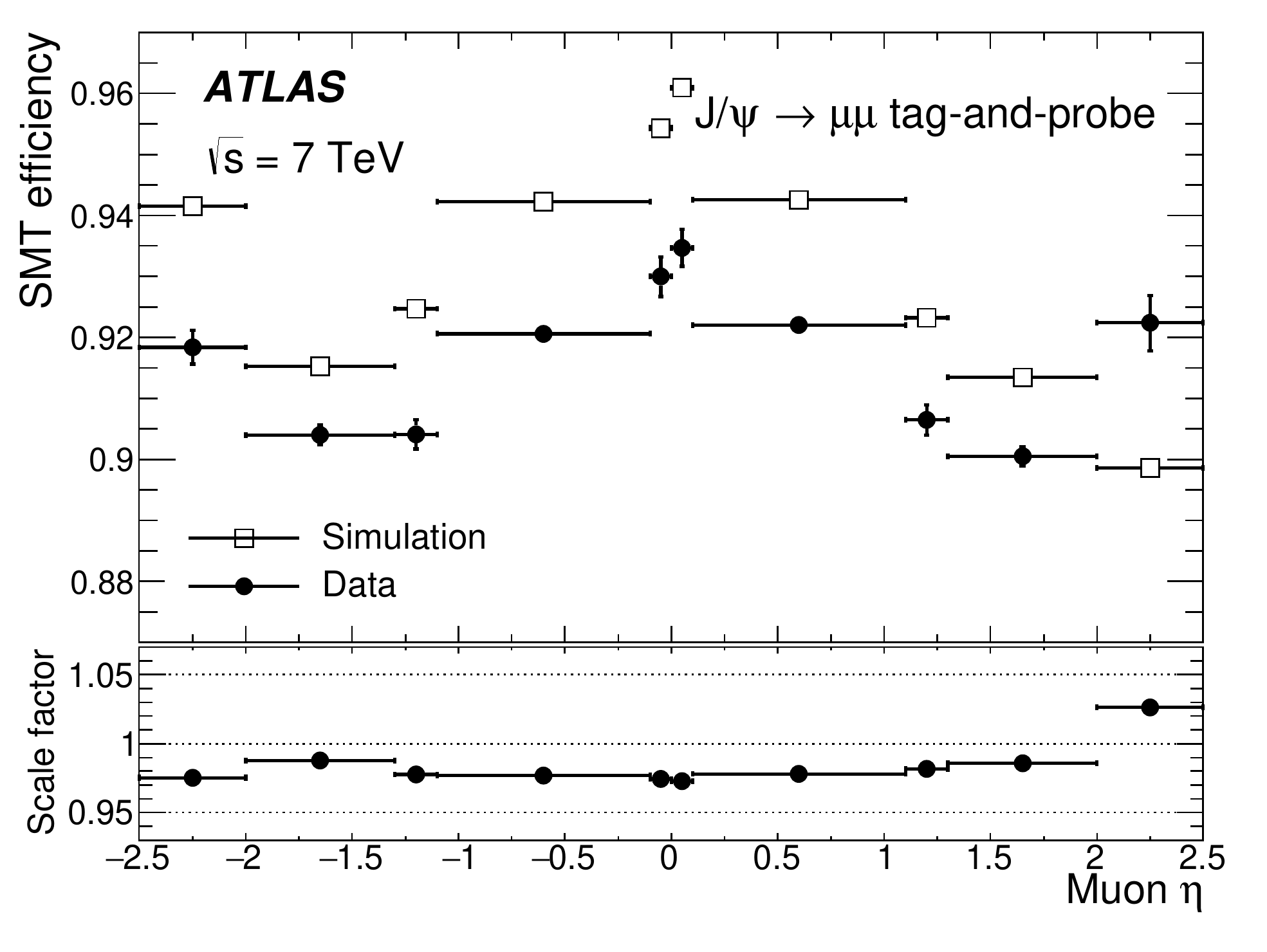}}
    \subfloat[]{
      \label{fig:AllJpsiEffSfPt}
      \includegraphics[width=0.49\textwidth]{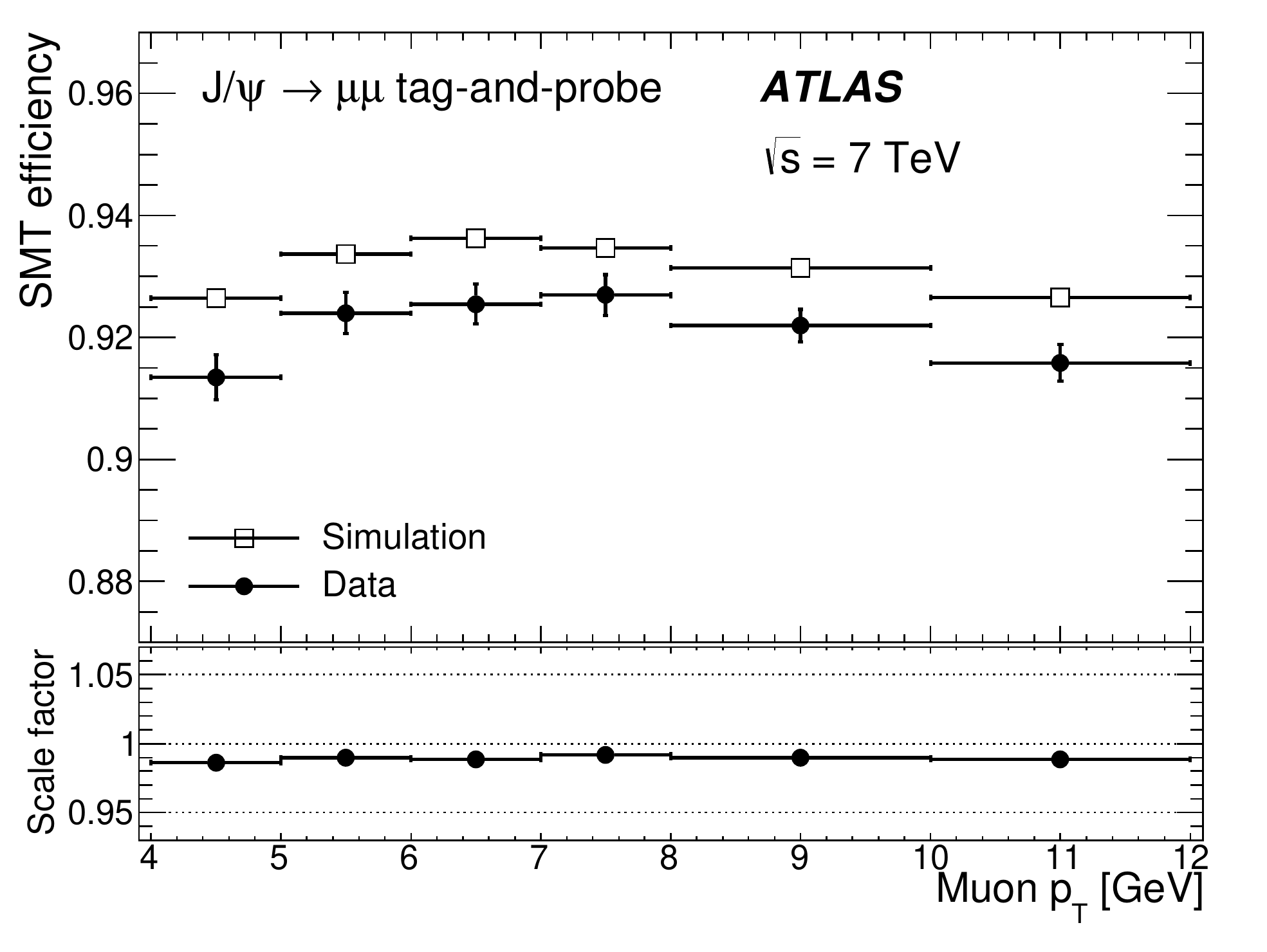}}
    \caption{SMT muon efficiencies and data-to-simulation scale factors from the $\Jpsi \to \mu\mu{}$ analysis 
      as a function of $\eta$ (a) and~\pt{} (b) of the probe muon. Uncertainties are statistical only.
      \label{fig:AllJpsiEffSf}}
  \end{center}
\end{figure}
\begin{figure}[htbp]
  \begin{center}
    \subfloat[]{
      \label{fig:AllZmumuEffSfEta}
      \includegraphics[width=0.49\textwidth]{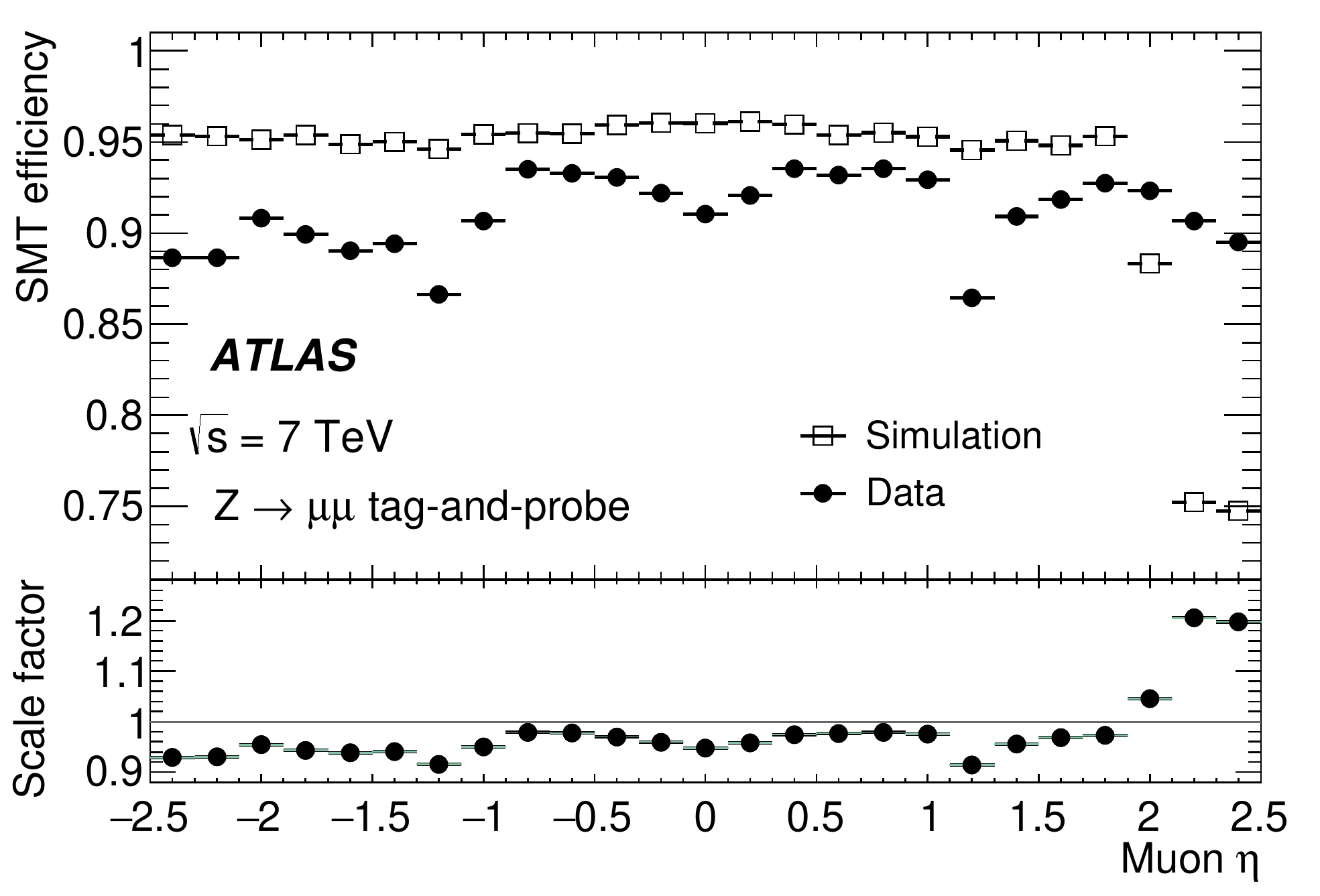}} 
    \subfloat[]{
      \label{fig:AllZmumuEffSfPt}
      \includegraphics[width=0.49\textwidth]{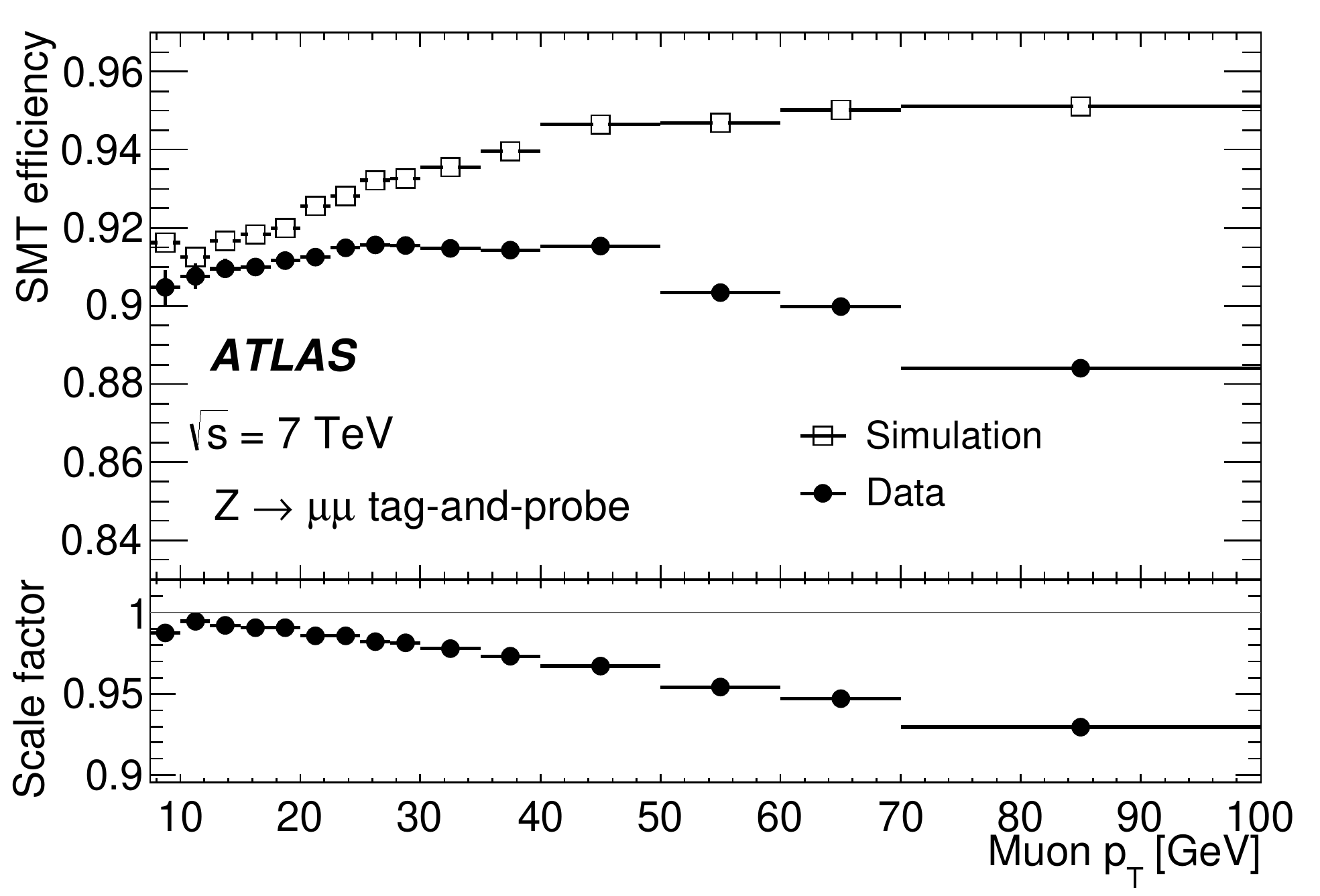}} 
    \caption{SMT muon efficiencies and data-to-simulation scale factors from the $Z \to \mu\mu$ analysis 
      as a function of $\eta$ (a) and~\pt{} (b) of the probe muon. Uncertainties are statistical only.}  
  \label{fig:AllZmumuEffSf}
  \end{center}
\end{figure}
\begin{figure}[htbp]
  \begin{center}
    \includegraphics[width=0.5\textwidth]{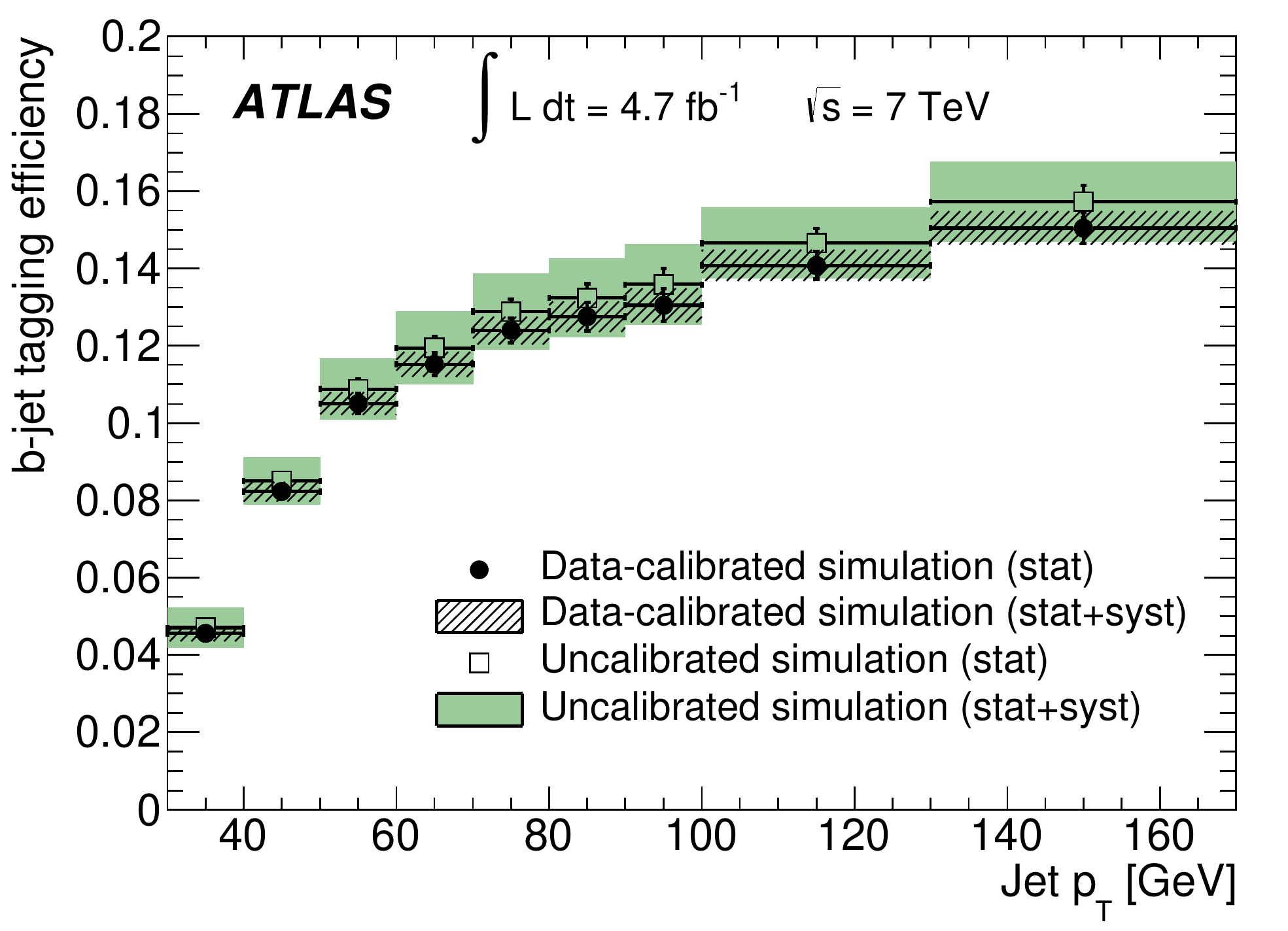}
    \caption{The $b$-jet tagging efficiency of the SMT algorithm in simulated \ttbar{} events, before (open squares) 
      and after (full black dots) applying the SMT muon efficiency and muon reconstruction efficiency 
      calibrations extracted from tag-and-probe methods.}
    \label{fig:smt_bjet_eff}
  \end{center}
\end{figure}

The \eta{} and \pt{} dependence of the SMT muon efficiency from the tag-and-probe method is shown in
Figs.~\ref{fig:AllJpsiEffSf} and~\ref{fig:AllZmumuEffSf} for the $\Jpsi \to \mu\mu{}$ and $Z \to \mu\mu$ analyses
respectively.
In most of the $\eta$ range the data efficiency is lower than the efficiency
predicted by simulation. Only in the range $\eta>2$ the simulation efficiency is
significantly lower; the reason is a misconfiguration in these simulations.
For the $Z \to \mu\mu$ sample the SMT efficiency in simulation is otherwise independent of $\eta$ while in data there are variations of up to 8\%. 
While the dependence on the transverse momentum in the $\Jpsi \to \mu\mu{}$ sample is weak, there is a 
a strong~\pt{} dependence of the data-to-simulation scale factors in the $Z \to \mu\mu$ sample 
which is caused by the sensitivity of the $\chi^{2}_{\rm match}$ cut to residual misalignments of the 
Muon Spectrometer relative to the inner tracker that are present in the data but not in the perfectly 
aligned simulation. The scale factors were found to exhibit no strong dependence on $\phi$.
The $b$-jet tagging efficiency of the SMT algorithm as a function of the jet \pt\ is shown in 
Fig.~\ref{fig:smt_bjet_eff}, where the simulation has been scaled with the muon reconstruction 
efficiency results from Refs.~\cite{ATLAS-CONF-2011-063,ATLAS-CONF-2012-125} and the SMT muon 
efficiency calibration results from Section~\ref{sec:SMT_TP}. For the latter scaling, 
the $J/\psi\to\mu\mu$ ($Z\to\mu\mu$) results are used for muon $\pt < 12\GeV$ ($\pt > 12\GeV$).
The green area indicates the statistical uncertainty summed in quadrature with the
systematic uncertainties from the $b$-jet modelling
presented in Section~\ref{sec:syst_smt_jet_extrapolation} (with the exception of the
uncertainty on the muon reconstruction efficiency) while the hashed area corresponds to
the quadrature sum of the statistical uncertainty and the systematic uncertaintes on
the muon reconstruction and SMT muon efficiencies. As analyses using the SMT tagging algorithm
use the muon-based scale factors rather than the jet-based ones, no jet-based 
data-to-simulation scale factors are presented here.

\section{$c$-jet tagging efficiency calibration using events with a $W$ boson produced in association with a $c$ quark}
\label{sec:ceff_Wc}

The efficiency with which a $b$-tagging algorithm tags $c$ jets is referred to as the \ctageff. 
The method to calibrate the \ctageff{} described in this section is based on the selection of a single $c$~jet
produced in association with a \Wboson~boson and identified by a soft muon stemming from the semileptonic decay of a $c$~hadron;
the \Wboson~boson is reconstructed via its decay into an electron and a neutrino.
In proton-proton collisions at a centre-of-mass energy of $\sqrt{s}=7$~\TeV, the dominant production mechanism is
$gs\rightarrow W^- c$ and $g\bar{s}\rightarrow W^+\bar{c}$, where the \Wboson~boson is always accompanied by a $c$~quark of opposite charge.
Given that the soft muon and $c$~quark charge signs are the same, requiring that the charge of the soft muon
and the charge of the electron from the \Wboson-boson decay should be of opposite sign selects \Wc{} events with high purity. 
Most of the background processes are evenly populated with events where the charges of the decay leptons are of opposite sign (OS) or of same sign (SS). 
Therefore, the number of \Wc{} signal events can be extracted as the difference between the numbers of events with opposite and with same charge leptons (OS-SS).
This fundamental strategy has already been exploited in several \Wc{} production cross section measurements~\cite{CDFII:Wc,CDFII:Wc2,D0:Wc,CMS-SMP-12-002,STDM-2012-14}. 
Since the present analysis was performed in the course of the recent \Wc{} cross section measurement~\cite{STDM-2012-14},
details about the extraction of the \Wc{} sample, i.e. the event selection and background estimations, can be found in this reference.
In the remainder of this section, jets that are soft-muon tagged (SMT) applying the algorithm described in Section~\ref{sec:smt_tagger}
are referred to as \smtjets~and a sample composed of such jets extracted as the number of OS-SS events is referred to as the \smtjetssample.

In a first step the \ctageff{} is measured using the \smtjetssample{} in data and simulation.
Following that, an extrapolation procedure is performed to derive data-to-simulation scale factors that are applicable to an unbiased, inclusive sample of $c$~jets. 
It should be noted that the analysis described here does not attempt to perform a measurement of the calibration factors in bins of $c$-jet transverse momentum,
because of the limited available number of events.

\subsection{Data and simulation samples}
\label{sec:dataMCSamples}

The signal process is defined as a \Wboson~boson produced in association with a single charm quark. 
\Wboson~bosons produced in association with light quarks or gluons, hereafter referred to as \Wlight,
as well as charm- or bottom-quark pairs are considered as backgrounds.
The contribution from \Wboson~bosons produced in association with a single bottom quark is negligible.
The background further includes the production of \Zboson/$\gamma^{*}$+jets, top-quark pairs, single top quarks, dibosons ($WW$, $WZ$ and $ZZ$) and \mj~events.
Because of the symmetry in the process of producing heavy charm- or bottom-quark pairs in \Wcc{} and \Wbb{} events, these are expected to show an
even population of OS and SS events, whereas this is not the case for the other processes.

The data sample used to perform the analysis is collected by a single-electron trigger.
The \Wc{} signal process is generated using \Alpgen\ 2.13~\cite{bib:alpgen}, 
where the showering and the hadronisation is done with \Pythia\ 6.423~\cite{pythia2}.
An additional signal sample is produced to study systematic uncertainties, where \Herwig{} 6.520~\cite{Corcella:2000bw} 
is used for the parton shower and \Jimmy\ 4.31~\cite{Butterworth:1996zw} for the underlying event. 
Samples with zero to four additional partons are used and the MLM~\cite{Mangano:2001xp} matching scheme is applied to
remove overlaps between events with a given parton multiplicity generated both by the matrix element and the parton shower.
Also overlaps with \Alpgen\ samples used to model background processes leading to \Wboson~bosons and heavy-flavour quarks are removed.
The CTEQ6L1 parton distribution function~\cite{bib:cteq6l} is used.

To improve on known shortcomings of the \Alpgen+\Pythia\ predictions and to minimise systematic uncertainties,
several $c$-quark fragmentation and $c$-hadron decay properties are corrected as explained in \mysec~\ref{sec:inclSF}.
In the following this corrected sample is referred to as the \Alpgen+\PythiaCorr~sample;
the sample without any of the fragmentation and decay corrections applied is called the \Alpgen+\PythiaDef~sample. 
To study $c$-hadron decay properties another signal sample is used which is also generated with \Alpgen\ and \Pythia,
but where the \EvtGen~\cite{Lange:2001uf} program is used to model the $c$-hadron decays.

Details on the simulated samples used to describe the \Wboson, \Zboson~and top background processes can be found in Ref.~\cite{STDM-2012-14}. 
Their contributions are normalised to NNLO predictions in case of the inclusive \Wboson, \Zboson~and 
\ttbar{} productions~\cite{Anastasiou:2003ds, bib:nnlocacciari} and to NLO predictions for the 
other processes~\cite{bib:dibXsec,Campbell:2004ch}.
The normalisation as well as other properties of the \mj~background are determined using data-driven techniques.

\subsection{Event selection}
\label{sec:eventSelection}

Only the most important steps of the event selection are mentioned here, while a more detailed description can be found in Ref.~\cite{STDM-2012-14}. 
Some information on the object definitions is also given in Section~\ref{sec:samples}.

\Wboson~bosons are reconstructed via their leptonic decay into an electron and a neutrino.
Electrons are required to have a transverse momentum $\pt>25$\,\GeV{} and a pseudorapidity range $|\eta|<2.47$, excluding the
calorimeter transition region $1.37<|\eta|<1.52$.
Electrons that fulfil the ``tight'' identification criteria described in Ref.~\cite{PERF-2010-04} and re-optimised for the 2011 data-taking conditions are selected.
In addition a calorimeter-based isolation requirement is applied:
the sum of transverse energies in calorimeter cells within a cone of radius $\Delta R<0.3$ around the electron direction, $\sum_{\Delta R < 0.3} E_{\rm T}^{\rm cells}$,
is required to be less than 3\,\GeV.
Only events with exactly one isolated electron are selected; events with additional electrons or isolated muons are
rejected to suppress events from $Z$ and $t\bar{t}$ background processes.
Events are required to have missing transverse momentum (\met) of at least $25\,\GeV$  accounting for the presence of the neutrino.
The transverse mass of the \Wboson~boson candidate reconstructed from the electron and neutrino candidates, $m_{\mathrm{T}}(l\nu)$, is required to exceed 40\,\GeV.
Events with exactly one jet with  $\pt>25$\,\GeV\ are selected. 
This single jet is moreover required to contain exactly one muon within a cone of radius $\Delta R = 0.5$ around the jet direction,
following the selection requirements of the SMT algorithm as described in Section~\ref{sec:smt_tagger}.

\subsection{Determination of the $W+c$ yield}
\label{subsec:backDet}

The yield of the \Wc{} signal process is determined exploiting its charge correlation, by subtracting the number of SS events from OS events, 
$\NOSmSS{}= \NOS{}-\NSS{}$. 
The remaining background, substantially reduced after the OS-SS subtraction, consists predominantly of  \Wlight~and to a lesser extent of \mj~events. 
Their contributions after the OS-SS subtraction are estimated using data-driven methods as sketched below and explained in more detail in Ref.~\cite{STDM-2012-14}. 
Smaller backgrounds from $Z/\gamma^*$+jets, top and diboson production are estimated from \mc~simulations.
Backgrounds from \Wbb{} and \Wcc{} events are negligible since they are expected to contain the same number of 
OS and SS events.

The number of OS-SS events of the  \Wlight~and \mj~backgrounds is obtained using the following relation

\begin{equation}
\NOSmSS{bkg} = \frac{2\cdot \Asym{bkg}{}}{1-\Asym{bkg}{}}\NSS{bkg},
\label{eq:OSmSS}
\end{equation}
where the number of background events in the SS sample, \NSS{bkg}, and the OS/SS asymmetry,  \Asym{bkg}, defined as

\begin{equation}
\Asym{bkg}{}=\frac{\NOS{bkg}-\NSS{bkg}}{\NOS{bkg}+\NSS{bkg}},
\label{equ:Asym}
\end{equation}
are determined independently.

For the \Wlight~background, an estimate of $N_{W+\text{light}}^{SS}$ is obtained from MC simulation, corrected for the rate 
of SMT light-flavour jets as described in Section~\ref{smt:mtag}.  
For the multijet background, $N_{\mj}^{SS}$ is determined using a data-driven method. 
In the SS sample, a binned maximum likelihood fit of two templates to the \met\ distribution in data is performed. 
One template represents the \mj~background and the other the contributions from all other sources (including the \Wc{} signal). 
While the former is extracted from a data control region selected by inverting some of the electron identification criteria 
as well as the electron isolation requirement, the latter is obtained from simulation.
Since the SS sample is mainly composed of  \Wlight~and \mj~events, the initial \Wlight~and  \mj~estimates are adjusted
by performing a constrained $\chi^2$ fit so that the sum of all backgrounds and the signal equals the number of data SS events.
The other backgrounds and the small \Wc{} signal contribution are fixed to their values predicted by simulations in this fit.

For both the \Wlight~and \mj~backgrounds, $\Asym{bkg}{}$ is obtained using data-driven methods. 
The asymmetry of the \mj~background is derived by
performing the template fit to the \met\ distribution in data, as described above, separately for the OS and SS samples. 
\Asym{\mj}{}, derived from the fit results according to \eq~\ref{equ:Asym}, is found to be consistent with zero within the assigned total uncertainties.
The total uncertainties are dominated by the statistical component. 

The asymmetry of the  \Wlight~background, \AWlight{}, is obtained from \mc~simulation,
but corrected by a factor derived in a data sample that is defined by omitting the identification requirements for the soft muon.
The correction factor is obtained by investigating the charge correlation of the electron from the \Wboson-boson decay and
generic tracks passing the soft-muon kinematic requirements and being associated with the selected jet. 
\AWlight{} is found to be approximately 10\%. The assigned total uncertainty is dominated by the statistical uncertainty
due to the limited size of the simulated sample. 

The selected numbers of OS and SS events in the data \smtjetssample\ are 7445 and 3125, respectively.
This results in a number of OS-SS data events of $4320\pm100\textrm{\,(stat.)}$
and an extracted number of SMT \Wc{} events of $3910 \pm 100\textrm{\,(stat.)} \pm 160\textrm{\,(syst.)}$. 
The estimated number of background events, the number of data events and the measured \Wc{} yield are summarised in Table~\ref{tab:finalYields}.
The only backgrounds exhibiting a significant asymmetry are from the single-top and diboson processes.

\begin{figure}							
  \subfloat[\label{fig:smtjetPt_rew}]{									
    \centering								
    \includegraphics[width=0.49\textwidth]{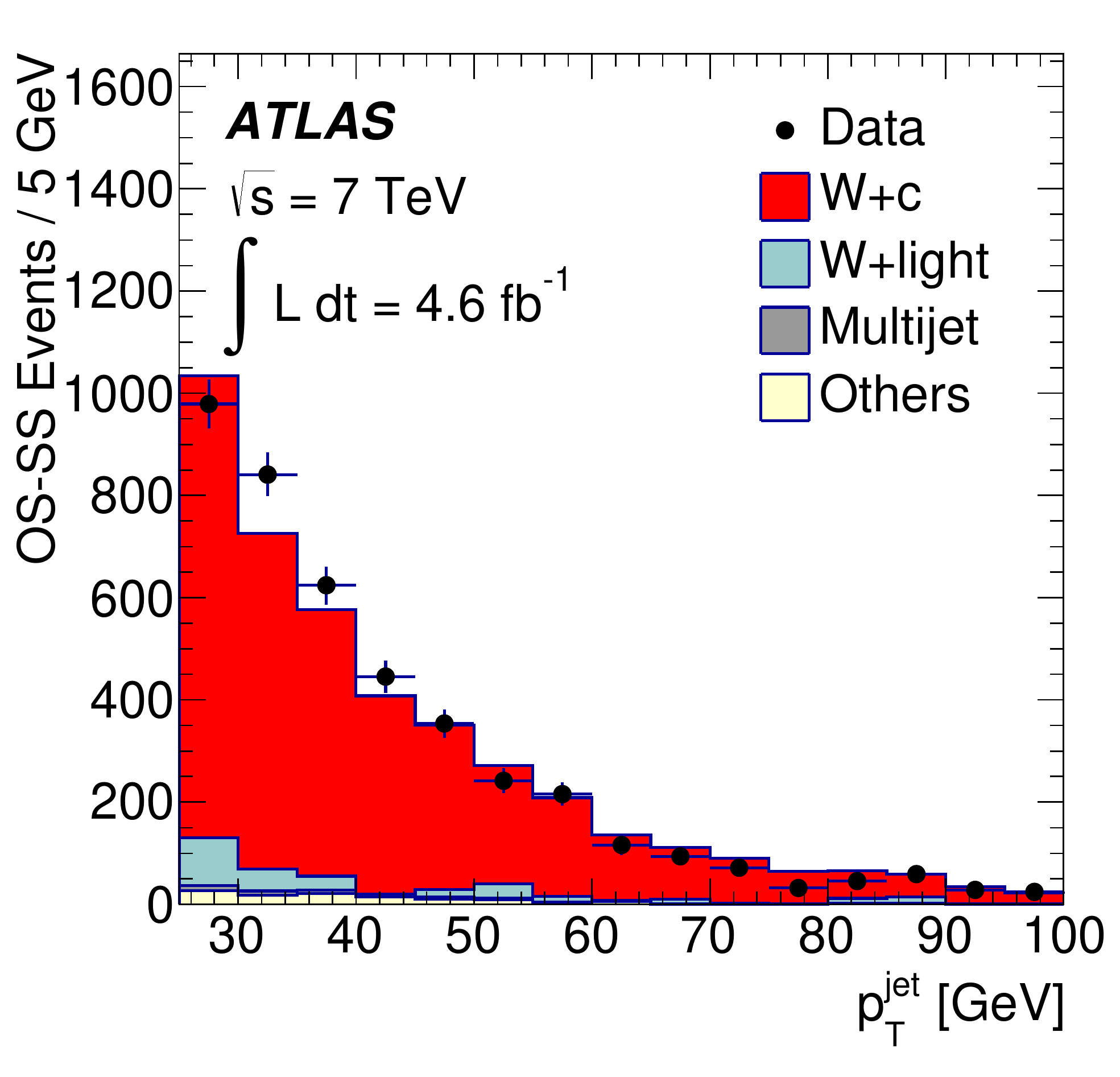}
  }\hfill										       									       
  \subfloat[\label{fig:taggingOutputs_rew}]{									
    \centering								
    \includegraphics[width=0.49\textwidth]{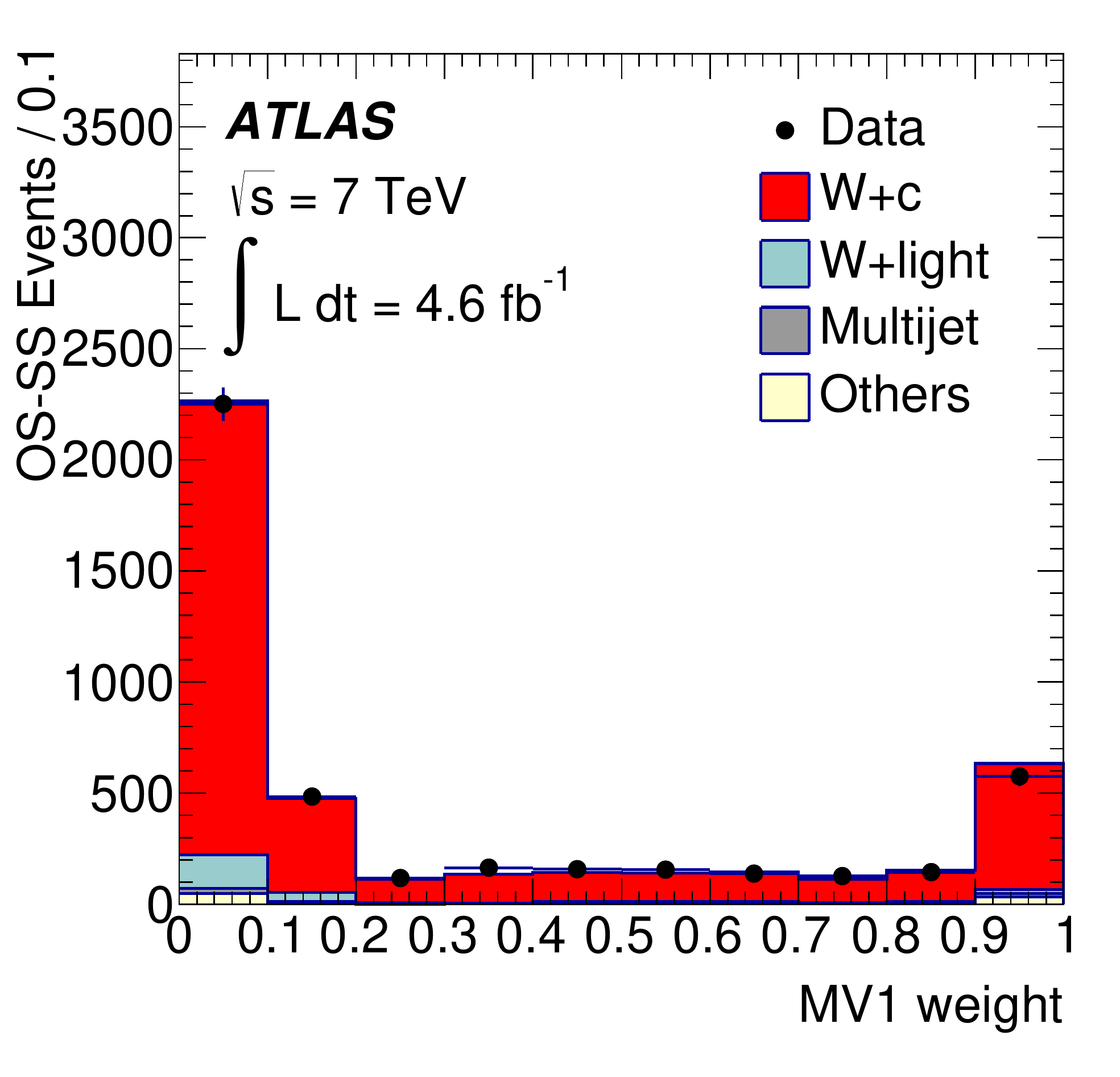}
  }\hfill
  \caption{The \pt~distribution (a) 
    and the output weight of the \MV~tagging
    algorithm (b) of \smtjets{} in a sample of events with opposite-sign lepton 
    charges from which the corresponding distribution of a sample with 
    same-sign lepton charges has been subtracted (``OS-SS events''). 
    The \Wc{} signal contribution is normalised to the measured yield, while the shapes
    are extracted from the \Alpgen+\PythiaCorr{} sample.
    The average $\pt^{\mathrm{jet}}$ in the data is about 42\,\GeV.}
  \label{fig:jetProb}										       
\end{figure}

Figure~\ref{fig:jetProb} shows the \pt\ distribution of $c$-jet candidates in OS-SS events as well as the output weight of the \MV~tagging algorithm
for which the \ctageff{} is calibrated as described in the following sections.
The signal contribution is normalised to the measured yield and the background contributions to the values listed in \tab~\ref{tab:finalYields}.
The \Wc{} signal shapes are derived from the \Alpgen+\PythiaCorr~simulated signal sample. 
The \mj{} shapes are extracted from the data control region used to derive the fit templates for determining \Asym{\mj}{}.

\begin{table}[t]
  \centering
  \scriptsize
  \begin{tabular}{lr@{ $\pm$ }l} \hline\hline
     \multicolumn{3}{c}{Number of events $N^{\rm OS-SS}$}\\
    \hline
    \Wlight       & 240  &110 \\
    \Mj           & 50   &130\\
    \ttbar        &  13  & 5\\
    \Singletop    &  62  & 10\\
    \Diboson      &  35  & 5\\
    \Zjets        &  6   & 14\\\hline
    Total background&410 &160 \\
    \Wc\,(meas.)  & 3910 & 190 \\\hline 
    Data          & 4320 & 100 \\
    \hline\hline
  \end{tabular}
  \caption{Number of OS-SS events of the different backgrounds and in data, as well as the measured \Wc{} yield. 
    The combined statistical and systematic uncertainties are quoted. Correlations between the uncertainties due 
    to exploiting the constraint in the SS sample are taken into account when computing the total background uncertainty.}
  \label{tab:finalYields}
\end{table}

\subsection{Measurement of the $c$-jet tagging efficiency of SMT $c$ jets}
\label{sec:effExtraction}

The output weight of the \MV~tagging algorithm for which the \ctageff{} is calibrated at operating points
corresponding to $b$-jet tagging efficiencies of 85\,\%, 75\,\%, 70\,\% and 60\,\%  in simulated \ttbar~events
is shown in \fig~\ref{fig:taggingOutputs_rew}. 
It should be noted that in what follows the number of events refers to the number of OS-SS subtracted events, unless indicated otherwise.

The \ctageff{} of \smtcjets{}, \effWc{data}, is derived as the fraction of \Wc{} events selected in data that pass a certain $b$-tagging requirement

\begin{equation}
  \effWc{data} =\frac{\NWctag{}}{\NWc{}},
\label{eq:effWcmu}
\end{equation}
where \NWc{} is the number of \Wc{} events  before applying the $b$-tagging requirement (hereinafter referred to as \textit{pre-tag} level or sample)
and \NWctag{} is the number of \Wc{} events passing the $b$-tagging requirement.
\NWc{} is derived as described in \mysec~\ref{subsec:backDet}.

The number of $b$-tagged signal events, \NWctag, is determined 
by subtracting from the number of all $b$-tagged events the expected number of $b$-tagged background events
taking into account the tagging rates of the different background components, which depend
on the jet flavour composition of the background component and their respective tagging 
efficiencies.
The \tagrates~of the \Wlight, \ttbar, \singletop, \diboson~and \Zjets~backgrounds are extracted using \mc~simulation
with the tagging efficiencies of the differently flavoured jets 
being corrected to match those in data by applying $b$-tagging scale 
factors (Sections~\ref{sec:combination} and \ref{sec:mistag})
and the corresponding uncertainties being taken into account.
The total uncertainties on the \tagrates{} are either dominated by or of the same order as the statistical uncertainties due to the limited size of the simulated samples.
\begin{figure}
  \subfloat[]{\label{fig:fitResult_a}
    \includegraphics[width=0.49\textwidth]{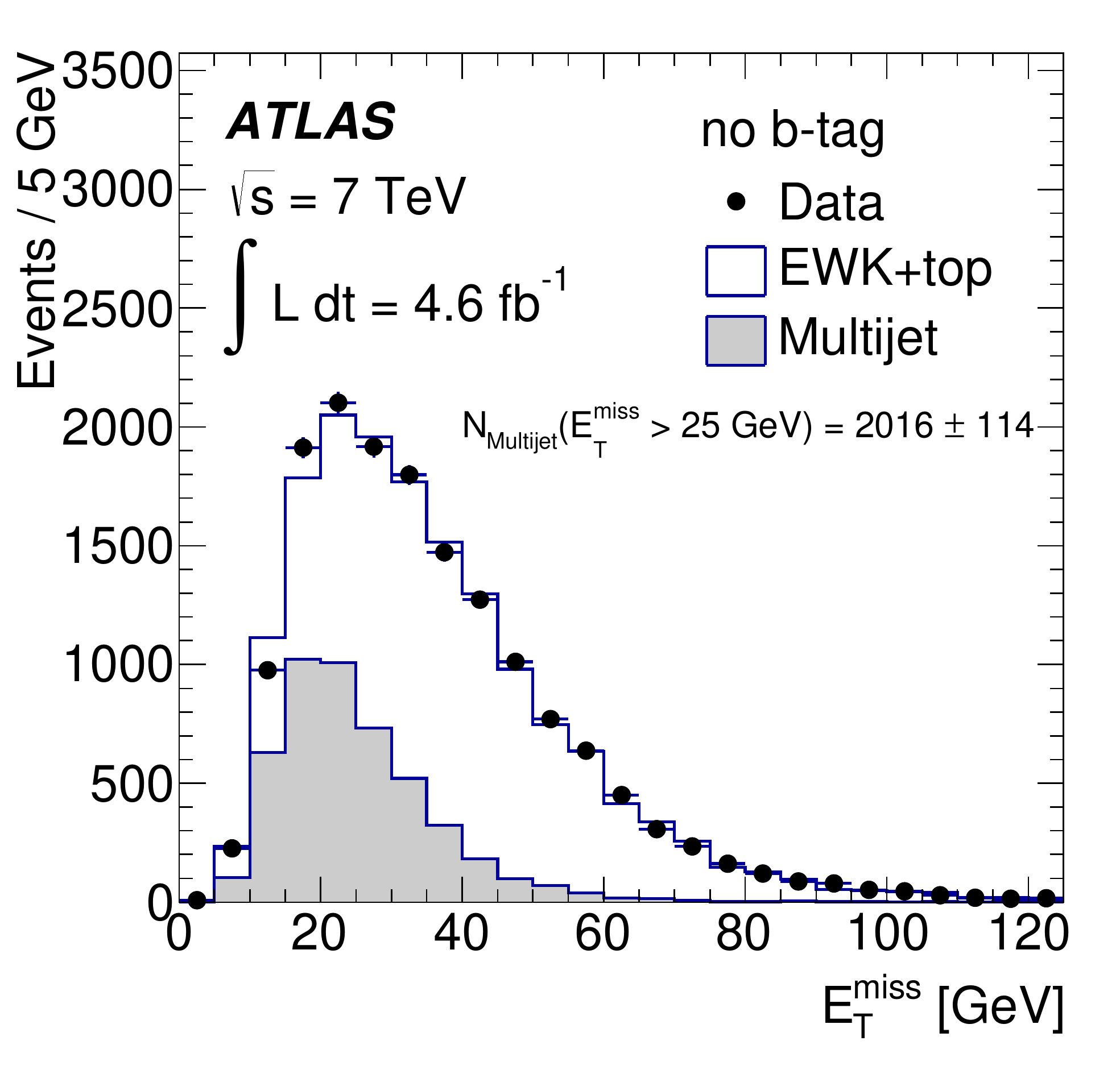}}
  \subfloat[]{\label{fig:fitResult_b}
    \includegraphics[width=0.49\textwidth]{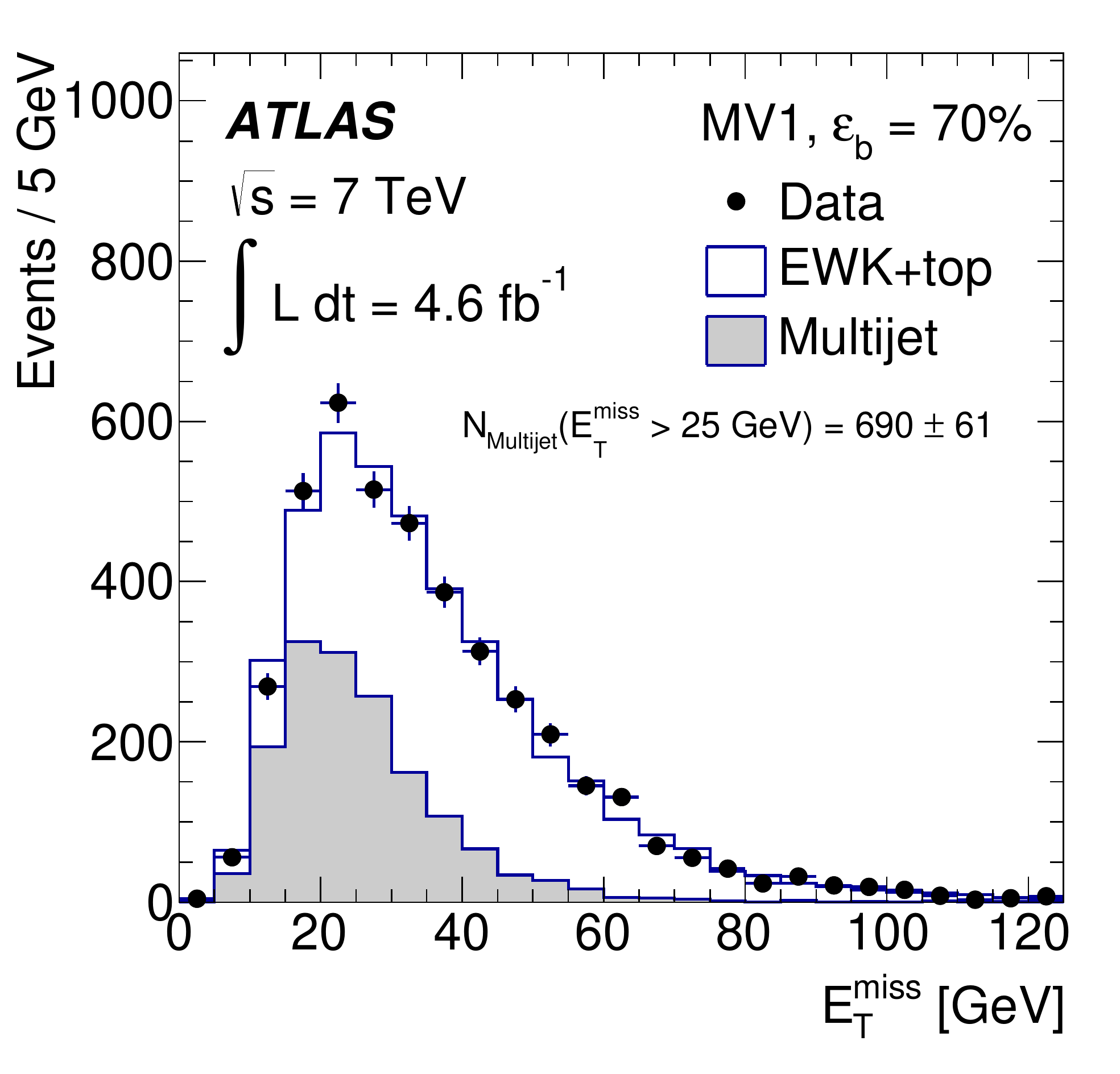}}
  \caption{Fits to the \met~distribution in the sum of the OS and SS samples used to determine the number of \mj~events  before (a) and 
    after (b) applying a cut on the \MV~weight corresponding to the $\bEff = 70\,\%$ operating point. 
    ``EWK+top'' corresponds to the production of $W$ and $Z$ bosons,
    the production of single top and top-quark pairs and diboson production.}
  \label{fig:fitResult}
\end{figure}

The \tagrate~of the \mj~background is estimated using a data-driven method which follows closely the
procedure to extract the OS/SS asymmetry at pre-tag level described in \mysec~\ref{subsec:backDet}: 
a binned maximum likelihood fit of templates to the \met~distribution in data before and after applying the \btagging~requirement is performed. 
The \mj~templates used to derive the number of $b$-tagged \mj~events are extracted from data control samples defined in \mysec~\ref{subsec:backDet}
with the additional requirement that the selected SMT jet is $b$-tagged.
Since the \mj~\tagrates~computed using the fit results for the OS and SS samples lead to compatible results,
the final \mj~\tagrate~is obtained from the fit results derived for the sum of the OS and SS samples.
The fit results before and after applying the MV1 $b$-tagging requirement corresponding to the \bEff=70\,\% operating point are shown in \fig~\ref{fig:fitResult}.
The \mj~\tagrates~for the different operating points vary between 26\,\% and 55\,\% indicating that the \mj~sample has a large heavy flavour component.
The relative total uncertainties range between 15\,\% and 23\,\% accounting for the dominating statistical uncertainties of the \met~fits and 
systematic uncertainties related to the choice of the fit range and variations of the shapes of 
the \mj~and non-\mj~templates, obtained by modifying the inverted electron identification requirements or the relative fractions of the 
different background processes, respectively.

\begin{figure}
  \centering
  \subfloat[\label{subfig:cEff}]{
    \includegraphics[width = 0.49\textwidth]{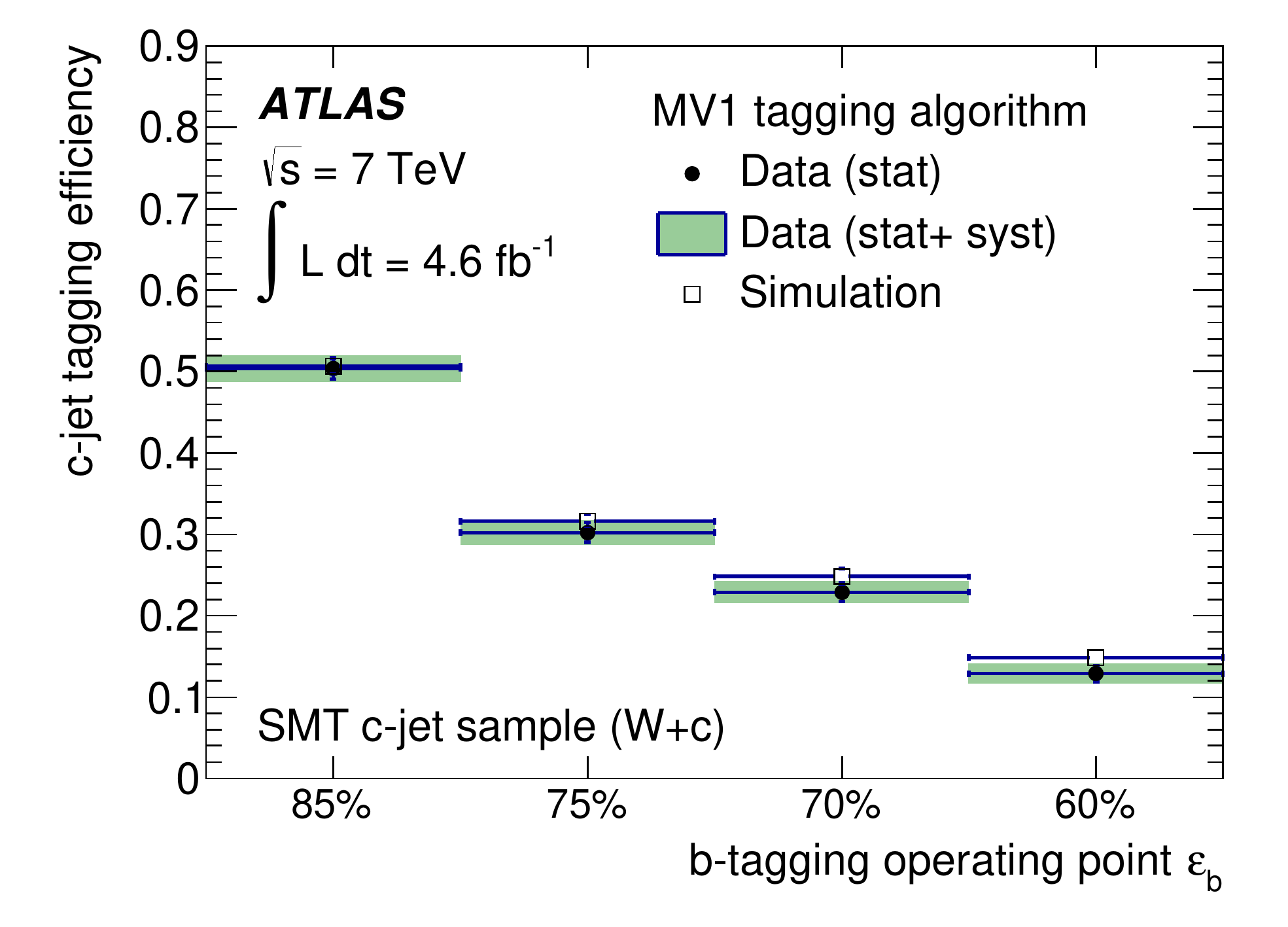}
  }
  \subfloat[\label{fig:smtSF}]{
    \includegraphics[width = 0.49\textwidth]{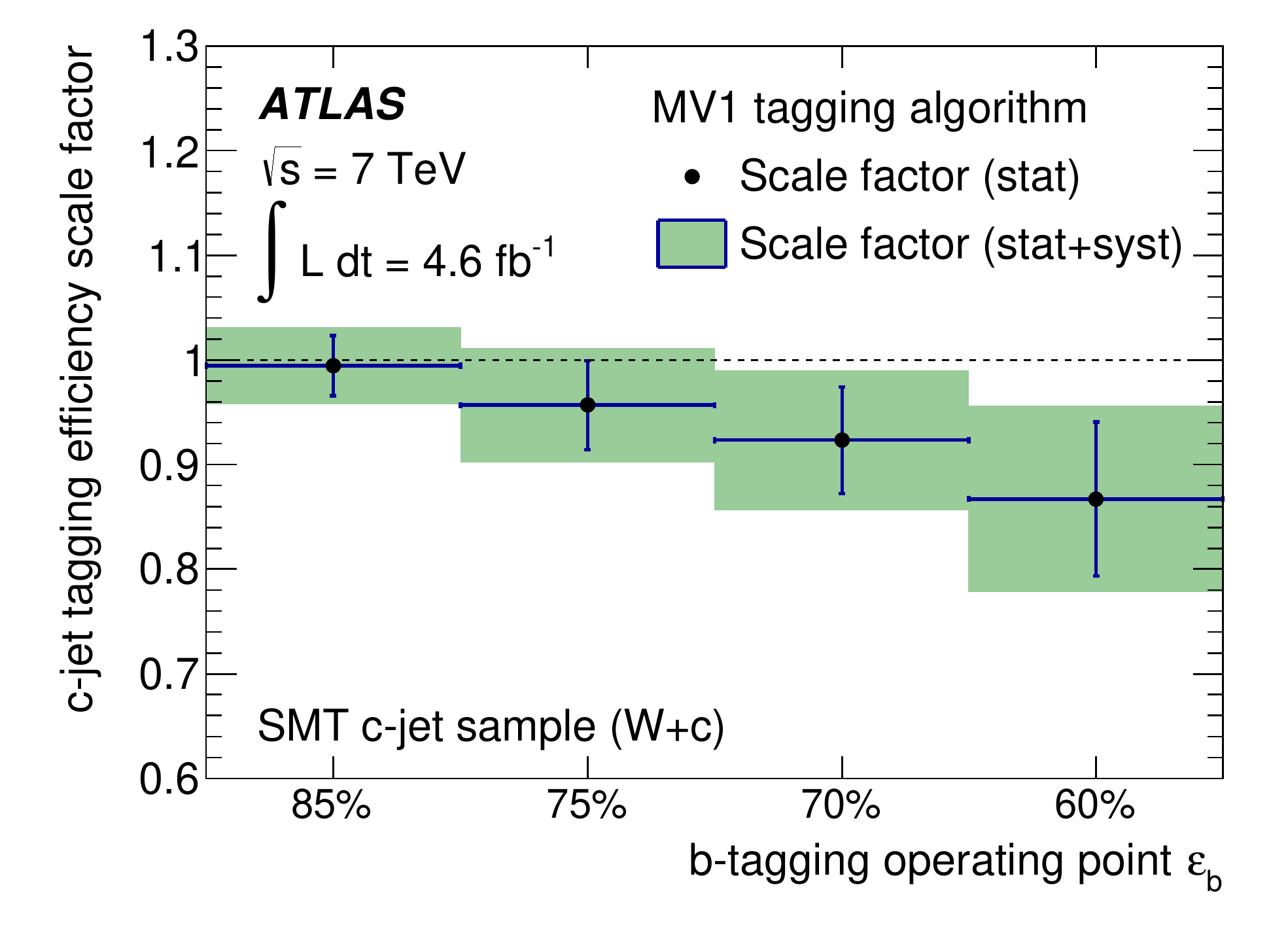}
  }
  \caption{Comparison of the \ctageffs{} for \smtcjets{} in data and the \Alpgen+\PythiaDef{} simulation (a) and the corresponding
    data-to-simulation \ctageff{} scale factors (b) for several operating points of the \MV~tagging algorithm derived using \Wc{} events.}
\end{figure}

The \ctageffs{}  of \smtcjets{}  in data \effWc{data} derived according to \eq~\ref{eq:effWcmu}
for the different operating points of the \MV~tagging algorithm are shown in \fig~\ref{subfig:cEff}. 
Their values decrease from 50\,\% to 13\,\% with increasing tightness of the operating point;
the corresponding relative total uncertainties increase from 3\,\% to 10\,\%. 
The systematic uncertainties are dominated by the precision on the \Wlight~and \mj~background yields at pre-tag level,
in particular on the data-driven OS/SS asymmetry estimates as mentioned in \mysec~\ref{subsec:backDet}, and on the \Wlight~\tagrate.
The statistical uncertainties are of the same order as the systematic uncertainties.
 
The expected \ctageff{}, \effWc{\rm \simu}, is defined as the fraction of \smtcjets{}
selected in simulated \Alpgen+\PythiaDef{} \Wc{} events that pass the $b$-tagging requirement. 
In Fig.~\ref{subfig:cEff} \effWc{\rm \simu} is compared to \effWc{\rm data} for the different $b$-tagging operating points.

The data-to-simulation scale factors for \smtcjets, $\sfcmu = {\effWc{\rm data}}/{\effWc{\rm \simu}}$, are shown in \fig~\ref{fig:smtSF}.
They decrease from 0.99 to 0.87 with increasing tightness of the operating points, while their total uncertainties increase from 4\,\% to 10\,\%.
The systematic uncertainties arise from the previously discussed background determinations
as well as from the \Wboson~boson reconstruction and the SMT $c$-jet identification efficiencies.
A more detailed discussion and a breakdown of the systematic uncertainties can be found in \mysec~\ref{sec:systUnc}.
The statistical uncertainties are of the same order as or larger than the systematic uncertainties. 

\subsection{Calibration of the $c$-jet tagging efficiency for inclusive $c$-jet samples}
\label{sec:inclSF}

Due to several differences between an inclusive sample of $c$~jets and a sample of \smtcjets{}
the derived \ctageff{} scale factors need to be extrapolated in order to be applicable also to the former. 

Selecting a sample of $c$~jets via semimuonic decays of $c$~hadrons leads to a $c$-hadron composition that is
different with regard to an inclusive $c$-jet sample,
given that the semileptonic branching fractions of the different $c$-hadron types vary significantly.
Since the $c$-hadron types differ also in several other characteristics relevant for the performance of 
$b$-tagging algorithms, e.g. the lifetime -- which is correlated with the semileptonic branching fraction -- or the charged particle decay multiplicity,
the tagging efficiencies of $c$~jets associated to different types differ considerably.
For instance, the tagging efficiencies of $c$-jets associated to the most prominent weakly decaying $c$-hadron types
for the \bEff=70\,\% operating point of the \MV~tagging algorithm  are 

\begin{equation*}
 \Dzero:\,\,0.157\pm0.001,\,\,\Dplus:\,\,0.280\pm0.002,\,\,\Ds: 0.152\pm0.003,\,\,\Lpc:\,\,0.041\pm0.002
\end{equation*}
as estimated from the \Alpgen+\PythiaCorr~sample, where the quoted uncertainties are statistical only.
Therefore, the overall tagging efficiency of a $c$-jet sample strongly depends on the admixture of different $c$ hadrons.
Furthermore, the tagging efficiencies for \smtcjets{} differ with regard to inclusive $c$-jet samples because of differences in the decay properties,
such as that requiring an associated muon guarantees at least one well reconstructed track stemming from a $c$-hadron decay.

The \ctageff{} of an inclusive sample of $c$~jets, \effWcIncl{}, can be obtained from the \ctageff{} of \smtcjets{}, \effWc{}, by applying an 
extrapolation factor \effCorr{}{}: 

\begin{equation}
 \effWcIncl{} = \effCorr{}{}\cdot \effWc{}.
\label{eq:inclEff}
\end{equation}
The comparison of the expected \effWcIncl{\simu} and \effWc{\simu} for several operating points of the \MV~tagging algorithm
derived using the \Wc{} sample simulated with \Alpgen+\PythiaDef~shows that \effWcIncl{\simu} is systematically 
lower than \effWc{\simu}, resulting in a correction factor \effCorr{\simu}{} of about 0.8 for the different operating points.

The \ctageff{} scale factor \sfc~for inclusive $c$~jets can be computed similarly from the measured \ctageff{} scale factor \sfcmu\ by applying a correction
factor \SFCorr

\begin{equation}
  \sfc = \frac{\effWcIncl{data}}{\effWcIncl{\simu}} = \frac{\effCorr{data}{}\cdot \effWc{data}}{\effCorr{\simu}{}\cdot \effWc{\simu}}
  = \frac{\effCorr{data}{}}{\effCorr{\simu}{}}\cdot\sfcmu = \SFCorr\cdot\sfcmu,
  \label{eq:inclSF}
\end{equation}
where \SFCorr~is expressed as the ratio of the efficiency extrapolation factors in data, \effCorr{data}{}, and simulation, \effCorr{\simu}{}.
Mismodelling of the differences between SMT $c$-jet and inclusive $c$-jet samples in \mc~simulation leads to different extrapolation factors \effCorr{}{} and 
thus to a ratio \SFCorr{} deviating from one.

The efficiency extrapolation factor \effCorr{data}{} is estimated using the simulated \Alpgen+\PythiaCorr{} sample 
where $c$-quark fragmentation and $c$-hadron decay properties are corrected to the current best knowledge 
as discussed in detail in the following, and yielding a corresponding extrapolation factor $\effCorr{\simu}{corr}$.
An estimate for the scale factor extrapolation factor \SFCorr~is then obtained from 

\begin{equation}
 \SFCorr \approx \frac{\effCorr{\simu}{corr}}{\effCorr{\simu}{}}.
\end{equation}
In order to describe the $c$-hadron composition of inclusive $c$-jet samples correctly, the fragmentation fractions
of the relevant weakly decaying $c$-hadron types in the \PythiaDef~sample are re-weighted to those obtained by combining the 
results of dedicated measurements performed in \ee~and \ep~collisions~\cite{heralep}. 
By correcting the semileptonic branching fractions of $c$~hadrons to match the world average values~\cite{PDG2014},
also the modelling of the $c$-hadron composition in the simulated \smtcjetssample{} is improved. 
Comparing the $c$-hadron fractions in the \smtcjetssample{} with the ones in the inclusive $c$-jet sample shows
that the \Dplus-meson component is strongly enhanced in the SMT $c$-jet sample due to its relatively large semileptonic branching fraction:
the \smtcjetssample{} consists of a similar amount ($\sim43\,\%$) of \Dzero~and \Dplus~mesons,
while in the inclusive $c$-jet sample the \Dzero~meson is clearly the dominant $c$-hadron type ($\sim60\,\%$).

Given that the $b$-tagging algorithms exploit track and vertex properties that can specifically be associated with $b$- and $c$-hadron decays,
it is important that these are well modelled by \mc~simulations.
This applies in particular to the charged particle decay multiplicity of $c$-hadron decays.
In order to improve its description in  the \PythiaDef~simulation the relative branching fractions of the dominant semileptonic decay channels of the 
abundant \Dplus~and \Dzero~mesons are corrected to match the world average values \cite{PDG2014}.
The less frequent decay channels of the two mesons, which are known to a lower precision, are adjusted to maintain the overall normalisation. 
It is found that the relative fractions of \Dplus~decays with one and three charged decay products have a noticeable impact on the \ctageff{} of \smtcjets.
Also the  hadronic $n$-prong branching fractions of $c$~hadrons in the \PythiaDef~sample are corrected
with a significant impact on the predicted \ctageff{} of an inclusive $c$-jet sample. 
While the corrections in case of the \Dzero~meson have been inferred from measured inclusive $n$-prong branching fractions \cite{PDG2014},
the hadronic $n$-prong branching fractions of the \Dplus~and \Ds~mesons as well as the \Lpc~baryon are re-weighted to the predictions of the \EvtGen~simulation. 
A comparison of the \PythiaDef~and \EvtGen~predictions, as well as the measured values in case of the \Dzero~meson,
reveals large differences in the hadronic $n$-prong distributions.
In particular the 2-to-0-prong ratio of the \Dzero~meson and the 3-to-1-prong ratio of the \Dplus~meson are found to have
a significant influence on the inclusive \ctageff.

Finally, the $b$-tagging performance also depends on the kinematic distributions of the $c$-hadron and its decay products:
first, the effect of any mismodelling of the momentum fraction of the $c$~hadron ($p_{\rm T}^{c~\rm hadron}/p_{\rm T}^{c~\rm jet}$),
which is sensitive to the $c$-quark fragmentation function, is evaluated by comparing different simulations;
second, the momentum of the decay muon in the rest frame of the $c$~hadron (\pstar) is re-weighted to agree with the \EvtGen~prediction.

The \Alpgen+\PythiaCorr{} simulation obtained by applying all corrections discussed above predicts a
significantly lower $b$-jet tagging efficiency for inclusive $c$-jet samples with
respect to the \Alpgen+\PythiaDef{} simulation, mainly due to the correction of the  hadronic $n$-prong decay branching fractions.
Since the impact of the corrections on the \ctageff{} predicted for a \smtcjetssample{} is very small ($<$2\% for all operating points), the efficiency correction factor computed 
using the \Alpgen+\PythiaCorr~sample $\effCorr{\simu}{corr} = 0.69-0.76$ is systematically 
lower than $\effCorr{\simu}{} = 0.79-0.83$ of the \Alpgen+\PythiaDef{} sample.
The resulting scale factor extrapolation factors \SFCorr{} are 0.86--0.95, systematically lower than unity with a decreasing trend towards tighter operating points. 
Their total systematic uncertainties, ranging between 3\,\% and 7\,\%, are due to the before-mentioned corrections and are discussed in detail in \mysec~\ref{sec:systUnc}.
Statistical uncertainties are neglected since the numerator and the denominator of $\SFCorr$ are computed using approximately the same simulated events.

\subsection{Systematic uncertainties}
\label{sec:systUnc}
\begin{table}
  \centering
  \scriptsize
  \begin{tabular}{lcccc}
    &\qquad\qquad\qquad&\qquad\qquad\qquad&\qquad\qquad\qquad&\qquad\qquad\qquad\\
    \hline\hline
    &\multicolumn{4}{c}{Operating points (\bEff) of the \MV~tagging algorithm}\\
    Source                         & 85\,\% &   75\,\%  &       70\,\%  &    60\,\%\\
    \hline
    Event reconstruction           &   1.4  &   2.1     &       3.4     &       3.4\\						  
    Background pre-tag yields      &   0.8  &	2.1	&	2.3	&	4.0\\							
    Background tagging rates       &   1.6  &	1.9	&	2.2	&	2.4\\
    $c$-quark fragmentation        &   0.7  &   0.7     &       0.9     &       1.0\\
    Hadronic $c$-hadron decays     &   2.1  &	3.7	&	4.8	&	6.3\\
    Semileptonic $c$-hadron decays &   2.1  &	2.9	&	2.5	&	3.3\\
    Simulated sample size          &   1.2  &	1.9	&	2.0	&	2.7\\
    \hline
    Total systematic uncertainty   &   4.0  &	6.2	&	7.4	&	9.6\\
    Statistical uncertainty        &   2.2  &	3.5	&	4.9	&	8.0\\
    \hline
    Total uncertainty              &   4.6  &	7.1	&	8.9	&        12\\
    \hline \hline
  \end{tabular}
  \caption{Summary of the systematic uncertainties on the \ctageff{} scale factor for inclusive $c$~jets \effSFWcIncl{}.
    The values are listed in percent.}
  \label{tab:syst}
\end{table}

Systematic uncertainties on the \ctageff{} scale factors arise from the \Wboson~boson reconstruction and SMT $c$-jet identification, the pre-tag yield and 
\tagrate~determination of the backgrounds as well as from the extrapolation procedure to correct the measured \ctageff{} scale factors for \smtcjets{}.
The different contributions are summarised in \tab~\ref{tab:syst} and discussed below.

\subsubsection*{Event reconstruction}
The \Wboson~boson reconstruction uncertainty arises from the electron trigger and reconstruction efficiencies,
the electron energy scale and resolution as well as the \met~reconstruction.
There are two main sources of uncertainty on the $c$-jet identification:
first the determination of the jet energy scale and resolution,
second the reconstruction efficiency and the energy resolution of the soft muon as well as the SMT tagging efficiency and mistag rate, respectively. 
The lepton uncertainties are assessed by varying each of the efficiencies, the mistag rate,
the energy scale and resolution in simulation within the range of the assigned uncertainties as determined from independent measurements and 
re-calculating the resulting \ctageff{}.
The uncertainties due to jet energy scale and resolution determinations are estimated in the same way.
The uncertainties on the lepton and jet energy scale and resolution are
additionally propagated to the reconstruction of the missing transverse momentum. 
Further systematic uncertainties that affect the \met~reconstruction,
but are not associated with reconstructed objects, are also accounted for.
The systematic uncertainties due to the jet energy scale and resolution calibrations dominate the event reconstruction uncertainties,
but are of the same order as the statistical uncertainty due to the 
limited size of the simulated signal sample.
A more detailed breakdown and discussion of the event reconstruction uncertainties can be found in Ref.~\cite{STDM-2012-14}.

\subsubsection*{Pre-tag yields and background tagging rates}

The determination of the OS-SS background yields at pre-tag level and the assessment of the corresponding uncertainties are discussed in \mysec~\ref{subsec:backDet}.
The main source of systematic uncertainties is the data-driven estimation of the OS/SS asymmetry of the \Wlight~and \mj~backgrounds. 
The uncertainty due to the background \tagrates~is dominated by the uncertainty on the \Wlight~\tagrate~mainly
because of the limited size of the simulated sample used to derive it, as discussed in \mysec~\ref{sec:effExtraction}.

\subsubsection*{Fragmentation and decay modelling}
The $c$-quark fragmentation and $c$-hadron decay properties are corrected to improve the modelling of the \Alpgen+\Pythia~signal sample
as described in \mysec~\ref{sec:inclSF}.
Whenever results from independent measurements are used to correct the \mc~description,
the uncertainties assigned to those results are propagated to the extrapolated scale factors. 
This is done for the fragmentation fractions and the semileptonic decay branching fractions of the prominent weakly decaying $c$-hadrons
as well as for the hadronic $n$-prong decay branching fractions of the \Dzero~meson.
Where corrections are derived from \mc~simulations because no measurements are available, the corresponding systematic uncertainties are assessed 
by comparing predictions from different \mc~generators.
Hence, the difference between the \Pythia{} and \Herwig~simulations is used to estimate the uncertainty due to the fragmentation function of $c$ quarks.
The systematic uncertainty due to a possible mismodelling of the \pstar{} distribution of the soft muon is evaluated
from the difference between the \EvtGen{} and \Pythia~simulations.
The largest difference between the \EvtGen~and either the \Pythia~or \Herwig~simulations is used to
estimate the uncertainties due to the  hadronic $n$-prong decay branching fractions of the \Dplus~and \Ds~mesons as well as the \Lpc~baryon.
The largest effect on the final scale factors computed for inclusive $c$~jets arises from the correction of the  $n$-prong decay branching fractions 
of hadronically decaying $c$~hadrons.
Since only semileptonically decaying $c$~hadrons are used in the data measurement, 
uncertainties on the properties of hadronically decaying $c$~hadrons propagate fully to the scale factors for inclusive $c$~jets.

\subsection{Results}
\label{sec:WcResults}

\begin{figure}
  \centering
  \includegraphics[width = 0.5\textwidth]{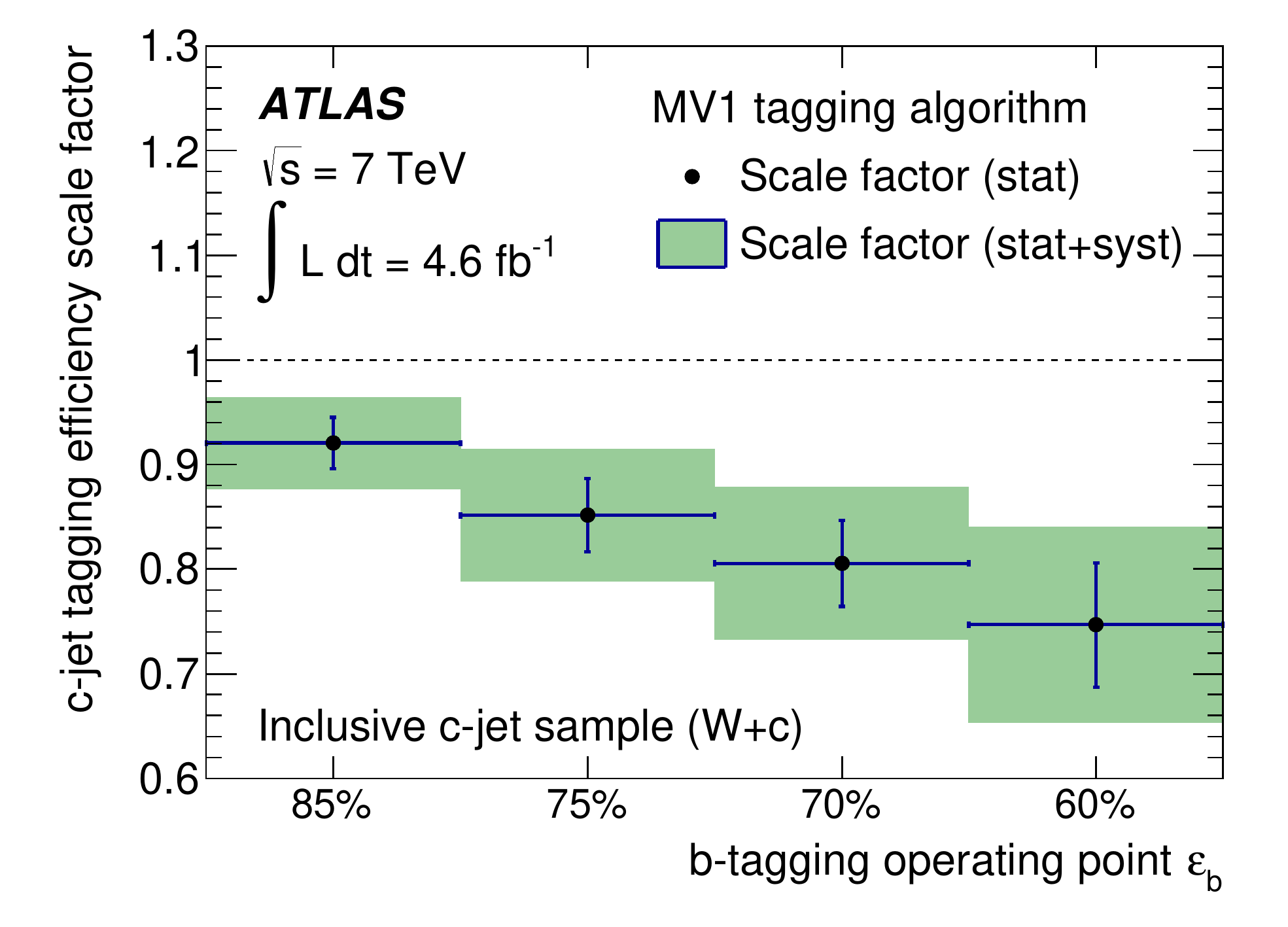}
  \caption{Data-to-simulation \ctageff{} scale factors for inclusive $c$~jets
    derived for the  \MV~tagging algorithm with respect to the \Alpgen+\PythiaDef{} sample.}
  \label{fig:finalResult}
\end{figure}

The data-to-simulation \ctageff{} scale factors for several operating points of
the \MV~tagging algorithm with respect to a \Wc{} sample simulated with
\Alpgen+\PythiaDef~are shown in \fig~\ref{fig:finalResult} for the different operating points.
Being applicable to inclusive samples of $c$~jets, these scale factors are derived from the measured \ctageff{}
scale factors for \smtcjets{} (see \mysec~\ref{sec:effExtraction}) by a simulation-based extrapolation procedure.
The results range between 0.75 and 0.92, decreasing with increasing tightness of the operating point,
while the assigned total uncertainties increase from 5\,\% to 13\,\%. 
There are three main sources of uncertainties that are of the same order: the statistical uncertainty,
the systematic uncertainty on the measured scale factors for \smtcjets{} and
the systematic uncertainty due to the extrapolation procedure described in \mysec~\ref{sec:inclSF}.
The main contribution to the latter is the limited knowledge of the charged particle 
multiplicity of $c$-hadron~decays.

\section{$c$-jet tagging efficiency calibration using the $D^{\star}$ method}
\label{sec:ceff_dstarbased}

In this section the $c$-jet tagging efficiency is measured using a sample of jets containing $\dstar$ mesons,
by comparing the yield of $\dstar$ mesons before and after the tagging requirement.
The measurement is based on the $D^{\star +} \to D^0(\rightarrow K^- \pi^+)\pi^+$ decay mode, and the contamination with $\dstar$ mesons that result from $b$-hadron decays is measured with a fit to the $D^0$ pseudo-proper time distribution.

\subsection{Data and simulation samples}
\label{sec:dstar_samples}

The data sample used in the $D^{\star}$ measurement was collected using a logical OR of inclusive jet triggers.
These triggers have been heavily prescaled to a constant bandwidth of about 0.5 Hz each and 
reach an efficiency of 99\% for events having the leading jet with an offline \pT{} higher than the 
corresponding trigger thresholds by a factor ranging between 1.5 and 2.
Events with at least one jet with \pT{} above a given threshold at the highest trigger level 
are selected, and using a combination of the inclusive jet triggers, the data set covers 
the entire 20--200~\GeV{} jet \pT{} range used in the analysis.

The analysis makes use of a Monte Carlo simulated sample of multijet
events. The samples used are equivalent to those used in the muon-based $b$-jet tagging efficiency measurement (see Section~\ref{sec:beff_mubased})
with the exception that each event in the sample used in the $D^{\star}$ analysis is required to contain a $\dstar$ meson,
in the decay mode $D^0(\rightarrow K^- \pi^+)\pi^+$.
Approximately one million events have been simulated per $\hat{p}_{\perp}$ bin.

As the trigger algorithms requiring a single jet with a \pT\ below approximately 250~GeV are prescaled
in data but not in simulated events, the \pT{} spectrum of  jets in the multijet samples is harder in data than in simulation.
Therefore the jet \pT{} distribution has been reweighted to match that observed in data.

\subsection{The $D^{\star}$ analysis}

\subsubsection*{\dstar{} selection}

\dstar\ mesons are reconstructed in the decay $\dstar \to D^0 \pi^+$, with $D^0\to K^-\pi^+$.
Pairs of oppositely charged tracks are considered for the $D^0$ candidates,
assigning both kaon and pion mass hypotheses to them. Studies on simulated data confirm that only the correct combination
of mass hypotheses produces a $D^0$ in the expected mass region. The $D^0$ candidates are then
combined with charged particle tracks with opposite sign to that of the kaon candidate, assigning the pion mass to them.

The \dstar{} candidates must satisfy the following criteria:
\begin{itemize}
  \item All tracks must have at least five hits in the silicon tracking detectors, 
        at least one of them in the pixel detector.
  \item The transverse momenta
        of the kaon and pion candidates from the $D^0$ decay candidate have to satisfy $\pT>1\GeV$.
  \item The reconstructed $D^0$ candidate mass $m_{K^-\pi^+}$ must satisfy $|m_{K^-\pi^+}-m_{D^0}|~<~40\MeV$,
    where $m_{D^0}$ is the world average $D^0$ mass, $m_{D^0} = 1864.83\pm 0.14\MeV$~\cite{pdg2010}.
  \item The transverse momentum of the \dstar\ candidate has to exceed $4.5\GeV$.
\end{itemize}
 The decay chain is fitted as follows: first the $D^0$ vertex is formed
 by fitting the kaon and pion candidates, and the resulting $D^0$ 
 direction is reconstructed by combining the kaon and pion four-momenta; the $D^0$ 
 direction is then extrapolated back and fitted with the pion candidate to form the $\dstar$ vertex.
 The decay chain is fitted with a tool allowing the simultaneous
 reconstruction and fit of both vertices.
 No requirements are made on the vertex fit $\chi^2$ probability 
 in order to minimise the bias on the $b$-tagging.

The $\dstar$ candidate is in turn associated with a reconstructed jet requiring its direction
to be within $\Delta R(\dstar,\mbox{jet})=0.3$ of the jet direction. 
Finally, to reduce the amount of combinatorial background, as well as the contribution from $b$ jets, 
the momentum of the $\dstar$ candidate projected along the jet direction
has to exceed 30\% of the jet energy.

\begin{figure}[ht]
  \centering
  \begin{tabular}{cc}
    \subfloat[]{\label{fig:dstar_a}\includegraphics[width=0.49\textwidth]{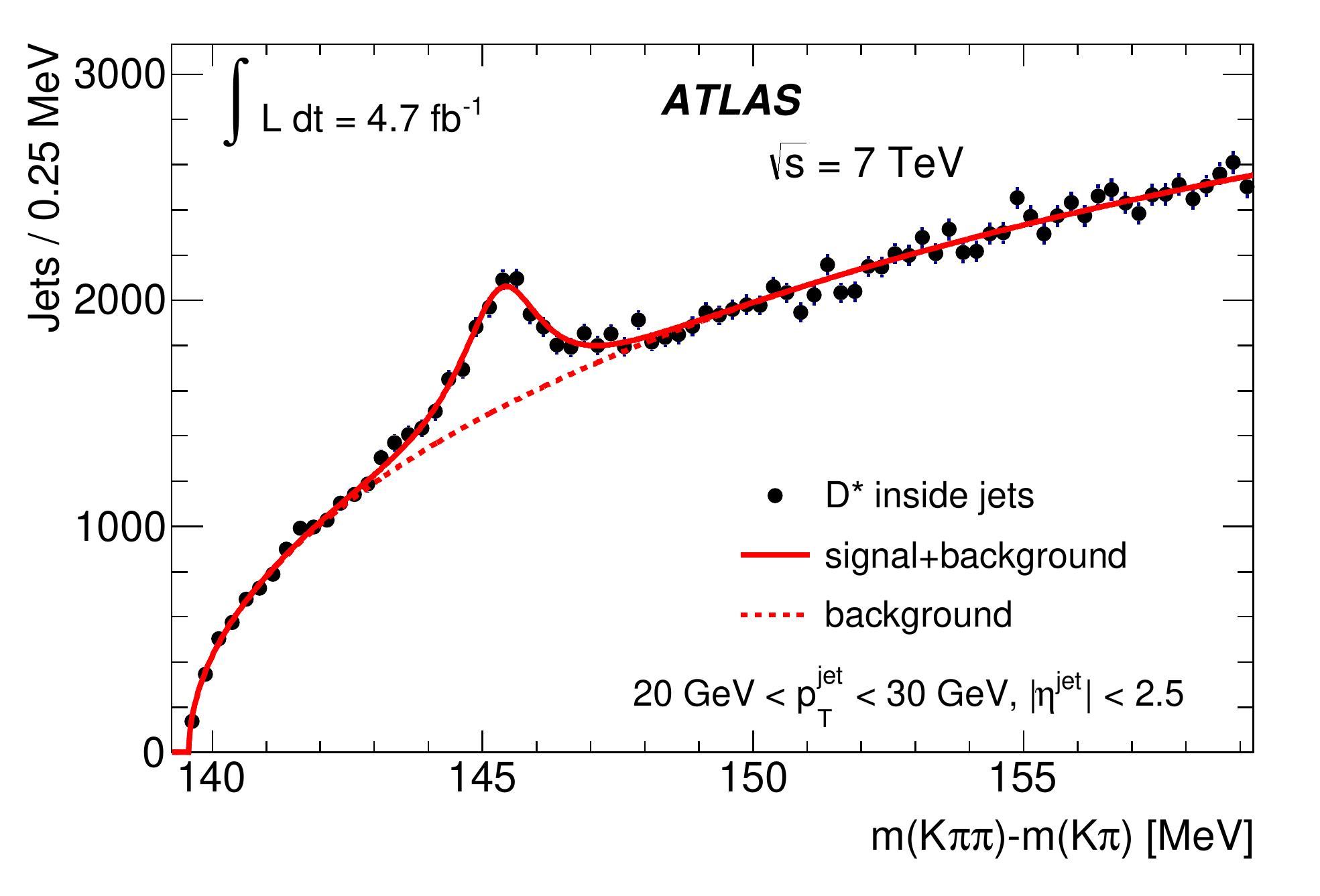}}
    \subfloat[]{\label{fig:dstar_b}\includegraphics[width=0.49\textwidth]{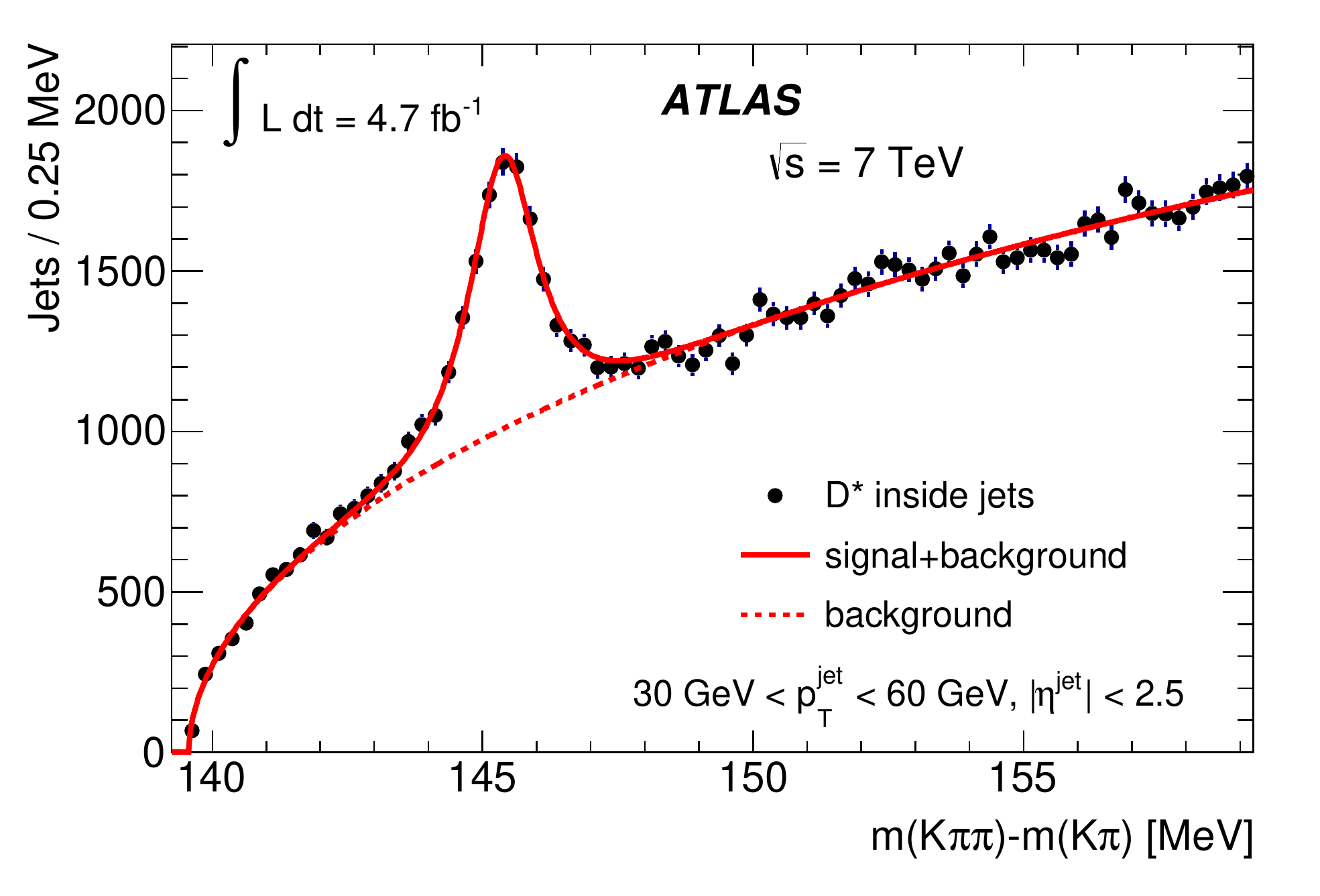}} \\
    \subfloat[]{\label{fig:dstar_c}\includegraphics[width=0.49\textwidth]{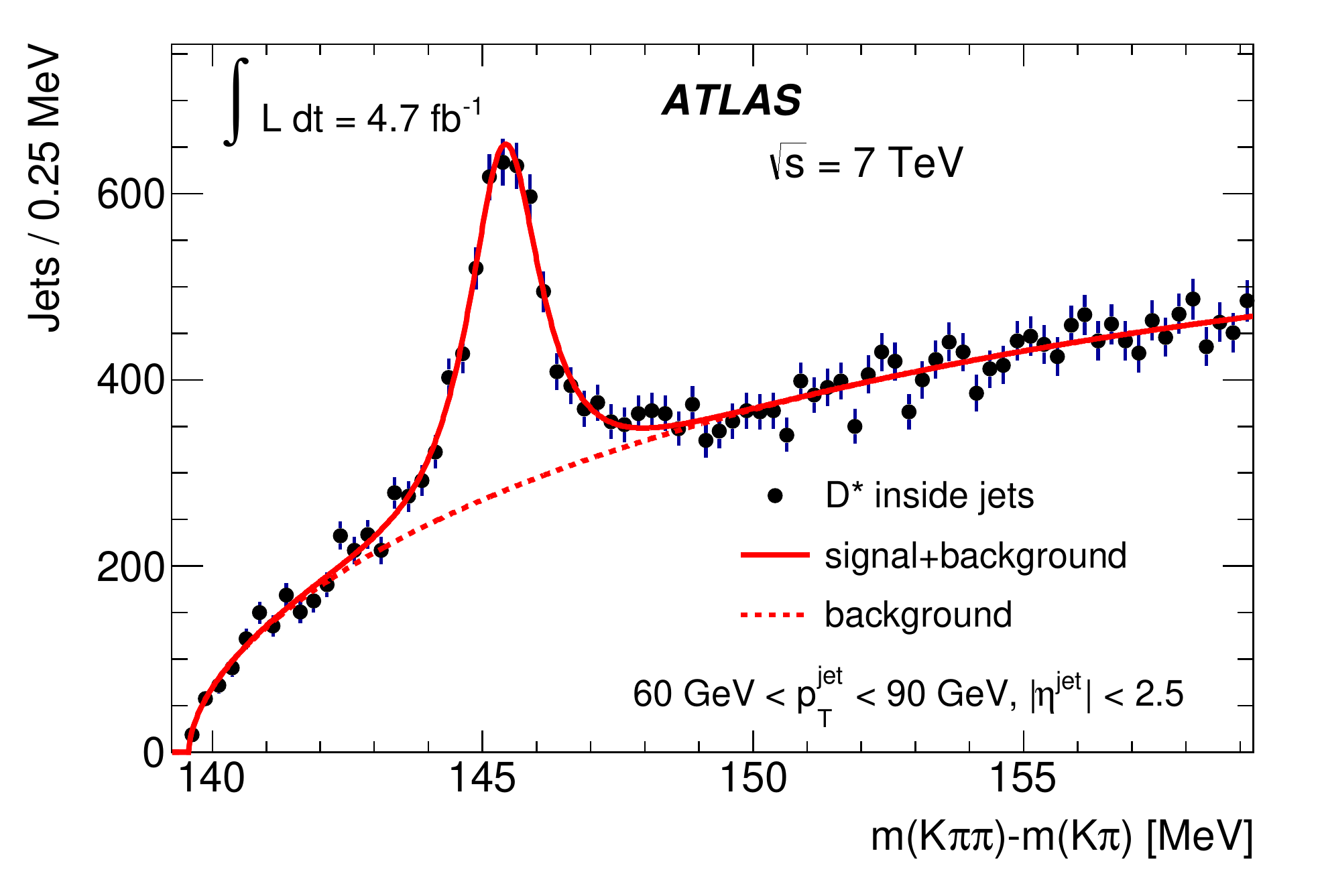}}
    \subfloat[]{\label{fig:dstar_d}\includegraphics[width=0.49\textwidth]{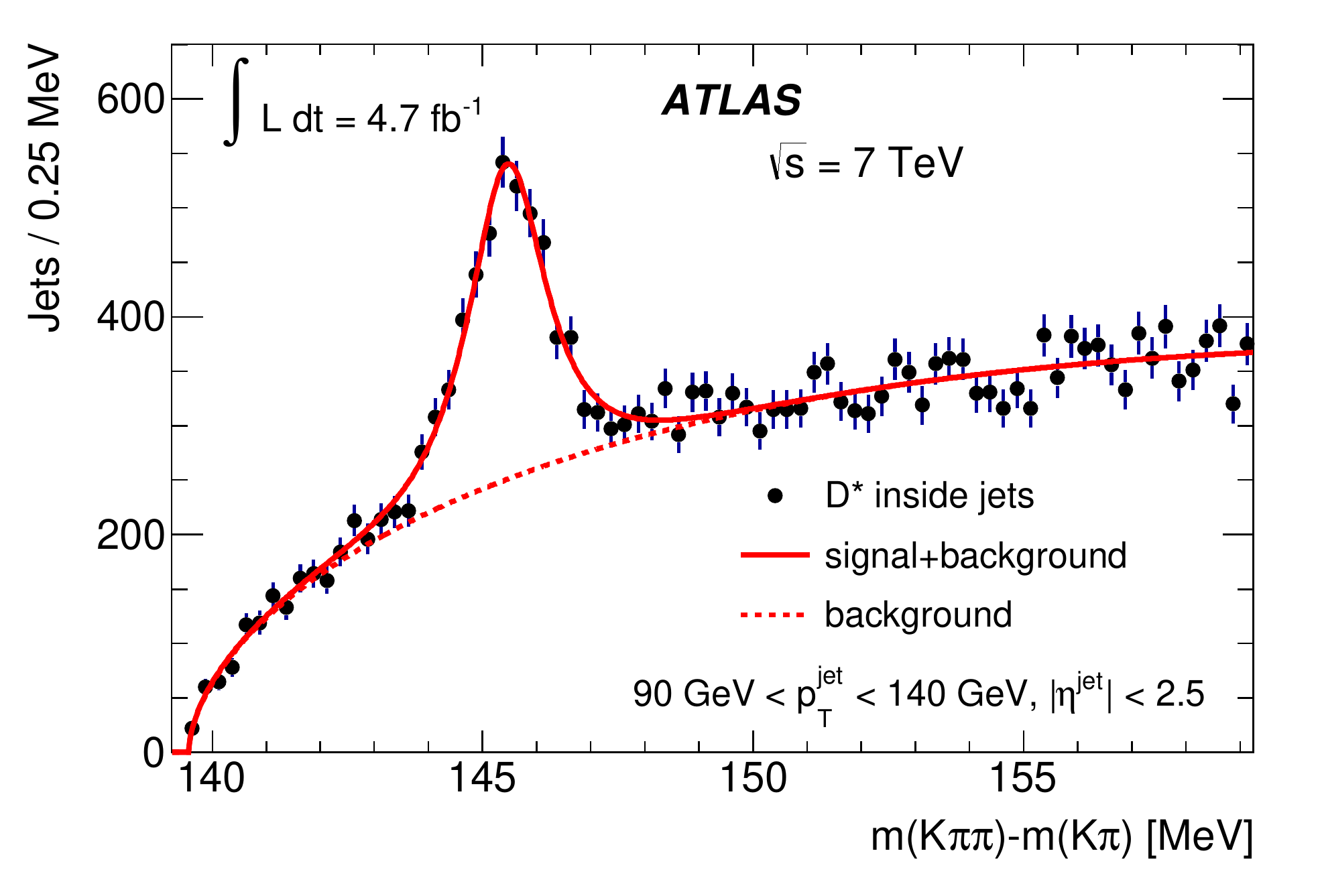}}
  \end{tabular}
  \caption{$\Delta m$ distribution of the $\dstar$ candidates associated with jets with $20\,\GeV < \pt^{\rm jet} < 30\,\GeV$ (a),
    $30\,\GeV < \pt^{\rm jet} < 60\,\GeV$ (b), $60\,\GeV < \pt^{\rm jet} < 90\,\GeV$ (c) and $90\,\GeV < \pt^{\rm jet} < 140\,\GeV$ (d).}
  \label{fig:dstar}
\end{figure}

The kinematics of the decay causes the $\dstar$
to release only a small fraction of energy to the prompt pion, 
usually called ``slow pion''; for this reason the $\dstar$
signal is commonly studied as a function of the mass difference $\Delta m$ between
the $\dstar$ and $D^0$ candidates. $\dstar$ mesons are expected to form a peak in the $\Delta m$ distribution around
145.4~\MeV, while the combinatorial background forms a rising distribution, starting at the pion mass.
Figure~\ref{fig:dstar} shows the distributions of the mass difference for the $\dstar$ pairs associated with a reconstructed jet
for four different jet \pT{} intervals: 20--30~\GeV, 30--60~\GeV, 60--90~\GeV, and 90--140~\GeV.

A fit of the $\Delta m$ distribution in each jet \pT{} interval is done in order to determine the yield of the $\dstar$ mesons.
The signal part $S$ of the $\Delta m$ distribution 
is fitted using a modified Gaussian ($\mbox{Gauss}^{\rm mod}$) function:
\begin{equation}
 S = \mbox{Gauss}^{\rm mod} \propto \mbox{exp}[-0.5\cdot x^{(1+\frac{1}{1+0.5x})}], \qquad x=|(\Delta m - \Delta m_0)/\sigma|,  
 \label{eq:S}
\end{equation}
which provides a better description of the signal tails than a simple Gaussian.
The mean and width of the $\Delta m$ peak, $\Delta m_0$ and $\sigma$, are the free parameters in the fit.
The combinatorial background $B$ is fitted with a power function multiplied by an exponential function:
\begin{equation}
 B \propto (\Delta m-m_{\pi})^{\alpha}e^{-\beta(\Delta m-m_{\pi})} ,  
 \label{eq:B}
\end{equation}
where $\alpha$ and $\beta$ are free fit parameters.

\subsubsection*{Background subtraction technique}
\label{bkg}
To allow for comparisons between data and simulation of observables related to the $\dstar$ mesons or jets,
for the mixture of $b$ and $c$ jets present in data, a background subtraction technique is used.
Signal and background regions are defined as the region within $3\sigma$
of the $\Delta m$ peak value and the region above 150~\MeV, respectively.
The choice of the $\Delta m$ intervals for the signal and background regions aims at including almost all the signal events
in the signal region and ensuring a negligible fraction of signal events in the background region.

For each observable, the data distribution extraction is carried out as follows: 
the distribution of events from the background region, normalised to the fitted background fraction in the signal region, 
is subtracted from the corresponding distribution in the signal region.
The procedure relies on the assumption that the distribution of the observable
of interest is the same for the combinatorial background under the peak 
and in the sidebands. This assumption has been verified to be valid in simulated events.
It is further supported by the observation in data that the distributions obtained from two different 
contiguous sideband regions ($\Delta m \in [150,160]\MeV$ and $\Delta m \in [160,168]\MeV$) 
are compatible with each other within their statistical uncertainty.

\subsubsection*{Measuring the flavour composition in the $\dstar$ sample}

The measurement of the flavour composition for the selected $\dstar$ sample is a
key ingredient for its use in $b$-tagging calibration studies. The
discriminating variable adopted in this paper to identify bottom and charm components is the $D^0$ pseudo-proper time defined as:
\begin{equation}
  t(D^0) = \mbox{sign}(\vec{L}_{xy}\cdot \vec{p}_{\rm{T}}(D^0)) \cdot m_{D^0}\cdot\frac{L_{xy}(D^0)}{\pt(D^0)},
  \label{lab:pseudopropertime}
\end{equation}
where $m_{D^0}$ is the $D^0$ meson mass, $\pT(D^0)$ is the transverse momentum of the reconstructed $D^0$ candidate
and $L_{xy}(D^0)$ is the distance,
in the transverse plane, between its decay vertex and the primary vertex in the event.

The first step of the flavour composition fit is the extraction of charm and bottom templates from simulated data:
\begin{itemize}
\item The resolution on the $D^0$ pseudo-proper time, $R(t)$, is described by the sum of a simple Gaussian and a modified Gaussian.
  As it has been verified with simulated events that the resolution does not depend on the $\dstar$ production mechanism,
  its parameters are fitted to the more abundant charm component.
\item The $D^0$ pseudo-proper time distribution for the charm component, $F_c(t)$,
  is modelled as a single exponential function with a time constant equal to the measured $D^0$ lifetime \cite{pdg2010},
  convolved with the pseudo-proper time resolution $R(t)$; no additional fits are needed to obtain the charm component model.
\item The model for the $D^0$ pseudo-proper time distribution for the bottom component, $F_b(t)$, cannot be easily inferred from physics arguments,
  since it depends on many variables, such as the bottom hadron and $D^0$ lifetimes, the momenta and the angle between their flight paths.
  Therefore, $F_b(t)$ is modelled as the convolution of two exponential functions, further convolved with the pseudo-proper time resolution $R(t)$;
  this empirical model provides a good agreement with the simulated distribution.
  The time constants of the two exponential functions are fitted using the simulated distribution for the bottom component.
\end{itemize}
Once the models for charm and bottom components are fixed, their sum is built as
\begin{align}
 F(t) = f_b \cdot F_b(t) + (1-f_b) \cdot F_c(t),
\end{align}
where $f_b$ is the fractional bottom abundance, and is used to fit the simulated or real background-subtracted data. 
A binned maximum likelihood fit is performed leaving the $f_b$ parameter free.

A validation of the fit procedure is performed by splitting the simulated inclusive sample
into 40 sub-samples and repeating the pseudo-proper time fit on each sub-sample. The pull distribution of the fitted purity
is found to be compatible with a Gaussian distribution centred on zero and with unit width, thus 
confirming that the fit results are unbiased and the uncertainties properly estimated.

\subsubsection*{Fit results}
\label{fitfb}
\begin{figure}[ht]
  \centering
    \subfloat[]{\label{fig:fitfb_a}\includegraphics[width=0.49\textwidth]{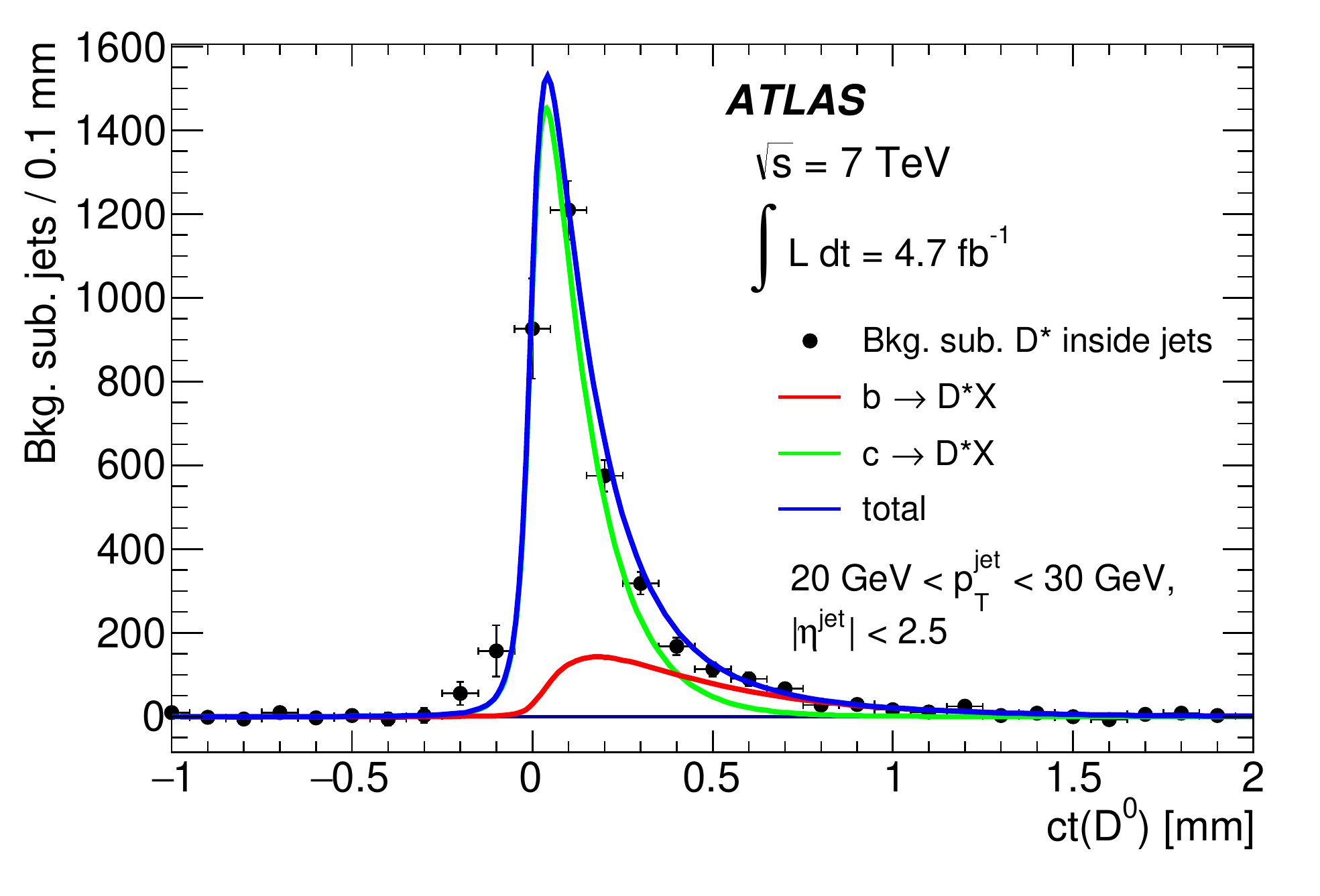}}
    \subfloat[]{\label{fig:fitfb_b}\includegraphics[width=0.49\textwidth]{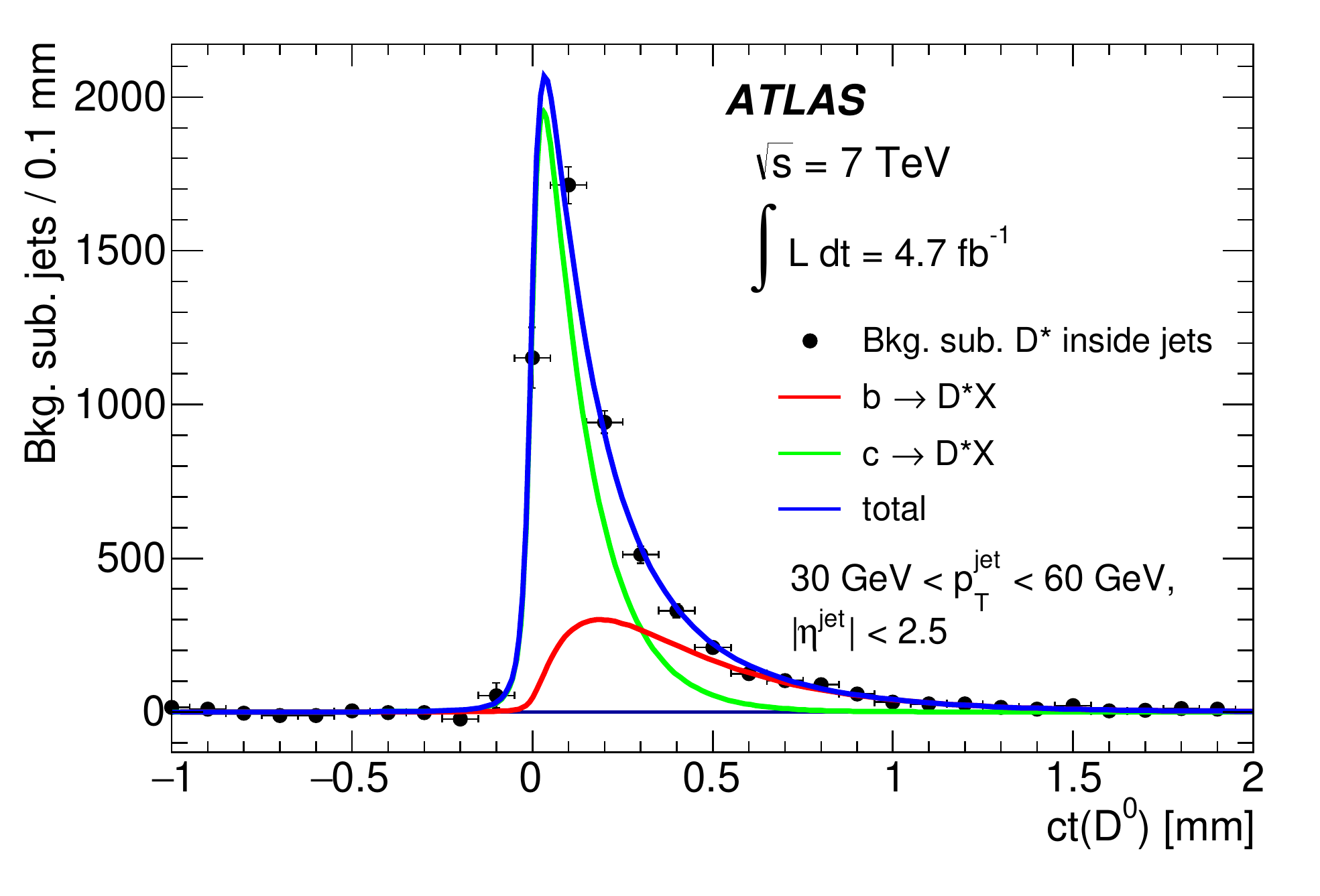}} \\ 
    \subfloat[]{\label{fig:fitfb_c}\includegraphics[width=0.49\textwidth]{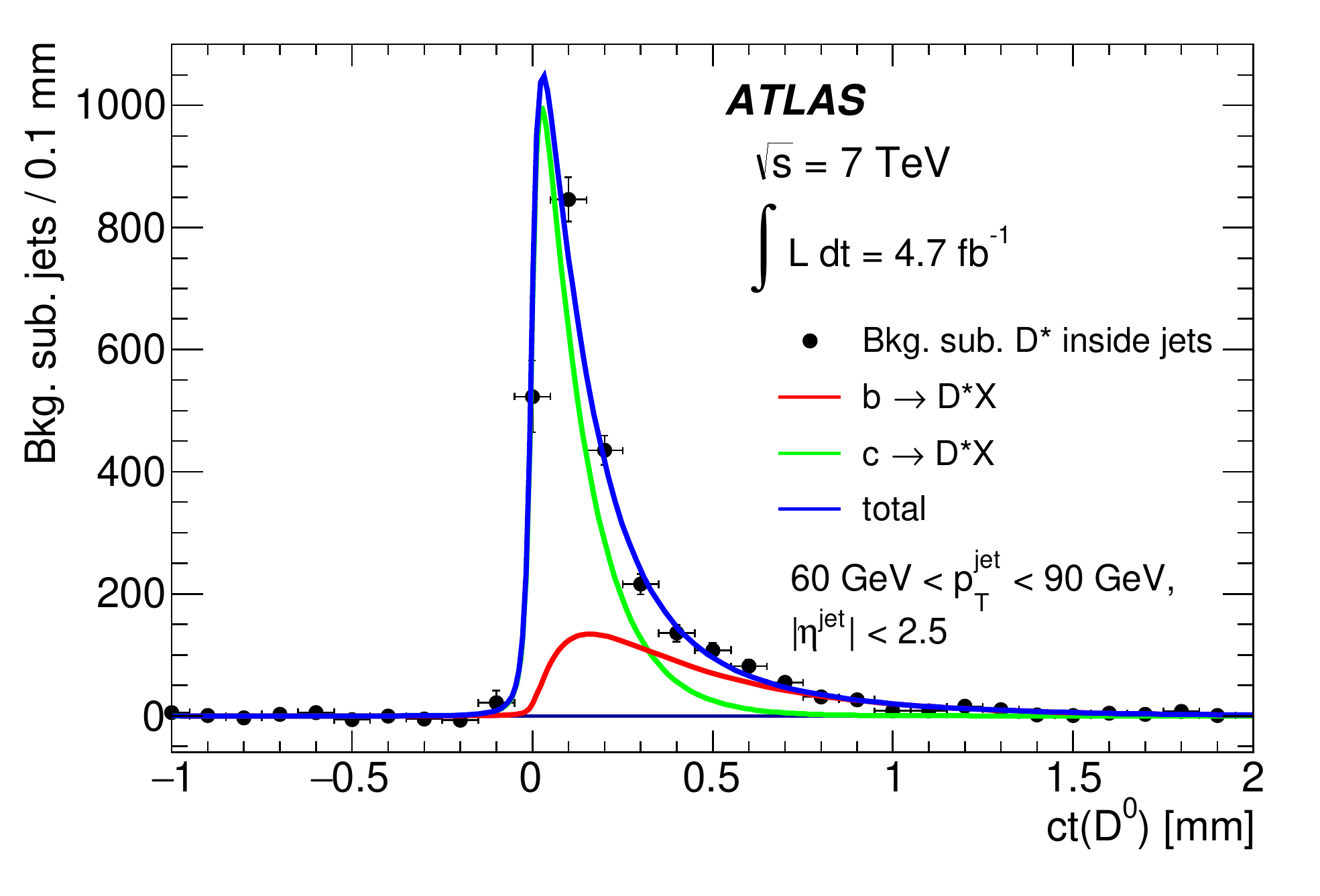}}
    \subfloat[]{\label{fig:fitfb_d}\includegraphics[width=0.49\textwidth]{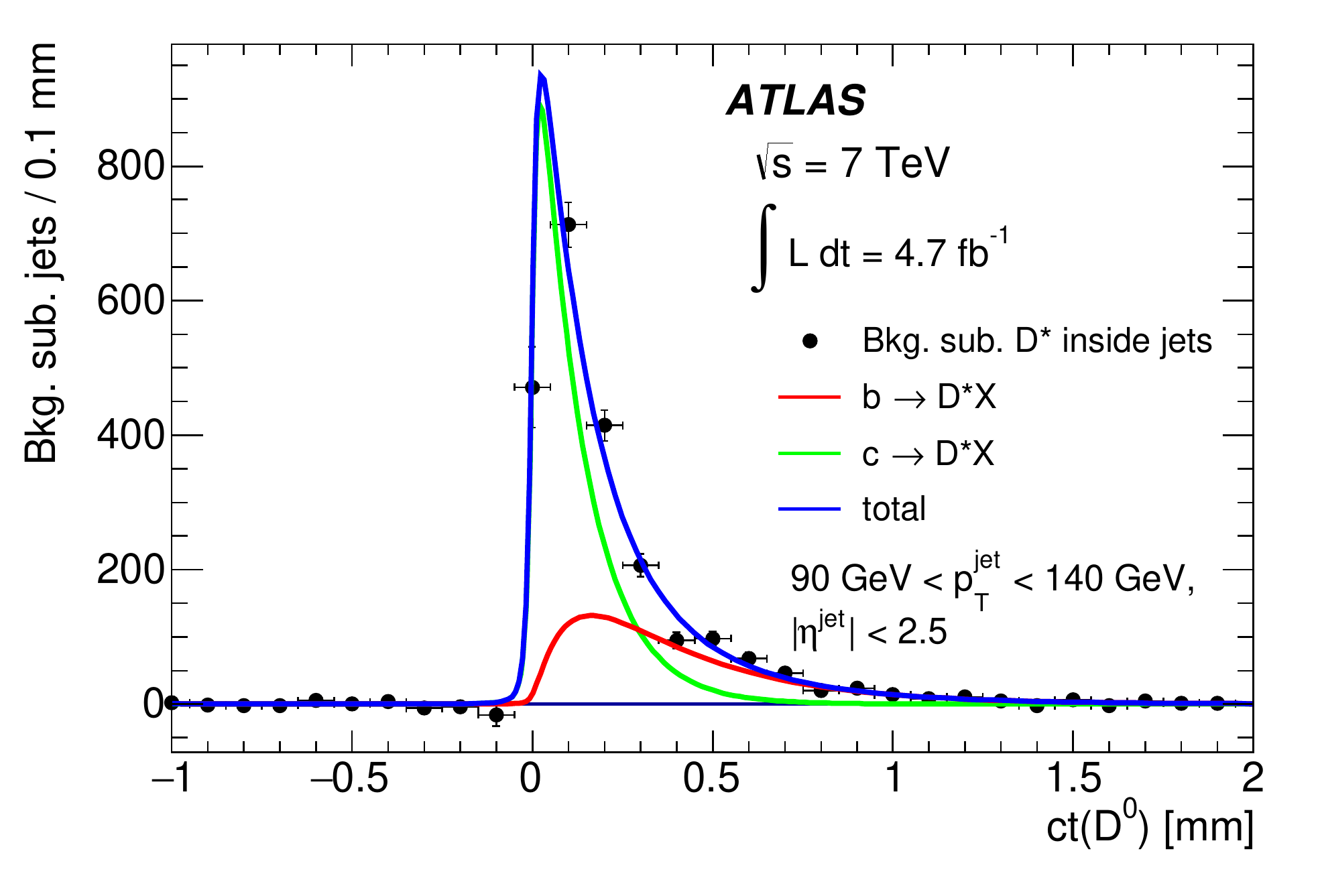}}  
  \caption{Fitted $D^0$ pseudo-proper time distributions to the background-subtracted $\dstar$ data samples of jets
    with $20\,\GeV < \pt^{\rm jet} < 30\,\GeV$ (a), $30\,\GeV < \pt^{\rm jet} < 60\,\GeV$ (b), $60\,\GeV < \pt^{\rm jet} < 90\,\GeV$ (c)
    and $90\,\GeV < \pt^{\rm jet} < 140\,\GeV$ (d).}
  \label{fig:fitfb}
\end{figure}

The fit is done using the $D^0$ pseudo-proper time defined in Eq.~\ref{lab:pseudopropertime},
in the range [-1, 2] mm.
The fit to background-subtracted real data, in the four bins of jet \pT{}, is shown in Fig.~\ref{fig:fitfb}. 
The bottom fractions as determined by these fits are summarised in Table~\ref{fitres_fb}.

\begin{table}[h]
  \centering
  \scriptsize
  \begin{tabular}{c|ccccc}
    \hline\hline
                      & \multicolumn{5}{c}{Jet \pt{} [\GeV]} \\
    & 20--30 & 30--60 & 60--90 & 90--140 & 20--140\\
    \hline
    Bottom fraction   & $0.212\pm0.010$ & $0.315\pm0.010$ & $0.303\pm0.015$ & $0.315\pm0.017$  & $0.286\pm0.006$ \\
    \hline \hline
  \end{tabular}
  \caption{Bottom fractions determined by fits to the data in four jet \pT{} bins,
    as well as in the full jet \pT{} range considered in the analysis.
    Only statistical uncertainties are shown.}
  \label{fitres_fb}
\end{table}

\begin{figure}[h!]
  \centering
  \begin{tabular}{c}
    \includegraphics[width=0.6\columnwidth]{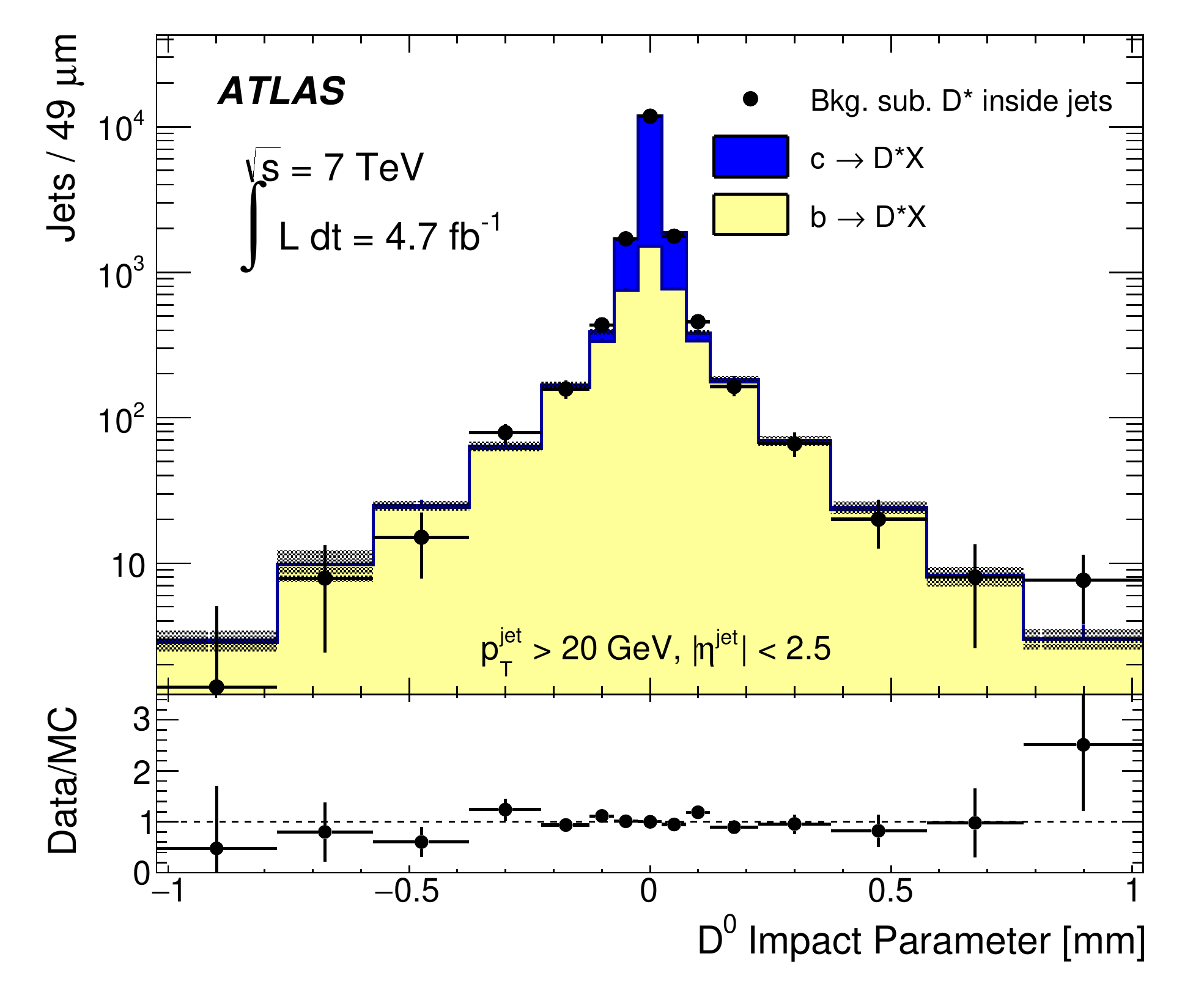}  
  \end{tabular}
  \caption{Comparison between the $D^0$ impact parameter in the background-subtracted $\dstar$ data samples with the corresponding simulated samples.
    The bottom fraction in the simulation is fixed to the value obtained by the pseudo-proper time fit in the full jet \pT{} range in data,
    given in Table~\protect\ref{fitres_fb}.
    The ratio between the two distributions is shown on the bottom of the plot.}
 \label{fig:dstar-distrib}
\end{figure}

In order to cross-check the fit results, the distribution of the $D^0$ impact parameter, a 
variable sensitive to the bottom component, is analysed.
Figure~\ref{fig:dstar-distrib} shows the comparison between the
background-subtracted data and the Monte Carlo simulation
for the impact parameter of the $D^0$ meson emerging from the $\dstar$ decay.
The distribution in simulated events is obtained by summing the bottom and charm
components according to the overall $f_b$ value given in Table~\ref{fitres_fb}.
Data and simulation distributions are found to be in reasonable agreement.

\begin{figure}[ht]
  \centering
    \subfloat[]{\label{fig:jetprob-tag_a}\includegraphics[width=0.49\textwidth]{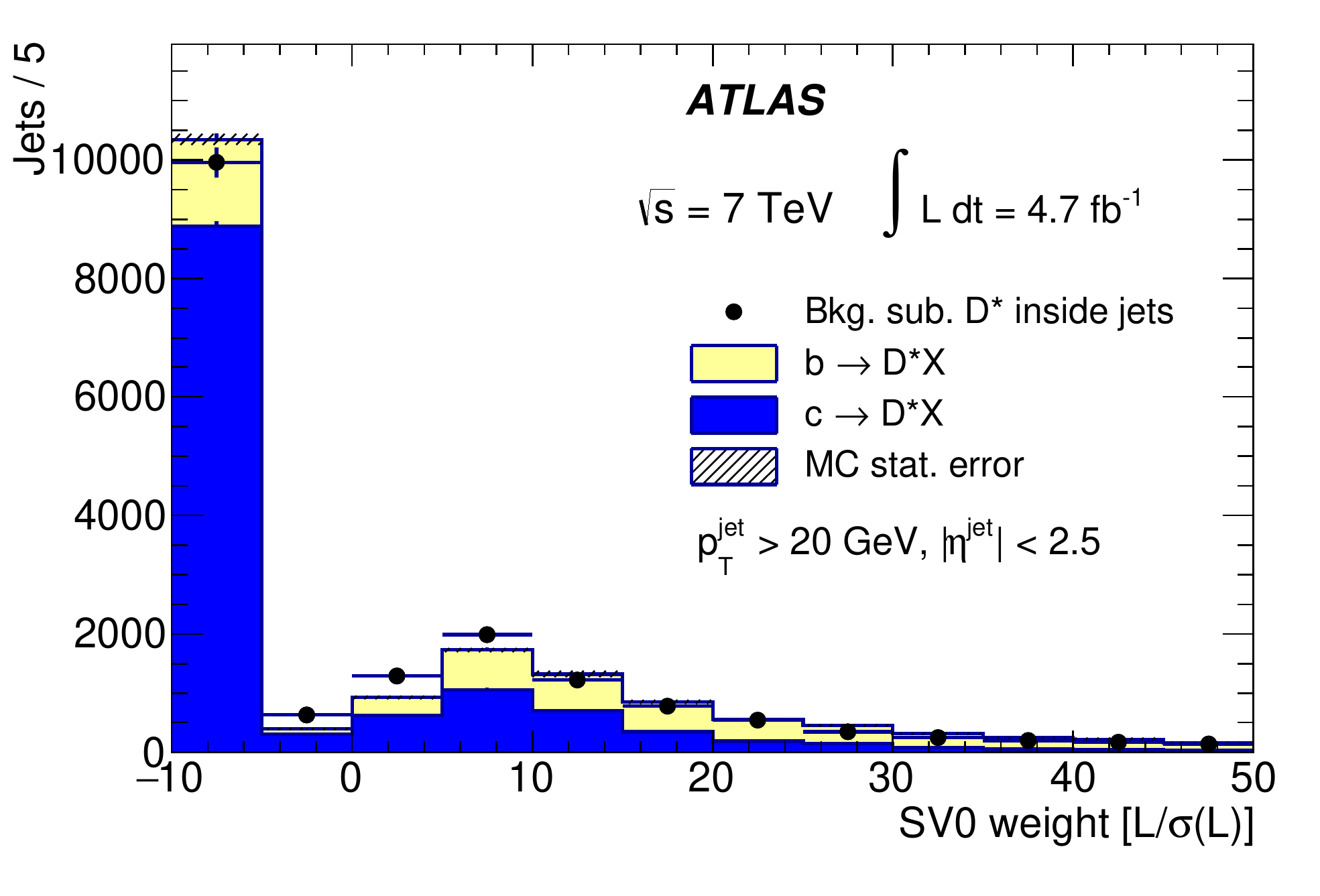}}
    \subfloat[]{\label{fig:jetprob-tag_b}\includegraphics[width=0.49\textwidth]{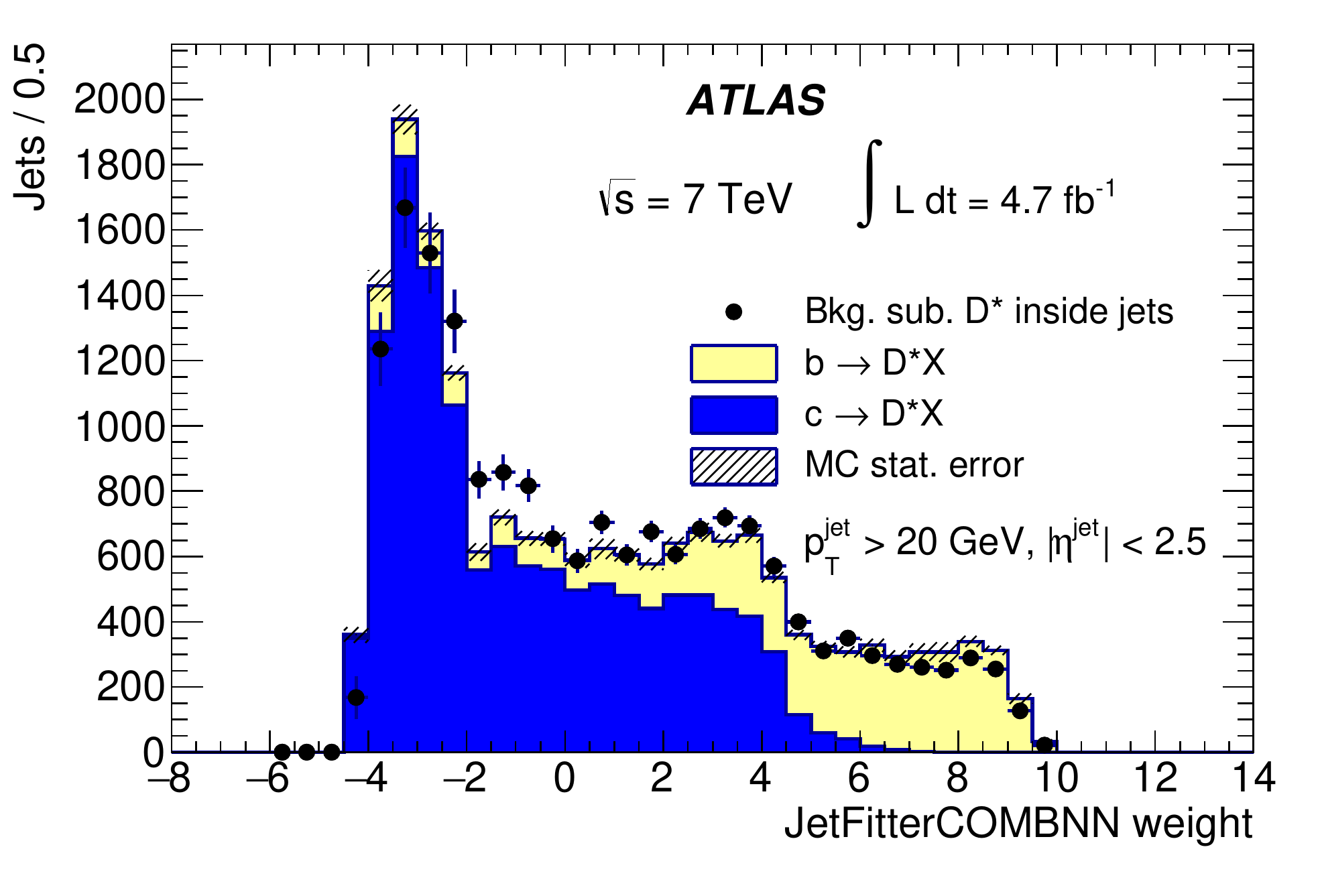}} \\ 
    \subfloat[]{\label{fig:jetprob-tag_c}\includegraphics[width=0.49\textwidth]{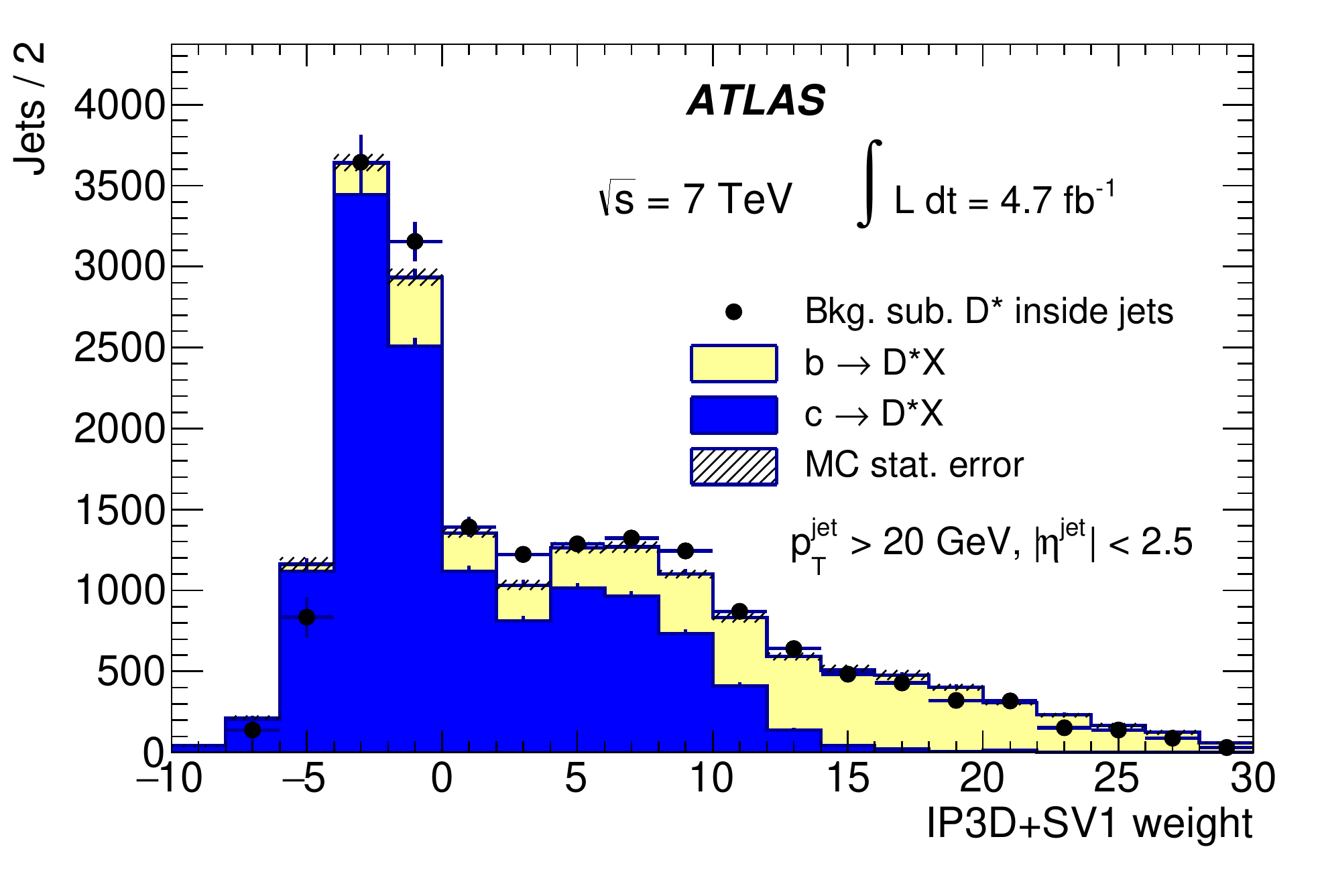}}
    \subfloat[]{\label{fig:jetprob-tag_d}\includegraphics[width=0.49\textwidth]{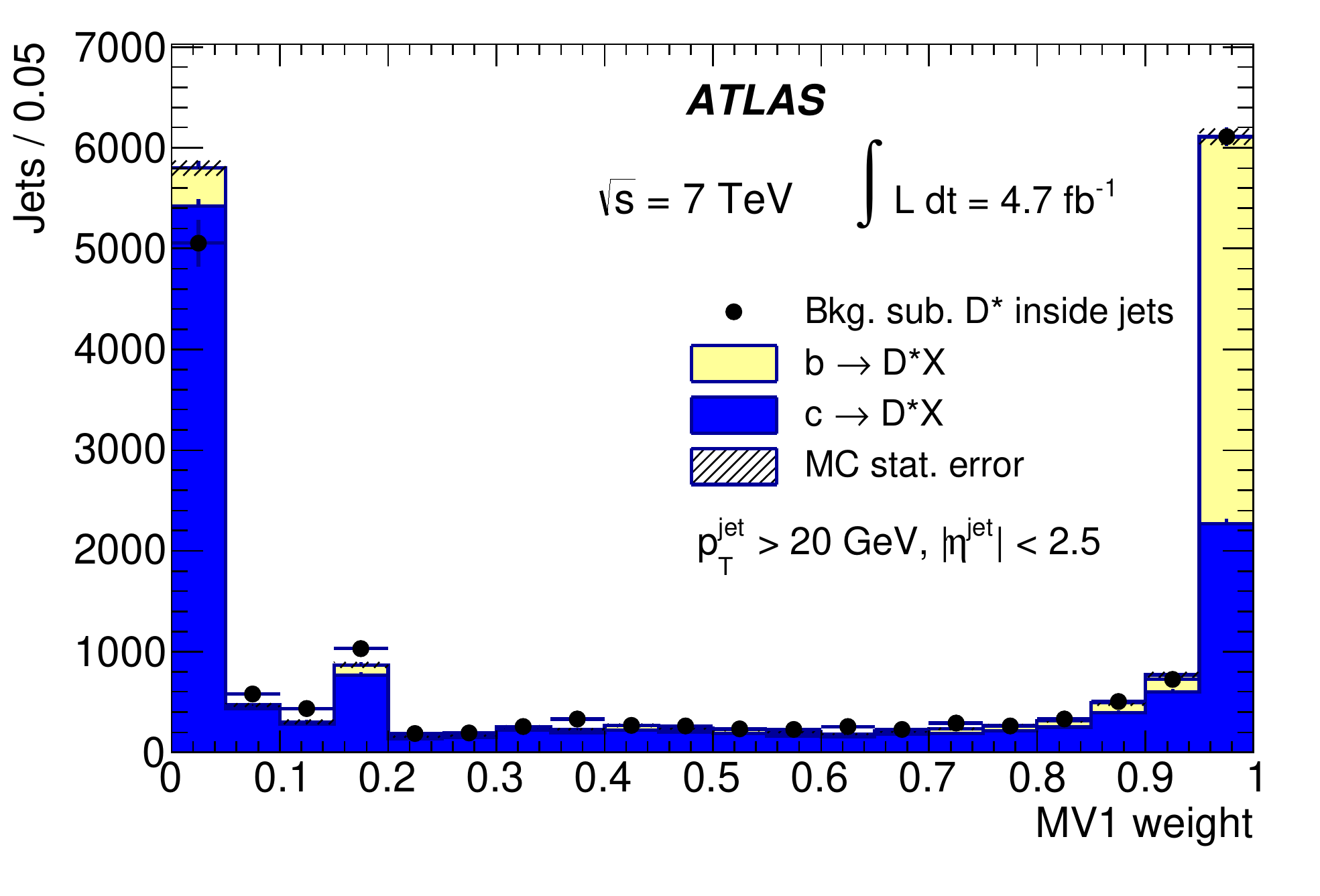}}
  \caption{Comparison between the weight distributions for the {SV0} (a), {IP3D+JetFitter} (b), {IP3D+SV1} (c) 
    and {MV1} (d) $b$-tagging algorithms in the background-subtracted $\dstar$ data sample and the corresponding simulated samples. The
    bottom fraction in the simulation is fixed to the value obtained by the pseudo-proper time fit in the full jet \pT{} range in data,
   given in Table~\protect\ref{fitres_fb}.}
  \label{fig:jetprob-tag}
\end{figure}

Using the background subtraction technique described above, the shape of any variable in data can be compared to that in simulation.
Figure~\ref{fig:jetprob-tag} shows the distributions of the {SV0} output weight,
namely the decay length significance,\footnote{The significance is set to $-$10 when no secondary vertex is found.}
the {IP3D+JetFitter} output weight, the {IP3D+SV1} output weight and the {MV1} output weight 
in the background-subtracted $\dstar$ sample. 
The discrepancies observed between the tag weight shapes in data and simulation will be reflected in the
data-to-simulation scale factors derived with the $D^{\star}$ method.

\subsubsection*{Measuring the $c$-jet tagging efficiency using $\dstar$ candidates}

The selected sample can be used to measure the $c$-jet tagging efficiency for jets associated with $\dstar$ candidates,
by performing a combined fit to the $\Delta m$ distributions for $\dstar$ mesons in jets before and after applying the $b$-tagging requirement.

The fit parameters describing the signal and the background shapes are required to be equal
for the two distributions and the combined fit only introduces the $\dstar$ tagging efficiency
$\epsilon_{\dstar}$ as an extra parameter accounting for the reduction in the $\dstar$ peak in the tagged jets.
The procedure was tested in simulation and it has been verified that the measured efficiency on jets associated with a $\dstar$ meson is unbiased.

Using this method it is possible to obtain the efficiency to tag jets associated with a $\dstar$ candidate.
This inclusive efficiency $\epsilon_{\dstar}$ is then decomposed into the efficiency for $b$ and $c$ jets using:
\begin{equation}
  \epsilon_{\dstar} = f_{b}\epsilon_b + (1-f_{b})\epsilon_{c},
\end{equation}
where $f_b$ is the fraction of $\dstar$ coming from bottom, before the
$b$-tagging selection, determined by the fit to the pseudo-proper lifetime.
The efficiency to tag a $b$ jet, $\epsilon_b$, is taken from simulation and
corrected by the data-to-simulation scale factors obtained by the \ptrel{} and system8 methods 
(the combination of individual calibration results is discussed in detail in Section~\ref{sec:combination}).
It is straightforward to solve this equation for $\epsilon_{c}$.

\subsubsection*{Extrapolation to inclusive charm}
\label{extrap}

The calibration procedure described above measures the $b$-tagging efficiency
for $c$ jets with an exclusively reconstructed $\dstar$ decay, $\dstar\rightarrow D^{0}(\to K^{-}\pi^{+})\pi^{+}$,
and hence the corresponding scale factor
$\sfcdstar = \epsilon^{\rm data}_{c(\dstar)}/\epsilon^{\rm MC}_{c(\dstar)}$.
To be applicable to an inclusive sample of $c$ jets, an extrapolation procedure
has to be applied to obtain the corresponding scale factor $\sfc$.
The extrapolation procedure follows closely the procedure described in 
Section~\ref{sec:inclSF} for the $c$-jet tagging efficiency calibration analysis
based on $W+c$ events (Section~\ref{sec:ceff_Wc}).                                                                                                                                             
The typical values of $\alpha$, defined following Eq.~\ref{eq:inclEff} as $\alpha = \epsilon_c / \epsilon_{c(\dstar)}$, 
evaluated for the MV1 algorithm, range between 0.5 and 0.7, depending on the working point.
Despite the fact that the weakly decaying $D^0$ meson has a significantly shorter lifetime than e.g. the $D^+$ meson,
$c$ jets containing an exclusively reconstructed $\dstar$ meson are tagged more often than generic $c$ jets,
explained by the requirement of having at least two reconstructed tracks from the weak decay of a $D^0$ meson.

To obtain the scale factor extrapolation factor $\delta$,
following Eq.~\ref{eq:inclSF} defined as $\delta = \sfc / \sfcdstar$,
the $c$-hadron fragmentation fractions of weakly decaying 
$c$ hadrons and the charged particle multiplicities in the decays of weakly decaying $c$ hadrons
have been corrected
as described in Section~\ref{sec:inclSF}.
The main discrepancies between the Monte Carlo simulation and the experimental knowledge are
the $\Lambda_c^+$ fragmentation fraction and the $D^0\to 0-$prong decay
branching ratio, which are both lower in the simulation.
Therefore the effect of the extrapolation procedure is to decrease the estimated inclusive $c$-jet tagging efficiency,
the extrapolation factor $\delta$
ranges between 0.82 and 0.92, depending on the
tagging working point, and is relatively independent of the jet \pt.

\subsection{Systematic uncertainties}

\label{sec:dstar-syst}

The dominant systematic uncertainties affecting the method presented in this paper are those related to the fit of the yield of $\dstar$ mesons,
to the extraction of the fraction of $\dstar$ mesons originating from bottom hadrons and to the extrapolation of the $c$-jet tagging efficiency scale factor
measured on jets associated with a $\dstar$ meson to that of an inclusive $c$-jet sample.

The systematic and statistical uncertainties on the $c$-jet tagging efficiency scale factors of the MV1 tagging algorithm at 70\% efficiency are
shown in Table~\ref{tab:systs_mv170}. Each source of systematic uncertainty listed in the table is explained below.

\begin{table}[htbp] 
  \begin{center}
    \scriptsize
    \begin{tabular}{l|cccc} 
      \hline 
      \hline 
                                              & \multicolumn{4}{c}{Jet $\pT{} \mathrm{[\GeV]}$} \\ 
      Source                                  & 20--30 & 30--60 & 60--90 & 90--140 \\
      \hline
      Bottom fraction fit                     & 2.8 & 2.3 & 2.1 & 2.4 \\
      $b$-tagging scale factor                & 4.3 & 5.2 & 4.9 & 7.6 \\
      Background parametrisation              & 0.8 & 0.9 & 0.7 & 1.8 \\
      $D^{*+}$ mass peak width                 &  16 & 4.3 & 4.7 & 10 \\
      Jet energy scale and resolution         & 2.6 & 1.7 & - & 0.5 \\
      Pile-up $\langle\mu\rangle$ reweighting & - & - & - & 0.5 \\
      Extrapolation $c\to D^0$                & 0.6 & 0.5 & 0.2 & - \\
      Extrapolation $c\to D^+$                & 1.3 & 1.1 & 0.5 & 0.6 \\
      Extrapolation $c\to D^+_s$              & - & 0.1 & - & - \\
      Extrapolation $c\to\Lambda_c$           & 2.2 & 1.8 & 0.9 & 1.4 \\
      Extrapolation $D_0\to 0$-prongs         & 2.8 & 3.2 & 2.4 & 2.9 \\
      Extrapolation $D^0\to 2$-prongs         & 0.5 & 0.4 & - & 0.4 \\
      Extrapolation $D^0\to 4$-prongs         & 0.2 & 0.2 & 0.2 & 0.2 \\
      Extrapolation $D^+\to$prongs            & 4.2 & 4.1 & 2.9 & 3.3 \\
      Extrapolation $D^+_s\to$prongs          & 0.7 & 0.9 & 0.9 & 1.1 \\
      Extrapolation $\Lambda_c\to$prongs      & 0.1 & 0.3 & 0.8 & 0.5 \\
      \hline
      Total systematic uncertainty            &  18 & 9.3 & 8.2 & 14 \\
      Statistical uncertainty                 &  13 & 9.2 &  11 & 12 \\
      \hline
      Total uncertainty                       &  22 &  13 &  14 & 19 \\
      \hline \hline
    \end{tabular} 
    \caption{\label{tab:systs_mv170}  Relative systematic and statistical
      uncertainties, in \%, on the data-to-simulation $c$-jet tagging efficiency
      scale factors for the MV1 tagging algorithm at 70\% efficiency.
      Negligibly small uncertainties are indicated by dashes.
    }
  \end{center} 
\end{table}

\subsubsection*{Bottom fraction fit}

To study the effects of imperfect modelling of the pseudo-proper time resolution in simulation and the
bottom lifetime uncertainty, the following procedure is adopted:
\begin{itemize}
\item Resolution systematics: the fit results have a weak dependence on the assumed resolution functions,
  and a conservative systematic uncertainty is assigned by fixing the Gaussian and modified Gaussian widths
  to 0.5 and 1.5 times the resolution fitted on the simulated sample. This mainly affects small pseudo-proper time
  values, while the fit results are mainly influenced by the bottom tails at high positive values.
\item Lifetime uncertainty: the lifetimes of the two exponentials used in modelling the bottom component are each varied
  by the fractional error on the inclusive $b$-hadron lifetime world average \cite{pdg2010}.
\end{itemize}
In both cases the maximum positive and negative variations in the bottom fraction central value 
are taken as an estimate of the corresponding systematic uncertainty.
The total uncertainty on the bottom fraction is calculated by combining the fit statistical error together 
with the resolution and lifetime systematics.

\subsubsection*{$b$-jet tagging efficiency scale factor}

The tagging efficiency for $b$ jets is evaluated by multiplying the value found in simulation
by the scale factor measured with the \ptrel{} and system8 methods, described in Section~\ref{sec:beff_mubased}.
The variation of this scale factor within its error is propagated to the final results as a systematic uncertainty.

\subsubsection*{$\dstar$ mass fit}

The systematic uncertainty in the mass fit is evaluated by removing the constraint that 
the width of the $\dstar$ mass peak and the parametrisation of the background shape are the same in the pre-tagged and tagged sample.
The fit is separately repeated with and without these assumptions and the efficiency variations are taken as two separate systematic uncertainties.
The obtained uncertainties, by definition single sided,
have been symmetrised assuming that a similar variation could have been observed also in the opposite direction.

\subsubsection*{Jet energy scale and resolution}

The systematic uncertainty originating from the jet energy scale is obtained by scaling the \pT{} of 
each jet in the simulation up and down by one standard deviation, according to the uncertainty of the jet energy scale~\cite{PERF-2011-03}.
This systematic uncertainty impacts both the true $c$-jet tagging efficiency  and the pseudo-proper time templates.
The effect of uncertainties on the jet energy resolution has been found to be negligible.

\subsubsection*{Pile-up $\langle\mu\rangle$ reweighting}

In principle, the $c$-jet tagging efficiency as well as its estimation using
reconstructed $\dstar$ candidates may be affected by event pile-up.
As for the $b$-jet efficiency measurements described in Sections~\ref{sec:beff_mubased} and~\ref{sec:beff_ttbarbased},
the distribution of the average number of interactions per bunch crossing,
$\langle \mu \rangle$, in simulated events is reweighted to agree with that in the data.
The evaluation of the corresponding systematic uncertainty follows the procedure described in Section~\ref{sec:syst}.

\subsubsection*{Extrapolation}

The uncertainties on the extrapolation factor are obtained by varying individually each charm fragmentation fraction
and the topological decay branching fractions of $c$ hadrons as described in Section~\ref{sec:inclSF}. 
Each variation is accounted for as a separate systematic uncertainty in Table~\ref{tab:systs_mv170}.

\subsection{Results}
\label{sec:dstar_results}

\begin{figure}[htb]
  \subfloat[]{\label{Dstar_MV170_a}\includegraphics[width=0.49\textwidth]{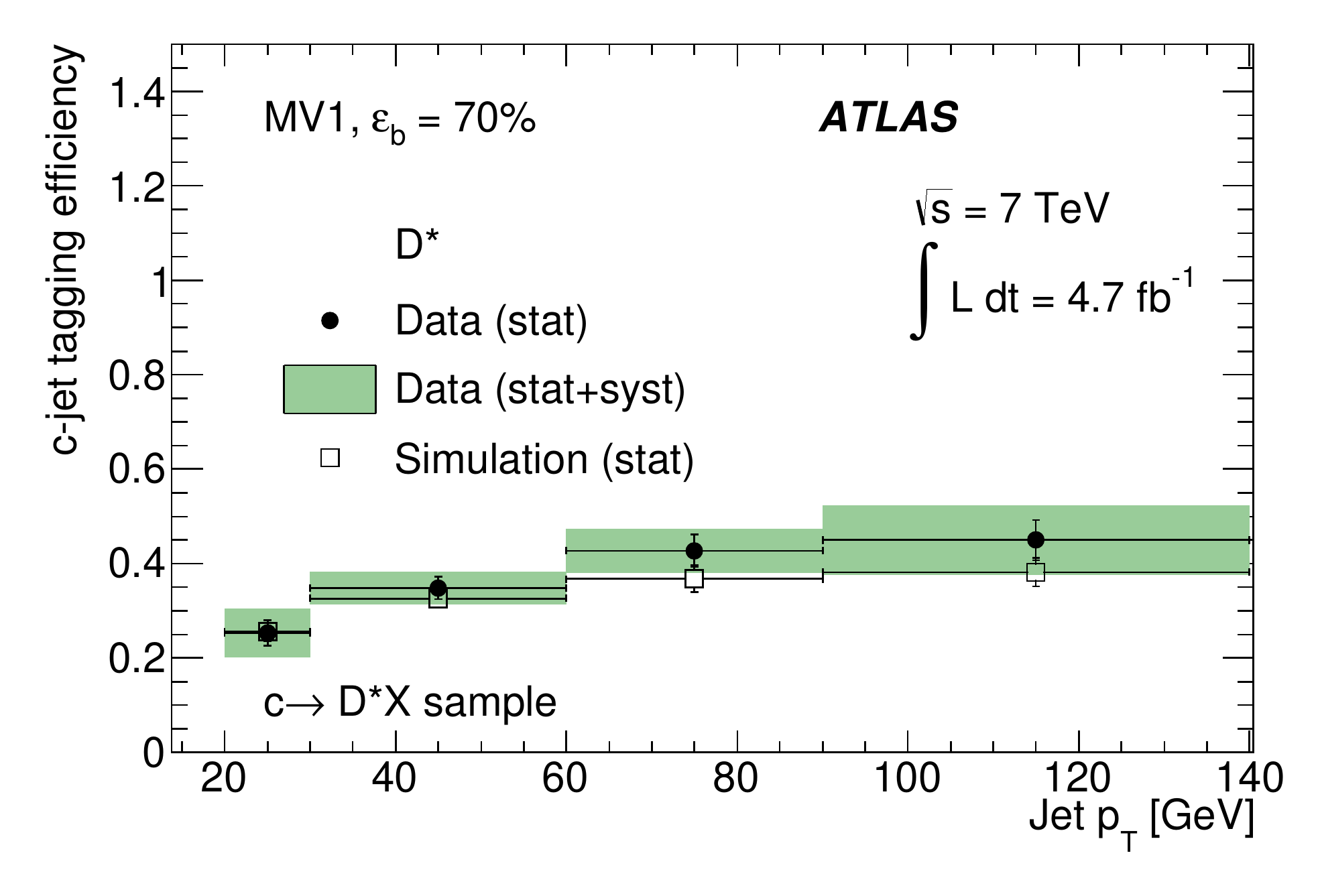}}
  \subfloat[]{\label{Dstar_MV170_b}\includegraphics[width=0.49\textwidth]{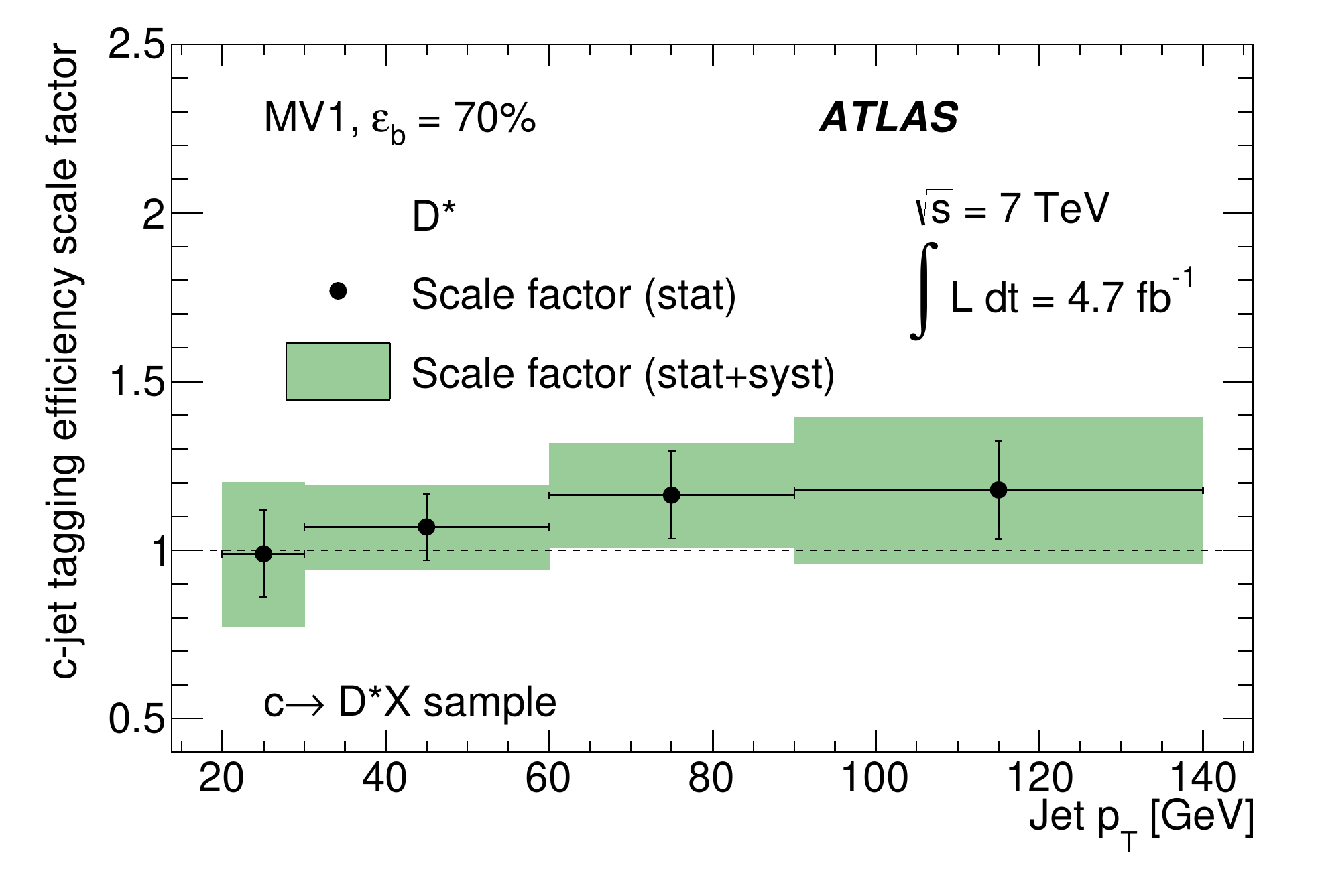}}
  \caption{The $c$-jet tagging efficiency in data and simulation (a) and the data-to-simulation scale factor (b)
    for jets containing a $D^{\star +}$ meson, for the 70\% operating point of the MV1 $b$-tagging algorithm.
    \label{Dstar_MV170}}
\end{figure}

\begin{figure}[htb]
  \centering
  \includegraphics[width=0.55\textwidth]{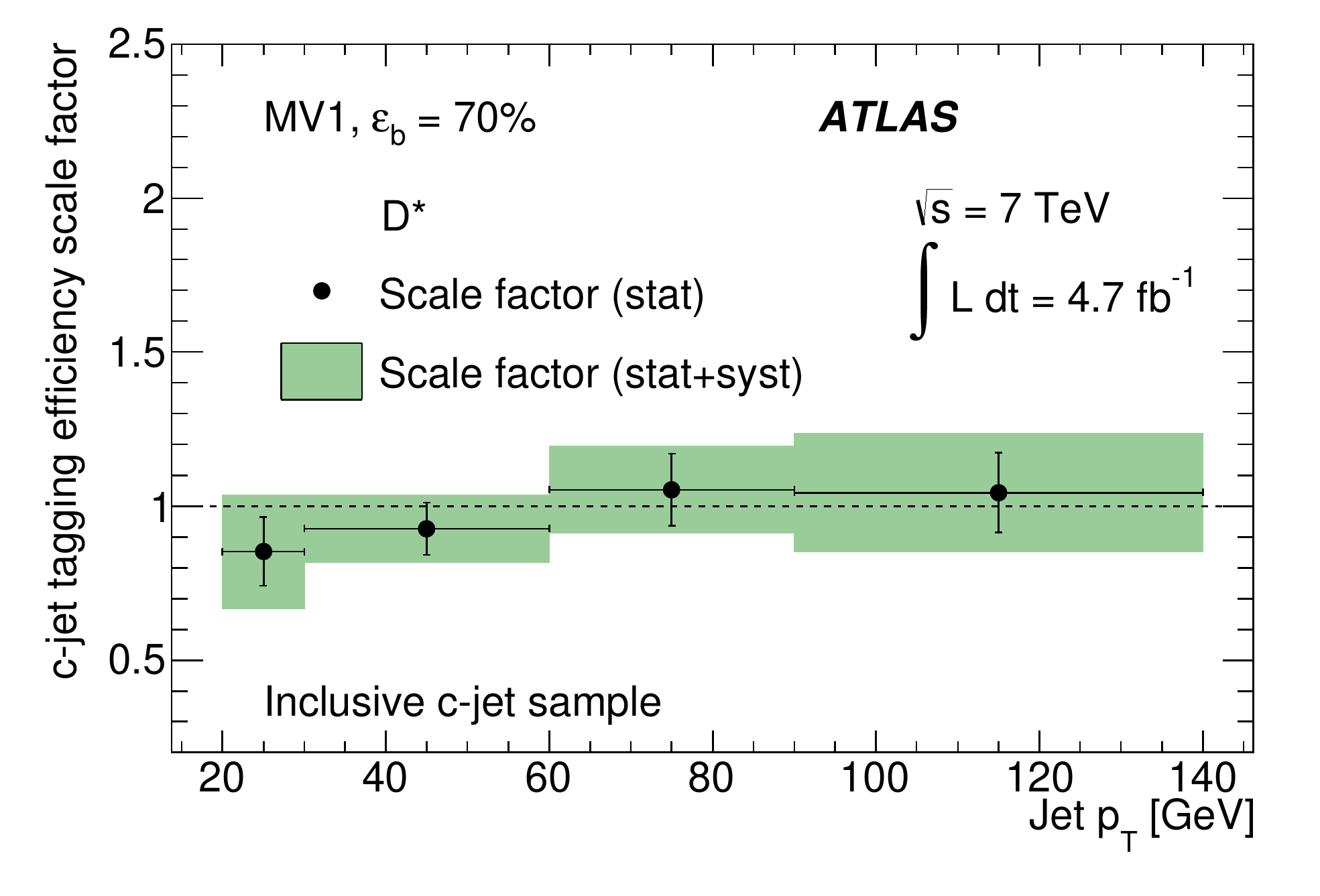}
  \caption{Data-to-simulation scale factor for inclusive $c$ jets, after the extrapolation procedure described in Section~\protect\ref{extrap},
    for the 70\% operating point of the MV1 $b$-tagging algorithm.
    \label{Incl_MV170}}
\end{figure}

The measured $c$-jet tagging efficiencies in data, the $c$-jet tagging efficiencies in simulation 
and the resulting data-to-simulation scale factors for the MV1 tagging algorithm at 70\% efficiency are shown in Fig.~\ref{Dstar_MV170}.
The corresponding data-to-simulation scale factors, after the extrapolation correction described in Section~\ref{extrap}, 
are shown in Fig.~\ref{Incl_MV170}.

The scale factors in all \pT{} bins are compatible with unity within uncertainties. No significant \pT{}
dependence of the scale factor is observed.
A comparison with the results of the $W+c$ analysis can be made by
weighting the results in the bins with $\pt < 90 \GeV$ by the expected jet \pt{}
spectrum in the $W+c$ analysis.
With conservative assumptions on the correlations of systematic uncertainties between the two analyses and bin-to-bin correlations within the $D^{\star}$ analysis,
the weighted scale factor of the $D^{\star}$ analysis is larger than the one of the $W+c$ analysis by about one standard deviation.

\section{Mistag rate calibration}
\label{sec:mistag}

The mistag rate is defined as the fraction of light-flavour jets that are tagged by a $b$-tagging algorithm.
The mistag rate is measured in data, using an inclusive sample of jets,
with the {\it negative tag} method which is described in the following.

\subsection{Data and simulation samples \label{sec:negtag_samples}}

The event sample for the mistag rate
measurement was collected using a logical OR of inclusive jet triggers, 
analogous to what was done in the $D^{\star}$ analysis (see Section~\ref{sec:dstar_samples}).

The analysis also makes use of simulated inclusive jet samples, similar to those used in the muon-based $b$-jet tagging 
efficiency measurements and the $D^{\star}$ $c$-jet tagging efficiency measurement, but without any muon or \dstar{} filter.
About 2.8 million events have been simulated per $\hat{p}_{\perp}$ bin.

\subsection{The negative tag method}
\label{sec:negtag}

Light-flavour jets are tagged as $b$ jets mainly because of the finite resolution of the Inner
Detector and the presence of tracks stemming from displaced vertices from long-lived particles or material
interactions.
Prompt tracks that are seemingly displaced, due to the finite resolution of the tracker, will as
often appear to originate from a point behind as in front of the primary vertex with respect to the jet axis.
In other words, the lifetime-signed impact parameter distribution of these tracks as well as the signed
decay length of vertices reconstructed with these tracks are expected to be symmetric about zero.

\begin{figure}[ht!]
  \begin{center}
    \subfloat[]{\label{fig:tagweight_a}\includegraphics[width=0.45\textwidth]{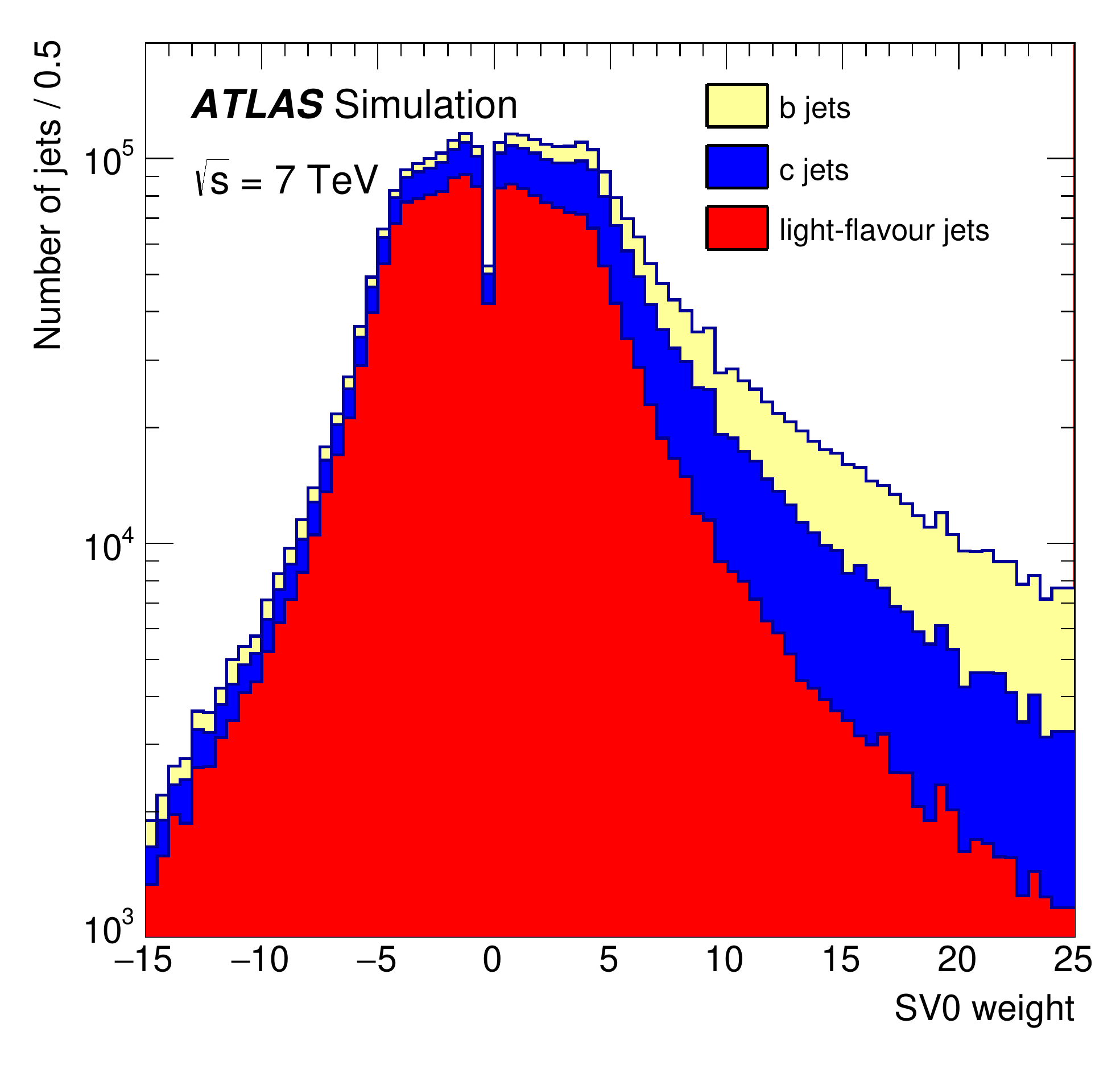}}
    \subfloat[]{\label{fig:tagweight_b}\includegraphics[width=0.45\textwidth]{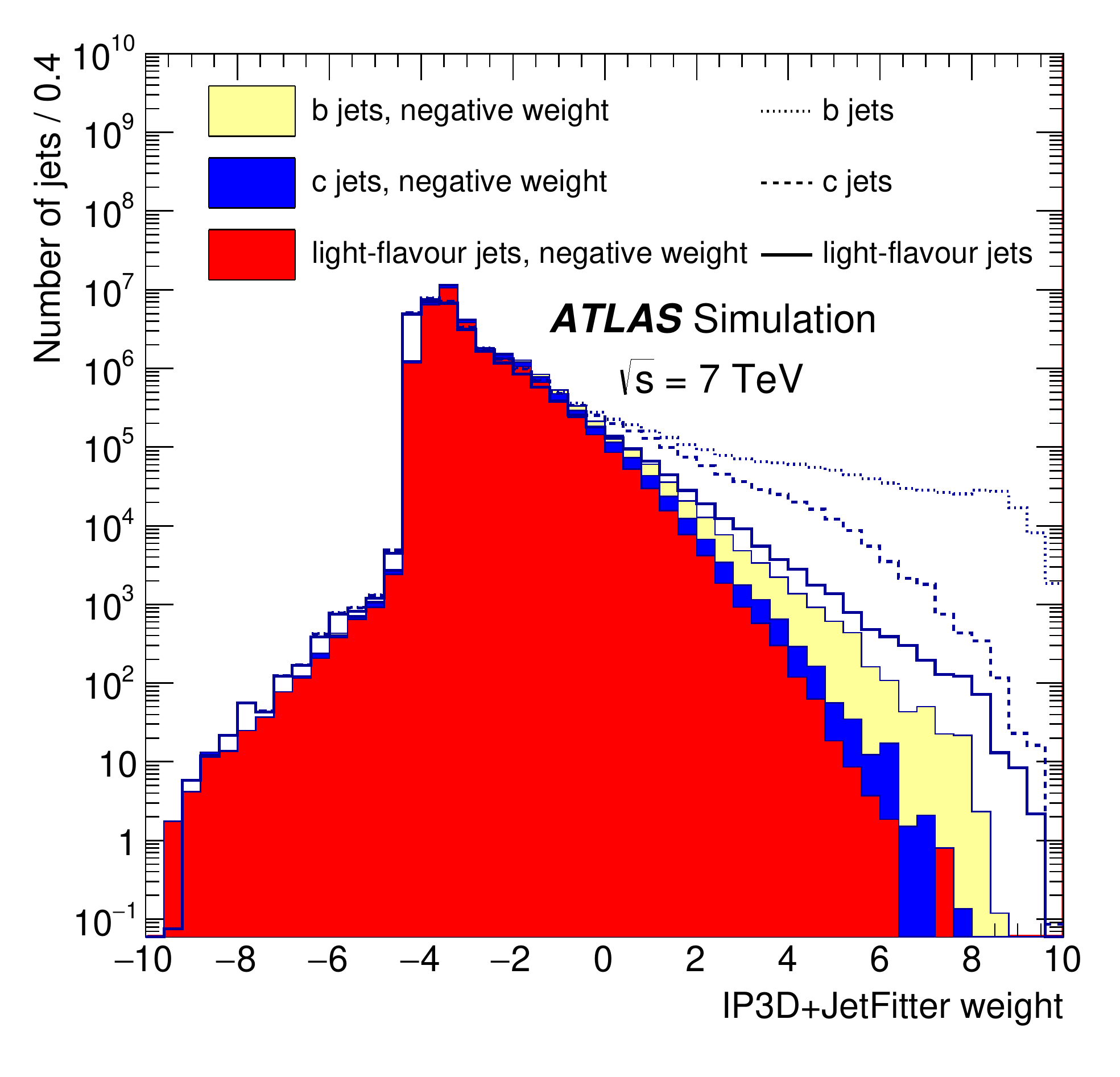}}
    \caption{The SV0 tag weight distributions (a) and the IP3D+JetFitter standard and negative tag 
      weight distributions (b) for $b$, $c$ and light-flavour jets in a simulated
      inclusive jet sample.
     }
    \label{fig:tagweight}
  \end{center}
\end{figure}

The inclusive tag rate obtained by reversing the impact parameter significance sign of tracks for
impact parameter based tagging algorithms, or reversing the decay length significance sign of secondary
vertices for secondary vertex based tagging algorithms, is therefore expected to be a good approximation
of the mistag rate due to resolution effects. For the SV0 algorithm, which is a basic secondary vertex based
algorithm where the tag weight $w$ is the signed decay length significance of the reconstructed secondary
vertex, a jet is considered negatively tagged if it contains a secondary vertex with decay length significance 
$w < -w_{\text{cut}}$ rather than decay length significance $w > w_{\text{cut}}$, where $w_{\text{cut}}$ is the
reference weight cut value for a particular
efficiency. For advanced tagging algorithms, based on likelihood ratios or neural networks, the negative
tag rate is instead computed in a more complex way, defining a negative version of the tagging algorithm
which internally reverses the impact parameter and the decay length selections. For such algorithms, a
jet is considered negatively tagged if it has a negative tag weight $w_{\text{\text{neg}}} > w_{\text{cut}}$ rather than 
the standard $w > w_{\text{cut}}$.
Figure~\ref{fig:tagweight} shows the tag weight distribution of the SV0 algorithm, as well as the standard and negative weight
distributions for the IP3D+JetFitter algorithm. For the SV0 algorithm and for
not too large weights (for reference a 50\% $b$-jet efficiency is obtained with a requirement
$w > 5.65$), the tag weight distribution is 
almost symmetric about zero for light-flavour jets, and the negative side of the tag weight distribution
is dominated by light-flavour jets. For the IP3D+JetFitter algorithm, the standard and negative tag
weight distributions for light-flavour jets are similar in shape,
while the tag weight distributions for $b$ and $c$ jets differ substantially. 
For reference the weight cut value $w$ for the IP3D+JetFitter algorithm, corresponding to a $b$-jet efficiency of 70\%, is 0.35.

The mistag rate $\varepsilon_l$ is then approximated by the negative tag rate of the inclusive jet sample,
$\varepsilon_{\text{inc}}^{\text{neg}}$. The approximation would be exact if the negative tag weight distribution is identical 
for all jet flavours and is identical to the normal tag weight distribution. In reality, two correction factors are 
applied to relate $\varepsilon_{\text{inc}}^{\text{neg}}$ to $\varepsilon_l$.

\begin{itemize}
\item The negative tag rate for heavy-flavour, $b$ and $c$ jets, differs from the negative tag rate for light-flavour jets.
  $b$ and $c$ jets are positively tagged mainly because of the measurable lifetimes of $b$ and $c$ hadrons, 
shifting the decay length significance distributions towards larger values.
  However, effects like the finite jet direction resolution can flip the sign of the discriminating variable, increasing
  significantly the negative tag rate for $b$ and $c$ jets.
  The correction factor $k_{\text{hf}}$ = $\varepsilon_l^{\text{neg}}$ / $\varepsilon_{\text{inc}}^{\text{neg}}$ is defined
  to account for this effect.
  Because of the effects described above and the relatively small fractions of $b$ and $c$ jets in the inclusive sample,
  $k_{\text{hf}}$ is typically smaller than, but close to unity.
\item A symmetric decay length or impact parameter significance distribution for light-flavour jets is
  only expected for fake secondary vertices arising e.g. from track reconstruction resolution effects.
  However, a significant fraction of reconstructed secondary vertices have their origin in charged
  particle tracks stemming from long-lived particles ($K_s^0$, $\Lambda^0$ etc.) or material interactions (hadronic
  interactions and photon conversions).
  These vertices will show up mainly at positive decay length significances and thus cause an asymmetry
  for the positive versus negative tag rate for light-flavour jets.
  The correction factor $k_{\text{ll}}$ = $\varepsilon_l$ / $\varepsilon_l^{\text{neg}}$ is defined to account for this effect.
  Because of the sources in light-flavour jets showing positive decay length, $k_{\text{ll}}$ is larger than unity.
  In particular $k_{\text{ll}}$ for the MV1 algorithm ranges, depending on jet \pt\ and on the operating point, between 1 and 13.
\end{itemize}

The measured negative tag rate value $\varepsilon_{\text{inc}}^{\text{neg}}$ is converted to the mistag rate $\varepsilon_l$ 
using the two correction factors $k_{\text{hf}}$ and $k_{\text{ll}}$ defined above: $\varepsilon_l = \varepsilon_{\text{inc}}^{\text{neg}}\,k_{\text{hf}}\,k_{\text{ll}}$.
Both correction factors are derived from simulated events.

\subsection{Systematic uncertainties}

The systematic uncertainties on the mistag rate scale factor $\sfl \equiv \epsilon_l^{\rm data}/\epsilon_l^{\rm sim}$ of the MV1 tagging
algorithm at 70\% efficiency are shown in Tables~\ref{tab:systs_neg_eta1} and~\ref{tab:systs_neg_eta2}, for jets 
with $|\eta|\in[0,1.2]$ and $|\eta|\in[1.2,2.5]$, respectively.
In cases where the number of simulated events is not sufficient to evaluate with sufficient precision the effect of a given 
systematic variation, the systematic uncertainty has been evaluated by merging two adjacent jet \pt\ bins.

\begin{table}[ptb]
  \begin{center}
    \scriptsize
    \begin{tabular}{l|cccccc}
      \hline\hline
      Source                                 & \multicolumn{6}{c}{Jet $\pt\ [\GeV]$} \\
                                             & 20--30 & 30--60 & 60--140 & 140--300 & 300--450 & 450--750\\
      \hline
      Simulation statistics                  & 3.1 & 2.5 & 2.2 & 1.8 & 1.4 & 1.6 \\
      $k_{\rm hf}$: $b$ fraction               & 0.7 & 1.0 & 2.0 & 3.0 & 2.8 & 3.0 \\
      $k_{\rm hf}$: $c$ fraction               & 1.3 & 3.1 & 5.2 & 6.5 & 6.3 & 5.3 \\
      $k_{\rm hf}$: $b$-jet tagging efficiency & 2.1 & 2.7 & 5.1 & 7.2 & 7.2 & 5.1 \\
      $k_{\rm hf}$: $c$-jet tagging efficiency & 3.8 & 6.4 & 9.2 & 11  &  11 & 8.2 \\
      $k_{\rm ll}$: long lived particles       & 0.7 & 0.6 & 0.8 & 1.3 & 0.5 & 0.5 \\
      $k_{\rm ll}$: fake tracks                & 0.3 & 0.7 & 0.8 & 2.6 & 2.7 & 1.7 \\
      $k_{\rm ll}$: hadronic interactions      & 1.3 & 2.2 & 3.8 & 7.3 & 2.5 & 1.1 \\
      $k_{\rm ll}$: photon conversions         & 0.5 & 0.2 & 1.1 & 0.1 & 1.5 & 1.1 \\
      Track multiplicity                     & 1.0 & 3.2 & 4.8 &  10 &  10 &  27 \\
      Impact parameter smearing              & 2.9 & 2.9 & 5.8 & 5.8 & 3.6 & 3.6 \\
      Data taking period dependence          & 2.1 & 1.1 & 2.2 & 5.6 & 3.8 & 2.3 \\
      Jet vertex fraction                    & 0.3 & - & 0.2 & - & 0.3 & 0.5 \\
      Jet energy scale                       & 2.1 & 0.3 & 0.4 & 0.3 & 0.2 & 1.7 \\
      Jet energy resolution                  & 5.3 & 0.4 & 2.3 & 0.7 & 0.2 & 0.2 \\
      Trigger bias                           & 4.1 & 3.6 &  10 & 6.2 & 7.0 & 12 \\
      \hline
      Total systematic uncertainty           & 9.8 &  10 &  18 &  22 &  20 &  32 \\
      Statistical uncertainty                & 1.5 & 0.8 & 0.5 & 0.3 & 0.4 & 0.8 \\
      \hline
      Total uncertainty                      & 9.9 &  10 &  18 &  22 &  20 &  32 \\
      \hline\hline
    \end{tabular}
    \caption{
      \label{tab:systs_neg_eta1}Relative systematic and statistical uncertainties, in \%, on the mistag rate 
      scale factor \sfl\ from the negative tag method for the MV1 tagging
      algorithm at 70\% efficiency for jets with $|\eta|\in[0,1.2]$.
      Negligibly small uncertainties are indicated by dashes.}
  \end{center}
\end{table}

\begin{table}[ptb]
  \begin{center}
    \scriptsize
    \begin{tabular}{l|cccccc}
      \hline\hline
      Source                                 & \multicolumn{6}{c}{Jet $\pt\ \mathrm{[GeV]}$} \\
                                             & 20--30 & 30--60 & 60--140 & 140--300 & 300--450 & 450--750\\
      \hline
      Simulation statistics                  & 3.7 & 3.7 & 2.9 & 2.1 & 2.2 & 6.0 \\
      $k_{\rm hf}$: $b$ fraction               & 0.3 & 0.4 & 1.0 & 1.7 & 1.5 & 2.3 \\
      $k_{\rm hf}$: $c$ fraction               & 0.5 & 1.0 & 2.9 & 3.8 & 4.0 & 5.2 \\
      $k_{\rm hf}$: $b$-jet tagging efficiency & 1.1 & 1.4 & 2.7 & 4.5 & 3.9 & 3.5 \\
      $k_{\rm hf}$: $c$-jet tagging efficiency & 2.8 & 3.6 & 6.2 & 7.0 & 7.4 & 7.3 \\
      $k_{\rm ll}$: long lived particles       & 1.4 & 0.4 & 0.4 & 0.2 & - & 2.0 \\
      $k_{\rm ll}$: fake tracks                & 0.2 & 1.8 & 0.3 & 0.9 & 0.7 & 1.0 \\
      $k_{\rm ll}$: hadronic interactions      & 1.8 & 2.8 & 3.6 & 6.6 & 3.9 & 2.6 \\
      $k_{\rm ll}$: photon conversions         & 0.1 & - & 0.4 & 1.6 & 1.1 & 1.4 \\
      Track multiplicity                     & 1.3 & 5.3 & 7.6 &  18 &  15 & 43 \\
      Impact parameter smearing              &  11 &  11 &  10 &  10 &  22 & 22 \\
      Data taking period dependence          & 1.2 & 7.4 & 2.7 & 5.7 & 9.0 & - \\
      Jet vertex fraction                    & - & 0.5 & 0.6 & 0.5 & 1.0 & 0.4 \\
      Jet energy scale                       & 1.1 & 1.8 & 1.5 & 1.6 & 1.2 & 2.3 \\
      Jet energy resolution                  & 2.9 & 5.5 & 2.0 & 3.8 & 1.3 & 3.6 \\
      Trigger bias                           & 2.5 & 4.1 &  10 & 8.0 & 8.1 & 2.6 \\
      \hline
      Total systematic uncertainty           &  13 &  17 &  19 &  26 &  31 &  50 \\
      Statistical uncertainty                & 2.2 & 1.1 & 0.8 & 0.6 & 0.9 & 2.3 \\
      \hline
      Total uncertainty                      &  13 &  17 &  19 &  26 &  31 & 50 \\
      \hline\hline
    \end{tabular}
    \caption{
      Relative systematic and statistical uncertainties, in \%, on the mistag rate
      scale factor \sfl\ from the negative tag method for the MV1 tagging algorithm at
      70\% efficiency for jets with $1.2 < |\eta| < 2.5$.
      Negligibly small uncertainties are indicated by dashes.}
    \label{tab:systs_neg_eta2}
  \end{center}
\end{table}

\subsubsection*{Simulation statistics}

The statistical uncertainties on $k_{\rm hf}$ and $k_{\rm ll}$ have been propagated through the
analysis. The resulting uncertainties range between 1 and 11\% (between 1\% and 6\% for the 70\%
operating point).

\subsubsection*{Data taking period dependence}

The negative tag analysis has been carried out in three different data taking
periods, and half of the largest difference between the results in different
data taking periods is assigned as an uncertainty.
These differences between periods (of up to 5.6\%) may be related to biases introduced by the trigger selection (the inclusive jet triggers used in this
analysis have undergone substantial changes in the prescale factors applied to them with evolving instantaneous 
luminosity) that have not been fully modelled in the selection of simulated events.

\subsubsection*{Jet vertex fraction}

A jet vertex fraction requirement is imposed for the measurement, which helps to suppress the 
jets originating from pile-up. The dependence on the cut is studied by removing the requirement and repeating the measurement.
The resulting uncertainties are below 1\%.

\subsubsection*{Jet energy scale and resolution}

A bias in the jet energy measurement in simulation compared to data will result in biases in the correction factors $k_{\rm hf}$ and $k_{\rm ll}$
if there is a correlation between the jet energy and these quantities.
As the mistag rate increases with the jet energy, a shift in the jet energy scale in simulated events will also
lead to an apparent mismatch between the mistag rate in data and simulated events.

To study this effect, the reconstructed jet energies were alternately shifted up and down 
by the uncertainty on the jet energy scale~\cite{PERF-2012-01}.
Half of the full difference of the corresponding shifts of the mistag rates is assigned as a systematic uncertainty.
The resulting uncertainties are below 2\%.

A bias of jet energy resolution between data and simulation is corrected with a smearing function applied to simulated events,
which leads to a migration of jets between neighbouring $\pT$ bins. The effective difference from nominal is below 5\%.

\subsubsection*{Trigger bias}

The negative tag analysis uses the two leading jets in each event and requires them to be in a back-to-back
configuration ($\Delta \phi > 2$). As generally only one of the two leading jets is in the jet $\pT$ 
region where the inclusive jet trigger used to select the events is fully efficient, any mismodelling
in the simulation of the trigger turn-on behaviour can lead to analysis biases.
For example, it is found that the number of tracks associated to the leading and sub-leading jets
in a given jet $\pT$ interval differs in data but not to the same extent in simulated events.

In order to account for possible trigger biases, the measurement has been repeated 
using only jets with sub-leading $\pT$. The variation in the mistag rate is taken as a systematic uncertainty.
The trigger bias systematic uncertainty is one of the most dominant in the negative tag analysis,
and is generally between 5 and 10\%.

\subsubsection*{Heavy flavour fractions}

The fractions of $b$ and $c$ jets enter directly into the correction factor $k_{\textrm{hf}}$ for the negative tag analysis.
Relative uncertainties on the $b$- and $c$-jet fractions of 10\% and 30\% have been propagated through the analysis,
resulting in uncertainties which are generally below 5\%.
These uncertainties are obtained from comparing these fractions as obtained from
simulation with their estimates obtained by fitting templates of the distributions of
the invariant mass of tracks significantly displaced from the primary vertex to the data.

\subsubsection*{Heavy flavour tagging efficiencies}

The negative-tagging efficiencies for $b$ and $c$ jets directly enter into the 
negative tag analysis through the correction factor $k_{\textrm{hf}}$. These efficiencies as obtained
from simulation have been varied by 20\% and 40\%, respectively. 
The variations used for the $b$- and $c$-jet tagging efficiencies have conservatively been doubled
compared to the uncertainties quoted in Sections~\ref{sec:beff_mubased} 
and~\ref{sec:ceff_dstarbased} to account for the extrapolation from the positive-tagging efficiency 
to the negative one. The resulting uncertainties are generally below 10\%.

\subsubsection*{Long-lived particle decays, material interactions, fake tracks}

The products from decays of long-lived particles, e.g. $K_s$, $\Lambda^0$, hadronic interactions 
or photon conversions in the detector material (mainly interactions in the first material layers of the 
detector), may cause reconstructed secondary vertices in light-flavour jets. While the secondary vertex 
based algorithms apply a veto to secondary vertices consistent with these decays or interactions, not all 
of them can be detected and there is a sizable fraction of vertices where one track arising from such 
decays or interactions is paired with a track from a different source into a vertex. Fake or badly measured 
tracks may also give rise to additional vertices. 

To estimate the resulting systematic uncertainty from an imperfect modelling of the rate of such vertices
in simulated events, the fraction of jets containing fake tracks, long-lived particles like 
$K_s$ and $\Lambda^0$ or material interactions have been varied based on
estimates in data, before the application of their suppression criteria.
The modelling of fake tracks is evaluated using the fraction of jets containing tracks with 
$\chi^2/N_{\rm dof} > 3$. The difference in the fraction of such tracks between data and 
simulated events is found to be 30\%, which is assigned as a systematic uncertainty.
The fraction of jets with $K_s$ or $\Lambda^0$ decays in data and simulation are compared
by counting the number of events in the $K_s$ and $\Lambda^0$ mass peaks.
As the fraction of reconstructed $K_s$ and $\Lambda^0$ candidates is consistent between 
data and simulation, the statistical uncertainty of the estimate, which is approximately 10\%, 
is used as a systematic uncertainty.
Finally, the uncertainty associated with jets with a hadronic interaction or photon conversion is estimated
in simulated events, considering jets containing a selected track
produced at a radial distance from the beam line $r > 25$~mm.
About 80\% of all jets have at least one such track, and a systematic uncertainty of 10\% is assigned
to this fraction, based on the precision with which the material in the detector is known.
The resulting uncertainties are up to 7\%, with the largest effects originating from hadronic interactions.

\subsubsection*{Track multiplicity}

The simulation does not properly reproduce the multiplicity of tracks associated with jets observed in
data. This could be due to imperfect modelling of fragmentation differences in the relative fraction of
quark and gluon jets in the light-flavour sample or differences between data and simulation in the track 
reconstruction efficiency in the core of jets where the track density is high. A higher track multiplicity 
implies a larger probability of accidentally
tagging a light-flavour jet as a $b$ jet for purely combinatorial reasons.
The systematic uncertainty in the negative tag analysis due to the track multiplicity is estimated by
reweighting the jet sample according to the ratio of distributions of the number of tracks associated to
jets in data and simulation.
The effect of the track multiplicity reweighting ranges between approximately 5\% at low jet $\pT$ and over 
40\% at high jet $\pT$ in the forward region. 
The track multiplicity systematic uncertainty affects the higher jet
$\pT$ bins more because the discrepancy between data and simulation is larger in this region,
presumably due to an imperfect modelling of track reconstruction in the core of high-\pt{} jets in simulated events
as well as to an imperfect description of the track multiplicities over a wide range of jet transverse momenta in the generator. 

\subsubsection*{Impact parameter resolution}

The secondary vertex reconstruction is very sensitive to the tracking resolution and the 
proper estimation of the track parameter errors, 
especially in light-flavour jets where a large contribution of fake vertices is present.
It has been shown in Section~\ref{sec:ipres} that the track impact parameter resolutions in simulation are slightly better than those in data.
Therefore, track impact parameters in the simulation have been smeared in order to bring data and simulation into better agreement.
The chosen smearing approach does not take into account correlated modifications of the impact parameters of tracks that pass through the same pixel module,
as would be needed to model residual misalignments in the Inner Detector.
The parameters for the smearing have been chosen to cover the observed discrepancies in the impact parameter resolution between data and simulation.
After having applied the track impact parameter smearing to the tracks in simulation,
the primary vertex reconstruction and $b$-tagging have been rerun and the whole analysis repeated.
The effect in the negative tag analysis is approximately 5\% in the central $\eta$ region
but can be as large as  22\% in the forward $\eta$ region where the modelling of the impact parameter resolution is worse.

Given that the impact parameter sign for tracks associated with a jet depends on the direction of the tracks relative to the jet direction
(unless a secondary vertex is found, as detailed in Section~\ref{sec:ip_algo}),
the finite jet angular resolution results in a degree of arbitrariness for tracks nearly aligned with the jet direction.
The factors $k_{\text{hf}}$ and $k_{\text{ll}}$ therefore are sensitive to this resolution,
and the uncertainty on the angular resolution in principle translates into uncertainties on $k_{\text{hf}}$ and $k_{\text{ll}}$.
In practice, however, smearing the jet directions in the simulation as done in Section~\ref{sec:ptrel} has a negligible effect on the impact parameter significance distributions, and consequently on the predicted mistag rate.

\subsection{Results}

\begin{figure}[!ht]
 \subfloat[]{\label{negativetag_MV170_eta1_eta2_a}\includegraphics[width=0.49\textwidth]{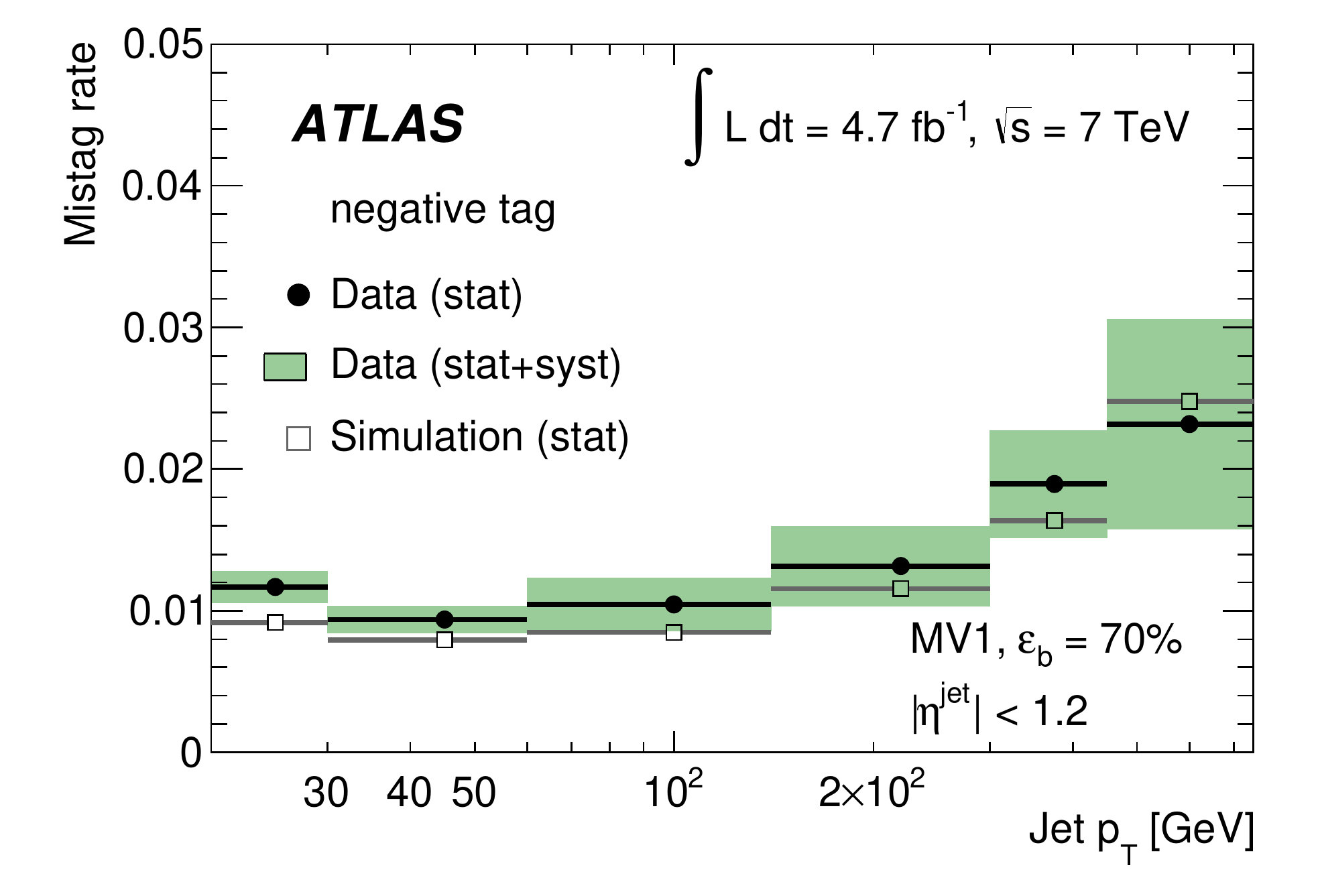}}
 \subfloat[]{\label{negativetag_MV170_eta1_eta2_b}\includegraphics[width=0.49\textwidth]{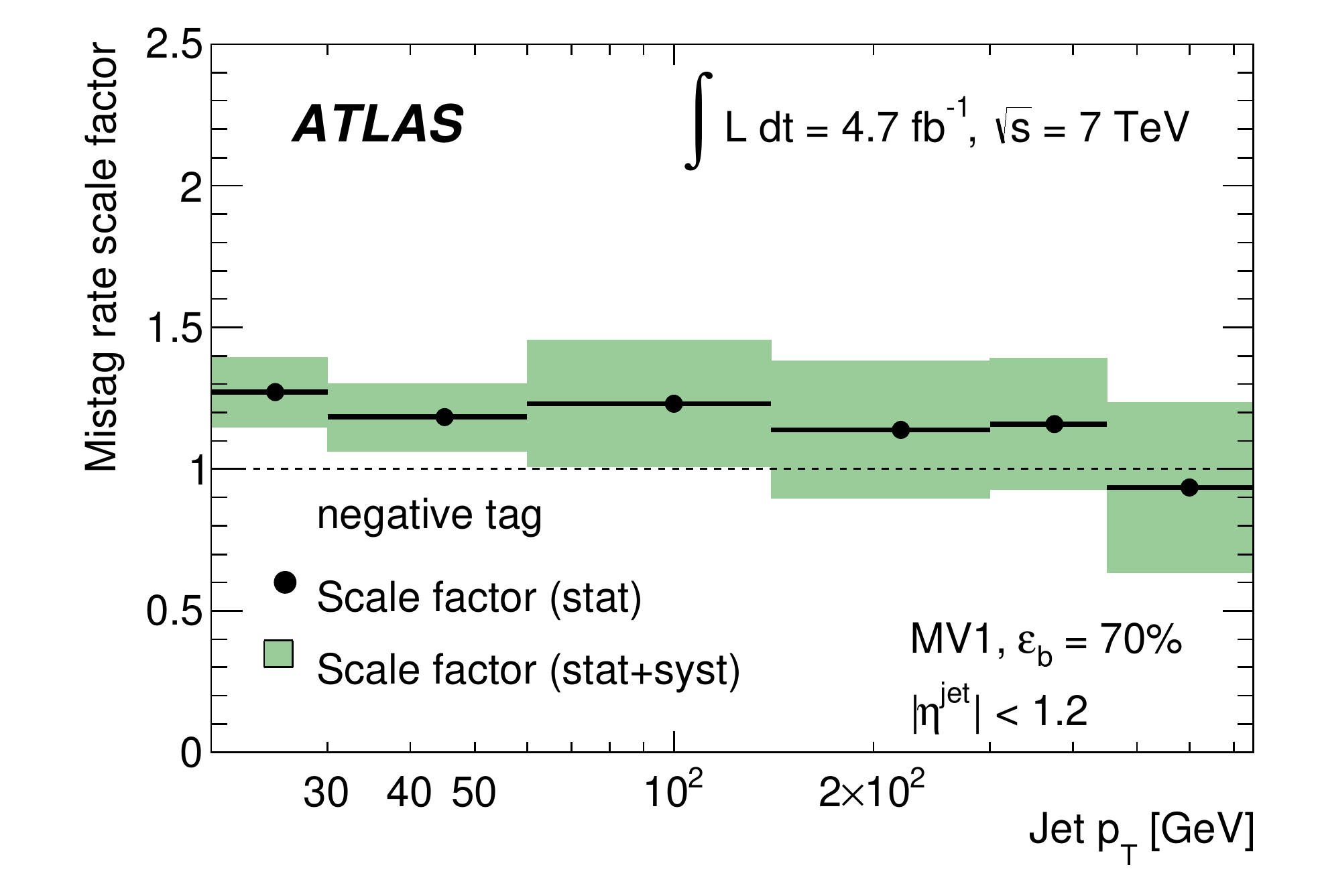}} \\
 \subfloat[]{\label{negativetag_MV170_eta1_eta2_c}\includegraphics[width=0.49\textwidth]{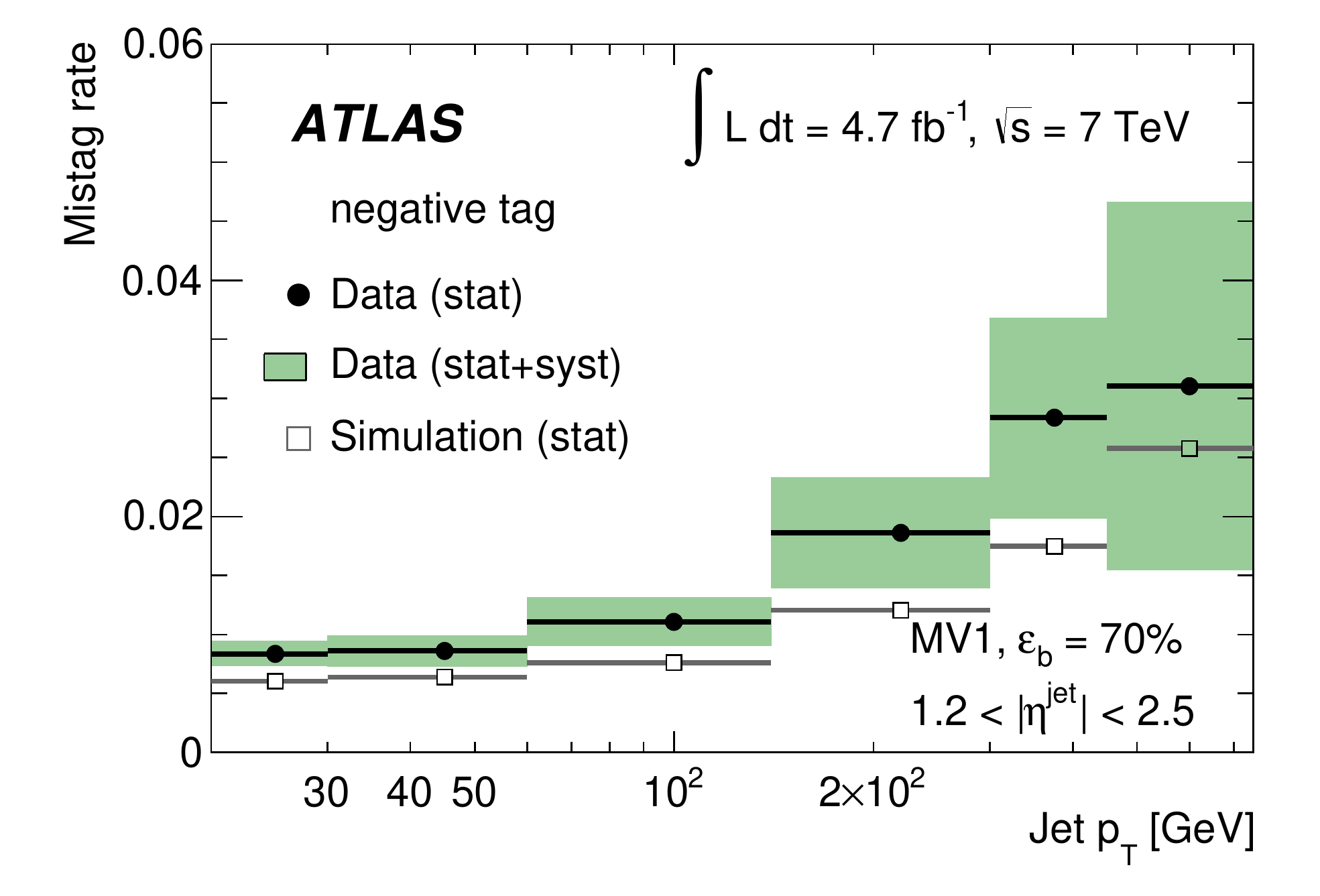}}
 \subfloat[]{\label{negativetag_MV170_eta1_eta2_d}\includegraphics[width=0.49\textwidth]{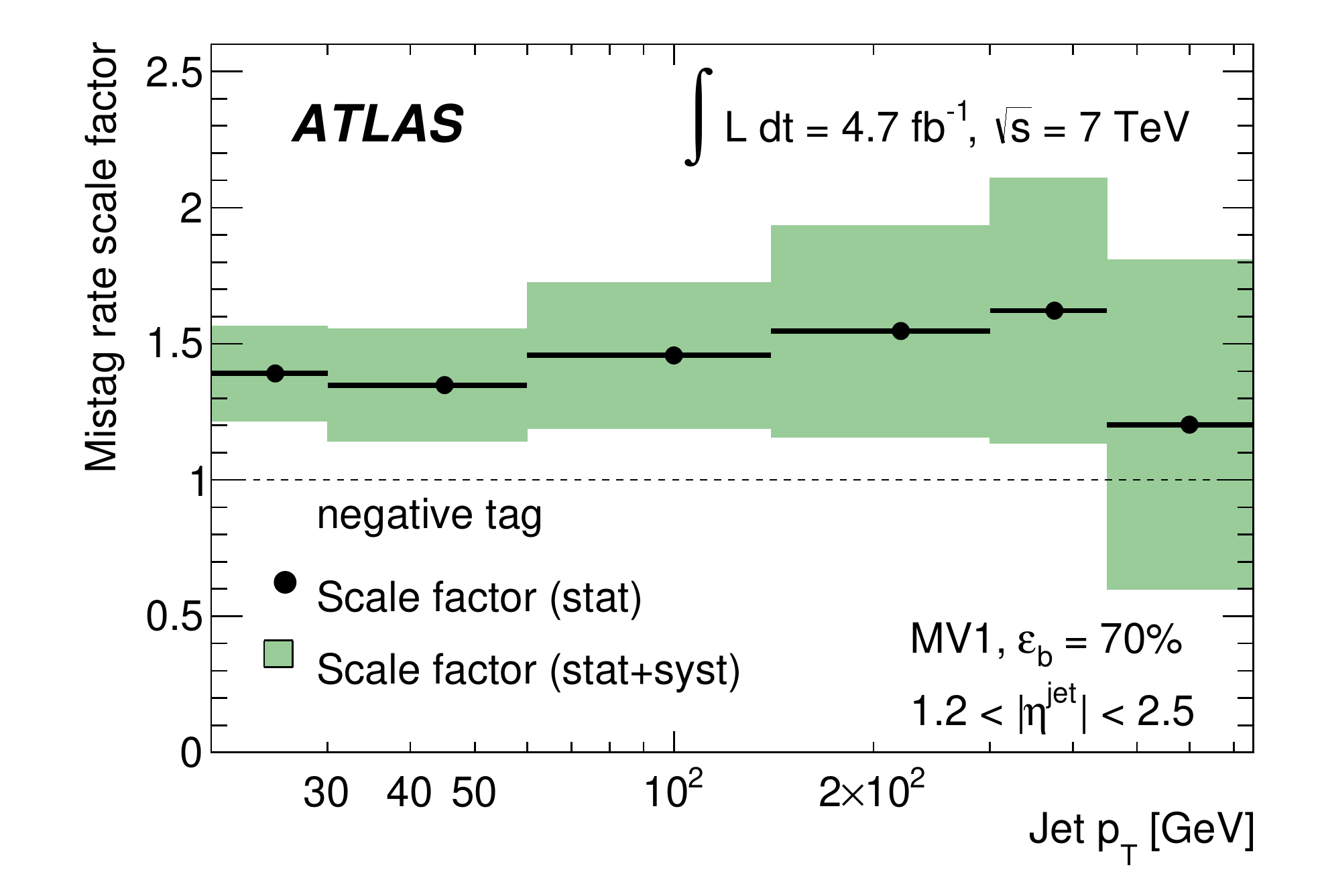}} \\
 \caption{The mistag rate in data and simulation (left) and the data-to-simulation scale factor (right)
   for the MV1 tagging algorithm at 70\% efficiency for jets with $|\eta|<1.2$ (top) and jets with $1.2<|\eta|<2.5$ (bottom).}
 \label{negativetag_MV170_eta1_eta2}
\end{figure}

The measured mistag rates in data, the mistag rates in simulation and the resulting data-to-simulation scale factors
for the MV1 tagging algorithm at 70\% efficiency are shown in Fig.~\ref{negativetag_MV170_eta1_eta2}
for two different regions of the jet pseudorapidity.

For the MV1 tagging algorithm at the 70\% efficiency operating point the efficiency in data tends to be higher than in simulation,
leading to data-to-simulation scale factors that are about 1.2 and 1.4 in the central and forward directions, respectively.

\section{Mistag rate calibration of the soft muon tagging algorithm}
\label{smt:mtag}

The probability with which a light-flavour jet is tagged by the SMT algorithm, referred to as the mistag rate,
is measured in data using an inclusive jet sample.
The method is designed to minimise biases from heavy-flavour jets and to make minimal use of information obtained from simulation.

\subsection{Data and simulation samples}

To cover a wide transverse momentum range, the events are required to pass one of several inclusive jet triggers, 
with \pT{} thresholds ranging between 10 and 40 \GeV. 
Only events in which the reconstructed primary vertex has at least five tracks associated to it are considered.

The data are compared to the same set of simulated QCD jet samples used in the negative tag analysis (see Section~\ref{sec:negtag_samples}).

\subsection{Mistag rate measurement} \label{sec:smt_mistagmethod}

The method to measure the mistag rate with collision data is based on a system of three equations and three
unknowns (among which the mistag rate). The IP3D+JetFitter lifetime tagging algorithm (see Section~\ref{sec:comb-vertex-algos})
is used as an auxiliary tagging algorithm to enhance the inclusive jet sample in light-flavour jets.
Two samples are selected from events with exactly two jets, one in which one jet is not tagged by
the IP3D+JetFitter algorithm (single-veto sample) and one in which neither jet is tagged by the IP3D+JetFitter algorithm (double-veto sample).
In the latter sample the fraction of heavy-flavour jets is expected to be considerably suppressed.
As the amount of heavy-flavour jets in the double-veto sample largely determines 
the uncertainty on the mistag rate measurement, the operating point of the IP3D+JetFitter algorithm is chosen to correspond 
to a high efficiency (80\% in simulated \ttbar\ events).

The number of jets in data which are tagged by the SMT in each of the two above samples is given by
\begin{eqnarray}
  N_{\rm SMT} &=& N \times [{(\varepsilon_{\rm HF}^{\rm SMT} \cdot f_{\rm HF}) + (\varepsilon_{\rm LF}^{\rm SMT} \cdot [1-f_{\rm HF}])}], \\
  N^{\prime}_{\rm SMT} &=& N^{\prime} \times [{(\varepsilon_{\rm HF}^{\rm SMT} \cdot f_{\rm HF}^{\prime}) + (\varepsilon_{\rm LF}^{\rm SMT} \cdot [1-f_{\rm HF}^{\prime}])}],
  \label{eqn:opt2_eqn}
\end{eqnarray}
where $N$ ($N^{\prime}$) is the number of selected jets in the single- (double-) veto sample in data,
$N_{\rm SMT}$ ($N^{\prime}_{\rm SMT}$) is the corresponding subset of jets which are also tagged by
the SMT, $f_{\rm HF}$ ($f_{\rm HF}^{\prime}$) is the fraction of heavy-flavour
jets in single- (double-) veto sample and $\varepsilon_{\rm LF}^{\rm SMT}$ is the SMT mistag rate.
Assuming that the single-veto sample is already dominated by light-flavour jets
and neglecting the effect of the second veto on this component (the inaccuracies
in these approximations have been verified to lead to measurement biases
negligible compared to the corresponding uncertainties), $f_{\rm HF}^{\prime}$
and $f_{\rm HF}$ are then related by
\begin{equation}
  f_{\rm HF}^{\prime} = f_{\rm HF} \cdot (1-\varepsilon_{\rm HF}),
  \label{eqn:opt2_eqn1}
\end{equation}
with $\varepsilon_{\rm HF}$ denoting the average of the $b$- and $c$-jet
lifetime tagging efficiencies weighted by their relative fractions in the single-veto sample.
Solving for $\varepsilon_{\rm LF}^{\rm SMT}$, one obtains
\begin{equation}
  \varepsilon_{\rm LF}^{\rm SMT} = \frac{1}{\varepsilon_{\rm HF}} \cdot [\frac{N^{\prime}_{\rm SMT}}{N^{\prime}} - (\frac{N_{\rm SMT}}{N} (1-\varepsilon_{\rm HF}))].
  \label{eqn:opt2_eqn2}
\end{equation}

The heavy-flavour efficiency of the IP3D+JetFitter algorithm, $\varepsilon_{\rm HF}$, is evaluated from true heavy-flavour jets in the simulated QCD jet sample described in Section~\ref{sec:smt_mc}, corrected using data-to-simulation scale factors
from the calibration methods described in Sections~\ref{sec:beff_mubased} and~\ref{sec:ceff_dstarbased}.
Since the ratio of the fractions of $b$ and $c$ jets in the single- and double-veto samples could be different in data and 
simulation, a systematic uncertainty is associated to variations of the $b$-to-$c$ ratio in the mistag rate estimate, as discussed in Section~\ref{sec:faker_systs}.

It has been observed that the estimation of the mistag rate bears a systematic bias with respect to the true value of the mistag rate for low jet transverse momenta.
This has been found looking at the true and estimated mistag rates in simulation: the method returns an estimate about 20\% to 40\% lower for jet \pt $<$ 40 \GeV.
This effect is due to the correlation between the $\chi^{2}_{\rm match}$ cut and the muon \pt,
which causes a migration of light-flavour jets towards higher values of the IP3D+JetFitter weights, rendering them more heavy-flavour-like.
This bias is taken into account in the treatment of the systematics uncertainties (Section~\ref{sec:faker_systs}).

\subsection{Systematic uncertainties}
\label{sec:faker_systs}

The systematic uncertainties considered for the mistag rate measurement in data are shown in
Tables~\ref{tab:faker_results} and \ref{tab:faker_results2} for
the two different jet pseudorapidity regions.

\begin{table}[!htbp]
  \begin{minipage}[t]{0.47\textwidth}
    \begin{center}
      \scriptsize
      \begin{tabular}{l|ccc}
        \hline
        \hline 
        Source                       & \multicolumn{2}{c}{Jet $\pt$ [\GeV]} \\
                                     & 30--60  & $> 60$ \\
        \hline
        IP3D+JetFitter calibration   &  0.4    &  2.5   \\
        Flavour composition          &  1.1    &  5.7   \\
        Method bias                  &   27    &  -   \\
        \hline
        Total systematic uncertainty &   27    &  6.2   \\
        Statistical uncertainty      &   22    &  25  \\
        \hline
        Total uncertainty            &   35    &  25 \\
        \hline
        \hline
      \end{tabular}
    \end{center}
    \caption{Relative systematic and statistical uncertainties, in \%, on the
      mistag rate for the SMT tagging algorithm for jets with $|\eta| < 1.2$.
      Negligibly small uncertainties are indicated by dashes.}
    \label{tab:faker_results}
  \end{minipage}
  \begin{minipage}[t]{0.05\textwidth}
    \mbox{}
  \end{minipage}
  \begin{minipage}[t]{0.47\textwidth}
    \begin{center}
      \scriptsize
      \begin{tabular}{l|ccc}
        \hline
        \hline 
        Source                       &  \multicolumn{3}{c}{Jet $\pt$ [\GeV]} \\
                                     &  30--60 & $> 60$ \\
        \hline
        IP3D+JetFitter calibration   &  3.9   &  0.7  \\
        Flavour composition          &  1.0   &  1.6  \\
        Method bias                  &   13   &  3.0  \\
        \hline
        Total systematic uncertainty &   14   &  3.5  \\
        Statistical uncertainty      &   48   &   26 \\
        \hline
        Total uncertainty            &   49   &   26 \\
        \hline
        \hline
      \end{tabular}
    \end{center}
    \caption{Relative systematic and statistical uncertainties, in \%, on the
      mistag rate for the SMT tagging algorithm for jets with $1.2 < |\eta| < 2.5$.}
    \label{tab:faker_results2}
  \end{minipage}
\end{table}

\subsubsection*{Calibration of the advanced tagging algorithm}

The data-to-simulation scale factors of the efficiency of the IP3D+JetFitter algorithm (which have been determined
in a way similar to that described in Sections~\ref{sec:beff_mubased}--\ref{sec:combination}) have been varied within
their uncertainties, which are also comparable to those derived for the MV1 tagging algorithm.

\subsubsection*{Flavour composition}

The ratio of the fractions of $b$ and $c$ jets in data can be different than that found in the simulated events
used to calculate $\varepsilon_{\rm HF}$. The systematic uncertainty associated to the limited knowledge of the 
$b$-to-$c$ composition in data is assessed by doubling the fraction of $b$ jets in the simulated QCD jet sample 
and re-deriving the heavy-flavour tagging efficiency for the IP3D+JetFitter algorithm.
This leads to a relative 3-4$\%$ difference in the mistag rate which is taken as a systematic uncertainty. 

\subsubsection*{Method bias}

The inaccurate estimation of the method with respect to the true mistag rate is evaluated in simulated events 
as a function of the jet \pt\ and $\eta$. The relative difference between the estimated and true mistag rates is 
summarised in Tables \ref{tab:faker_results} and \ref{tab:faker_results2}.
The bias is found to originate from the correlation between $\chi^2_{\rm match}$~and muon (or jet) \pt, 
resulting in a slightly harder jet \pt\ spectrum for jets tagged by the SMT.
The difference is assigned as a systematic uncertainty on the estimation of the mistag rate.

\subsubsection*{Muon momentum corrections}

The effect of the muon momentum corrections on the acceptance of the SMT algorithm to light-flavour jets 
(from the $\pt(\mu)>4\GeV$ cut) has been studied and found to be negligible in all cases.

\subsubsection*{Pile-up dependence}

The mistag rate is studied as a function of the number of additional minimum-bias interactions. As no dependence 
is observed, no systematic uncertainty has been assigned. However, the available statistics does not 
allow to be conclusive on the matter.

\subsection{Results} \label{sec:faker_results}

\begin{figure}[!htbp]
  \centering
  \subfloat[]{\label{fig:faker_results_a}\includegraphics[width=0.49\textwidth]{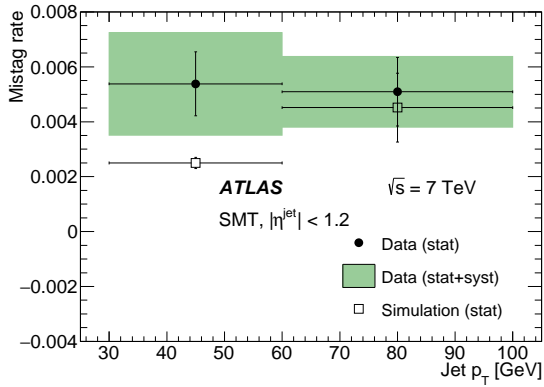}}
  \subfloat[]{\label{fig:faker_results_c}\includegraphics[width=0.49\textwidth]{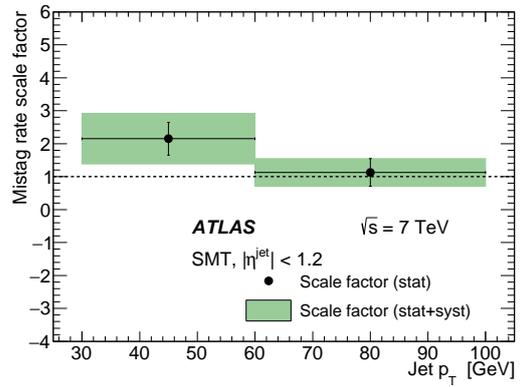}} \\
  \subfloat[]{\label{fig:faker_results_b}\includegraphics[width=0.49\textwidth]{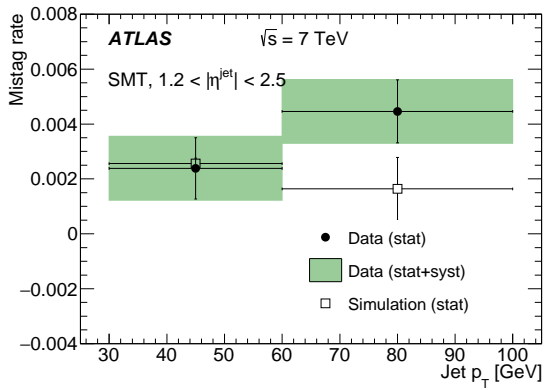}}
  \subfloat[]{\label{fig:faker_results_d}\includegraphics[width=0.49\textwidth]{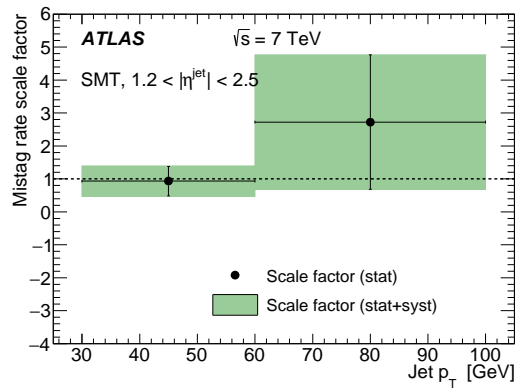}}
  \caption{The mistag rate in data and simulation (left) and the data-to-simulation scale factor (right)
    for the SMT algorithm, for jets with $|\eta|<1.2$ (top) and jets with $1.2<|\eta|<2.5$ (bottom).
    The last bin also includes jets with $\pt > 100\GeV$.}
  \label{fig:faker_results}
\end{figure}

The mistag rate $\varepsilon_{\rm LF}^{\rm SMT,data}$, measured in data using Eq.~\ref{eqn:opt2_eqn2},
together with the mistag rate in simulated events, $\varepsilon_{\rm LF}^{\rm SMT,sim}$, and the data-to-simulation
scale factor $\kappa_{\rm LF}^{\rm SMT} = \varepsilon_{\rm LF}^{\rm SMT,data}/\varepsilon_{\rm LF}^{\rm SMT,sim}$
are displayed in Fig.~\ref{fig:faker_results}. 
The statistical uncertainties for jets with $\pt < 30\GeV$ are too large to
allow for a meaningful measurement; 
the results for higher jet $\pt$ values indicate a scale factor $\kappa_{\rm LF}^{\rm SMT}$ compatible with unity within one standard deviation.

\section{Conclusions}

Several $b$-tagging algorithms to identify jets arising from the hadronisation 
of $b$ quarks have been developed in the ATLAS collaboration.
The most powerful $b$-tagging algorithms are based on the lifetime of
$b$ hadrons leaving detectable signatures of charged particle tracks 
significantly displaced from the primary event vertex or secondary
decay vertices in the detector.
Whereas relatively simple and robust algorithms have been used already in 
analyses based on data collected in the very early periods of LHC running,
more advanced algorithms based on sophisticated reconstruction techniques
and multivariate combination methods have been provided for the analyses
based on data collected from 2011 onwards.  
The most performant single algorithm is based on the complete reconstruction
of the $b$-hadron decay chain involving secondary and tertiary decay vertices.
This technique is used for the first time at a hadron collider experiment.
To obtain the best possible performance, several $b$-tagging algorithms have been combined using multivariate analysis techniques like neural
networks and boosted decision trees.     
The choice of a certain working point corresponding to a certain 
$b$-jet tagging efficiency and rejection of non-$b$ jets allows to adapt
the use of the $b$-tagging information to the needs of specific physics analyses.
For an efficiency to identify $b$ jets of 70\%, the MV1 algorithm -- the main algorithm used in ATLAS 
to analyse the 2011 and 2012 data --
achieves rejection rates for light-flavour jets of about 100 and for $c$ jets of about five,
as estimated using simulated $t\bar{t}$ events.
For kinematic properties of jets that are particularly favourable for $b$-tagging -- in the central
region of the detector and for jet transverse momenta around 80--150~\gev{} --
the rejection of light-flavour jets reaches values above 300.  

In the 2011 data a significant number of additional pile-up events, leading to additional vertices along the beam line, were present.
It has been shown that this level of pile-up leads to only a minor degradation of the $b$-tagging performance. 

To increase the efficiency to trigger mainly pure hadronic event topologies without leptons in the final state,
$b$-tagging is also applied at the software-based trigger levels allowing to significantly decrease the trigger thresholds for these events. 

In addition to the algorithms based on the lifetime of $b$ hadrons, the presence of a muon from a semileptonic $b$-hadron decay in a jet is used
in a dedicated $b$-tagging algorithm.

Since $b$-tagging algorithms rely critically on the reconstruction of
charged particle tracks and the determination of their properties, dedicated studies have been performed.
The impact parameter resolution of charged particle tracks -- an important 
quantity driving the performance of $b$-tagging algorithms -- has been measured
over a wide kinematic range, after deconvoluting the pure track impact parameter
resolution from the contribution of the primary vertex resolution.
The simulation describes the data well, especially in the region of low track transverse
momenta where the multiple scattering contribution dominates, pointing to an excellent
description of the ID material. Differences at high track transverse
momenta can be attributed to some ID residual misalignments.

To obtain a sample where particles originating from $b$-hadron decays and jet fragmentation
can be cleanly separated, the decay $B^{\pm} \to \Jpsi (\mu^+\mu^-)K^{\pm}$ has been reconstructed.
This allows to validate properties of $b$-hadron and fragmentation tracks separately as well as
details of the $b$-tagging algorithms like the association of the different
types of particles to reconstructed primary and secondary vertices.
Very good agreement has been found between data and simulation.   

To make use of $b$-tagging algorithms in data analyses and fully specify 
their associated systematic uncertainties, these algorithms have been
calibrated using the data themselves. The comparison to the expectation
from simulation is achieved through data-to-simulation scale factors 
for the tagging efficiencies of the different kinds of jets ($b$, $c$ and light-flavour jets).
These scale factors are then applied as corrections in physics analyses and their
uncertainties are propagated to the final result.

The efficiency to tag $b$ jets with the muon-based tagging algorithm
has been calibrated using $\Jpsi \to \mu\mu$ and $Z \to \mu\mu$ events
while the rate at which light-flavour jets are misidentified as $b$ jets 
has been calibrated with an inclusive jet sample.

To calibrate the efficiency of the lifetime-based tagging algorithms 
to tag $b$ jets two classes of events have been
used. The first one is composed of QCD jet events containing a jet with an identified
muon inside. Such a sample is enriched with $b$ jets in which a semileptonic
$b$-hadron decay has occurred. Two methods, \ptrel{} and system8, allow a measurement
of the $b$-jet tagging efficiency of lifetime-based algorithms up to jet transverse momenta
of 200~\gev\ with statistical (systematic) uncertainties in the range of 1.5\% to 4\%
(5\% to 7\%).
The second class of events are selected $t\bar{t}$ events, which naturally have
a high $b$-jet content, containing either 
one or two isolated leptons from the decays of $W$ bosons. Several calibration methods 
are exploited for these samples. The tag counting method is based on the number of identified $b$ jets
in each event, while the kinematic selection method exploits the fraction of jets in 
a $b$-enriched sample that are identified as $b$ jets.
The combinatorial likelihood method increases the precision by 
exploiting the kinematic correlations between the jets while the kinematic fit method
makes use of a kinematic fit to the $t\bar{t}$ system to identify the $b$ jets. 
The \ttbar-based measurements allow to extend the calibration analyses to jet transverse momenta of 300~\GeV. 
Typical statistical (systematic) uncertainties for the combinatorial likelihood method range 
from $2$\% to $8$\% ($2$\% to $8$\%).
The \ptrel{} method has also been applied to $t\bar{t}$ events, yielding
results compatible with those from the method applied to dijet events, albeit
with larger statistical and systematic uncertainties.

To obtain the best possible accuracy for the $b$-jet efficiency data-to-simulation scale factors, three individual measurements 
have been combined based on statistical methods taking into account the correlations in statistical and systematic uncertainties.
This combined fit shows good consistency of the different measurements and results in uncertainties between $2$\% and $4$\%
for transverse momenta between 20 and 200~\gev, rising to $12$\% for jets with transverse momenta between 200 and 300~\GeV.

Two novel methods have been developed to measure the efficiency to tag $c$ jets.
The first one is based on a sample where a $c$ jet -- identified through a semileptonic decay
of a $c$ hadron into a muon -- is produced in association with a $W$ boson. Exploiting the correlation 
of the charges between the muon in the $c$ jet and the $W$ boson provides a $c$-jet sample with very
high purity. The resulting $c$-jet tagging efficiency scale factors have uncertainties between
5\% and 13\%, depending on the chosen $b$-tagging operating point. 
The second method is based on the exclusive reconstruction of the decay $\dstar\to D^0(K^-\pi^+)\pi^+$ 
which allows to define a sample of $c$ jets after the subtraction of the
$b$-jet contribution. The statistical and systematic uncertainties on the data-to-simulation 
scale factors are about 10\% and between 10\% and 20\%, respectively.
Since both methods are based on sub-samples of specific $c$-hadron decays, a consistent
procedure has been developed to obtain results valid for an inclusive sample of $c$ jets.
These two methods have been adopted for the first time to measure the efficiencies of $b$-tagging algorithms for $c$ jets. 

The rate to misidentify light-flavour jets as $b$ jets has been measured
on a sample of QCD jet events using the negative tag method. This
method is based on modified versions of the algorithms where the signs of 
quantities sensitive to $b$-hadron lifetimes have been inverted.
The uncertainties for the individual measurements extending up to jet transverse
momenta of 750~\gev\ are typically in the range from 20\% to 50\% with a close
to negligible statistical contribution.   
 
The $b$-tagging algorithms discussed in this paper and their data-to-simulation scale factors derived in the calibration analyses
have been applied in many ATLAS physics analyses covering a wide range of physics processes.

\section*{Acknowledgements}

We thank CERN for the very successful operation of the LHC, as well as the
support staff from our institutions without whom ATLAS could not be
operated efficiently.

We acknowledge the support of ANPCyT, Argentina; YerPhI, Armenia; ARC, Australia; BMWFW and FWF, Austria; ANAS, Azerbaijan; SSTC, Belarus; CNPq and FAPESP, Brazil; NSERC, NRC and CFI, Canada; CERN; CONICYT, Chile; CAS, MOST and NSFC, China; COLCIENCIAS, Colombia; MSMT CR, MPO CR and VSC CR, Czech Republic; DNRF, DNSRC and Lundbeck Foundation, Denmark; IN2P3-CNRS, CEA-DSM/IRFU, France; GNSF, Georgia; BMBF, HGF, and MPG, Germany; GSRT, Greece; RGC, Hong Kong SAR, China; ISF, I-CORE and Benoziyo Center, Israel; INFN, Italy; MEXT and JSPS, Japan; CNRST, Morocco; FOM and NWO, Netherlands; RCN, Norway; MNiSW and NCN, Poland; FCT, Portugal; MNE/IFA, Romania; MES of Russia and NRC KI, Russian Federation; JINR; MESTD, Serbia; MSSR, Slovakia; ARRS and MIZ\v{S}, Slovenia; DST/NRF, South Africa; MINECO, Spain; SRC and Wallenberg Foundation, Sweden; SERI, SNSF and Cantons of Bern and Geneva, Switzerland; MOST, Taiwan; TAEK, Turkey; STFC, United Kingdom; DOE and NSF, United States of America. In addition, individual groups and members have received support from BCKDF, the Canada Council, CANARIE, CRC, Compute Canada, FQRNT, and the Ontario Innovation Trust, Canada; EPLANET, ERC, FP7, Horizon 2020 and Marie Skłodowska-Curie Actions, European Union; Investissements d'Avenir Labex and Idex, ANR, Region Auvergne and Fondation Partager le Savoir, France; DFG and AvH Foundation, Germany; Herakleitos, Thales and Aristeia programmes co-financed by EU-ESF and the Greek NSRF; BSF, GIF and Minerva, Israel; BRF, Norway; the Royal Society and Leverhulme Trust, United Kingdom.

The crucial computing support from all WLCG partners is acknowledged
gratefully, in particular from CERN and the ATLAS Tier-1 facilities at
TRIUMF (Canada), NDGF (Denmark, Norway, Sweden), CC-IN2P3 (France),
KIT/GridKA (Germany), INFN-CNAF (Italy), NL-T1 (Netherlands), PIC (Spain),
ASGC (Taiwan), RAL (UK) and BNL (USA) and in the Tier-2 facilities
worldwide.

\bibliographystyle{atlasBibStyleWithTitle}
\bibliography{paper}

\newpage
\begin{flushleft}
{\Large The ATLAS Collaboration}

\bigskip

G.~Aad$^\textrm{\scriptsize 85}$,
B.~Abbott$^\textrm{\scriptsize 113}$,
J.~Abdallah$^\textrm{\scriptsize 151}$,
O.~Abdinov$^\textrm{\scriptsize 11}$,
R.~Aben$^\textrm{\scriptsize 107}$,
M.~Abolins$^\textrm{\scriptsize 90}$,
O.S.~AbouZeid$^\textrm{\scriptsize 158}$,
H.~Abramowicz$^\textrm{\scriptsize 153}$,
H.~Abreu$^\textrm{\scriptsize 152}$,
R.~Abreu$^\textrm{\scriptsize 30}$
% ...
\bigskip
\\
% ...
$^{11}$ Institute of Physics, Azerbaijan Academy of Sciences, Baku, Azerbaijan\\
% ...
$^{30}$ CERN, Geneva, Switzerland\\
% ...
$^{85}$ CPPM, Aix-Marseille Universit{\'e} and CNRS/IN2P3, Marseille, France\\
% ...
$^{90}$ Department of Physics and Astronomy, Michigan State University, East Lansing MI, United States of America\\
% ...
$^{107}$ Nikhef National Institute for Subatomic Physics and University of Amsterdam, Amsterdam, Netherlands\\
% ...
$^{113}$ Homer L. Dodge Department of Physics and Astronomy, University of Oklahoma, Norman OK, United States of America\\
% ...
$^{151}$ Institute of Physics, Academia Sinica, Taipei, Taiwan\\
$^{152}$ Department of Physics, Technion: Israel Institute of Technology, Haifa, Israel\\
$^{153}$ Raymond and Beverly Sackler School of Physics and Astronomy, Tel Aviv University, Tel Aviv, Israel\\
% ...
$^{158}$ Department of Physics, University of Toronto, Toronto ON, Canada\\
% ...
\end{flushleft}


\end{document}